%% file: main.tex
\documentclass[twoside,12pt]{article}

\usepackage{epsfig}
\usepackage[numbers,sort&compress]{natbib}
\usepackage{slashed}

\def\p{\vec p}

\newcommand{\be}{\begin{equation}}

\newcommand{\ee}{\end{equation}}

\newcommand{\bea}{\begin{eqnarray}}

\newcommand{\eea}{\end{eqnarray}}

\newcommand{\kp}{\ensuremath{K^{+}}} 
\newcommand{\km}{\ensuremath{K^{-}}} 
\newcommand{\photon}{\ensuremath{\gamma}} 

\newcommand{\Lagr}{\mathcal{L}}
\newcommand{\no}{\nonumber}

\newcommand{\ls}{\left(}
\newcommand{\rs}{\right)}

\begin{document}

\title{ \vspace{1cm} Strangeness in Nuclei and Neutron Stars}

\author{L.\ Tolos,$^{1,2,3,4}$ L.\ Fabbietti,$^5$ 
\\
$^1$  Institut f\"ur Theoretische Physik, University of Frankfurt,\\ Max-von-Laue-Str. 1, 60438 Frankfurt am Main, Germany \\
$^2$ Frankfurt Institute for Advanced Studies,  University of Frankfurt, \\Ruth-Moufang-Str. 1, 60438 Frankfurt am Main, Germany \\
$^3$ Institute of Space Sciences (ICE, CSIC), \\Campus UAB, Carrer de Can Magrans, 08193, Barcelona, Spain \\
$^4$ Institut d'Estudis Espacials de Catalunya (IEEC), 08034 Barcelona, Spain \\
$^5$ Technische Universit\"at M\"unchen, Germany}
\maketitle

\begin{abstract} 
We review the present status of the experimental and theoretical developments in the field of strangeness in nuclei and neutron stars. We start by discussing the $\bar K N$ interaction, that is governed by the presence of the $\Lambda(1405)$. We continue by showing the two-pole nature of the $\Lambda(1405)$, and the production mechanisms in photon-, pion-, kaon-induced reactions as well as proton-proton collisions, while discussing the formation of $\bar K NN$ bound states. We then move to the theoretical and experimental analysis of the properties of kaons and antikaons in dense nuclear matter, paying a special attention to kaonic atoms and the analysis of strangeness creation and propagation in nuclear collisions. Next, we examine the $\phi$ meson and the advances in photoproduction, proton-induced and pion-induced reactions, so as to understand its properties in dense matter. Finally, we address the dynamics of hyperons with nucleons and nuclear matter, and the connection to the phases of dense matter with strangeness in the interior of neutron stars.

\end{abstract}

\eject
\tableofcontents
\setcounter{page}{2}

\input{Content/Introduction.tex}

\input{Content/KaonAntikaon.tex}

\input{Content/Phi.tex}

\input{Content/Hyperon.tex}

\input{Content/EOSandNeutronStars.tex}

\input{Content/Conclusions.tex}

\newenvironment{acknowledgement}{\relax}{\relax}
\begin{acknowledgement}
\section*{Acknowledgements}
L.F. and L.T. warmly thank Albert Feijoo, Avraham Gal, Manuel Lorenz, Eulogio Oset, Assumpta Parre\~no, Angels Ramos and Isaac Vida\~na for the careful reading of the manuscript and very constructive comments. L.F. and L.T. thank Alexandre Obertelli for the continuous support during the writing of this review, and $\Lambda(1405)$ for interesting discussions. L.T. acknowledges support from Deutsche Forschungsgemeinschaft under Project Nr. 383452331 (Heisenberg Programme) and Project Nr. 411563442.  This
research is supported by the Spanish Ministerio de
Ciencia, Innovacion y Universidades  under
contract FPA2016-81114-P; the THOR COST Action CA15213 and PHAROS COST Action CA16214; and by the EU STRONG-2020 project under the program H2020-INFRAIA-2018-1, grant agreement no. 824093. L.F. acknowledges support from MLL M\"unchen, DFG  EXC 2094 – 390783311 ORIGINS, GSI TMLRG1316F, BmBF 05P15WOFCA, SFB 1258, DFG FAB898/2-2.

\end{acknowledgement}

\bibliographystyle{elsarticle-num}
\bibliography{Content/Bibliography.bib}


\end{document}

%% file: Content/Introduction.tex
\section{Introduction}

Understanding the dynamics of hadrons with strangeness has received a lot attention over the past decades in connection with the study of exotic atoms \cite{Friedman:2007zza},  the analysis of strangeness production and propagation in particle and nuclear research facilities \cite{Fuchs:2005zg,Hartnack:2011cn}, and the investigation of the possible "strange" phases  in the interior of neutron stars \cite{Watts:2016uzu}. 

Kaons (antikaons) are the lightest mesons with strangeness that are made of one antistrange (strange) quark, and one up or down quark (antiquark).  Considering the fact that kaons are Goldstone  bosons, a quantitative study of their interaction with other hadrons and, in particular, nucleons allows for the exploration of the non-perturbative character of QCD at low energies. This allows to test the scales and symmetries of QCD in this energy regime, such as chiral symmetry and its partial restoration in dense and/or hot matter \cite{Lutz:1994cf}. Also, it permits a better understanding of the nature of newly discovered states with the strange degree of freedom, whether they can be understood as dynamically generated states via hadron-hadron scattering processes. 

In particular, the $\bar KN$ scattering amplitude has been extensively studied, showing a repulsive behaviour at energies close to the $\bar KN$ threshold. This is due to the presence of the $\Lambda(1405)$ resonance, located about 30 MeV below threshold \cite{Tanabashi:2018oca}. The nature of this baryonic resonant state has been analyzed experimentally in photon-, pion-, kaon-induced reactions as well as proton-proton collisions, in particular after the emergence of theoretical predictions describing this state as the superposition of two poles \cite{Oller:2000fj,Jido:2003cb,Hyodo:2007jq}. The theoretical and experimental determination of the two-pole nature of the $\Lambda(1405)$ moreover indicates that the $\bar K N$ interaction could be attractive enough to produce bound states. This finding has triggered the analysis of the possible formation of bound states with one $\bar K$ and one or more nucleons, such as the $\bar KNN$ state (see Ref.~\cite{Nagae:2016cbm} and references therein).

The properties of (anti-)kaons in dense nuclear matter have also prompted the interest of the scientific community, specially after the detection of kaonic atoms \cite{Friedman:2007zza}, the theoretical predictions of (anti-)kaon condensation in neutron stars \cite{Kaplan:1986yq}, and the analysis of strangeness creation and propagation in nuclear collisions \cite{Fuchs:2005zg,Hartnack:2011cn}. 
Indeed, kaons are considered suited probes to study the dense and hot nuclear matter formed in heavy-ion collisions. They are created in the early stages of the reaction and, even as they undergo scattering in the hot and dense medium, they mostly emerge intact thus carrying the information on the properties of hot dense matter until their detection. 

Moreover, the study of the hidden strange $\phi$ meson (together with $\omega$ and $\rho$) in dense nuclear matter has become a matter of great interest, since they are also probes for the chiral restoration \cite{Rapp:1999ej}. The study of the light vector mesons is indeed very appealing, as the dileptonic decay offers clean information on nuclear matter. The $\phi$, in particular, is very narrow in vacuum and is well separated from the $\rho$ and the $\omega$ mesons, allowing for the measurement of any modifications of its mass or width in matter.  To that end, several experiments have been carried out, such for photoproduction  \cite{Ishikawa:2004id,Wood:2010ei}, for proton-induced \cite{PhysRevC.85.035206,Polyanskiy:2010tj} and pion-induced \cite{Adamczewski-Musch:2018eik} reactions, an effort that has been accompanied by important theoretical developments to understand the modifications of the $\phi$ meson in dense nuclear matter.

Another venue of interest in the field of strangeness is the study of strange baryons, the so-called hyperons, and their dynamics with nucleons and nuclear matter. Theoretical studies have gone hand in hand with scattering experiments employing secondary hyperon beams \cite{Thomas:1973uh} or, more recently, using femtoscopy techniques \cite{FemtopXi}. Also, the possible formation of nuclei with one or more hyperons inside the nucleus, the so-called hypernuclei \cite{Gal:2016boi,Tamura:2013lwa,Feliciello:2015dua}, has triggered a lot of theoretical advances. The aim is to understand how hyperons interact with nucleons and hyperons in vacuum and as a function of density.

And last but not least, understanding the behaviour of strange mesons and baryons in the presence of a surrounding dense medium is of particular interest to determine the features of the possible phases of dense matter in compact astrophysical objects, such as neutron stars. Neutron stars are one of the most compact astrophysical objects in the universe and, therefore, serve as a unique laboratory for testing matter with strangeness under strong  gravitational and magnetic fields, as well as extreme conditions of density, isospin asymmetry and temperature \cite{Lattimer:2006xb,Watts:2016uzu,Watts:2018iom}. 

In this review we aim at discussing the present experimental and theoretical status in the field of strangeness in nuclei and neutron stars. We start by reviewing in Sec.~\ref{sec:kaon-antikaon} the kaon and antikaon interaction with nucleons, paying a special attention to the role of the $\Lambda(1405)$ in the determination of the $\bar K N$ interaction, in Sec.~\ref{sec:Lambda1405}, and the formation of bound states, such as $\bar KNN$ in Sec.~\ref{sec:KNN}. Moreover, kaons and antikaons in matter are reviewed in Sec.~\ref{sec:kaons-matter}, with a special emphasis on the production and propagation of strangeness in hadron-hadron collisions in view of the present and forthcoming  experimental programs on strangeness. 
Next, in Sec.~\ref{sec:phi} we review the experimental and theoretical advances on the study of the $\phi$ meson behaviour in nuclear matter, whereas the hyperon-nucleon and hyperon-hyperon interactions are discussed in Sec.~\ref{sec:YN-YY}. Among the experimental searches, we clearly identified those performed in the field of hypernuclei (Sec.~\ref{sec:hypernuclei}), whereas we address the hyperon properties in a dense medium in Sec.~\ref{sec:hyperon-dense}. Indeed, investigating the properties of strange hadrons in a dense medium is of extreme relevance for the physics of neutron stars, as discussed in Sec.~\ref{sec:NS}. In particular, the Equation of State of nuclear matter in the high density regime is discussed in Secs.~\ref{sec:nucEoS} and \ref{sec:constraints}. There, we also point out the changes induced in the Equation of State of dense matter once strangeness is considered (Sec.~\ref{sec:eos-strange}). 
Our review finishes in Sec.~\ref{sec:conclusions} with some conclusions and future perspectives, specially focusing on future experiments.

%% file: Content/KaonAntikaon.tex
\section{Kaon(Antikaon)-Nucleon Interaction }
\label{sec:kaon-antikaon}
\subsection{$\bar K N$ interaction: the  $\Lambda(1405)$}
\label{sec:Lambda1405}
\subsubsection{Experiments for $\bar KN$}
\label{sec:KNexperiment}

Scattering data of low-energy charged kaons with proton or deuteron targets, performed in Bubble-Chamber experiments or with emulsions, exist for kaon incident momenta below~350 MeV/c~\cite{Humphrey:1962zz,Watson:1963zz,Mast:1975pv,Nowak:1978au,Ciborowski:1982et}. Such data represented so far the only reference for the understanding of the  low energy  kaon-nucleon interaction, since the works provide precise cross section measurements and, with the help of scattering theory scattering, parameters can be extracted. The typical cross sections are small (of the order of 10~mb) and are dominated by the elastic channel with a small contribution from charge-exchange processes. 

The \kp\ and \km\ behavior in the scattering process is very different and markedly depends on the meson strangeness content. The $\kp N$ interaction is well established due to the lack of coupled-channels, that is, there are no meson-baryon pairs with the same quantum numbers that couple to the $\kp N$ final state. 
The interaction is moderately repulsive, due to the strong and the Coulomb interactions \cite{Hadjimichef:2002xe}. The \kp $p$ system is a pure isospin I~=~1 state, and its S-wave scattering length has been determined with good (1\%) precision ~\cite{Dover:1982zh, Hyslop:1992cs}. The two isospin states of the $KN$ interaction in the $t$-channel, $I = 0,1$, are both
contributing to the \kp $n$ interaction. The $I=0$ part of the $KN$ amplitude can be extracted by using \kp $d$ in combination with \kp $p$ data.
 Last but not least, \kp $N$ data can be employed to assess the strangeness content of the nucleon ~\cite{Gasser:1999df,Jaffe:1987sw}. This aspect plays an important role in the determination of the chiral potential that drives the behaviour of kaons within nuclear matter.

For antikaons, the behaviour above threshold is only scarcely constrained by scattering experiments due to the low available statistics close to the $\bar{K} N$ threshold. But of late, new measurements above threshold have been made available exploiting correlation techniques to \km-p pairs \cite{Acharya:2019bsa}, that are measured in pp collisions at the LHC by the ALICE collaboration.

The correlation function between two hadrons is defined 
as the ratio of the distribution of relative
momenta for correlated and uncorrelated baryon pairs.
Experimentally, one considers particle pairs emitted in the same collision and in different collisions, where correlations are absent (mixed event method). This results in the correlation function $C(k^*)=\frac{N_{\mathrm{same}}(k^*)}{N_{\mathrm{mixed}}(k^*)}$, that is, the ratio of the number of pairs with a given reduced relative momentum in the pair rest frame $k^*$ ($k^*=|\vec{p_1}-\vec{p_2}|/2$) measured
in the same collision, divided by the number of pairs with the same relative momentum
obtained combining particles from different collisions. 

This correlation function is equal to one in absence of any correlation for the pair of
interest. Correlations can occur either because of quantum-mechanical interference or final state interactions.
In the case of an attractive interaction, the correlation will be larger than one. In the case of a repulsive interaction or in the presence of a bound state, the correlation will be between zero and one.
The measured correlation function can be expressed theoretically  as 
$C(k^*)=\int d^3\vec{r}\, S(r) \cdot|\psi(k^*,r)|^2$ \cite{Pratt:1986cc,Lisa:2005dd}, where $S(r)$ is
the distribution at the distance $r$ at which particles are emitted (source) and $|\psi(k^*,r)|$ represents the relative wave function of the
pair of interest.

The particle emitting source is found to be very small in pp and p+Pb collisions at the LHC \cite{Mihaylov:2018rva,Acharya:2018gyz}, that is of the order of 1 fm.
These values are evaluated by using known interactions,  with  pp and $K^+ p$ as benchmarks \cite{Acharya:2019bsa}.

\begin{figure}[htb]
\centering
\includegraphics[width=5cm]{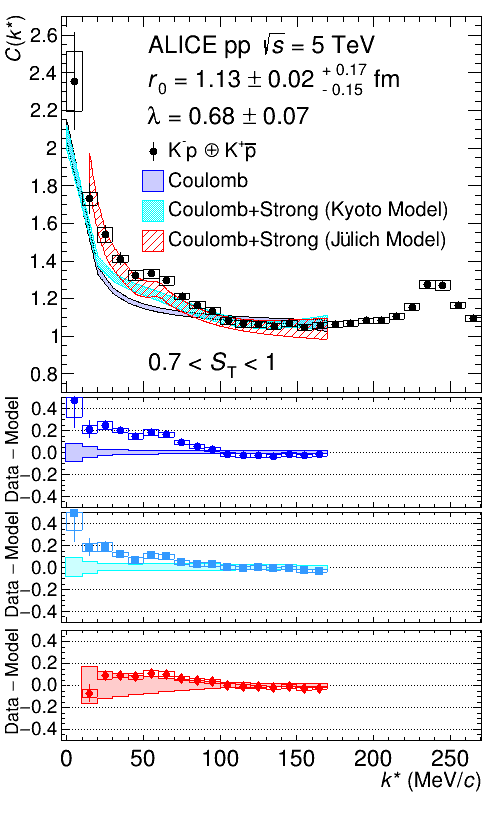}
\includegraphics[width=5cm]{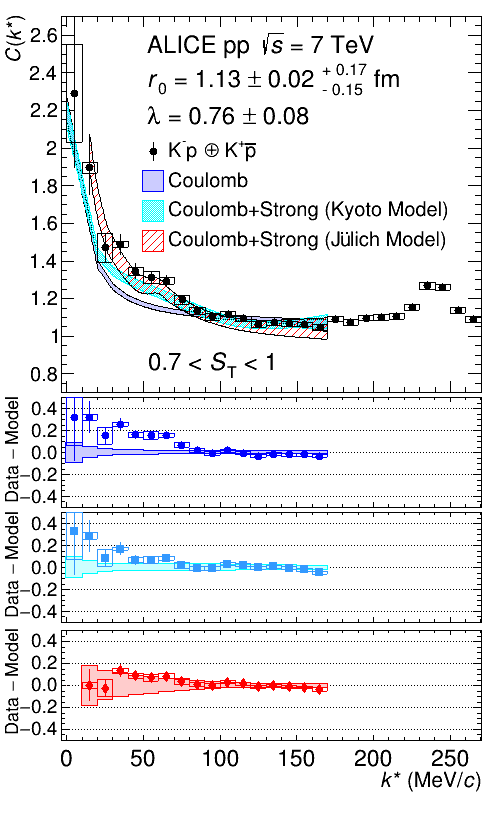}
\includegraphics[width=5cm]{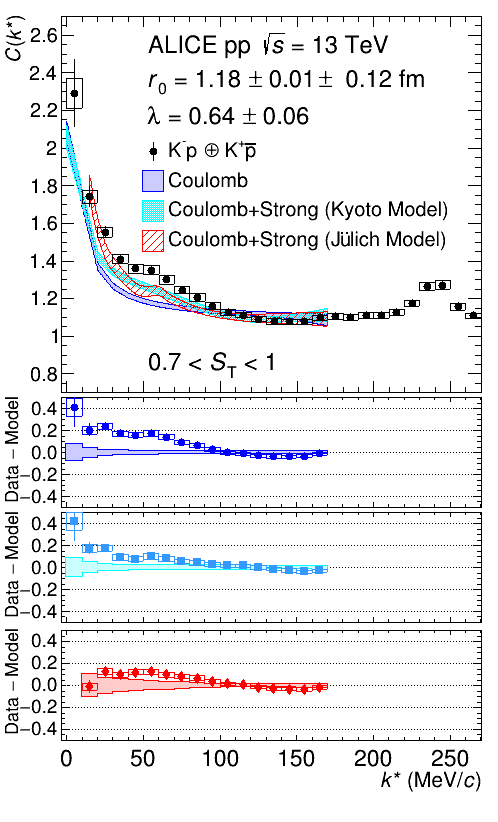}
\caption{\label{fig:k-p} 
\textit{(Color online)} $K^- p$ correlation functions obtained from $pp$ collisions at 5~TeV (left), 7 TeV (middle) and 13 TeV (right). Three different potentials were considered: Coulomb potential (blue band), Kyoto model and J\"ulich model \cite{Acharya:2019bsa}. In the bottom panels, differences between the data and the model are shown.}
\end{figure}

Figure \ref{fig:k-p} shows the correlation functions measured for $K^- p$ pairs in $pp$ collisions at 5, 7 and  13 TeV by the ALICE collaboration \cite{Acharya:2019bsa}. The fact that the correlation function is larger than one indicates the overall attractive nature of the interaction. However, the assumption that only an attractive Coulomb interaction drives the correlation is not sufficient to explain  the measurements (see blue histograms in Fig.~\ref{fig:k-p}). The cyan and red histograms show the prediction considering,  additionally to the Coulomb interaction, also a strong attractive interaction based on \cite{Ikeda:2012au,Haidenbauer:2018jvl}. 
One can  see that the agreement improves, but also that some deviations are still present.
One of the aspects that should be considered in the comparison is the fact that in the femtoscopy studies only the final state is fixed, while different initial states can be created.
In the specific case  of the $K^-p$ final state, the coupling to a $\pi\Sigma$ initial state is possible. This process would manifest itself in an increased correlation for low values of $k^*$ for the $K^-p$ pair. Recent predictions \cite{Kamiya:2019uiw}, that take into account this coupled channel, achieve a much improved agreement with the experimental correlation function. A better improved statistics (factor 50 more) is expected in the upcoming Run3 and Run4 at the LHC, and this will allow to better constrain existing models of the $K^-p$ interaction.

The scattering parameters that can be extracted from both experimental methods above and at the $\bar{K}N$ threshold would suggest a repulsive interaction. In reality, the interaction is attractive because a 
sign change of the scattering length occurs slightly below threshold 
due to the presence of the $\Lambda(1405)$ resonance. The latter is located $27$ MeV below the $\bar{K} N$ threshold.
Precise data above and at the threshold are mandatory to constrain the properties of this resonance and pin down the $\bar{K}N$ interaction quantitatively.

A crucial measurement is the evaluation of the scattering parameters exactly at the threshold and this is made 
possible by studying the kaonic hydrogen atom.
This is a system composed of a proton
and a negatively charged kaon bound by the Coulomb
force, where the kaon cascades down to the atomic $1s$ level. The energy levels are shifted from their
pure electromagnetic values and have a finite absorption width because of the effect of the strong $\bar{K} N$ interaction. The measurement of the K-series X-rays can be related to the real and imaginary parts of the scattering amplitude for the $\bar{K} p$ system.

 Kaonic hydrogen atoms were first investigated at KEK by the KpX experiment \cite{Iwasaki:1997wf,Ito:1998yi}, where negatively charged kaon beams with a momentum of $600$ MeV/c were first slowed down and then employed for the kaonic atom creation. Later, kaonic hydrogen atoms were investigated at Da$\phi$NE e$^+$e$^-$ collider in Frascati by the DEAR \cite{Beer:2005qi} and SIDDHARTA \cite{Bazzi:2011zj,Bazzi:2012eq} collaborations.  These experiments presented the advantage of a slower beam as negative kaons
 were stemming from the $\phi$ decay almost at rest, hence, giving rise to large cross sections for the creation of the exotic atoms.
 
The so far most precise measurement of the scattering lengths was obtained by SIDDHARTA experiment, due to the excellent time and energy resolution of the large area silicon drift detectors that have been employed for the first time by this collaboration. The collaboration obtained a shift and width of the $1s$ state equal to $\epsilon _{1s}=\,-283\pm36(stat)\pm 6(syst)$ and $\Gamma_{1s}=\, 541\pm 89(stat)\pm 22 (syst)$.
These numbers can be converted into the real and imaginary parts of the isopin-averaged $\bar{K} N$ scattering amplitude. In order to extract the isospin dependent scattering parameters, the measurement of kaonic deuterium has been proposed, since bound systems of negative charged kaons and neutrons can not be built.
These experiments are envisaged in the next two years at the J-PARC \cite{Zmeskal:2015efj} and Da$\phi$NE \cite{Curceanu:2017jyd} facilities.

\subsubsection{Theory for the $\bar KN$ interaction}
\label{sec:theoryKN}
As previously mentioned, the $\bar K N$ scattering in the $I=0$ channel is governed by the presence of the $J^P=1/2^-$ $S=-1$ $\Lambda(1405)$. The dynamical origin of the $\Lambda(1405)$ as a molecule was predicted more than 50 years ago by Dalitz and Tuan  \cite{Dalitz:1959dn,Dalitz:1959dq} and, up to now, is the only accepted molecular state. Over the past years, the molecular nature of the $\Lambda(1405)$ has been revisited. On the one hand, lattice QCD simulations \cite{Hall:2014uca} have determined the molecular bound $\bar K N$ nature of the $\Lambda(1405)$ due to the vanishing strange quark contribution to the magnetic moment and by the dominance of the $\bar K N$ component in the finite-volume treatment. 

On the other hand, a lot of effort has been invested within coupled-channel unitarized theories. The coupled-channel unitarized schemes result from solving the Bethe-Salpeter equation (or the three-dimensional Lippman-Schwinger reduction) 
\begin{equation}
T=V+VGT ,
\label{bethe-free}
\end{equation}
with $G$ being the meson-baryon propagator, and $V$ standing for the meson-baryon interaction kernel. The coupled-channel structure implies that all meson-baryon pairs from the pseudoscalar-meson and baryon octets with the same quantum numbers as the $\bar K N$ have to be considered in the solution of the Bethe-Salpeter equation. The unitarization comes from the non-pertubative behaviour of the Bethe-Salpeter equation that implies an infinite resummation of two-body diagrams. The Bethe-Salpeter equation, moreover, has to be renormalized to avoid logarithmically ultraviolet divergences by means of a subtraction scheme. Note that a commonly used approach for solving the $\bar KN$ Bethe-Salpeter equation (mainly in $S$-wave) relies on the on-shell factorization \cite{Oset:1997it,Oller:2000fj,Jido:2003cb,Nieves:1999bx}. In this case, two different subtraction schemes are usually considered, the cutoff regulator or the dimensional regularization.

The crucial ingredient for a reliable evaluation of the $\bar K N$ scattering is the determination of the microscopic interactions of $\bar K N$ and associated meson-baryon channels. There have been attempts within meson-exchange models that extended pion-nucleon meson-exchange schemes by incorporating the strangeness degree of freedom \cite{MuellerGroeling:1990cw,Haidenbauer:2010ch}.  Also, there has been a lot of effort within meson-baryon chiral effective field theories ($\chi EFT$) \cite{Kaiser:1995eg, Oset:1997it,Oller:2000fj,Lutz:2001yb,GarciaRecio:2002td,Jido:2003cb,Borasoy:2005ie,Oller:2006jw,Feijoo:2018den}.

These latter works on meson-baryon $\chi EFT$ are based on the SU(3) chiral effective Lagrangian that reads
\begin{equation}
\Lagr_{\phi B}^{eff}=\Lagr_{\phi B}^{(1)}+\Lagr_{\phi B}^{(2)}  \ ,
\end{equation}
being $\Lagr_{\phi B}^{(1)}$ and $\Lagr_{\phi B}^{(2)}$ the most general form of the leading-order (LO), and contact next-to-leading order (NLO) (relevant in $S$-wave) contributions to the meson-baryon interaction Lagrangian, respectively. Those read   
\begin{eqnarray}
\Lagr_{\phi B}^{(1)} & = & i \langle \bar{B} \gamma_{\mu} [D^{\mu},B] \rangle
                            - M_0 \langle \bar{B}B \rangle  
                           - \frac{1}{2} D \langle \bar{B} \gamma_{\mu} 
                             \gamma_5 \{u^{\mu},B\} \rangle \no \\
                  & &      - \frac{1}{2} F \langle \bar{B} \gamma_{\mu} 
                               \gamma_5 [u^{\mu},B] \rangle \ ,
\label{LagrphiB1} 
\end{eqnarray}
\begin{eqnarray}
    \Lagr_{\phi B}^{(2)}& = & b_D \langle \bar{B} \{\chi_+,B\} \rangle
                             + b_F \langle \bar{B} [\chi_+,B] \rangle
                             + b_0 \langle \bar{B} B \rangle \langle \chi_+ \rangle \no \\ 
                     &  & + d_1 \langle \bar{B} \{u_{\mu},[u^{\mu},B]\} \rangle 
                            + d_2 \langle \bar{B} [u_{\mu},[u^{\mu},B]] \rangle    \no \\
                    &  &  + d_3 \langle \bar{B} u_{\mu} \rangle \langle u^{\mu} B \rangle
                            + d_4 \langle \bar{B} B \rangle \langle u^{\mu} u_{\mu} \rangle \ .
\label{LagrphiB2}
\end{eqnarray}
In Eqs.~(\ref{LagrphiB1},\ref{LagrphiB2}) $B$ is a $3\times3$ matrix that contains the fundamental baryon octet $(N,\Lambda,\Sigma,\Xi)$. The pseudoscalar meson octet $\phi=$($\pi,K,\eta$) enters through $u_\mu = i u^\dagger \partial_\mu U u^\dagger$, with $U(\phi) = u^2(\phi) = \exp{\left( \sqrt{2} {\rm i} \phi/f \right)} $, where $f$ is the meson decay constant. In Eq.~(\ref{LagrphiB1}) $M_0$ is the common baryon octet mass in the chiral limit, whereas the SU(3) axial vector constants  $D$ and $F$ are subject to the constraint $g_A=D+F=1.26$.
 Moreover, in Eq.~(\ref{LagrphiB2}), the term $\chi_+$ = $2 B_0 (u^\dagger \mathcal{M} u^\dagger + u \mathcal{M} u)$  breaks chiral symmetry explicitly via the quark mass matrix  $\mathcal{M} = {\rm diag}(m_u, m_d, m_s)$, where $B_0 = - < 0 | \bar{q} q | 0 > / f^2$ is related to the order parameter of spontaneously broken chiral symmetry. The coefficients $b_D$, $b_F$, $b_0$ and $d_i$ $(i=1,\dots,4)$ are the corresponding low energy constants at NLO, that need to be determined from experiment.  The details of the different terms can be found, for example, in Ref.~\cite{Feijoo:2018den}.
 

\begin{figure}[t]
\begin{center}
\centering
\includegraphics[width=\textwidth]{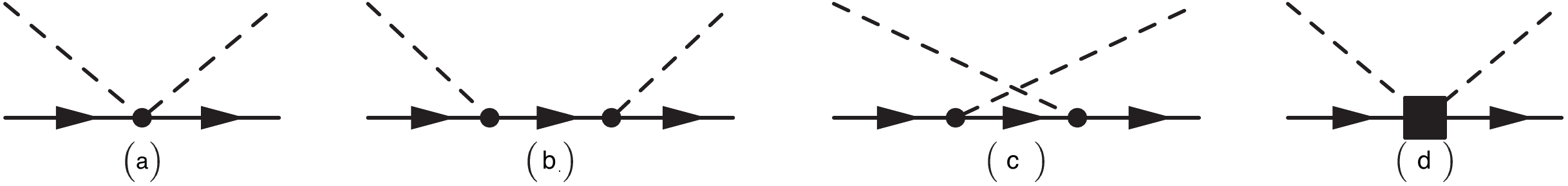}
\caption{\label{fig:epsart} Different contributions from the chiral meson-baryon Lagrangian: Weinberg-Tomozawa term (a), direct and crossed Born terms (b) and (c), and NLO terms (d). The dashed (solid) lines indicate the pseudoscalar octet mesons (octet baryons).}
\end{center}
\end{figure}

The meson-baryon interaction kernel in momentum space is  derived from both Eqs.~(\ref{LagrphiB1}) and (\ref{LagrphiB2}). In particular, the Weinberg-Tomozawa (WT) contribution corresponds to the contact diagram (a) in Fig.~\ref{fig:epsart}. Next, the vertices of diagrams (b) and (c) from the direct and crossed Born contributions in Fig.~\ref{fig:epsart} are obtained from the third and fourth terms of  Eq.~(\ref{LagrphiB1}), whereas the NLO contact term is directly extracted from Eq.~(\ref{LagrphiB2}), that is,  diagram (d) of  Fig.~\ref{fig:epsart}.

The different works on $\chi EFT$ have analyzed the effects of including a complete basis of meson-baryon channels, studied the differences in the regularization scheme, included $s$- and $u$-channel Born terms in the Lagrangian, implemented NLO contributions, and so on. These analyses have arrived to the conclusion that the significant contribution to the $\bar K N$ scattering that allows for a good fit to the experimental data comes from the WT interaction, with the known exception of $K^- p \rightarrow \eta \Lambda$ \cite{Guo:2012vv}. The inclusion of 
direct and crossed Born terms, and NLO terms fine tune the fitting to the experimental data, becoming more important for scattering processes that do not have a direct contribution from the WT term \cite{Feijoo:2015yja}. 

\begin{figure}[htb]
    \centering
\includegraphics[width=0.49\linewidth]{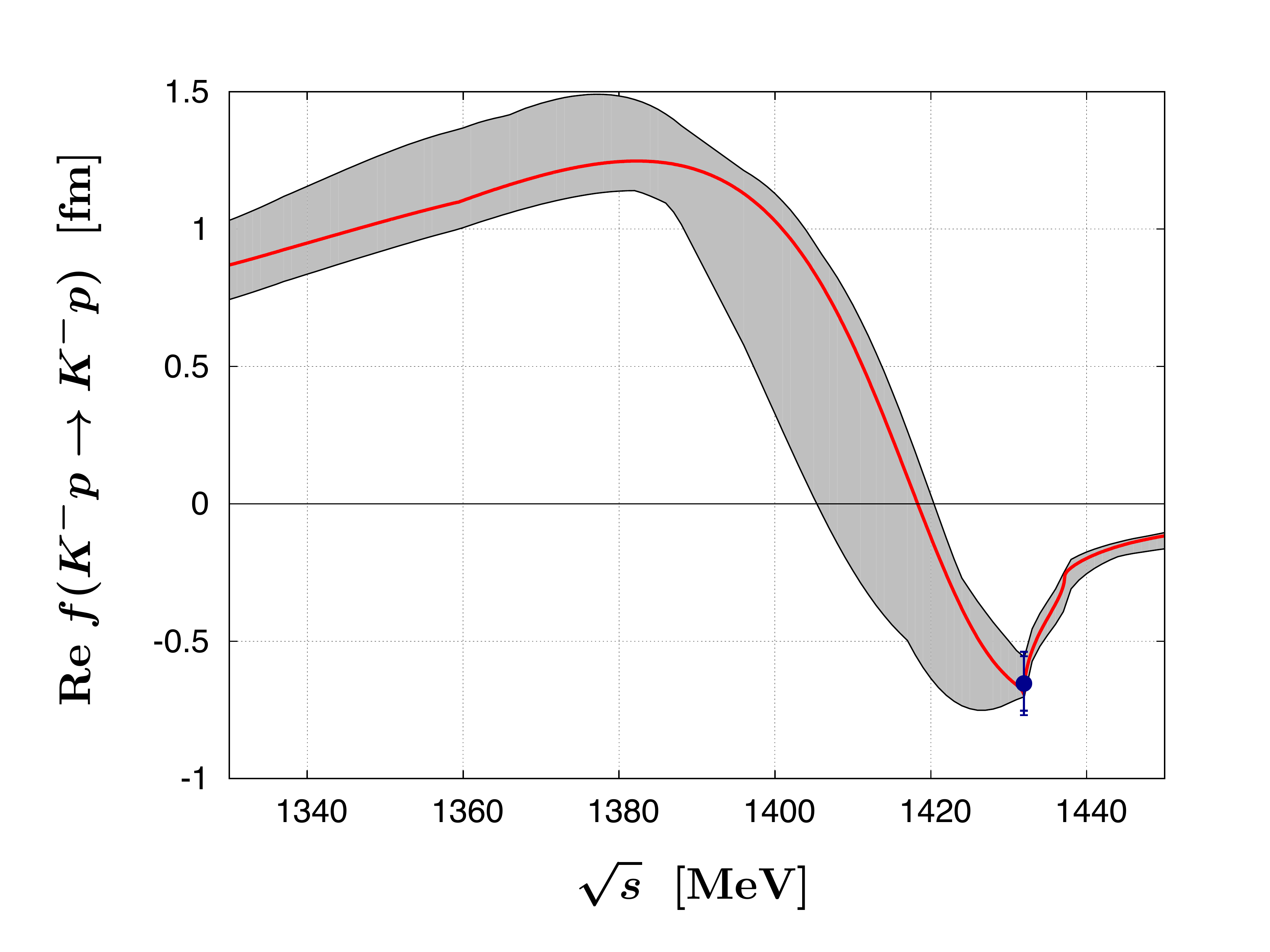}
  \hfill
\includegraphics[width=0.49\linewidth]{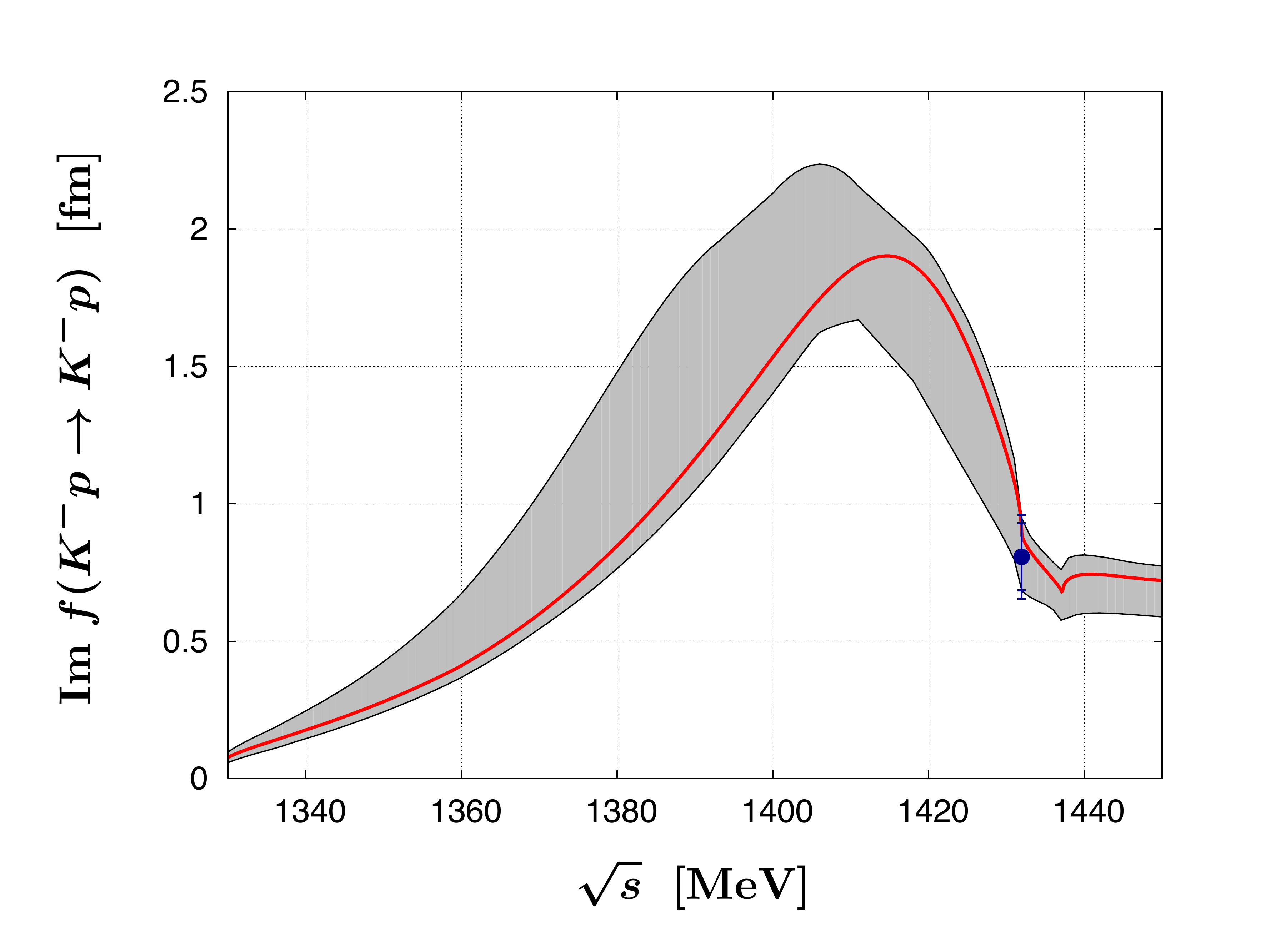}
    \caption{\textit{(Color online)} The real   (left)  and  imaginary   (right)  parts  of  the $K^- p \rightarrow K^- p$  scattering  amplitude  obtained  from the NLO calculation and extrapolated to the subthreshold region. The real and imaginary parts of the $K^- p$ scattering length from SIDDHARTA experiment are indicated by the dots including the statistical and systematic errors. Figures taken from Ref.~\cite{Ikeda:2012au}.}
    \label{fig:amplitude-weise}
\end{figure}

The fine tuning is indeed important for the correct description of the shift and width of the $1s$ state kaonic hydrogen obtained by the SIDDHARTA collaboration, as discussed in Sec.~\ref{sec:KNexperiment}. The $K^-p$ scattering length $a(K^-p)$ can be related to the shift and width of the $1s$ state of kaonic hydrogen atom using a Deser-type relation with isospin-breaking corrections as \cite{Meissner:2004jr}:
\begin{eqnarray}
\Delta E - i \Gamma/2 = -2 \alpha^3 \mu_r^2 a(K^-p) [1+ 2 \alpha \mu_r (1- {\rm ln} \alpha) a(K^-p)],
\end{eqnarray}
where $\alpha$ is the fine-structure constant and $\mu_r$ is the $K^-p$ reduced mass.
The SIDDHARTA results on kaonic hydrogen as well as the earlier results on total cross sections and threshold branching ratios can be successfully described considering all contributions up to NLO meson-baryon contributions \cite{Ikeda:2012au,Guo:2012vv,Mai:2012dt}. Indeed, in  Fig.~\ref{fig:amplitude-weise} the subthreshold extrapolation of the real and imaginary parts of the $K^- p \rightarrow K^- p$ amplitude are shown by the best-fit NLO scheme of Ref.~\cite{Ikeda:2012au}. The amplitude displays the structure of the $\Lambda(1405)$, resulting from the strong attraction in the $I=0$ sector. As mentioned in Sec.~\ref{sec:KNexperiment}, it is expected that the future measurement of kaonic deuterium in J-PARC \cite{Zmeskal:2015efj} and DA$\phi$NE \cite{Curceanu:2017jyd} facilities will further constrain the subthreshold behaviour of the $\bar K N$ amplitude. 

Another interesting conclusion of all these works is that the dynamics of the $\Lambda(1405)$ is described by the superposition of two poles of the scattering matrix, between the $\bar K N$ and $\pi \Sigma$ thresholds \cite{Oller:2000fj,Jido:2003cb,Hyodo:2007jq}.
Experimentally, the $\Lambda(1405)$ is seen as one resonance shape, while the existence of two poles is supported in  reaction-dependent line shapes \cite{Jido:2003cb}. This two-pole structure has its origin in the two attractive channels of the LO WT interaction in the SU(3) basis (singlet and octet) \cite{Jido:2003cb} and in the $\bar K N$ and $\pi \Sigma$ isospin basis \cite{Hyodo:2007jq}. 

Recent results from NLO chiral unitary coupled-channel approaches that include the SIDDHARTA constraint are given in the PDG \cite{Tanabashi:2018oca} and reproduced in Table~\ref{tab:poles}. In spite of the differences (coming from the various approximations to the Bethe-Salpeter equation, the fitting procedure, the inclusion of SU(3) breaking effects and others), there is a consensus that the dominant contribution to the $\Lambda(1405)$ stems from the state located near the $\bar K N$ threshold, with a strong coupling to that meson-baryon system, whereas the pole at lower energies with a larger width strongly couples to the $\pi \Sigma$ state.

\begin{table}[]
    \centering
    \begin{tabular}{c|c|c}
         {\rm Model} & {\rm First Pole [MeV]} & {\rm Second Pole [MeV]} \\
         \hline
         \hline
          {\rm NLO} \cite{Ikeda:2012au} & $1424^{+ 7}_{-23} - i \, 26 ^{+3}_{-14}$ & $1381^{+ 18}_{-6} - i \, 81 ^{+19}_{-8}$\\
         \hline
          {\rm Fit II} \cite{Guo:2012vv} & $1421^{+ 3}_{-2} - i \, 19 ^{+8}_{-5}$ & $1388^{+ 9}_{-9} - i \, 114 ^{+24}_{-25}$\\
          \hline 
          {\rm Solution Nr. 2} \cite{Mai:2014xna} & $1434^{+ 2}_{-2} - i \, 10 ^{+2}_{-1}$ & $1330^{+ 4}_{-5} - i \, 56 ^{+17}_{-11}$\\
         \hline
         {\rm Solution Nr. 4} \cite{Mai:2014xna} & $1429^{+ 8}_{-7} - i \, 12 ^{+2}_{-3}$ & $1325^{+ 15}_{-15} - i \, 90 ^{+12}_{-18}$\\
         \hline
    \end{tabular}
    \caption{Pole positions for the $\Lambda(1405)$ coming from recent chiral effective models including the SIDDHARTA constraint. Table adapted from \cite{Tanabashi:2018oca}. }
    \label{tab:poles}
\end{table}


\subsubsection{Experimental measurements of the $\Lambda(1405)$ }
\label{sec:expLambda1405}

The region below the $\bar{K} N$ threshold and, hence, the spectral shape of the $\Lambda(1405)$ can be studied experimentally by investigating its strong decays, that is, $\Lambda(1405)\rightarrow \Sigma\pi$. The measurement of these 
final states is non-trivial, since they all contain a neutral particle and, for this reason, the data set is rather limited.

\begin{figure*}[hbt]
  \centering
  \includegraphics[width=0.49\linewidth]{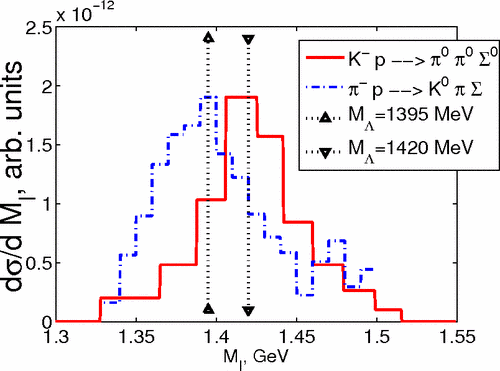}
  \hfill
  \includegraphics[width=0.49\linewidth]{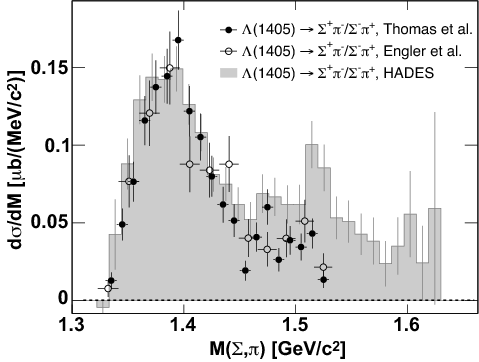}
\caption{\textit{(Color online)}  Left panel: Comparison of the $\Lambda(1405)$ spectral function measured in  kaon- and pion-induced reactions \cite{Magas:2005vu}. Right panel: Comparison of the  pion-induced data with the measurement of the reaction $p+p\rightarrow \Lambda(1405)K^+p \rightarrow \Sigma^{\pm}\pi^{\mp}K^+p$ \cite{Agakishiev:2012xk,Siebenson:2013rpa}.
  }
  \label{fig:L1405exp1}
\end{figure*}
Motivated by the theoretical work of R. Dalitz \cite{Dalitz:1959dn,Dalitz:1959dq}, first experimental efforts were carried out  in the sixties at BNL \cite{Thomas:1973uh} and in the eighties at CERN \cite{Hemingway:1984pz}, employing pion and kaon beams, with a momentum of $1.69$ and $4.2$ GeV/c, respectively, and hydrogen Bubble Chambers for the detection. Despite of the limited statistics, the clear detection procedure that was possible with the photographic plates allowed to tag reactions as $\pi^- + p \rightarrow \Lambda(1405)K^0 \rightarrow \Sigma^{\pm} \pi^{\mp}K^0$, and $K^- + p \rightarrow \Sigma^+(1660)\pi^-\rightarrow \Lambda(1405)\pi^+\pi^- \rightarrow \Sigma^{\pm} \pi^{\mp}\pi^+\pi^-$. The study of the $\Sigma^{\pm}\pi^{\mp}$ invariant mass allowed to determine the $\Lambda(1405)$ spectral shape and measure the angular distribution of the resonance to extract its spin. 

In both analyses, the $\Lambda(1405)$ spin was found to be equal to $1/2$ and the contribution of the non-resonant final state was not isolated from the resonance. Interferences were also neglected.
The left panel of Fig.~\ref{fig:L1405exp1} shows the comparison of the measured spectral functions of the $\Sigma^{\pm}\pi^{\mp}$ final state associated to the $\Lambda(1405)$ for kaon- and pion-induced reactions. One can see that the maxima of the two distributions are shifted with respect to each other.

Although the statistics is limited, the different spectral function obtained for two different initial states supports the molecular interpretation of the $\Lambda(1405)$: the strength of the $\bar K N$ and $\pi \Sigma$ poles depends on the initial state. 
The right panel of Fig.~\ref{fig:L1405exp1} shows the comparison of the pion-induced data with the measurement of the reaction $p+p\rightarrow \Lambda(1405)K^+p \rightarrow \Sigma^{\pm}\pi^{\mp}K^+p$ \cite{Agakishiev:2012xk}. Although the initial states are very different, the obtained spectral shapes are in good agreement.
If one takes the nominal mass of the resonance at $1405$ MeV, one observes a shift of about 20 MeV towards lower masses for both pion- and proton-induced reactions, while a shift towards higher masses is measured in kaon-induced reactions.
One has to point out that the contributions of nearby resonances, as the $\Sigma(1385)$, has been quantified in these analyses and found to be rather negligible with respect to the total yield. The production of $\Lambda(1405)$ in proton-proton induced reactions has
been also investigated by the ANKE collaboration \cite{Zychor:2007gf}, exploiting the $\Sigma^0\pi^0$ decay channel and the results are consistent within errors.

It is clear that in reactions as $p+p$ several rather broad baryonic resonances can be excited in the final states and, hence, interferences can take place. Different theoretical calculations have been developed to 
explain the apparent shift of the $\Lambda(1405)$ peak in p+p collisions and possible explanations are discussed in Sec. \ref{sec:1405production}.

\begin{figure*}[hbt]
  \centering
\includegraphics[width=\linewidth,height=90mm]{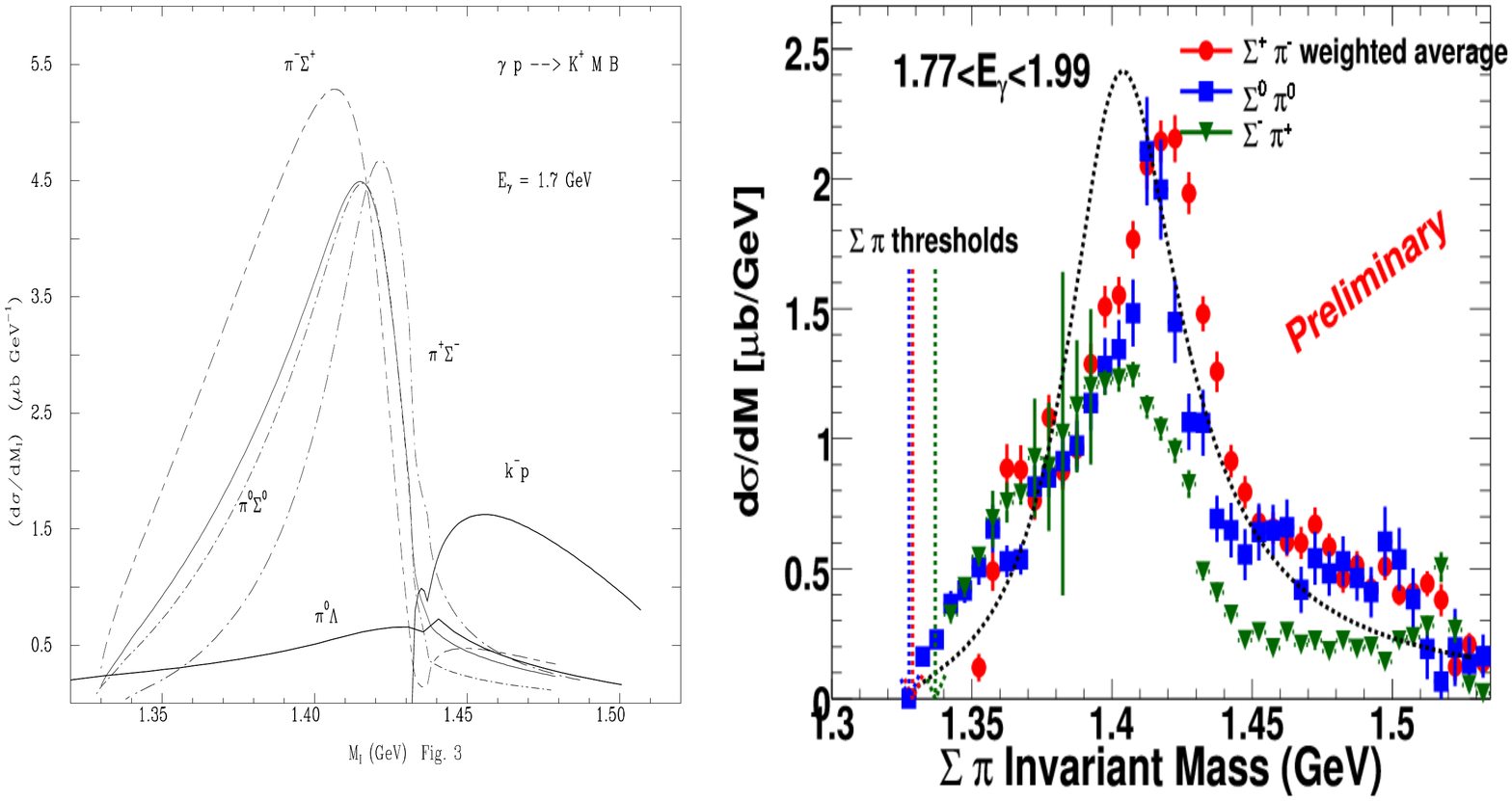}
  \hfill
 \caption{\textit{(Color online)} Left panel: Predicted $\Lambda(1405)$ spectral function for the decays into $\Sigma^{\pm}\pi^{\mp}$ and $\Sigma^0\pi^0$ \cite{Nacher:1998mi}. Right panel: Spectral function of the $\Lambda(1405)$ in the three $\Sigma\pi$ final states measured by the CLAS experiment with photon beams  \cite{Moriya:2009mx}.
 }
  \label{fig:L1405exp2}
\end{figure*}

A break-through was achieved by the CLAS collaboration only in recent years by employing photon beams \cite{Moriya:2013eb}. All the charged combinations of the $\Lambda(1405)$ decays could be measured for the reaction $\gamma +p \rightarrow \Lambda(1405)K^+\rightarrow (\Sigma\pi)^0K^+$ at center-of-mass energies in the range $1.95 <\, W <\,2.85$ GeV.
Different spectral functions had been predicted \cite{Nacher:1999ni} for the three decays of the $\Lambda(1405) \rightarrow \Sigma^+\pi^-/\Sigma^-\pi^+/\Sigma^0\pi^0 $ due to the interference of isospin components. Experimentally, a difference was measured, but the 
surprising fact is that the predicted shapes do not reproduce at all the measured hierarchy.
Figure \ref{fig:L1405exp2} shows the comparison of the theoretical predictions and experimental measurement for 
the three decays and the inverted hierarchy is evident.

One of the attempts to understand the behaviour of the $\Lambda(1405)$ in the CLAS data consisted in a parameterization with momentum dependent spectral functions of three different poles \cite{Moriya:2013eb}. Despite of the fact that the theory predicts only $I=0$ states for the $\Lambda(1405)$, the three poles considered in this analysis were one rather broad $I=0$ state ($M=\,1338\pm10$ MeV/c$^2$, $\Gamma= \,85\pm10$ MeV/c$^2$) and two $I=1$ states ($M=\,1431\pm10$ MeV/c$^2$, $\Gamma= \,52\pm10$ MeV/c$^2$ and $M=\,1394\pm20$ MeV/c$^2$, $\Gamma= \,149\pm40$ MeV/c$^2$). Sizable contributions from interferences between the poles were found as well in \cite{Mai:2014xna}. 

The $\Lambda(1405)$ plays also an important role in reactions at the LHC, where the LHCb collaboration claimed the existence of several pentaquarks looking at reactions as $\Lambda_b\rightarrow J/\psi K^- p$ \cite{Aaij:2015tga}. In this type of analysis, it is important to precisely know the background stemming from different resonances, as the coupling $\Lambda(1405) \rightarrow \bar{K}N$ and the information retrieved from low energy experiments have been employed to provide predictions for such mechanisms at higher energies \cite{Roca:2015tea}.

\subsubsection{Theoretical models for the $\Lambda(1405)$ production}
\label{sec:1405production}

From the theoretical point of view, several approaches have addressed the production of the $\Lambda(1405)$ related to  photon-induced, pion-induced and kaon-induced reactions as well as proton-proton collisions and heavy-meson decays.

The work on the $\Lambda(1405)$ photoproduction on proton and nuclei of Nacher et al. \cite{Nacher:1998mi} predicted that the detection of the $K^+$ would be sufficient to determine the shape and strength of the $\Lambda(1405)$ resonance, whereas the detection of the resonance decay channels ($\pi \Sigma$) was needed in order to pin down its properties in nuclei. This is due to the fact that the Fermi motion, because of the large momentum of the original photon, would be blurring any trace of the resonance, if only the $K^+$ is detected. As a consequence, in a latter work \cite{Nacher:1999ni} the crossed channel $K^- p \rightarrow \Lambda(1405) \gamma$ had been proposed, so as to detect a particle that interacts weakly with the nucleus, allowing to inspect the invariant mass region below the $K^-p$ threshold, where the $\Lambda(1405)$ is present. In this reaction, the smaller energy of the $K^-$ and the exiting photon would lead to a suppression of the Fermi motion effects in nuclei, so as to detect the shape of the $\Lambda(1405)$ resonance.

\begin{figure*}[hbt]
  \centering
  \includegraphics[width=0.8\linewidth]{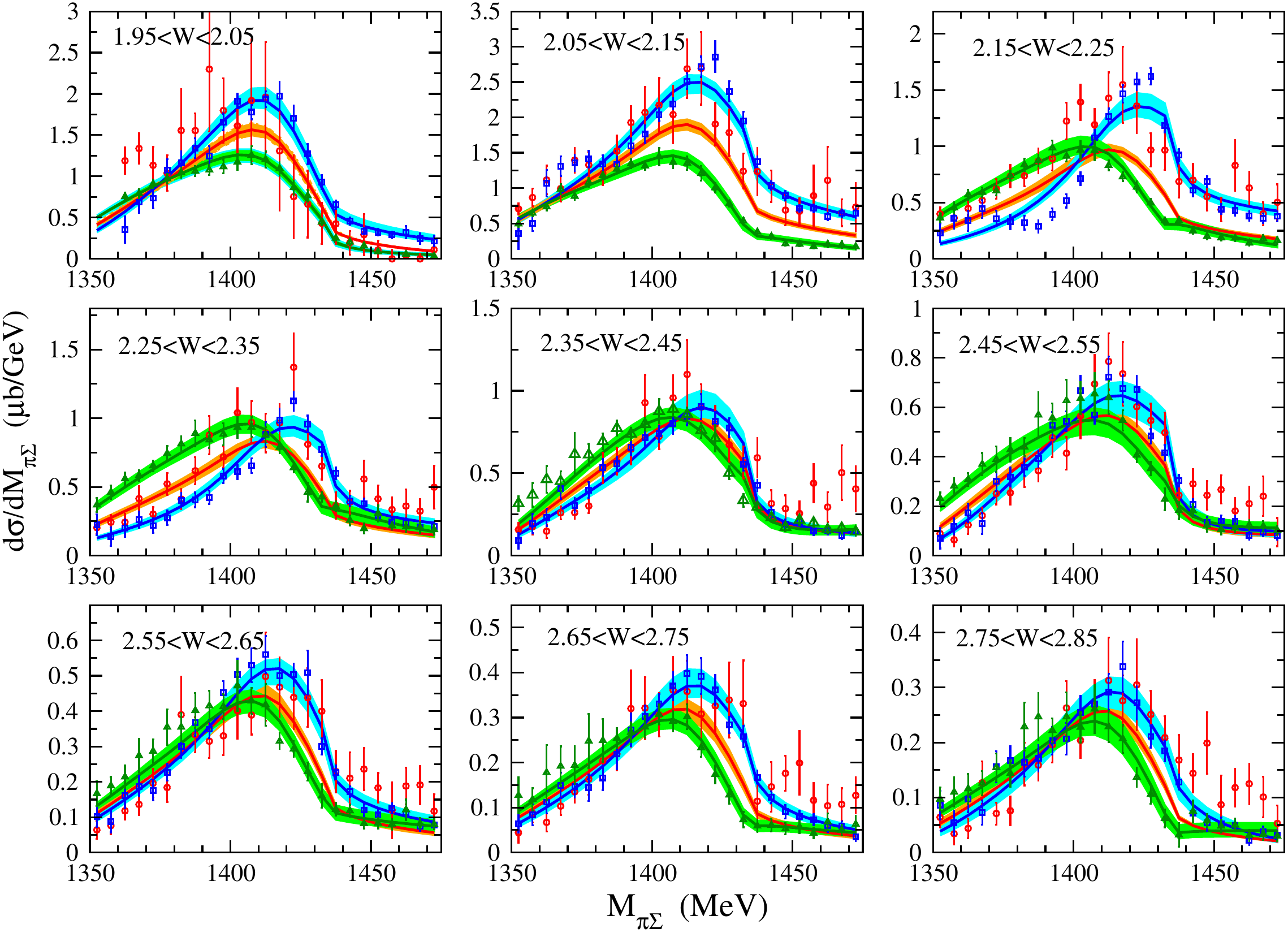}
\caption{\textit{(Color online)} Fit to photoproduction data of Ref.~\cite{Roca:2013cca}, where $\pi^0 \Sigma^0$ is in red, $\pi^- \Sigma^+$ in blue and $\pi^+ \Sigma^-$ in green. Experimental data taken from Ref.~\cite{Moriya:2013eb}.}
  \label{fig:1405oset}
\end{figure*}

The theoretical interest on $\Lambda(1405)$ photoproduction increased after the  CLAS measurements of the reaction $\gamma p \rightarrow K^+ \Sigma \pi$ had been performed at Jlab, as discussed in Sec.~\ref{sec:expLambda1405}. The predictions of Nacher et al. \cite{Nacher:1998mi,Nacher:1999ni} did not explain the experiment, thus the posterior theoretical effort was concentrated in reproducing the experimental data. The authors of \cite{Roca:2013cca} constructed a  model for the photoproduction amplitude, where the mechanism for the reaction $\gamma p \rightarrow K^+ \pi \Sigma$ is decomposed into two parts. On the one hand, the photoproduction part, $\gamma p \rightarrow K^+ M B$, where the meson-baryon ($MB$) with strangeness $S = -1$ is described by energy-dependent coupling constants. On the other hand, the final state interaction $M B \rightarrow \pi \Sigma$ is calculated solving the Bethe-Salpeter equation in coupled channels from the LO chiral unitary approach. Figure \ref{fig:1405oset} shows the measured $\Sigma\pi$ invariant mass distributions for different photon energies together with the calculations from the aforementioned model. A good agreement over the whole energy range for all decay channels is achieved. 
In this theoretical work, additionally to the two poles for the $\Lambda(1405)$ resonance, the possible existence of an isospin $I=1$ resonance in the vicinity of the $K^-N$ threshold was put forward. 

In Ref.~\cite{Nakamura:2013boa} the $\photon p \rightarrow K^+ \pi \Sigma$ reaction was also decomposed into photoproduction and final state interaction parts. While the final-state interaction term is described using the LO chiral Lagrangian, similarly to \cite{Roca:2013cca}, the photoproduction part is constructed in a gauge-invariant manner coupling the photon to the WT, the Born and the vector-meson exchange diagrams, allowing for tree-level contributions in contrast to  \cite{Roca:2013cca}. The experimental data on the line shapes of the three $\pi \Sigma$ states were reproduced and the $K^+$ angular distribution was calculated. 

The analysis of the $\Lambda(1405)$ was later carried out in \cite{Mai:2014xna}, using a simple semi-phenomenological model for the photoproduction process that combines the precise NLO description of the $K^- p$  scattering data and the kaonic hydrogen SIDDHARTA data  with a simple polynomial and energy-dependent ansatz for the process $\photon p \rightarrow K^+ M B$, as similary done in Ref.~\cite{Roca:2013cca}. The corresponding energy- and channel-dependent constants were fitted to the CLAS data, constraining the hadronic solutions that were allowed by the CLAS results.

The $\Lambda(1405)$ photoproduction reactions are most sensitive to the high-energy pole of the $\Lambda(1405)$ \cite{Roca:2013cca}. As determined from the theoretical models using the experimental data in Fig.~\ref{fig:L1405exp1}, kaon-induced reactions such as $K^-p \rightarrow \pi^0 \pi^0 \Sigma^0$ are largely dominated by a mechanism in which a $\pi^0$ is emitted prior to the $K^-p \rightarrow \pi^0 \Sigma^0$ amplitude. This amplitude is contributing to the narrower state at higher energy \cite{Magas:2005vu}. The pion induced reaction $\pi^- p \rightarrow K^0(\Sigma \pi)^0$, on the other hand, is complementary, as it  gives more weight to the low-energy pole, exhibiting a peak around 1390 MeV in the $\pi \Sigma$ invariant-mass
distributions \cite{Hyodo:2003jw}. Thus, the quite different shapes of the $\Lambda(1405)$ resonance seen in these experiments are evidences of the existence of the two-pole structure of the $\Lambda(1405)$.

 The kaon-induced reaction in deuteron $K^- d \rightarrow \pi \Sigma n$ has been also of prime interest, given the proposed experiment at J-PARC (E31) with a 1 GeV $K^-$ beam \cite{2012PTEP.2012bB013N}. Several theoretical models with comparable kinematics were performed \cite{Jido:2009jf,Miyagawa:2012xz,Jido:2012cy},  trying to extract the information of the subthreshold $K^-N$ interaction from the experimental spectrum. In a more recent paper \cite{Ohnishi:2015iaq},  a full three-body calculation of the $\bar K NN- \pi Y N$ amplitudes on the physical real energy axis has been performed and investigated how the signature of the $\Lambda(1405)$ appears in the cross sections of the reaction.  
 
 Proton-proton collisions have been also a matter of theoretical analysis in view of the  $pp \rightarrow p K^+ \Lambda(1405)$ reaction investigated at ANKE using a 3.65 GeV/c \cite{Zychor:2007gf}. In Ref.~\cite{Geng:2007vm}, a chiral unitary model including three different mechanisms was constructed to analyze the two-pole structure of the $\Lambda(1405)$. The mechanisms include single-kaon, single-pion, and single-rho exchanges. It was shown that the kaon exchange mechanism is mostly sensitive to the high-energy pole of the $\Lambda(1405)$, whereas its combination with the pion exchange mechanism has the effect of producing strength in the low invariant-mass part, resulting in a broadening of the invariant-mass distribution and a better agreement with experiment. The shape for the sum of the three contributions as well as the total cross section are consistent
with the ANKE data within experimental and theoretical uncertainties, if one reduces their contribution with the use of adequate form factors in the meson-baryon vertices. 

 It is also worth mentioning the role of the triangle singularity (TS) (a kinematical effect that leads to a peak structure) in the production of the $\Lambda(1405)$, that improves the previous calculation of \cite{Geng:2007vm}. In Ref.~\cite{Wang:2016dtb}, the effect of the TS on the angle and the energy dependence of $\Lambda(1405)$ photoproduction was studied. More recently, in pion-induced and proton-proton collisions  a clear peak around 2100 MeV in the $K \Lambda(1405)$ invariant mass has been found, due to the resonance peak of a $N^*$ state that plays a crucial role in the $K^* \Sigma$ production \cite{Bayar:2017svj}. This mechanism  produces the peak of the $\Lambda(1405)$ around or below 1400 MeV, as it was seen in the HADES experiment $p p \rightarrow p K^+ \pi \Sigma$ \cite{Agakishiev:2012xk}. The TS appears with a $\pi \Sigma$ in the intermediate state, which gives a dominant role to the low energy $\Lambda(1405)$ pole.

Finally, heavy decays such as $\Lambda_c \rightarrow \pi^+ M B$  can also give some light into the structure of the $\Lambda(1405)$. Calculating the final-state interaction using a chiral unitary approach, it was found that the $\pi \Sigma$ invariant mass distributions have the same peak structure in  the different $\pi \Sigma$ charge combinations that are related to the higher pole of the $\Lambda(1405)$ \cite{Miyahara:2015cja}. 

\subsection{$\bar K NN$}
\label{sec:KNN}
\subsubsection{Experimental determination of $\bar KNN$}
\label{sec:expKNN}

\begin{figure*}[hbt]
  \centering
  \includegraphics[width=0.48\linewidth,height=70mm]{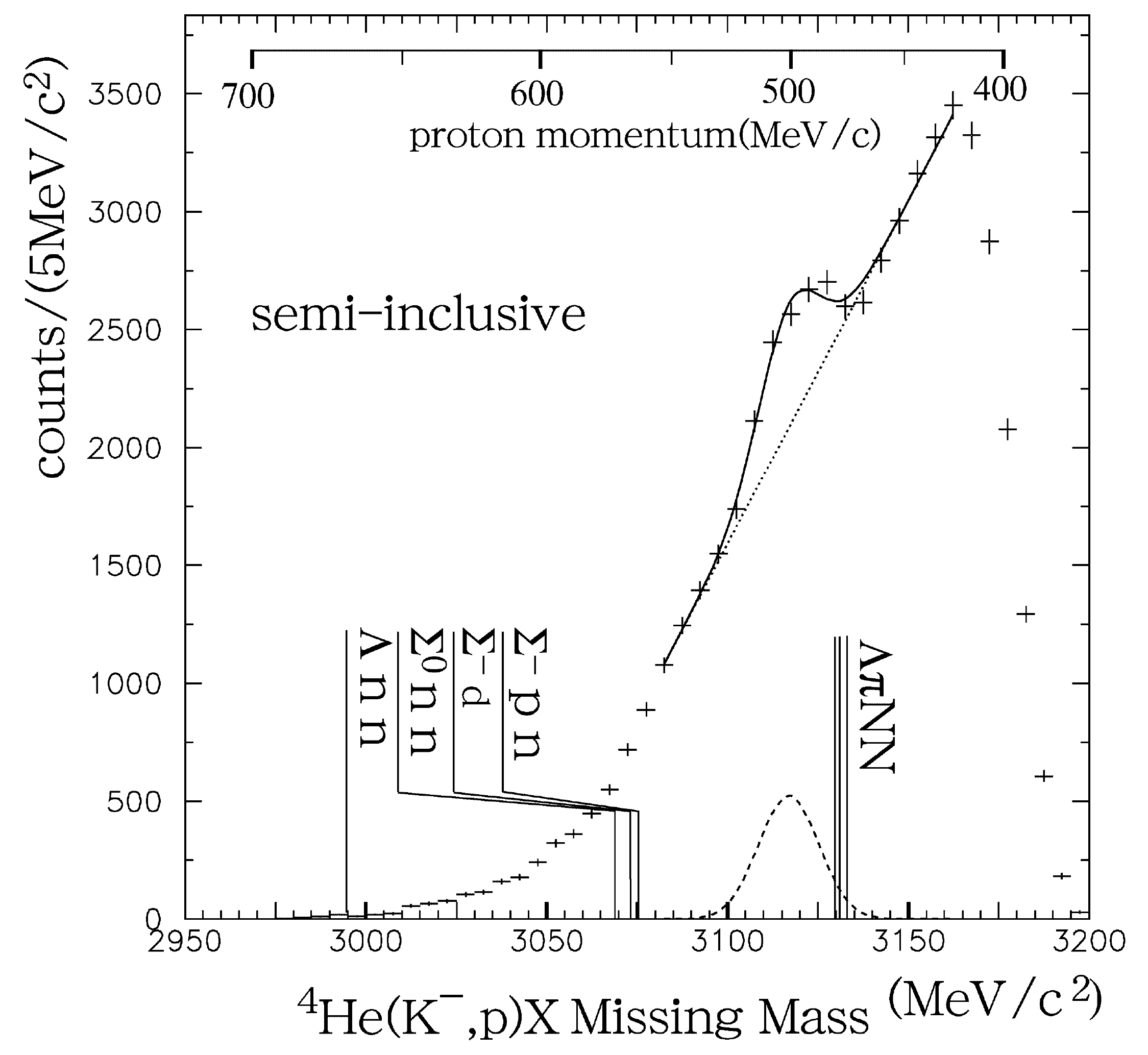}
  \hfill
  \includegraphics[width=0.49\linewidth,height=70mm]{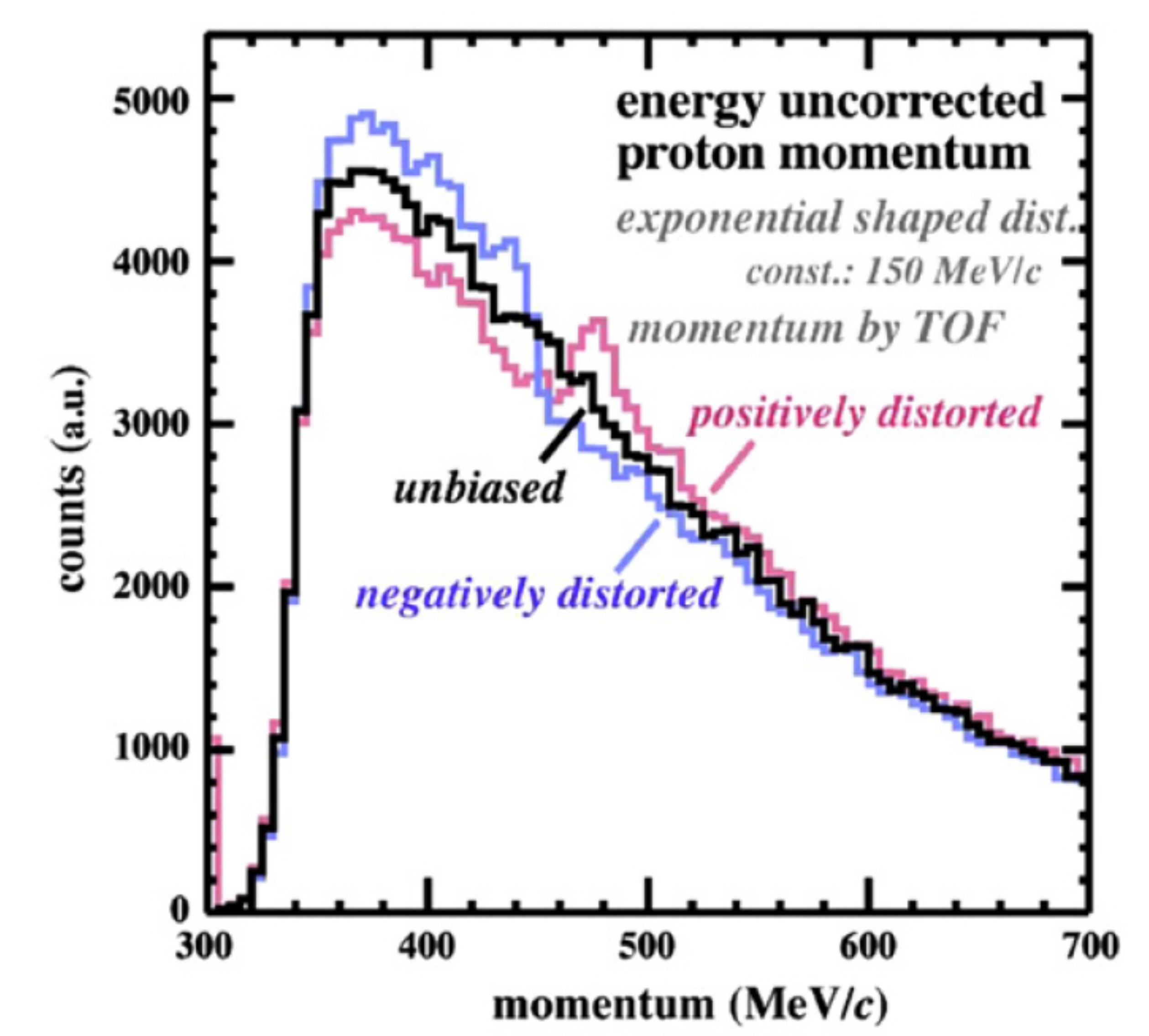}
\caption{\textit{(Color online)}~
  Proton missing mass spectrum (left panel) measured in the reaction $K^-(^4{\rm He},p)$ used for the claim of the S$^0(3115)$ discovery \cite{SUZUKI2004263}.
  Simulation (right panel) of the measured proton momentum assuming different distortion effects in the signal used for the velocity measurement of the spectator nucleon  \cite{Iwasaki:2008zza}.}
  \label{fig:S03115}
\end{figure*}
Following the interpretation of the $\Lambda(1405)$ as a molecular state, kaonic bound states were searched in several experiments.

Kaon-induced experiments performed at KEK were pioneers in the field 
exploiting the following chain of processes: i) the incoming $K^-$ beam at 600 MeV/c is stopped in an atomic state (typically $^4$He or $^3$He), and then decays first electromagnetically and eventually forms a deeply bound nuclear state, ii) a kaonic bound state is
formed in a nucleus with one nucleon less than the
target nucleus, iii) a monochromatic nucleon is emitted by the nuclear Auger process and serves as a spectator of the state formed. If 
kaonic bound states such as $K^-npp$ or $K^-nnp$ are formed, they can be tagged by the identification of a monochromatic neutron or proton.

\begin{figure*}[hbt]
  \centering
  \includegraphics[width=0.49\linewidth]{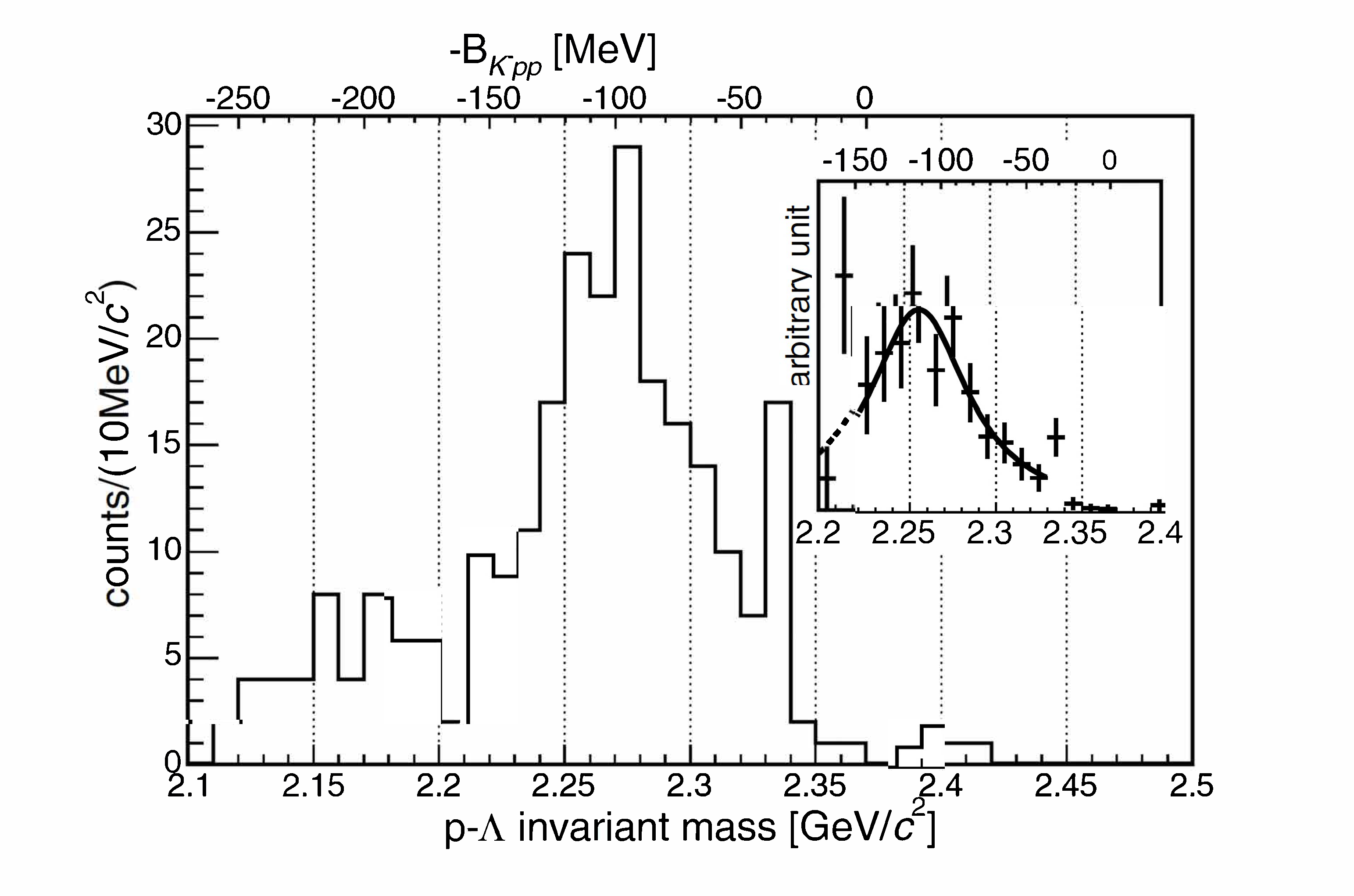}
  \hfill
  \includegraphics[width=0.48\linewidth]{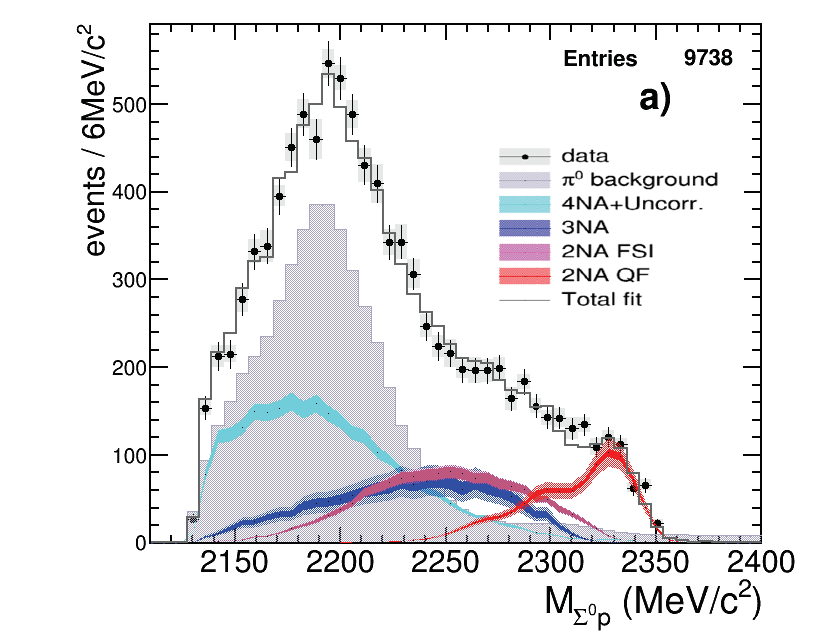}
\caption{\textit{(Color online)} The $\Lambda p$ invariant mass spectrum (left panel) measured by FINUDA in $K^-$ absorption on light nuclear targets associated to the kaonic bound state $K^-pp$ \cite{Agnello:2005qj}. The $\Sigma^0 p$ invariant mass measured by AMADEUS in $K^-$ absorption on carbon target compared to simulations of different absorption processes \cite{Doce:2015ust}.}

  \label{fig:ppK1}
\end{figure*}
\begin{figure*}[hbt]
  \centering
  \includegraphics[width=9cm,height=8cm]{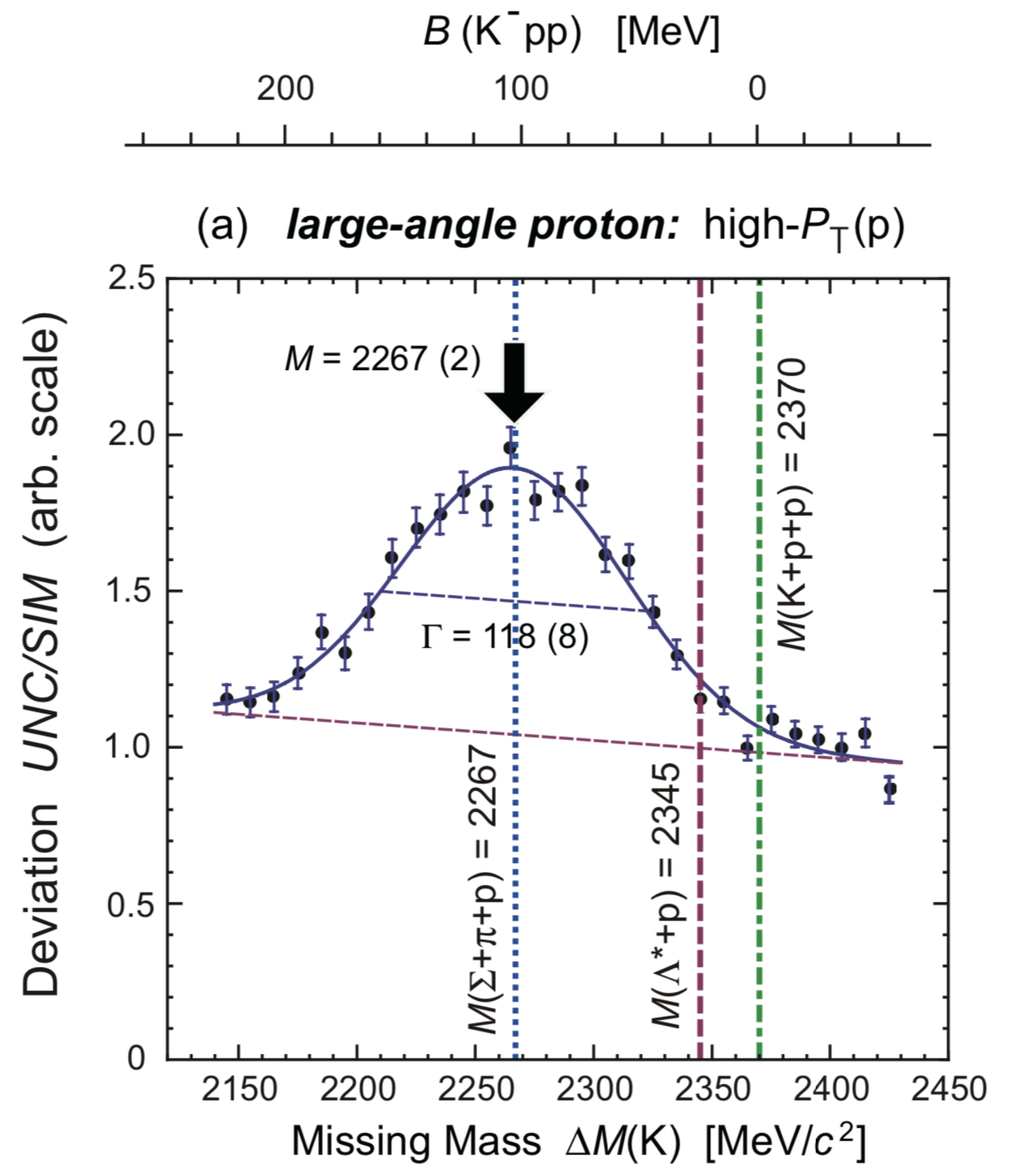}
  \hfill
  \includegraphics[width=9cm,height=8cm]{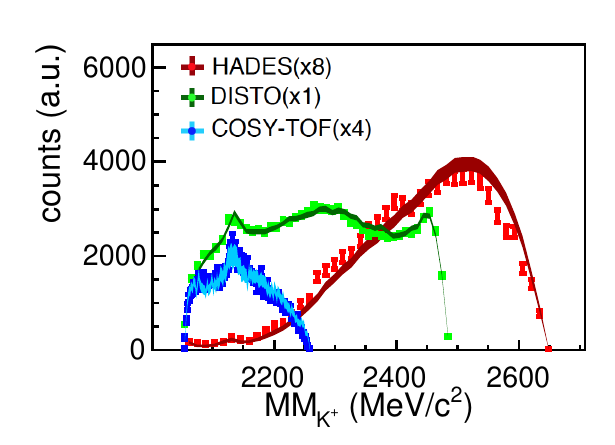}
\caption{\textit{(Color online)} Left panel: $K^+$ missing mass to the $pp$ initial state in $p+p \rightarrow p+K^+ + \Lambda$ reactions at 2.85 GeV, measured by DISTO, divided by the phase-space simulation of the same final state \cite{Yamazaki:2010mu}. Right panel: $K^+$ missing mass to the $pp$ initial state in $p+p \rightarrow p+K^+ + \Lambda$ reactions, measured by HADES, DISTO and COSY-TOF at kinetic energies of 3.5, 2.85 and 2.16 GeV, respectively, and partial-wave analysis results \cite{Munzer:2017hbl}. }
  \label{fig:ppK3}
\end{figure*}

The KEK-PS E471 experiment was optimized such to tag precisely stopped 
$K^-$ beams, to measure the proton or neutron velocity accurately and also to tag the production of such spectator nucleons with an additional pion emission.
Assuming a strong decay of the $K^-npp$ or $K^-nnp$ into a $\Sigma NN$ final state, a fast pion stemming from the weak decay of the $\Sigma$ could help in selecting the events of interest.

Two claims of the observation of kaonic bound states were put forward by the E471 collaboration: the $S^0(3115)$ \cite{SUZUKI2004263} and the $S^+(3140)$ \cite{Suzuki:2005bj,Iwasaki:2003kj}. Figure \ref{fig:S03115} shows on the left panel the missing momentum of the spectator proton in the reaction ${K^-} {}^4{He} \rightarrow p (\pi) +X$, where the peak at around 500 MeV/c was interpreted as the evidence of the $S^0(3115)$, a $K^- npp$ bound state with a width  of $<$ 20 MeV.

The right panel of Fig.~\ref{fig:S03115} shows the simulation of \cite{Iwasaki:2008zza}, that presents the effect of a wrong calibration of the time-of-flight detectors on the momentum spectrum of the spectator proton. Indeed, a new experiment (E549) was carried out with 10 times more statistics than the previous one, and no clear signature of the $S^0(3115)$ could be observed. A reanalysis of the data evidenced that the claimed kaonic bound state was due to a systematic distortion in the spectrum \cite{Iwasaki:2008zza}.

These first experiments came to the conclusion that 
inclusive measurements are not very well suited to investigate kaonic bound states and, later on, the attention of experimentalists  was focused on the search of the smallest of the kaonic bound states, the $K^-pp$,  by means of the strong decay into $\Lambda p$ pairs.

The first experimental claim about the existence of the $K^- pp$ bound state was reported by the FINUDA collaboration
\cite{Agnello:2005qj} in the stopped $K^-$ absorption reactions on $^6$Li, $^7$Li, and $^{12}$C targets. By focusing on $\Lambda$p pairs emitted back-to-back, the resulting invariant mass, prior to acceptance and efficiency corrections, is shown in the left panel of Fig.~\ref{fig:ppK1}. Since a peak is visible below the $K^- pp$ mass threshold, this was interpreted as the signature of the creation of
a $K^-pp$ bound state decaying into the $\Lambda p$ final state. 
A binding energy of $115 +6-5 (stat.) +3 -4 (syst.)$ MeV and a decay width of $\Gamma= 67 +14
-11 (stat.) +2
-3 (syst.)$ MeV were associated to this state.

Theoretical calculations \cite{Ramos:2008zza} proposed an alternative interpretation of the measured $\Lambda p$ peak as stemming from absorption of antikaons on two nucleons.
This effect was, hence, further investigated experimentally, first by the E549 collaboration \cite{Suzuki:2007pd} and then, by the AMADEUS collaboration \cite{DelGrande:2018sbv,Doce:2015ust}. The AMADEUS collaboration explained the $\Lambda p$ and $\Sigma^0 p$ kinematic distributions, resulting from low momentum $K^-$ absorbed in carbon targets, in terms of two and three-nucleons absorption including final state interactions. The right panel of Fig.~\ref{fig:ppK1} shows the $\Sigma^0 p$ measured invariant mass in $K^- +^{12}C$ reactions together with the results from full-scale simulations considering absorption on two-, three- and four- nucleons and possible final state interactions. One can see that the spectrum can be described without advocating the existence of a kaonic bound state.

\begin{figure*}[hbt]
  \centering
  \includegraphics[width=9cm,height=8cm]{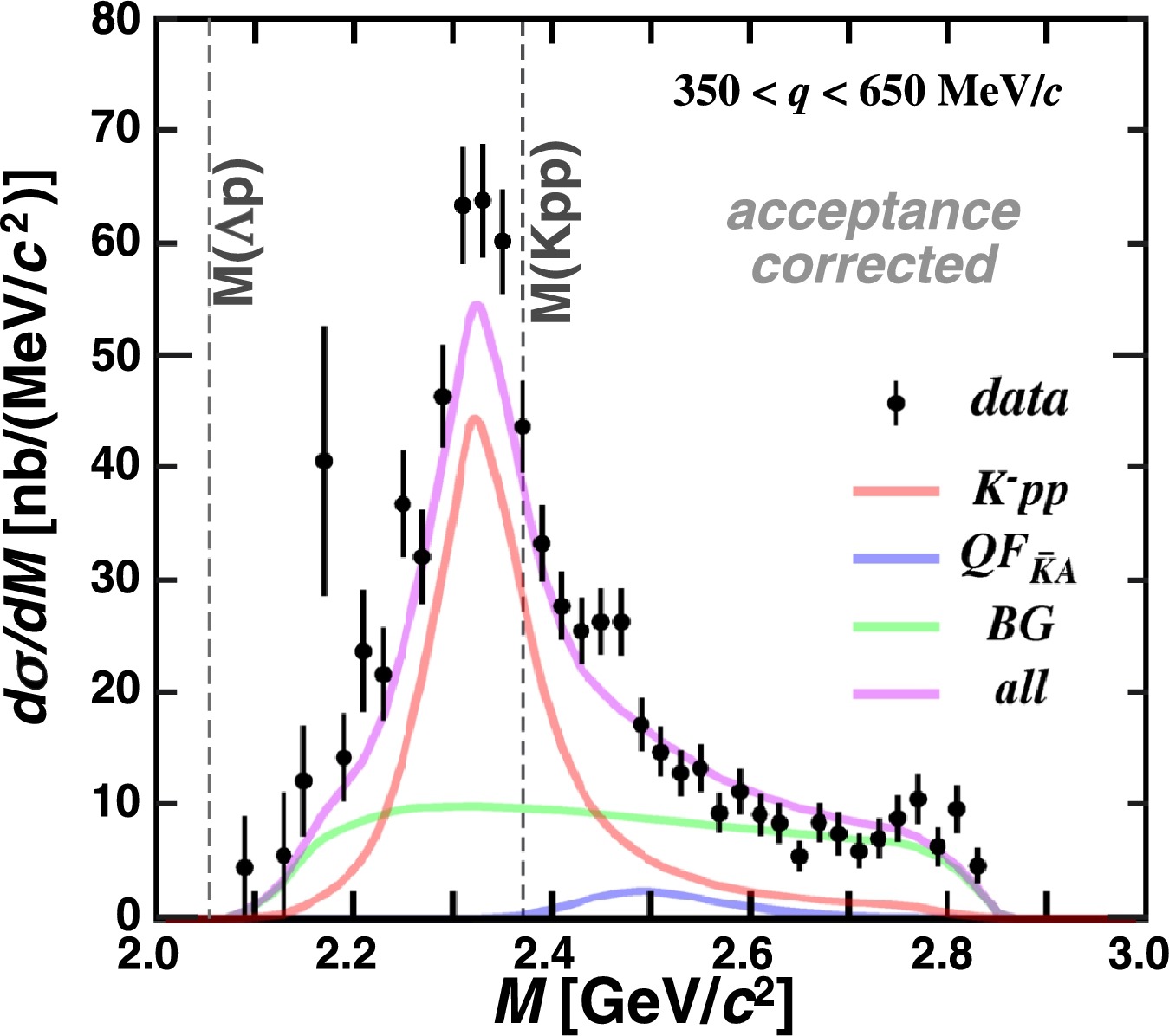}
\caption{\textit{(Color online)} $\Lambda p$ invariant mass spectrum measured by the E15 experiment in $^3{\rm He}(K^-,\Lambda p)n$ \cite{Ajimura:2018iyx}.}
  \label{fig:ppK4}
\end{figure*}

Another claim of the existence of the $K^-pp$
bound state was reported by the DISTO collaboration \cite{Yamazaki:2010mu} in the analysis of the reaction $p+p \rightarrow p+K^+ + \Lambda$ in a fixed target experiment at a beam energy of 2.85 GeV. 
 In this case, the claim of the observation of a $K^-pp$ bound state was based on a deviation spectra that resulted from dividing the measured $\Lambda p$ invariant mass and the $K^+$ missing mass by full-scale simulations. These simulations were obtained assuming pure phase-space production of the final state after selections on the event topology, meant to enhance the acceptance of monochromatic kaons in the final state.
 A binding energy of $103 \pm 3(stat.) \pm 5(syst.)$ MeV and a width of $118 \pm 8(stat.) \pm 10(syst.)$ MeV were associated to the resulting deviation spectrum. 
This interpretation was however strongly criticized \cite{Epple:2015fna}, since the deviation spectrum normalized to a phase-space simulation does not take into account the contributions of baryonic resonances decaying into the $K^+\Lambda$ final state. Indeed, detailed partial-wave analyses of the same reaction $p+p \rightarrow p+K^+ + \Lambda$ at kinetic energies between 2 and 4.5 GeV \cite{Agakishiev:2014dha,Munzer:2017hbl} demonstrated the contribution of $N^*$ resonances with masses between $1650$ and $1950$ MeV and the role played
by interferences among these resonances.

The left panel of Fig.~\ref{fig:ppK3} shows the DISTO deviation spectrum. The spectrum presents a wide peak-structure, that has been interpreted as a bound state. The energy scale shown on the top of the histogram indicates the binding energy. The right panel of Fig.~\ref{fig:ppK3} shows the experimental $K^+$ missing mass spectrum measured by three different experiments (HADES, DISTO and COSY-TOF) with different beam energies (3.5, 2.85 and 2.16 GeV), and 
within different geometrical acceptance. One can see that the three $K^+$ missing mass distributions can be perfectly reproduced by a partial-wave analysis including only the contributions of N$^*$ resonances and no kaonic bound states. The direct comparison of the HADES and DISTO analyses demonstrates the difficulty in drawing solid conclusions on the matter.

The $K^- pp$ bound state was also searched by means of the 
$\gamma +d\rightarrow K^+ \pi^- X$ reaction
with E$_{\gamma}$ = 1.5 - 2.4 GeV \cite{Tokiyasu:2013mwa}.  
Using the inclusive missing mass of the $K^+\pi^-$ to the 
colliding system, structures were searched and not found, 
whereas upper limits for the production of a $K^-pp$ bound 
state in the order of 1 $\mu$b were extracted.

A drastic improvement in the search of kaonic bound state has been achieved only recently by the E15 experiment, where a distinct peak in the $\Lambda p$ invariant mass spectrum, measured for the reaction $^3{\rm He}(K^-,\Lambda p)n$, has been found below the $m_K + 2m_p$ threshold, after a selection on the neutron momentum to enhance events with a large momentum-transfer ($q= 360-650$ MeV/c) \cite{Ajimura:2018iyx}. The analysis carried out by the E15 experiment shows that the peak below threshold appears in correspondence with a large momentum for the detected neutron. This implies that the formation of the $K^-pp$ bound state occurs associated to a large momentum transfer of the incoming kaon beam to the target. This is the first analysis that succeeds in measuring the whole event kinematics and, hence, the claimed observation is based on a more solid ground. The binding energy and width associated to the peak shown in Fig.~\ref{fig:ppK4} are found to be equal to $B_{Kpp}=\, 47 \pm 3(stat.)^{+3}_{-6}(syst.)$ and $\Gamma_{Kpp}=\, 115 \pm 7 (stat.)^{+10}_{-20}(sys.)$ MeV. 
In the next Sec.~\ref{sec:KNN-theory}, the theoretical interpretation of this measurement is discussed.

\subsubsection{Theoretical predictions for $\bar KNN$}
\label{sec:KNN-theory}

\begin{table}[]
    \centering
    \begin{tabular}{c|c|c|c|c}
        {\rm Work} & {\rm B [MeV]} & {$\Gamma$ [MeV]}  & {\rm Method} & {\rm Type of potential}\\
         \hline
         \hline
         {\rm Barnea et al. \cite{Barnea:2012qa}} & 16 & 41 & {\rm Variational} & {\rm chiral} \\
         \hline
          {\rm Dote et al. \cite{Dote:2008hw}} & 17-23 & 40-70 & {\rm Variational } & {\rm chiral} \\
         \hline
          {\rm Dote et al. \cite{Dote:2017wkk}} & 14-50 & 16-38 & {\rm ccCSM } & {\rm chiral} \\
         \hline
          {\rm Ikeda et al. \cite{Ikeda:2010tk}} & 9-16 & 34-46 & {\rm Faddeev } & {\rm chiral} \\
         \hline
          {\rm Bayar et al. \cite{Bayar:2012hn}} & 15-30 & 75-80 & {\rm Faddeev } & {\rm chiral} \\
           \hline
          {\rm Sekihara et al. \cite{Sekihara:2016vyd}} & 15-20 & 70-80 & {\rm Faddeev } & {\rm chiral} \\
          \hline
          {\rm Yamazaki et al. \cite{Yamazaki:2002uh}} & 48 & 61 & {\rm Variational } & {\rm phenomenological} \\
         \hline
          {\rm Shevchenko et al. \cite{Shevchenko:2006xy}} & 50-70 & 90-110 & {\rm Faddeev} & {\rm phenomenological} \\
         \hline 
         {\rm Ikeda et al. \cite{Ikeda:2007nz}} & 60-95 & 45-80 & {\rm Faddeev } & {\rm phenomenological} \\
         \hline
          {\rm Wycech et al. \cite{Wycech:2008wf}} & 40-80 & 40-85 & {\rm Variational } & {\rm phenomenological} \\
          \hline
          {\rm Dote et al. \cite{Dote:2017veg}} & 51 & 32 & {\rm ccCSM  } & {\rm phenomenological} \\
          \hline
           {\rm Revai et al. \cite{Revai:2014twa}} & 32/ 47-54 & 50-65 & {\rm Faddeev} & {\rm chiral/phenomenological} \\
        \end{tabular}
    \caption{ Binding energy and width of $K^-pp$ for different chiral and phenomenological calculations using variational, Faddeev or ccCSM+Feshbach methods. }
    \label{tab:KNN}
\end{table}

The dynamical generation of the two-pole structure of the $\Lambda(1405)$ indicates that the $\bar K N$ interaction might be attractive enough to produce bound states. As discussed in Sec.~\ref{sec:expKNN}, $\bar K$-nuclear clusters may form, such as the $\bar K NN$ in isospin $I=1/2$ and $J^{\pi}=0^-$. From the early predictions of Ref.~\cite{Akaishi:2002bg}, the $I=1/2$ $\bar K NN$ state has been extensively studied, both theoretically and experimentally (see Ref.~\cite{Nagae:2016cbm} and references therein). 

From the theoretical point of view, several works have addressed the existence of this state together with its mass and width. In Table \ref{tab:KNN} we show a summary of the most recent results on $K^- pp$ binding energies $(B)$ and widths $(\Gamma)$ resulting from these models. In this table, we differentiate between variational, three-body Faddeev calculations or the more recent coupled-channel Complex Scaling Method (ccCSM). For each of them, we distinguish between those where the two-body interactions are based on $\chi EFT$ theory or a phenomenological model, usually energy independent.  We observe that there is a disparity of values for the binding energy and the width of the $K^-pp$ bound state, with binding energies ranging between 9 and 95 MeV, while the decay widths cover values between 16 and 110 MeV. 

The variety of values for the binding energy and width is due to different sources. One of the reasons is the uncertainties in the subthreshold extrapolation of the $\bar KN$ interaction. While in the vicinity of the $\bar K N$ threshold the phenomenological models and chiral effective approaches roughly agree, as they fit the available $\bar KN$ data close to threshold, the effective interaction deduced from chiral SU(3) dynamics turns out to be less attractive than the phenomenological potentials in the subthreshold region. As a result, the chiral energy dependent interactions give binding energies considerably lower than those of static phenomenological interactions. The trend is that the $\bar K N$ $I=0$ interaction gives rise to a quasi-bound state at higher energies in the chiral approaches as compared to the phenomenological ones. The use of variational or Faddeev calculations leads to certain approximations that explain additional differences between the approaches. Variational calculations use effective $\bar K N$ interactions,  where the effect of the $\pi \Sigma$ channels are introduced in the non-locality and energy dependence of the off-shell $\bar K N$ amplitude. In this way, they do not account for the specific features of the full three-body $\bar K NN - \pi YN$ coupled-channel problem, which is handled correctly in the Faddeev calculations. However, although dealing with the full $\bar KNN$ dynamics, the Faddeev approaches have to deal with separable two-body interactions. 
More recently, the ccCSM allows for the treatment of the full coupled channels and resonant states to study $K^-pp$, thus involving the merits of the variational and Faddeev approaches, but with a high computational cost. However, 
the two-nucleon absorption $\bar K N N - YN$ is not considered yet, in spite of claims of having a sizable contribution to the $K^-pp$ width \cite{Bayar:2012hn}.


Certain experimental claims of the formation of $K^-pp$ have been re-interpreted in terms of conventional mechanisms, as discussed in Sec.~\ref{sec:expKNN} for the FINUDA, AMADEUS and COSY-TOF experiments, and following the discussion in Refs.~\cite{Gal:2013vx,Ramos:2008zza,Epple:2015fna,Munzer:2017hbl,Doce:2015ust}. 
On the other hand, the J-PARC E15 experiment for the in-flight $^3{\rm He}(K^-, \Lambda p)n$ reaction is free from final state interactions, leading to a much cleaner signal of the possible $K^-pp$ bound system. The authors of Ref.~\cite{Sekihara:2016vyd}
have evaluated the $\Lambda p$ invariant mass spectrum considering two possible scenarios, in order to interpret the observed experimental peak. The first one assumes that the $\Lambda(1405)$ is generated after the emission of an energetic neutron from the absorption of the initial $K^-$, but not giving rise to a bound state with the remaining proton. This system decays in the final $\Lambda p$ afterwards. The second scenario implies the formation of a $\bar K NN$ bound state after the emission of the energetic neutron, finally decaying into $\Lambda p$. The authors obtain a two-peak structure of the mass spectrum near the $K^-NN$ threshold and interpret the peak below the threshold as the signal of the $\bar KNN$ bound state with binding energy of 15-20 MeV and width of 70-80 MeV (see Table \ref{tab:KNN}), while the peak above threshold originates from the quasi-elastic scattering of the kaon in the first collision emitting a fast nucleon. 


\subsection{Kaons and Antikaons in Matter}
\label{sec:kaons-matter}

\subsubsection{Theoretical approaches for kaons in matter}
\label{subtheorykaonsmatter}

 As we mentioned in the discussion of the two-body $KN$ interaction in Sec.~\ref{sec:KNexperiment}, the latter is considered to be smooth since no baryonic resonances with positive strangeness would exist. Hence, the $K$ single-particle potential, $U_{K}$, that results from the interaction with the surrounding nuclear medium can be described at low densities by
\begin{equation}
U_{K} \sim T_{KN-KN} \ \rho ,
\end{equation}
using the so-called low-density theorem or, in other words, within the $T \rho$ approximation. The low-density theorem is based on the fact that the two-body $KN$ interaction in dense nuclear matter can be approximated by the free $KN$ two-body scattering (see Eq.~\ref{bethe-free}), as it is not expected that the propagation of kaons in a nuclear medium will be much different than the one in free space. 

Early results were based on a Nambu and Jona-Lasinio (NJL) model within the low-density approximation \cite{Lutz:1994cf}. This is an effective chiral model for nucleons and mesons that realizes the spontaneous chiral symmetry breaking mechanism, and it is constructed in a similar manner as Cooper pairs of electrons in the BCS theory of superconductivity. Within this model, the authors obtained a maximum change of 10$\%$ in the $K^+$ mass for nuclear matter saturation density ($\rho_0 \sim 0.16 {\rm fm}^{-3}$). In Ref.~\cite{Schaffner:1996kv} a similar mass change was calculated within the relativistic-mean-field approach (RMF), that is, within a covariant effective model where baryons interact through the exchange of mesons, and the mesons are replaced by their respective mean-field expectation values. This approach used, as starting effective interaction, either a one-boson model (OBM) or chiral perturbation theory ($\chi$PT) for the $KN$ interaction. This result is indeed expected as the behavior in matter follows from the low-density theorem, with the coupling constants fixed to the $KN$ scattering lengths. Differences, though, arise for densities around and above 2$\rho_0$, when further medium corrections to the effective $K^+$ mass should be considered, such as the Sigma  term contribution, $\Sigma_{KN}$ \cite{Schaffner:1996kv}. This term is related to the explicit chiral symmetry breaking. Also, a calculation based on the quark-meson coupling (QMC) was performed in \cite{Tsushima:1997df}, where the interaction between kaons and nucleons is determined by the exchange of mesons between the quarks in a bag that conformed the kaons and nucleons. Within this framework, a repulsive potential for kaons with a value of less than 20 MeV was found, if a rescaling of the coupling of the $\omega$ meson to the non-strange quark in the kaon is not performed.

The  $KN$ effective interaction has also been studied within unitarized coupled-channel approaches, taking the SU(3) chiral effective Lagrangian as kernel of the interaction \cite{Kaiser:1995eg,Oset:1997it},  which is the Lagrangian of Eqs.(\ref{LagrphiB1},\ref{LagrphiB2}), introduced in Sec.~\ref{sec:theoryKN}. The inclusion of medium corrections on the $KN$ amplitude beyond the $T \rho$ approximation, such as Pauli blocking, Fermi motion in the $K$ dispersion relation or a self-consistent treatment, has given rise to small changes of about 10$\%$ or less in the $K$ mass \cite{Waas:1996fy,Korpa:2004ae,Tolos:2005jg}.

However, the theoretical models of the kaon optical potential based on the low-density approximation fail systematically in reproducing $K^+$-nucleus total and reaction cross sections, underestimating the data by about 10-15\% \cite{Weiss:1994kt}. Although several
mechanisms have been explored (such as swelling of the nucleon, meson-exchange currents, a smaller mass for the exchanged vector mesons than the nominal ones), there is no satisfactory solution to this problem (see  Ref.~\cite{Friedman:2007zza} and references therein). The approach of Ref.~\cite{GarciaRecio:1994cn} considered the interaction of the $K^+$ with the pion cloud of the nucleus, leading to a significant improvement of the differential and total $K^+$ nuclear cross sections.  Alternatively, substantially improved fits to the data were achieved by incorporating the absorption of $K^+$ by nucleon pairs via the inclusion of the  $\Theta^+(1540)$ pentaquark \cite{PhysRevLett.94.072301,PhysRevC.73.015208,Tolos:2005jg}, although the evidence of the  pentaquark has faded away in the last years. 

An alternative way to determine the kaon optical potential is to analyze the formation and propagation of kaons in the dense medium created in pion-nucleus, proton-nucleus or nucleus-nucleus collisions for intermediate beam kinetic energies (GeV). As we will discuss in the next Sec.~\ref{kaonInMatterExp}, several experiments have been performed in the recent years and there is a strong need for understanding the properties of $K$ (and $K^*$) in dense matter.  A detailed analysis of the theoretical $KN$ (and $K^*N$) cross sections and scattering lengths has been performed recently, based on a unitarized coupled-channel model involving both pseudoscalar and vector mesons with baryons in the strangeness S$=+$1 sector \cite{Khemchandani:2014ria}, with the motivation of testing these improved cross sections in proton- and nucleus-nucleus collisions.

The analysis of heavy-ion experiments (HICs) for the extraction of the kaon optical potential requires, however, the use of transport models.  Transport models provide the link between the experiments and the underlying physical processes, as they take into account the dynamics of the production and propagation of all kind of particles and, in particular, of strange mesons (see Ref.~\cite{Hartnack:2011cn} for a recent review on strangeness production). Those models are based on the solution of Boltzmann-type semi-classical transport equations, that can be derived from non-equilibrium quantum field theory. 

Such hadronic transport models propagate one-body phase space distributions in self-consistent potentials, and account for elastic and inelastic two-body scattering processes. The various hadron species are coupled via production and absorption processes, and by their mean fields. Transport models are generally mean-field approaches which rely on the quasiparticle approximation. Those include the Boltzmann-Uehling-Uhlenbeck (BUU) approach \cite{Cassing:1990dr} or the quantum-molecular-dynamics (QMD) approach \cite{Aichelin:1991xy}. Among the latter, we note the recent  hadronic transport approach of SMASH \cite{Weil:2016zrk}. Whereas kaons can be treated  as quasiparticles also in a dense environment due to its weak interaction with the surrounding medium, antikaons require to account for off-shell effects beyond the quasiparticle approximation. Therefore, a transport model for off-shell propagation of particles beyond the quasiparticle approach has been developed, the Hadron-String-Dynamics model (HSD) \cite{Cassing:1999es}.

\begin{table}
\begin{center}
\begin{tabular}{|c|c|c|}
\hline
 $ K^+$  & $K^-$  & type\\ 
\hline\hline
$NN \longrightarrow NYK^+$ & $NN \longrightarrow NNK^+K^-$ &
strangeness production\\
 &   & primary/secondary \\
\hline
$\pi N \longrightarrow YK^+$ & $\pi N \longrightarrow NK^+K^-$ &
strangeness production\\
 &   & secondary \\
\hline
 & $N Y \longrightarrow NN K^-$  & strangeness exchange\\
\hline
 & $\pi Y \longleftrightarrow N K^-$  & strangeness exchange\\
\hline
$YK^+ \longrightarrow \pi N$  &   & strangeness absorption\\
\hline
$NK^+ \longleftrightarrow N K^{+/0}$ & 
$NK^- \longleftrightarrow N K^- (\overline{K}^{0})$  & elastic/charge exchange\\
\hline
\end{tabular}
\end{center}
\caption{
Elementary hadronic reactions which are relevant for kaon and antikaon dynamics 
at intermediate energies. $N$ stands for nucleons or nucleon 
resonances, and $Y$ for hyperons ($\Lambda ,
\Sigma$). Table adapted from \cite{Fuchs:2005zg}. 
}
\label{tab_cross1}
\end{table}

Coming back to kaons, those are produced at supra-normal densities and during the early stages of HiCs  (e.g.
for incident energy of 1 AGeV they reach chemical freeze-out after 15 fm/c). However, at such timescales, the surrounding nuclear environment is still in a non-equilibrium state. The various channels for kaon production are shown in the first column of Table~\ref{tab_cross1}, adapted
from Ref.~\cite{Fuchs:2005zg}. For kaons, those include primary and secondary strangeness production reactions, strangeness absorption mechanisms and elastic/charge exchange reactions. Only strangeness exchange mechanisms can happen for antikaons. Strangeness production can be classified into nucleon induced ($NN \longrightarrow NYK^+$, $NN \longrightarrow NNK^+K^-$) and pion induced ($\pi N \longrightarrow YK^+$,  $\pi N \longrightarrow NK^+K^-$) 
reactions, where $N$ indicates a nucleon or a nucleon resonance, and $Y$ stands for hyperons ($\Lambda ,
\Sigma$). Primary reactions are those reactions with two nucleons in the entrance channel, whereas secondary reactions have at least one nucleon resonance in the entrance channel. Meson induced reactions are of secondary type. Note that primary reactions are dominant for kaon (and antikaon) production at higher energies, and thus contributing dominantly to the high momentum part of the spectra, whereas at subthreshold energies the production is dominated by the secondary ones.

In order to derive the kinetic equation describing the production and propagation of kaons, one should start from the non-equilibrium real time Green's function formalism. After truncation of the Green's function hierarchy at the two-body level, one gets the Kadanoff-Baym equations for the one-body Green's function. After a gradient expansion of Wigner transforms, one arrives finally at the semi-classical kinetic equation (see Ref.~\cite{Fuchs:2005zg} for a brief description of the relevant steps and details). In the Hartree approximation, that is neglecting the momentum dependence on the self-energy \footnote{The $K$ self-energy is the potential felt by the $K$ meson due to the interaction with the surrounding medium. In the non-relativistic approximation, it is referred to the $K$ single-particle potential.}, it reads \cite{Fuchs:2005zg}
\begin{eqnarray}
&&  \left[k^{*\mu} \partial_{\mu}^x  + \left( k^{*}_{\nu} F^{\mu\nu}     
+ m^* \partial^{\mu}_x m^* \right) 
\partial^{k^*}_{\mu} \right] (af) (x,k^* ) 
\nonumber\\
&=&  \frac{1}{2} \int \frac{d^4 k_{2}}{(2\pi)^4} \frac{d^4 k_{3}}{(2\pi)^4}
             \frac{d^4 k_{4}}{(2\pi)^4} 
             a(x,k)  a(x,k_2)  a(x,k_3)  a(x,k_4) W(k k_2|k_3 k_4) 
\nonumber \\    
&\times& (2\pi)^4 \delta^4 \left(k + k_{2} -k_{3} - k_{4} \right)    
\left[ f(x,k_3) f(x,k_4) \ls 1\mp  f(x,k) \rs \ls 1\mp f(x,k_2) \rs \right.  
\nonumber \\  
&-&   \left. f(x,k) f(x,k_2) 
\ls 1 \mp  f(x,k_3) \rs \ls 1\mp f(x,k_4) \rs \right]    
\quad ,
\label{TP5}
\end{eqnarray}
with $a$ the scalar spectral function, and $f$ the scalar
distribution function of kaons in our case. The left-hand side of Eq.~(\ref{TP5}) is given by a drift term (the Vlasov term) driven by the mean field via the kinetic momenta $k^{*}$, the field strength tensor 
$F^{\mu\nu} (x) = \p^{\nu}_x  Re\Sigma^{+\mu}_H (x) 
                -\p^{\mu}_x  Re\Sigma^{+\nu}_H (x) $, 
with $\Sigma^+$ the retarded self-energy of kaons in the Hartree approximation, and the effective mass $m^*$. The right-hand side of Eq.~(\ref{TP5}) is a collision integral, that contains the transition rate $W$ or the in-medium cross section given by 
$(k^* + k_{2}^*)^2 d\sigma /d\Omega(k,k_2) = W(kk_2|k_3 k_4)$. The collision term includes Pauli-blocking or Bose enhancement factors $(1\mp f)$ for the final states.  Note that dealing with different interacting particle species  leads to  a coupled-channel transport model, where
one has to solve a separate transport equation for each degree of freedom. 

\begin{figure}[htb]
\begin{center}
\includegraphics[width=0.6\textwidth]{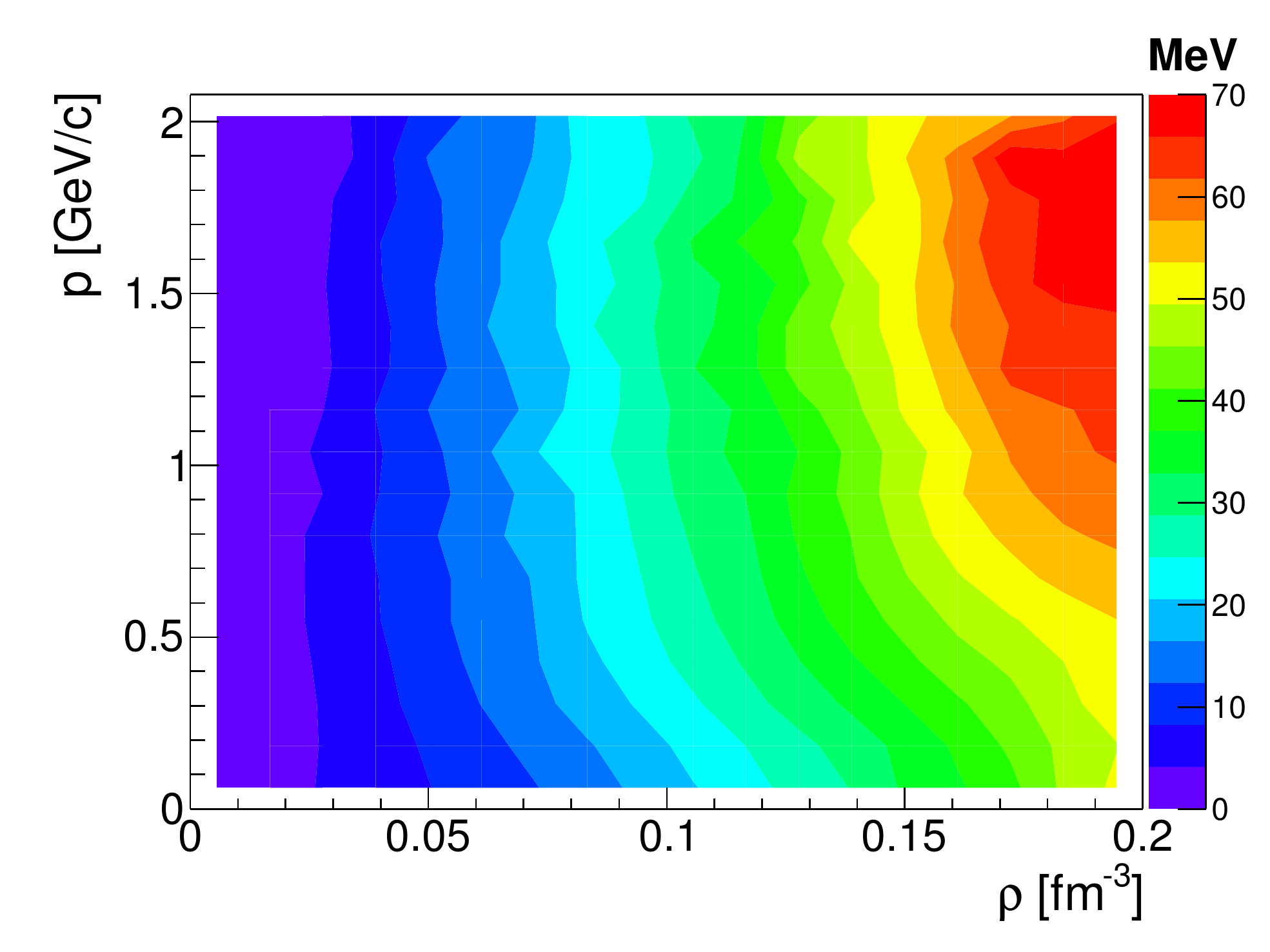}
\caption{\label{fig:U_ChPT_rho_p} \textit{(Color online)} In-medium $\chi$PT kaon potential $U = E^{*} - E$ (in MeV) as a function of the baryonic density and the kaon momentum, where $E$ is the free spectrum. Figure taken from \cite{Agakishiev:2014moo}.}
\label{fig:ChiralKPot}
\end{center}
\end{figure}

The theoretical predictions for the collective flow pattern and spectra as well as abundances of $K^+$ thus result from the solution of the kinetic equation for the $K^+$ in proton-proton and proton-nucleus reactions, and HICs, as shown in  Sec.~\ref{kaonInMatterExp}. The crucial ingredient is the $K^+$ self-energy or the non-relativistic kaon potential. In Fig.~\ref{fig:U_ChPT_rho_p} we show the repulsive kaon potential coming from $\chi$PT, used in BUU simulations of the Giessen group (GiBUU) for proton-induced $K^0$ production in $p+p$ and $p+{\rm Nb}$ collisions, measured with the HADES detector at a beam kinetic energy of 3.5 GeV \cite{Agakishiev:2014moo}. The $\chi$PT  optical potential shows a non-linear behaviour with density and an explicit momentum dependence, especially at high energies, resulting from the combined effect of the WT interaction and the $\Sigma_{KN}$ term contribution. For the kaon at rest and at normal nuclear density, the $\chi$PT potential amounts to $\sim 35 \ {\rm MeV}$.

\subsubsection{Experimental measurements of kaons in matter}
\label{kaonInMatterExp}
Precise measurements of the total, inelastic and quasi-elastic cross sections
for charged kaons impinging on different nuclear targets were carried out 
in the nineties at BEVALAC
\cite{Weiss:1994kt,PhysRevC.51.669,MICHAEL199629}.
There, secondary $K^+$ beams at momenta varying between 400-700 MeV/c were
impinged on different targets ($D,C,Li,Ca$), and the measured cross sections were compared
to state-of-the-art calculations based on the low-density approximation.
The ratio of the total reaction cross sections off different targets 
with respect to the deuterium target as a function of the kaon momentum was found much larger than the
calculation. This effect was interpreted as due to the enhanced medium modification of kaon properties
within nuclei \cite{Weiss:1994kt}.
 More exclusive studies allowed to isolate 
the elastic cross section, where the ratio of the cross section for different targets
as a function of the emission angle also shows larger values than predicted by the theory
\cite{MICHAEL199629}. This further disagreement was interpreted as caused by
the limitation of the optical models in the data interpretation.

\begin{figure*}[hbt]
  \centering
 \includegraphics[width=9cm,height=9cm]{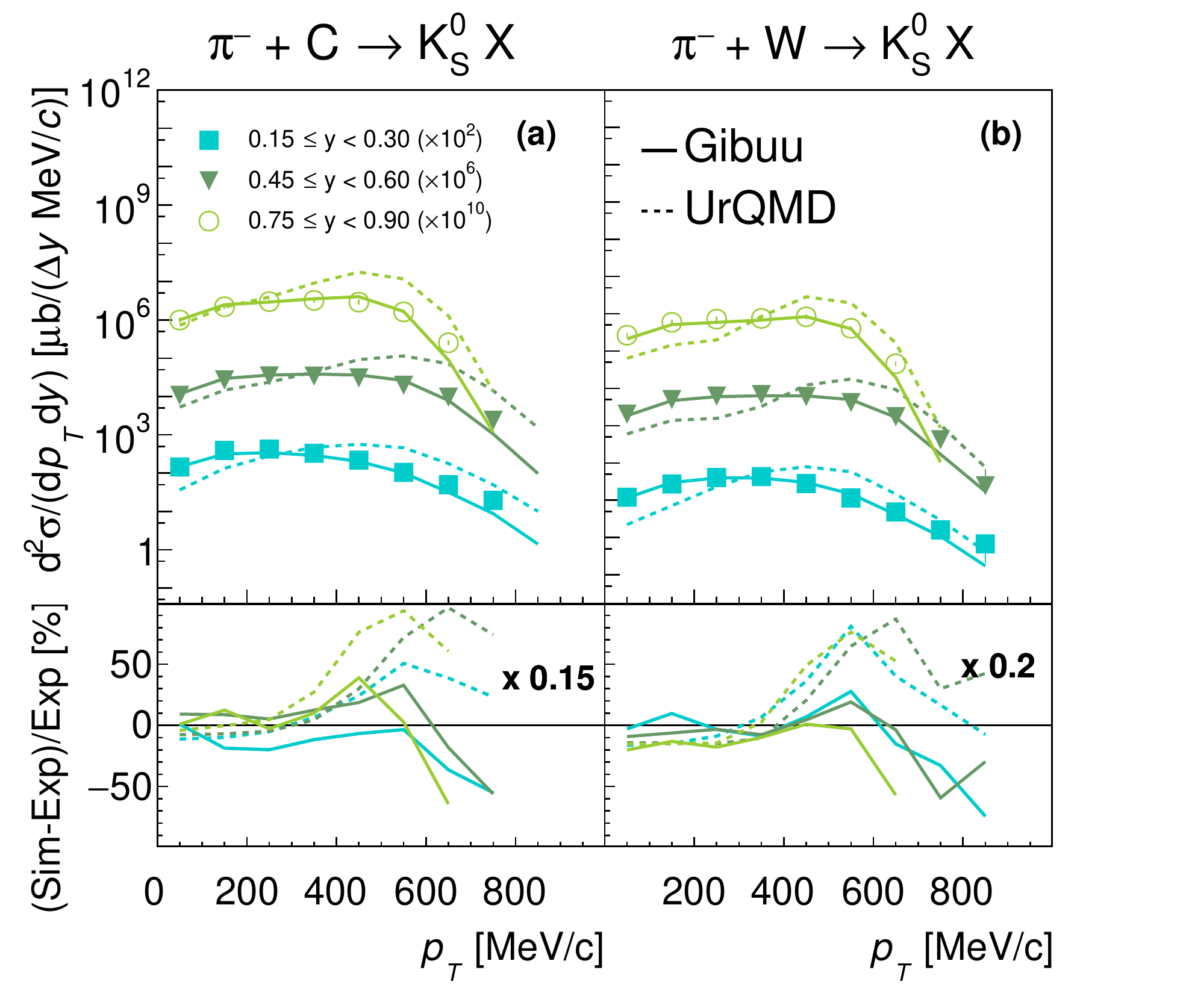}
  \hfill
  \includegraphics[width=9cm,height=9cm]{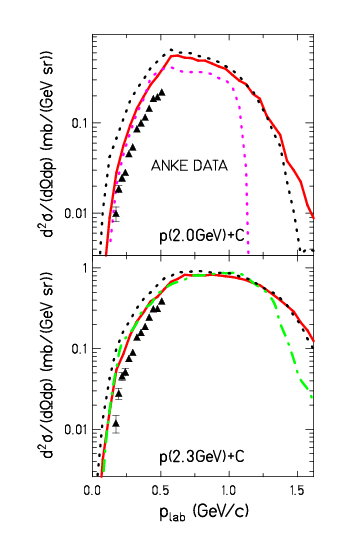}
\caption{\textit{(Color online)} Left upper panels: Differential cross sections of $K_S^0$ (points) as a function of transverse momentum for $\pi^-C$ (a) and $\pi^-W$ (b) reactions at 1.7 GeV for three rapidity bins and comparisons to GiBUU (solid curve) and UrQMD (dotted curve).
  Lower left panels: Deviation of the two transport models from to the cross section of $K_S^0$ as a function of rapidity. All deviations for the UrQMD model are scaled by the factor indicated in black. Right panel: Calculated $K^+$ spectra measured in $p+C$ reactions at energies above the threshold as compared to the experimental results of the ANKE Collaboration \cite{Buescher:2001qu}. Full red line: Standard IQMD version. Dotted black line: Standard without $K^+$ nucleus potential. Green dashed-dotted: Standard without
Fermi motion. Dotted magenta: Superposition of $pp$ cross section. Taken from \cite{Hartnack:2011cn}.}
  \label{fig:KaonMat0}
\end{figure*}

When one considers kaons produced in proton-, photon-, pion-induced or in HICs several mechanisms contribute to the final spectra  and the in-medium properties of kaons become more challenging to isolate.

Pion-nucleon and nucleon-nucleon collisions have been used to extract
the elementary production cross sections for fixed target experiments with beam energies between 1 and 3 GeV \cite{Hartnack:2011cn}. But, already when considering $\pi^-$+A reactions, transport models can not 
 perfectly reproduce the experimental data. 
 
 The left upper panels of Fig.~\ref{fig:KaonMat0} show the differential cross sections for $K^0_S$ produced in $\pi^-+C$ and $\pi^-+W$ reactions with a beam kinetic energy of 1.7 GeV measured by the HADES experiment, compared to several transport models. The lower left panels 
 display the relative difference between the data and the transport distributions, and sizable differences can be recognized. The transport predictions have been calculated without any in-medium potential. A similar trend is observed for the measured $K^+$ spectra in the same colliding system.
 The differences can be due to the incomplete implementation of
 intermediate resonances for the kaon production, that do play
 a role in reactions with nuclear targets, since the Fermi momentum
 of the nucleons within the target allows for intermediate resonance excitations. 
 
 The situation becomes even more entangled if 
 proton-induced reactions are considered. The right panel of Fig.~\ref{fig:KaonMat0} shows the measured and calculated $K^+$ differential spectra in $p+C$ reactions at energies below and above the $NN$ threshold for $K^+$ production measured by the ANKE collaboration \cite{Buescher:2001qu}. Different versions of the model are considered, but none delivers a good agreement with the experimental data. 
 
 \begin{figure*}[hbt]
  \centering
 \includegraphics[width=8cm,height=8cm]{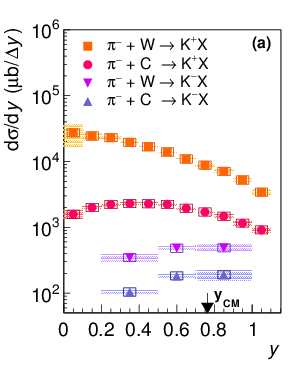}
  \hfill
  \includegraphics[width=9cm,height=9cm]{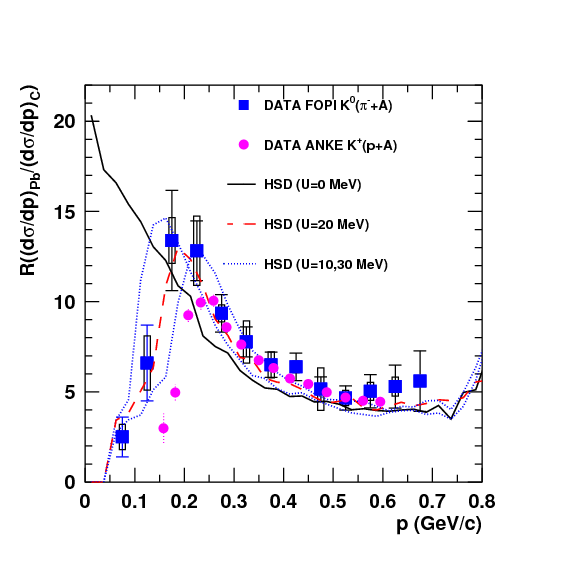}
\caption{\textit{(Color online)} Left panel: Cross sections of K$^+$ and K$^-$ in $\pi^- +C$ and $\pi^- +W$
 collisions as a function of rapidity measured with the HADES spectrometer. The shaded bands denote the systematic errors. The open boxes indicate the normalization error. The statistical uncertainties are smaller than the symbol size. The arrow indicates the rapidity \cite{Adamczewski-Musch:2018eik}. Right panel: Ratio of the $K^0$/$K^+$ yields produced by pions (protons) on heavy and light targets plotted as a function of the momentum and comparing to theoretical predictions assuming different in-medium potentials \cite{Benabderrahmane:2008qs}. The measurements have been carried out by the FOPI and ANKE collaborations. See text for details.}
  \label{fig:KaonMat1}
\end{figure*}

\begin{figure*}[hbt]
  \centering
  \includegraphics[width=9cm,height=9cm]{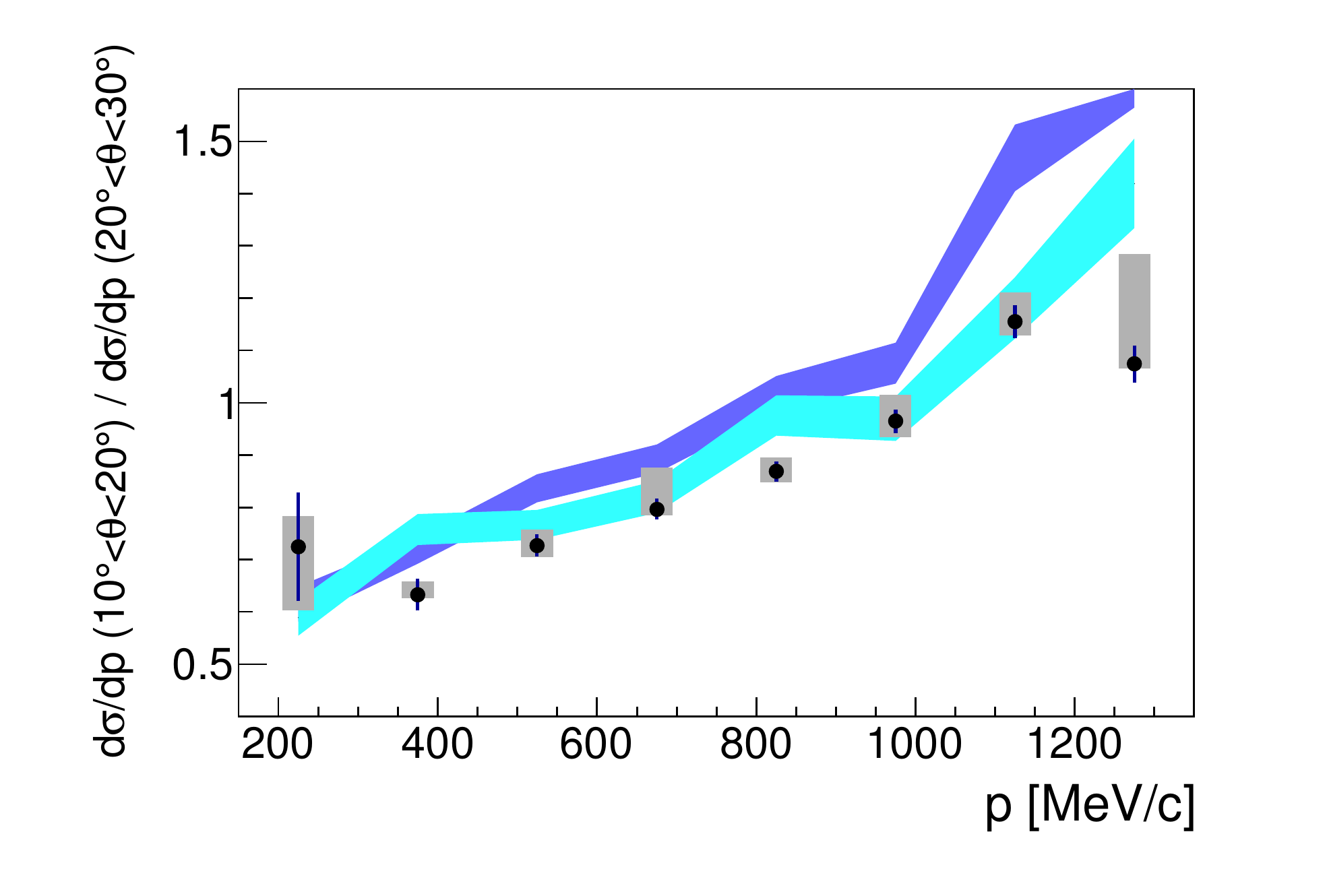}
 \includegraphics[width=9cm,height=9cm]{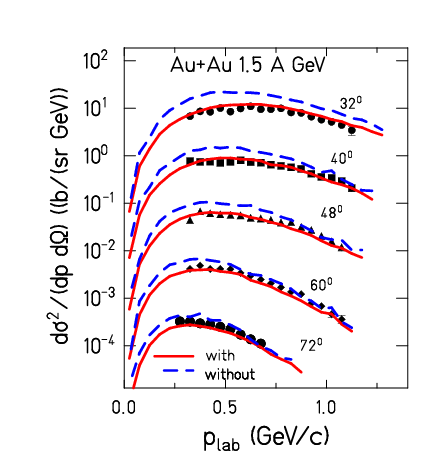} 
\caption{\textit{(Color online)} Left panel:  Ratio of the experimental momentum distributions in two neighbouring rapidity bins together with the prediction by the GiBUU model including the chiral potential (cyan histogram) and without any potential (blue histogram) \cite{Agakishiev:2014moo}. Right panel: Comparison of the experimental laboratory momentum spectra with IQMD calculations \cite{Hartnack:2011cn},  including and excluding the \kp nucleus potential. The data are from the KaoS Collaboration~\cite{Forster:2007qk}.}
  \label{fig:KaonMat2}
\end{figure*}

In $p+A$ collisions secondary and tertiary reactions driven by either pions
 or intermediate resonances contribute to the kaon yield \cite{Munzer:2017hbl}, and the shown comparison to the experimental data prove that all these effects are yet not under control.

 Also, in-medium modifications of the kaon properties could modify
 the measured kaon yields, since a repulsive interaction within nuclear matter would decrease the kaon production close to the threshold. These effects could show up already in $p+A$ and
 $\pi^-+A$ reactions, since the target nuclei provide a finite density  although smaller than the saturation density.

In order to quantify the in-medium effects in an alternative way,
the rapidity and momentum distribution of $K^0_S$ or/and $K^+$ have been analyzed by several experiments: KaoS  \cite{Scheinast:2005xs}, FOPI \cite{Benabderrahmane:2008qs} and HADES 
\cite{Agakishiev:2014moo,Adamczewski-Musch:2018eik} at GSI.

We consider first pion-induced reactions measured at GSI. These experiments are rather challenging due to the fact that secondary pion-beams have large emittance and limited intensities. 
The left panel of Fig. \ref{fig:KaonMat1} shows the rapidity distributions of $K^+$ and $K^-$ produced in $\pi^- +C$ and $\pi^-+W$ reactions with an incident beam momentum of 1.65 GeV/c measured by the HADES spectrometer at GSI \cite{Adamczewski-Musch:2018eik}. 
One can see that the $K^+$ rapidity distributions differ strongly between the two targets, and that for the heavier target the kaons are predominantly emitted in the backward direction (the arrow represents the center of mass rapidity of the $\pi N$ system). This clearly indicates the effect of multiple scattering within larger nuclei. 
A direct comparison of the rapidity distributions to transport models to check whether this effect is due to an enlarged scattering cross section in the medium with respect to the vacuum $KN$ scattering is, unfortunately, not straightforward. Indeed, this comparison depends on the production cross sections included in the transport codes. Still, the data are now available and can be used in the future as reference.

If we consider the momentum distributions of kaons in $\pi^-$-induced reactions, the right panel of Fig.~\ref{fig:KaonMat1} shows how neutral kaons have been measured by the FOPI collaboration \cite{Benabderrahmane:2008qs}.
This plot presents the comparison of the $K^0_S$ momentum measured in $\pi^-+C$ and $\pi^-+Pb$ collisions at a kinetic energy of 1.15 GeV to different versions of the transport code HSD that include an in-medium repulsive potential for the K$^0_S$ of varying strength (10, 20, 30 MeV). The figure shows that, within this model, the effect of a repulsive potential pushes kaons to higher momenta creating a depletion at low momenta. One can also see that the assumption of a repulsive potential
of 20 MeV provides the best agreement between the transport calculations and the data.
Possible effects at momenta higher than 700 MeV/c can not be verified here, since the phase space of the experimental measurement is limited.
One should also note that the involved transport model do not reproduce the inclusive kaon spectra for the same reaction.

Figure~\ref{fig:KaonMat2} shows the ratio of the momentum distribution of $K^0_S$ produced in $p+Nb$ reactions at a kinetic energy of 3.5 GeV for two different bins in the polar emission angle $\theta$ in the laboratory reference system \cite{Agakishiev:2014moo}. Forward emission angles in the laboratory frame are selected, since for this phase space 
region the effect of the potential is more evident due to the longer average path of kaons within the nucleus.

\begin{figure*}[hbt]
  \centering
  \includegraphics[width=10cm,height=7.5cm]{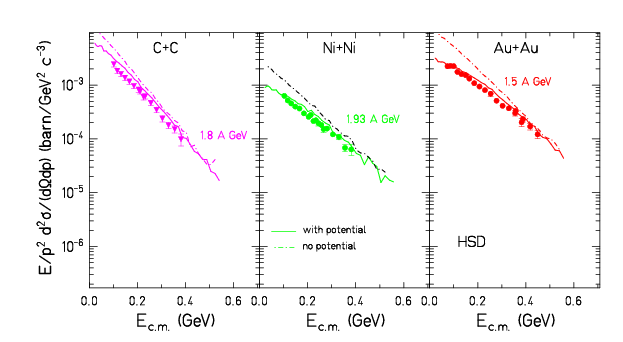}
   \includegraphics[width=7cm,height=8cm]{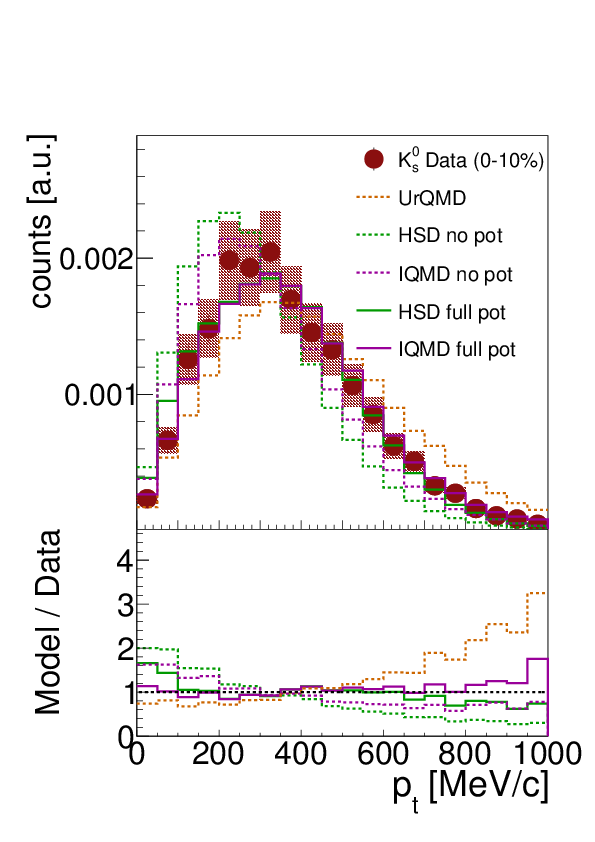}
\caption{\textit{(Color online)} Left panel: Inclusive invariant cross sections of \kp  at mid-rapidity as a function of the kinetic energy 
  for $C+C$ (left), $Ni+Ni$ (middle) and $Au+Au$ (right) reactions at various beam energies measured by the KaoS collaboration~\cite{Hartnack:2011cn}. The calculations refer to the HSD model \cite{Mishra:2004te}. The dashed lines refer to the option without \kp nucleus potential, the solid ones to that with \kp nucleus potential. Right panel: Transverse momentum of $K^0_S$ is measured in $Au+Au$ collisions at 1.23 AGeV by the HADES collaboration \cite{ Adamczewski-Musch:2018xwg}.}
  \label{fig:KaonMat3}
\end{figure*}

The experimental ratio is compared to the predictions by the GiBUU transport model including a chiral interaction potential
for the kaons within the nucleus (cyan histogram), or without any potential (blue histogram). In this comparison, 
the clearer effect of the in-medium kaon scattering is seen for larger momenta, starting from 800 MeV/c, and the scenario
with the potential is preferred by the data.
The very low momenta region ($p<\,200$ MeV/c) is not accessible in this experiment.
The potential used in the transport model for this comparison is the same shown in Fig. \ref{fig:ChiralKPot} and discussed in Sec. \ref{subtheorykaonsmatter}.

If one combines the findings presented in \cite{Benabderrahmane:2008qs} and \cite{Agakishiev:2014moo}, one can see that
some evidence of a repulsive interaction between 20 and 40 MeV at saturation density for neutral kaons within nuclei is visible. The effect is present at both small and large momenta, but the conclusions are still extremely model dependent.
The results are although obtained with different implementations of the kaon-nucleus potential since the HSD employed in \cite{Benabderrahmane:2008qs} contains only a density dependent interaction, while in \cite{Agakishiev:2014moo} the chiral potential has been employed in the GiBUU code for the comparison.

When going to HICs, the production of kaons is mainly mediated by intermediate resonances \cite{Hartnack:2011cn} in reactions as $N\Delta$ or even $\Delta\Delta$. 
This makes the uncertainties on the kaon production cross section rather large.
A lot of work has been invested in transport models to include all contributing production channels, and implement optical potentials to
describe the propagation of kaon within nuclear matter produced in HICs. Also, in this case, the measured momentum and rapidity distributions have been compared to transport code predictions.

The right panel of Fig.~\ref{fig:KaonMat2} shows the $K^+$ momentum distribution measured by the KaoS collaboration in Au+Au collisions at 1.5 AGeV \cite{Forster:2007qk}, in comparison with IQMD calculations with or without a repulsive and density dependent kaon-nucleus potential  \cite{Hartnack:2011cn}. One can see that the calculation including the repulsive potential are in better agreement with the experimental data, and that the effect of the potential is more pronounced for small
kaon momenta, but visible over the whole momentum range. 

The left panel of Fig.~\ref{fig:KaonMat3} shows the inclusive invariant cross sections for $K^+$ at mid-rapidity\footnote{Rapidity region perpendicular to the beam axis in the center-of-mass reference system} as a function of the collision energy in the centre-of-mass system for three different colliding systems ($C+C$, $Ni+Ni$ and $Au+Au$). In this case, the KaoS data are compared to the predictions by the HSD model \cite{Mishra:2004te}. Also in this case, the calculations with and without potential are shown. One can see that the contributions of the repulsive kaon potential is visible already in $C+C$ collisions, but the effect 
becomes more evident in larger colliding systems, since higher baryonic densities can be reached. The data seem in good agreement with the theoretical models.

The right panel of Fig.~\ref{fig:KaonMat3} shows the transverse momentum distribution for $K^0_S$ measured in $Au+Au$ collisions at a kinetic energy below the production threshold, 1.23 AGeV, by the HADES collaboration \cite{Adamczewski-Musch:2018xwg}. The data
are compared to transport model predictions normalized to the total experimental yield, and including calculations with or without a repulsive kaon-nucleus potential.
This comparison is more qualitative than the one carried out for reactions above threshold as all of the models over-predict the kaon production yields by more than 20\%, and does not deliver a consistent picture for all the transport models. On the other hand, 
an energy scan of the kaon production in HICs is  necessary to draw solid conclusions. When considering the momentum spectra for different colliding systems and their comparison to transport models, we find that any transport model can to this end reproduce all hadron spectra in several colliding systems with similar energies. 

An alternative method to study the in-medium properties of kaons exploits the investigation of the elliptic and radial flow of kaons.
The concept of flow comes from the fact that the azimuthal distribution of the emitted particles is often parameterized in a Fourier
series as
\begin{equation}
\frac{dN(y)}{d\phi}=  C[1+2v_1(y)cos(\phi) +2v_2(y)cos(2\phi)+..] ,
\end{equation}
with $y$ being the rapidity, $\phi$ the  particle azimuthal angle with respect to the reaction plane\footnote{The plane spanned by the  direction of the beam particle and impact parameter of the two nuclei at the collision point},  and $v_1$ and $v_2$ the parameters to be determined. 
The parameter $v_1$ is called 
radial or in-plane flow. It assumes positive values when particles are emitted in the same direction as the projectile on the reaction plane, or negative values when they follow the target.
The parameter $v_2$ is called elliptic or out-of-plane flow. It indicates if particles are emitted more in a direction perpendicular ($v_2<0$) or parallel ($v_2>0$) to the reaction plane. These observables can be experimentally measured for all particles, including kaons, and reflect the interaction of tagged particle with the nuclear medium created in the collision. The repulsive $K^+$-nucleus potential accelerates the $K^+$ to the side opposite to that of the projectile/target remnant, but, on the other hand, rescattering collisions have the opposite effect by aligning the $K^+$ to the flow of the nucleons. The final distribution is, hence, a combination of both effects and is sensitive to the number of rescattering collisions as well as to the $K^+$ nucleus potential.

\begin{figure*}[hbt]
  \centering
  \includegraphics[width=0.7\textwidth]{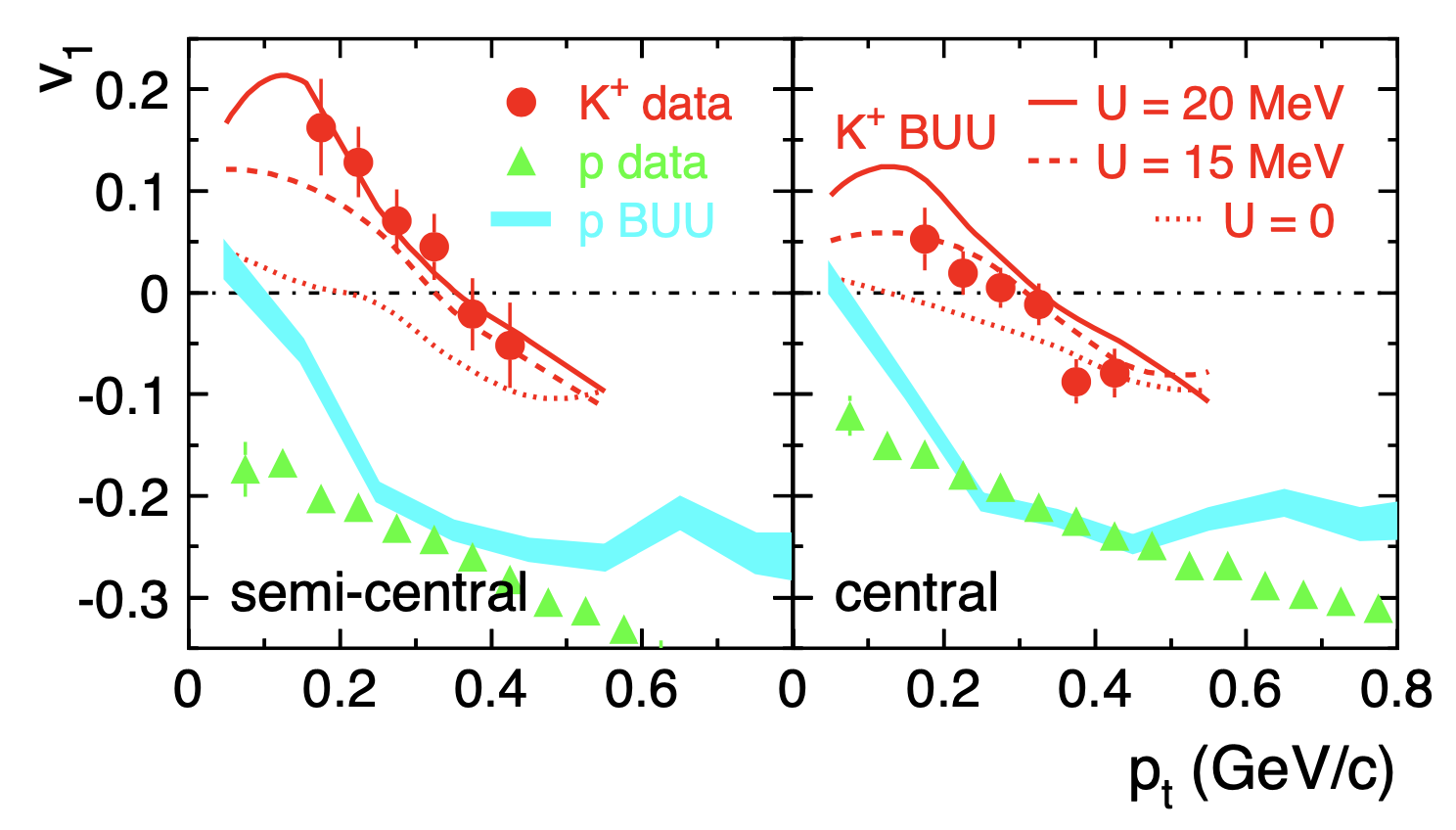}
\caption{\textit{(Color online)} $v_1$ versus $p_T$ for protons (triangles) and $K^+$ (dots) for semi-central (left) and central (right) Ru+Ru reactions at 1.69A GeV \cite{Crochet:2000fz}.  The curves and shaded area show the predictions of the RBUU model for different interacting potentials.}
  \label{fig:KaonMat4}
\end{figure*}

Figure \ref{fig:KaonMat4} shows the parameter $v_1$ extracted from experimental data by analysing the azimuthal angle distributions of protons and $K^+$ as a function of the particle transverse momentum, measured in semi-central (left) and central (right) $Ru+Ru$ reactions at 1.69 AGeV.
The experimental data are shown together with the predictions by the transport model RBUU \cite{Cassing:1999es}, with different assumptions on the $K^+$-nucleus potential. One 
can notice that the prediction of the sensitivity of the $v_1$ parameter to the potential is rather significant, specially for 
semi-central collisions.
This data are best reproduced assuming a repulsive potential with a value of $20$ MeV for kaons at rest, and for a density equal to saturation density.

This result has been confirmed also in other studies of kaon flow, carried out by the FOPI \cite{Zinyuk:2014zor} and KaoS \cite{Forster:2007qk} collaborations. One can, hence, summarize the findings about the properties of kaons in medium by saying that a repulsive potential between 20 and 40 MeV was measured in different experiments.

However, these interpretations do not go without criticisms, since the so far summarized conclusions are all rather model-dependent. 
Any modification of the production cross sections, scattering probability and inelastic cross sections would modify the data description. On the other hand, the experimental searches did exploit all the possible observables and a rather consistent picture was obtained for kaons. Antikaons, on the other hand, are much more challenging to describe, as we will see in Sec.~\ref{AntiKaonMatterExp}.

\begin{figure*}[hbt]
  \centering
  \includegraphics[width=7cm,height=10cm]{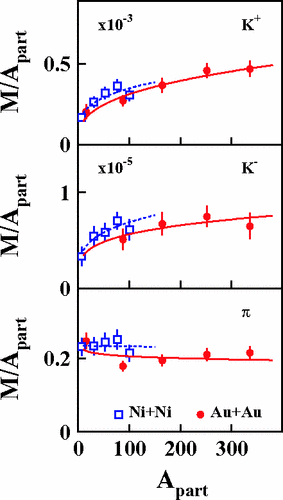}
   \includegraphics[width=7cm,height=10cm]{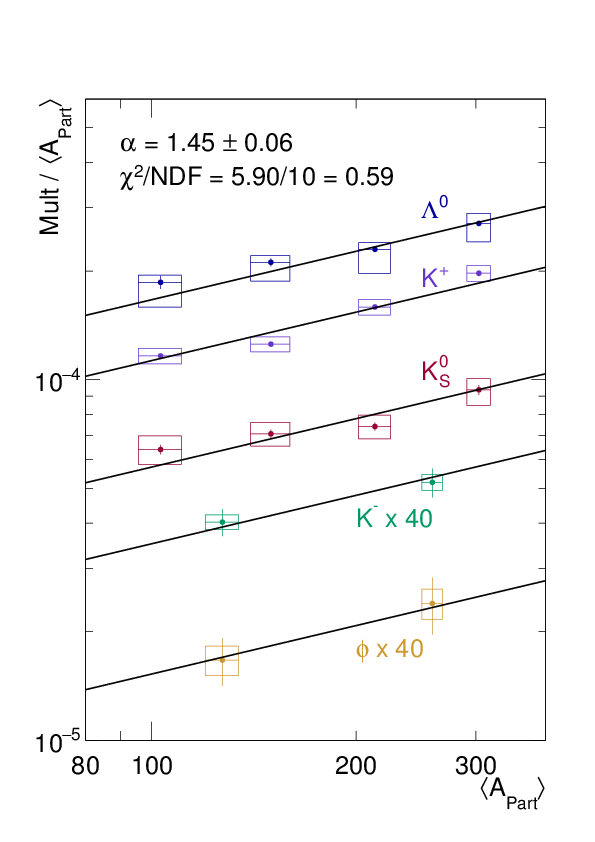}
\caption{\textit{(Color online)} Left panel: Dependence of the multiplicities of $K^+$ (upper panel) and of $K^-$ mesons (middle panel) as well as
of pions (lower panel) on A$_{\rm part}$ \cite{Forster:2007qk}. Full symbols denote Au+Au,
open symbols $Ni+Ni$, both at 1.5 AGeV. Right panel: Strange hadron multiplicities divided by the average 
participant number A$_{\rm part}$ as a function of A$_{\rm part}$, measured in $Au+Au$ collisions at 1.23 AGeV
by the  HADES collaboration \cite{Adamczewski-Musch:2018xwg}.}
  \label{fig:KaonMat5}
\end{figure*}

If we consider that the $NN$ production threshold for 
$K^+$ and $K^-$ lay at 1.58 GeV and 2.5 GeV, respectively, it is clear that 
most of the HICs measured by the KaoS, FOPI and HADES
collaboration concern sub-threshold production of antikaons and often also 
of kaons. The assumption made in some of the analyses 
of kaon and antikaon production \cite{Forster:2007qk,Hartnack:2011cn} is that single $NN$  collisions serve as reservoir to collect the necessary energy to produce kaons. The intermediate excitation of baryonic resonances would then play a major 
role.

This working hypothesis might be completely made obsolete by the fact that a 
sort of universal scaling for the measured multiplicity of strange particles 
as a function of the number of participating nucleons in the reactions is 
observed. Already the KaoS collaboration 
made this observation for $K^+$ and $K^-$ produced in $Ni+Ni$ and $Au+Au$ 
collisions, as seen  in the left panel of Fig.~\ref{fig:KaonMat5} by showing the $K^+$ and $K^-$ multiplicities divided by number of 
participants as a function of the number of participants.

Further, measurements in $Au+Au$ collisions at 1.23 AGeV, hence
below the $NN$ threshold for both kaon species, confirm this universal scaling as a function
of A$_{\rm part}$ for all strange hadrons. 
This effect was expected for $K^+$ and $\Lambda$ due to associated production, and for $K^-$ due to coupling by strangeness exchange to the former ones, but not at all for the $\phi$
The right panel of Fig.~\ref{fig:KaonMat5} demonstrates
this universal scaling. This observation does not support the interpretation of HICs at intermediate energies as a superposition of primary and secondary $NN$ collisions, but
indicates  the presence of an intermediate pre-hadronization phase common to all hadrons,
despite of the different mass and quark content.
Clearly effects as re-scattering, mean potential and absorption can be also present, but they will be difficult to disentangle.

\subsubsection{Theoretical approaches for antikaons in matter}
\label{sec:antikaon-matter}

\begin{figure*}[t]
  \centering
 \includegraphics[width=0.5\textwidth,angle=-90]{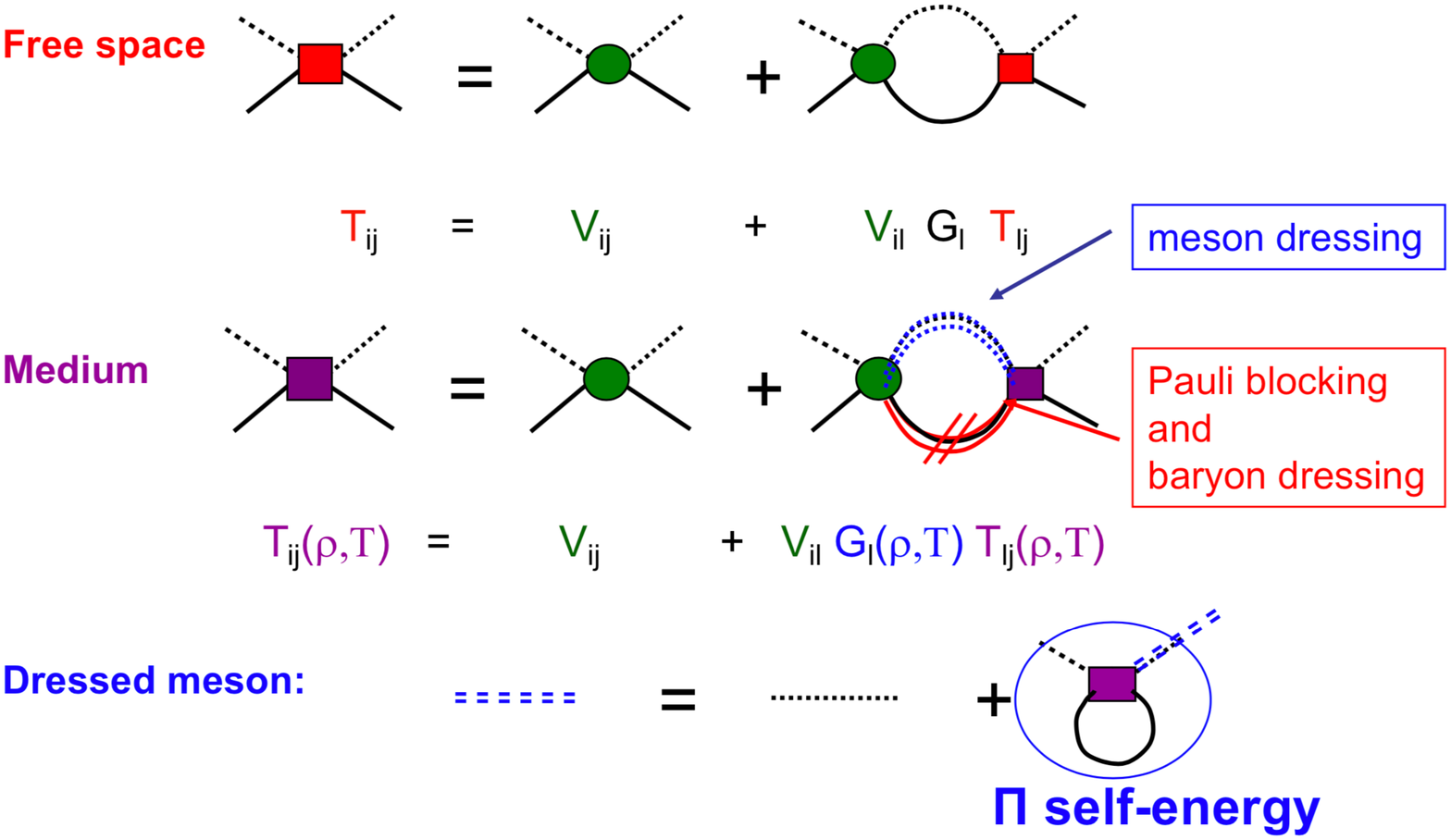}
\caption{\textit{(Color online)} Diagrams showing the effective $\bar K N$ interaction in vacuum via the Bethe-Salpeter equation (upper panel), the effective $\bar K N$ interaction in matter (middle panel) and the Dyson-Schwinger equation for the self-energy (lower panel).}
\label{bethe-medium}
\end{figure*}

\begin{figure*}[t]
  \centering
 \includegraphics[width=0.7\textwidth,angle=-90]{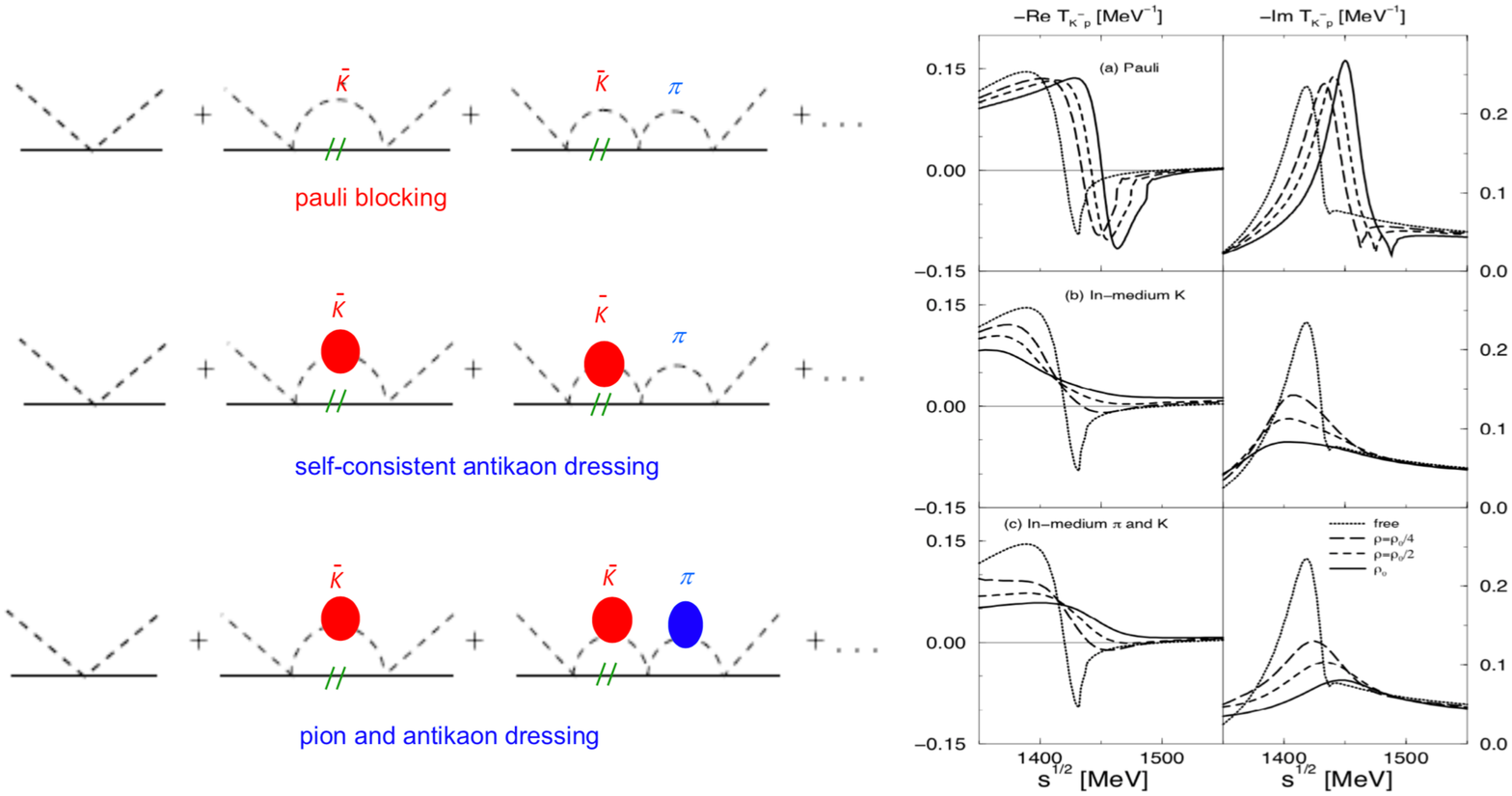}
\caption{\textit{(Color online)} Left panel: Diagrams for the three in-medium corrections: incorporation of Pauli blocking (upper panel), self-consistent inclusion of antikaon dressing (middle panel) and implementation of of pion and antikaon dressing (lower panel). Right panel: Effects of the previous in-medium corrections on the behaviour of the $\Lambda(1405)$ in matter, taken from \cite{Ramos:1999ku}. }
\label{diagram-medium}
\end{figure*}

\begin{figure*}[t]
  \centering
 \includegraphics[width=0.48\textwidth, height=0.4\textwidth]{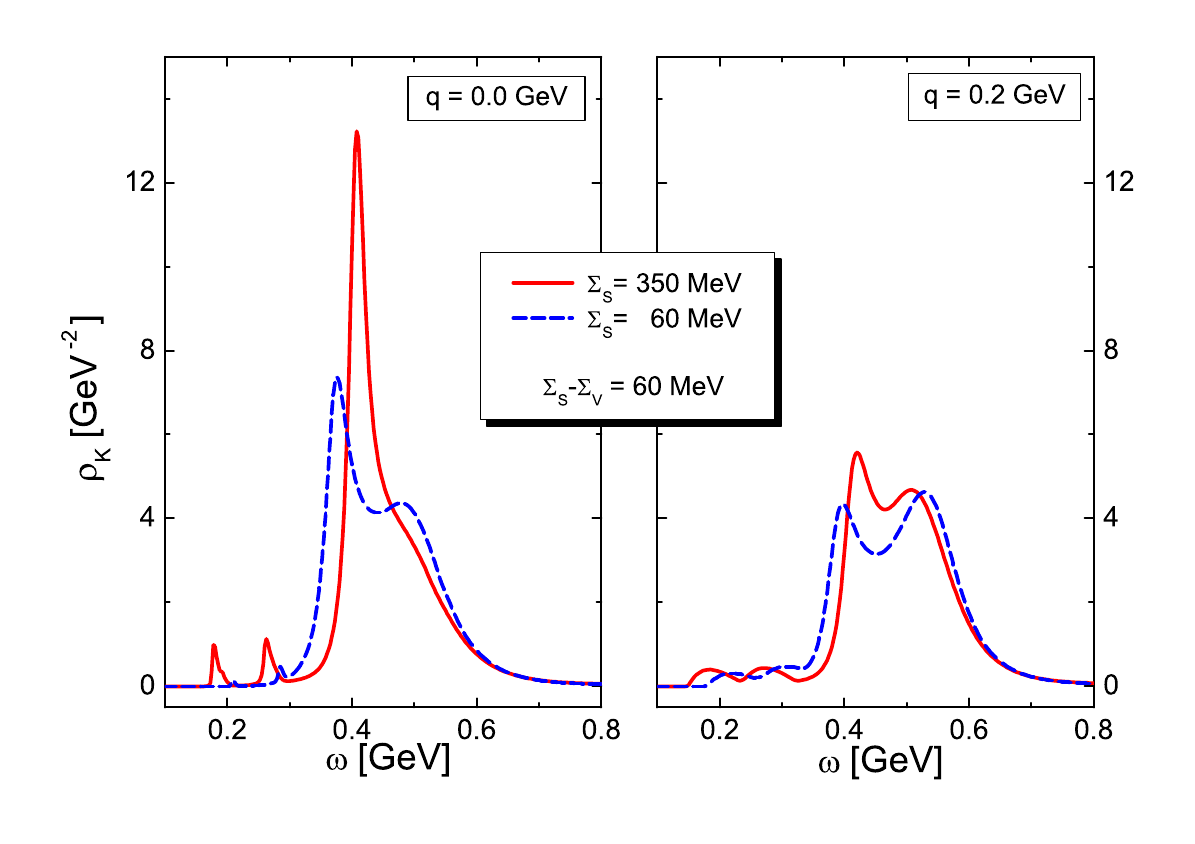}
 \includegraphics[width=0.48\textwidth, height=0.4\textwidth]{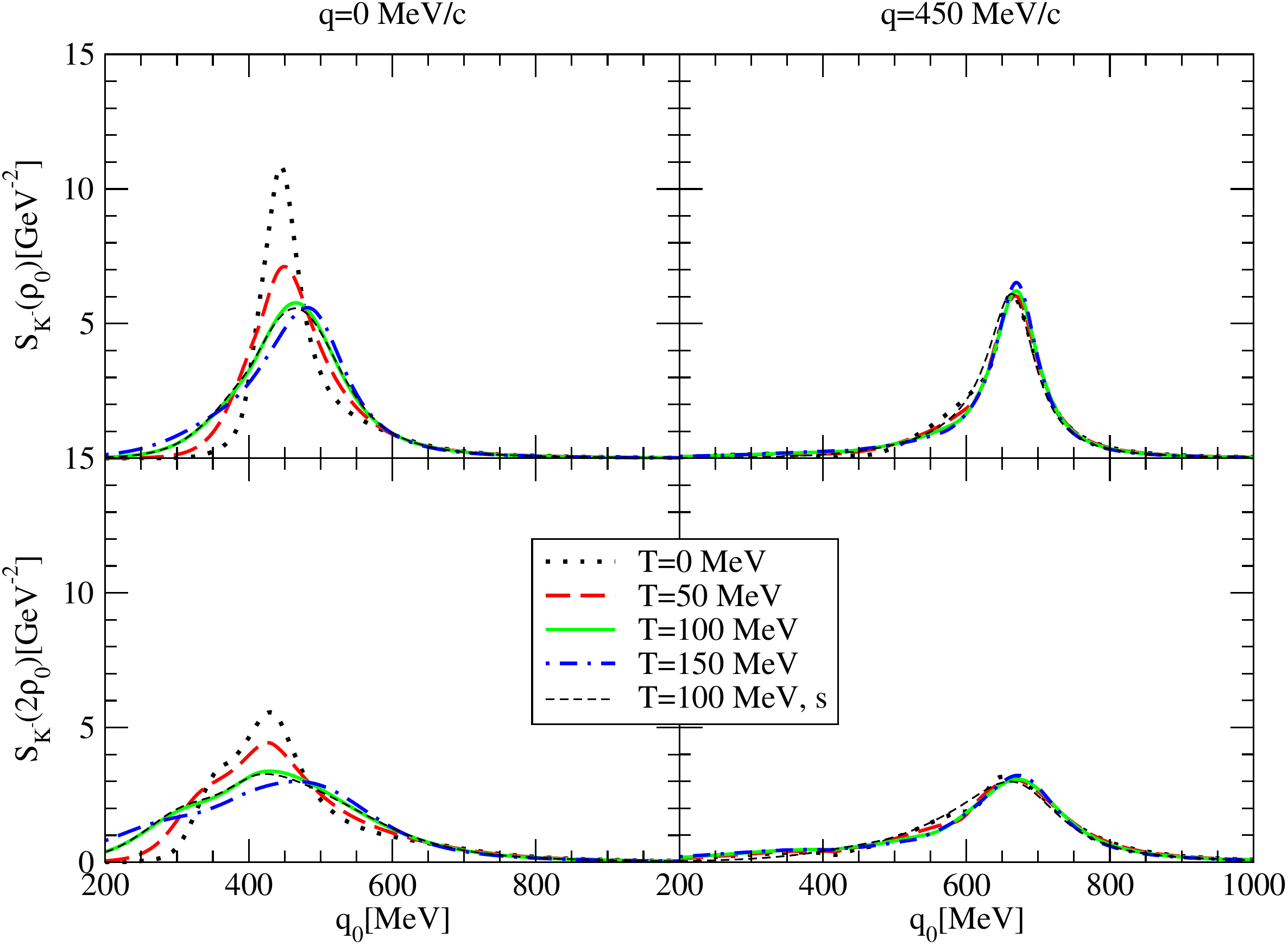}
\caption{\textit{(Color online)} $\bar K$ spectral functions for two chiral approaches (left figure taken from \cite{Lutz:2007bh} and right figure from \cite{Tolos:2008di}). }
\label{Kbarspectral}
\end{figure*}

Antikaons in matter have been extensively studied from the theoretical point of view. Early works on RMF (based on OBM or $\chi$PT \cite{Schaffner:1996kv}) or QMC schemes \cite{Tsushima:1997df} obtained very deep potentials of a few hundreds of MeVs at $\rho_0$. However, most of these models have assumed that the low-density theorem is valid for the calculation of the $\bar K$ optical potential in dense matter. As compared to the $KN$ interaction, the applicability of the low-density theorem for the $\bar K$ in matter is doubtful, since the behaviour of the $\bar KN$ interaction close to threshold is dominated by the presence of the $\Lambda(1405)$, as discussed in Sec.~\ref{sec:theoryKN}. Thus, it is of fundamental importance to study the properties of the $\Lambda(1405)$ in matter and the consequences on the $\bar K N$ scattering amplitude and, thus, on the $\bar K$ optical potential.

One way to proceed is to use unitarized theories in coupled channels in dense matter, starting either from $\chi EFT$ with strangeness \cite{Waas:1996fy,Lutz:1997wt,Ramos:1999ku} or from meson-exchange models \cite{Tolos:2000fj,Tolos:2002ud}. The idea is to address the meson-baryon scattering in matter by solving an equivalent equation to the Bethe-Salpeter one (see Eq.~\ref{bethe-free}), introducing medium corrections in the intermediate coupled-channel meson-baryon propagators. This is shown in Fig.~\ref{bethe-medium}. In the upper panel, we display the diagram for the Bethe-Salpeter equation, where the effective interaction is indicated with a red filled square and the microscopic meson-baryon interaction (or potential coming from chiral dynamics or meson-exchange models) is shown with a green filled circle. The middle panel represents the in-medium effective interaction (violet filled square), where the dressed meson is indicated with a dotted double line. The dressed baryon propagator takes into account, on the one hand, Pauli blocking so as to forbid access to already occupied Fermi levels (two tilted vertical lines), and, on the other hand, the modification of the baryon spectra due to the medium (solid double red line). 

In the lower panel of Fig.~\ref{bethe-medium}, we also show the Dyson-Schwinger equation for the calculation of the meson self-energy (at the lowest order). Indeed, the determination of the in-medium $\bar KN$ effective interaction demands that the Dyson-Schwinger equation for the $\bar K$ self-energy and the equation for the effective $\bar KN$ interaction have to be solved simultaneously.  This is due to the fact that the dressed $\bar K$ propagator and, hence, the $\bar K$ self-energy enter in the meson-baryon intermediate loop when $\bar K$ and nucleons are considered, and this self-energy, in turn, depends on the in-medium  $\bar KN$ effective interaction, as seen in the lower panel. Hence, a self-consistent procedure is required in order to finally obtain both the in-medium $\bar KN$ effective interaction and the $\bar K$ self-energy.

As previously discussed in Sec.~\ref{sec:theoryKN}, the $\bar K N$ amplitude close to the $\bar KN$ threshold (for $S$-wave) is determined by the behaviour of the $\Lambda(1405)$. Thus, any change in the behaviour of the $\Lambda(1405)$ induced by the presence of matter will affect the in-medium $\bar KN$ and, hence, the value of the antikaon potential. The attractive antikaon potential within unitarized coupled-channel models is a consequence of the modified $S$-wave $\Lambda(1405)$ resonance in the medium due to several effects: a) the consideration of Pauli blocking on the baryon states in the intermediate meson-baryon propagator \cite{Koch:1994mj}; b) the inclusion of the $\bar K$ self-energy \cite{Lutz:1997wt} in the propagation of $\bar K$ in matter, that is, the "dressing" of a $\bar K$ due its interaction with the surrounding nucleons; and  c) the implementation of self-energies of the other mesons and baryons in the intermediate states \cite{Ramos:1999ku}. These effects are pictured in the left panel of Fig.~\ref{diagram-medium}, whereas the effect on the behaviour of the $\Lambda(1405)$ is seen in the right panel. 

The right panel of Fig.~\ref{diagram-medium} displays the real part (on the left) and imaginary part (on the right) of the $K^-p \to K^-p$ scattering amplitude for a total momentum $\mid \vec{p}_K + \vec{p}_N \mid =0$ as a function of $\sqrt{s}$. The results are shown for several densities and the inclusion of the different medium modifications earlier discussed. The Pauli blocking effects on the nucleonic propagator (upper left panel) moves the $\Lambda(1405)$ to higher energies, thus changing its position from 27 MeV below
the $K^- p$ threshold to above it at finite density. The resonance shape remains unaltered, whereas appearing, for a given density, below the new
threshold imposed by the Pauli principle, i.e. $m_K + M +
p_F^2/2 M$, where $m_K$ is the antikaon mass and $M$  the nucleon mass. When including the Pauli blocking as well as the self-energy of the ${\bar K}$ meson (middle left panel), the $\Lambda(1405)$ stays close to the free space value smearing out with density. This is due to the cancellation between the Pauli blocking effect and the attraction felt by the $\bar K$ meson. The inclusion of self-energies of the mesons and baryons in the intermediate states, and, in particular, of pions (lower left panel) induces a further broadening of the $\Lambda(1405)$. Indeed, new channels appear when pions are dressed in the medium, such as $\bar K NN \rightarrow Y N $ or $\bar K N N \rightarrow Y \Delta $, with $Y=\Lambda,\Sigma$. 

Once the $\bar K N$ effective interaction in matter and, hence, the $\bar K$ self-energy are determined, it is possible to obtain the spectral representation of the $\bar K$ propagator as
\begin{equation}
S_{\bar K} (\omega,{\vec q})= -\frac{1}{\pi} {\rm Im}\, D_{\bar K}(\omega,{\vec q})
= -\frac{1}{\pi}\frac{{\rm Im}\, \Pi_{\bar K}(\omega,\vec{q})}{\mid
\omega^2-\vec{q}\,^2-m_{\bar K}^2- \Pi_{\bar K}(\omega,\vec{q}) \mid^2 } \ ,
\label{eq:spec}
\end{equation}
where $D_{\bar K}(\omega,{\vec q}\,)$ stands for the propagator and $\Pi_{\bar K}(\omega,\vec{q})$ is the $\bar K$ self-energy. In both expressions, the energy ($\omega$) and momentum ($\vec{q}$) of the $\bar K$ are treated separately, as the energy and momentum of a particle in the medium are not correlated.

The $\bar K$ self-energy is shown in Fig.~\ref{Kbarspectral} for two different chiral models, that mainly differ on the renormalization scheme and some technicalities, such as whether the angle-average is used for the calculation of the meson-baryon loop function. In the left panel of Fig.~\ref{Kbarspectral}, the antikaon spectral function as a function of energy and momentum at nuclear saturation density is shown, where not only the $S$-wave $\bar KN$ interaction is considered, but also $P$- and $D$-waves \cite{Lutz:2007bh}. In this figure, two possible different self-energies for the nucleons in matter are also included (red solid and dashed blue lines). On the right panel of Fig.~\ref{Kbarspectral}, the $\bar K$ spectral function is explored including $S$- and $P$-wave contributions for different temperatures (different lines) for two densities ($\rho_0$ in the upper panels and 2$\rho_0$ in the lower panels) and two momenta (left and right) within the chiral scheme of \cite{Tolos:2008di}. The $\bar K$ in matter feels a slight attraction (by shifting the peak of the spectral function to lower energies), while acquiring a remarkable width, that close to $\bar KN$ threshold is dominated by the presence of the $S$-wave $\Lambda(1405)$.

With these figures we also see the importance of considering higher partial waves as well as analysing the effect of finite temperature corrections for the realistic determination of the properties of $\bar K$ in dense matter \cite{Tolos:2008di,Tolos:2006ny,Lutz:2007bh,Cabrera:2009qr,Cabrera:2014lca}. In particular, the knowledge of higher-partial waves beyond $S$-wave  as well as finite temperature corrections become essential for analyzing the results of HiCs at beam energies below 2 AGeV \cite{Cassing:2003vz,Tolos:2003qj,Hartnack:2011cn}.  

As discussed in the previous Sec.~\ref{kaonInMatterExp} and the next Sec.~\ref{AntiKaonMatterExp}, the in-medium modification of kaon and antikaon properties have been explored experimentally close to threshold in HICs (see \cite{Hartnack:2011cn} for a recent review). The various channels for antikaon production are shown in the middle column of Table~\ref{tab_cross1}. As compared to kaons, strangeness exchange reactions are also taking place, whereas strangeness absorption mechanisms are not possible for antikaons. With the help of microscopic transport models \cite{Fuchs:2005zg,Hartnack:2011cn}, the creation and transport of kaons/antikaons have been studied. The first transport calculations for antikaon observables in nuclear matter were performed neglecting the finite width of the antikaon spectral function \cite{Cassing:1999es,Hartnack:2001zs}, as explained in Sec.~\ref{subtheorykaonsmatter} when discussing transport models (see Eq.~(\ref{TP5}) and following discussion). Some years later, antikaon production were studied using off-shell dynamics with in-medium spectral functions in the HSD transport model, where the $\bar K$ self-energy enters in both the drift term and the collision term of Eq.~(\ref{TP5}) \cite{Cassing:2003vz}. In this case,  the J\"ulich meson-exchange model \cite{Tolos:2000fj,Tolos:2002ud} was employed as the effective $\bar KN$ interaction in matter. 

The analysis of experimental data in conjunction with microscopic transport models have revealed a complicated interaction scenario for antikaons, as it will be discussed in Sec.~\ref{AntiKaonMatterExp}. Whereas a kaon behaves almost as a good quasiparticle with a narrow spectral function, for antikaons the situation is much more unclear due to several reasons. First, antikaons have a broad spectral function due to the strong interaction with the medium. Second, the $T \rho$ approximation for the antikaon optical potential is not valid due to the presence of the $\Lambda(1405)$ resonance. Third, the $\pi Y \to {\bar K}N$ strangeness exchange reaction is the dominant channel for antikaon production in HICs, as hyperons are more abundantly produced together with kaons, and this reaction is substantially modified in the hot dense medium. Thus, in spite of all the effort, the analysis of all experimental antikaon observables has not allowed so far for a consensus on the antikaon cross sections and optical potential (cf. the recent review \cite{Hartnack:2011cn}).

 \begin{figure}[htb]
 \begin{center}
\includegraphics[width=0.38\linewidth,height=12.5cm]{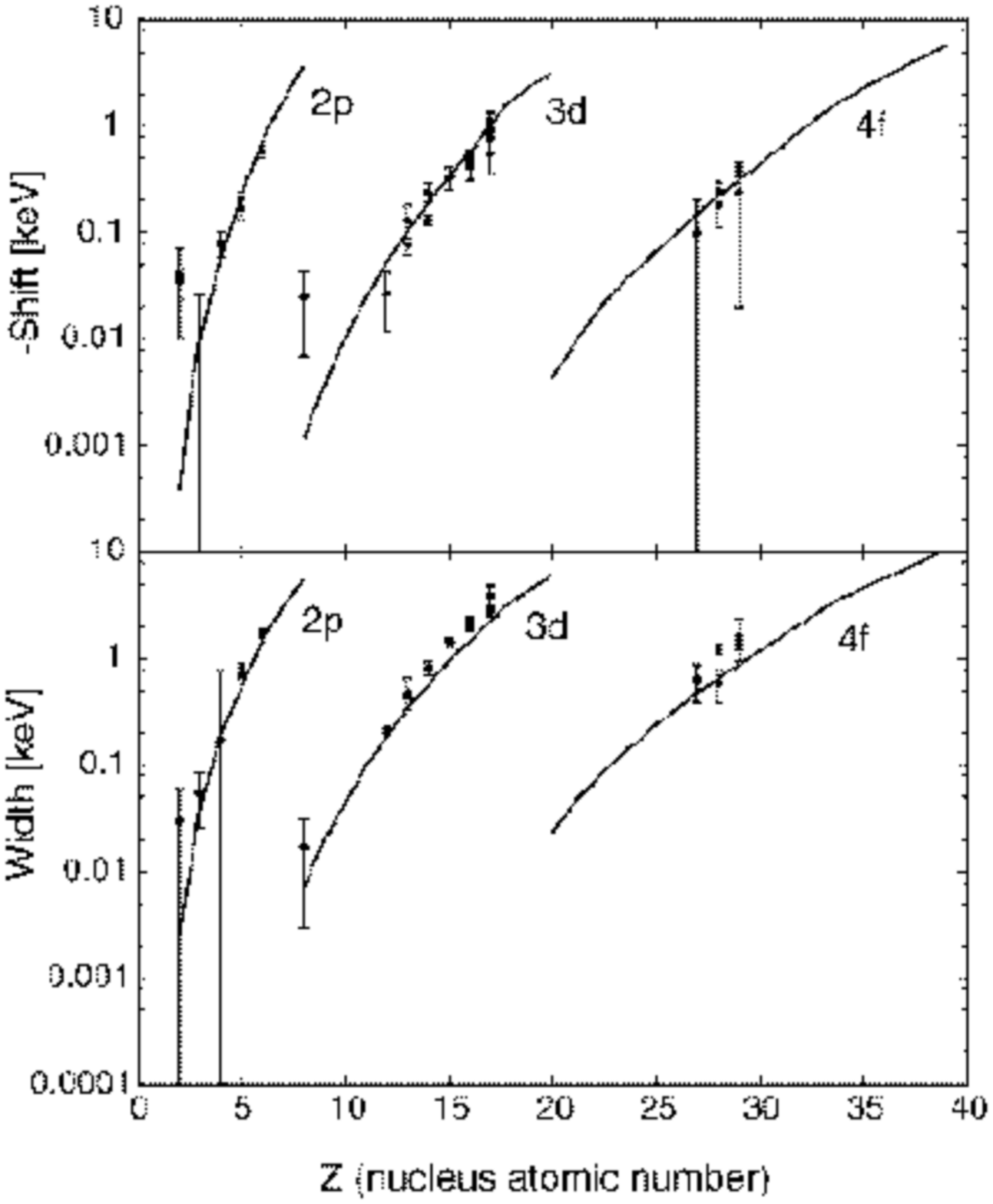}
\includegraphics[width=0.55\linewidth, height=13cm]{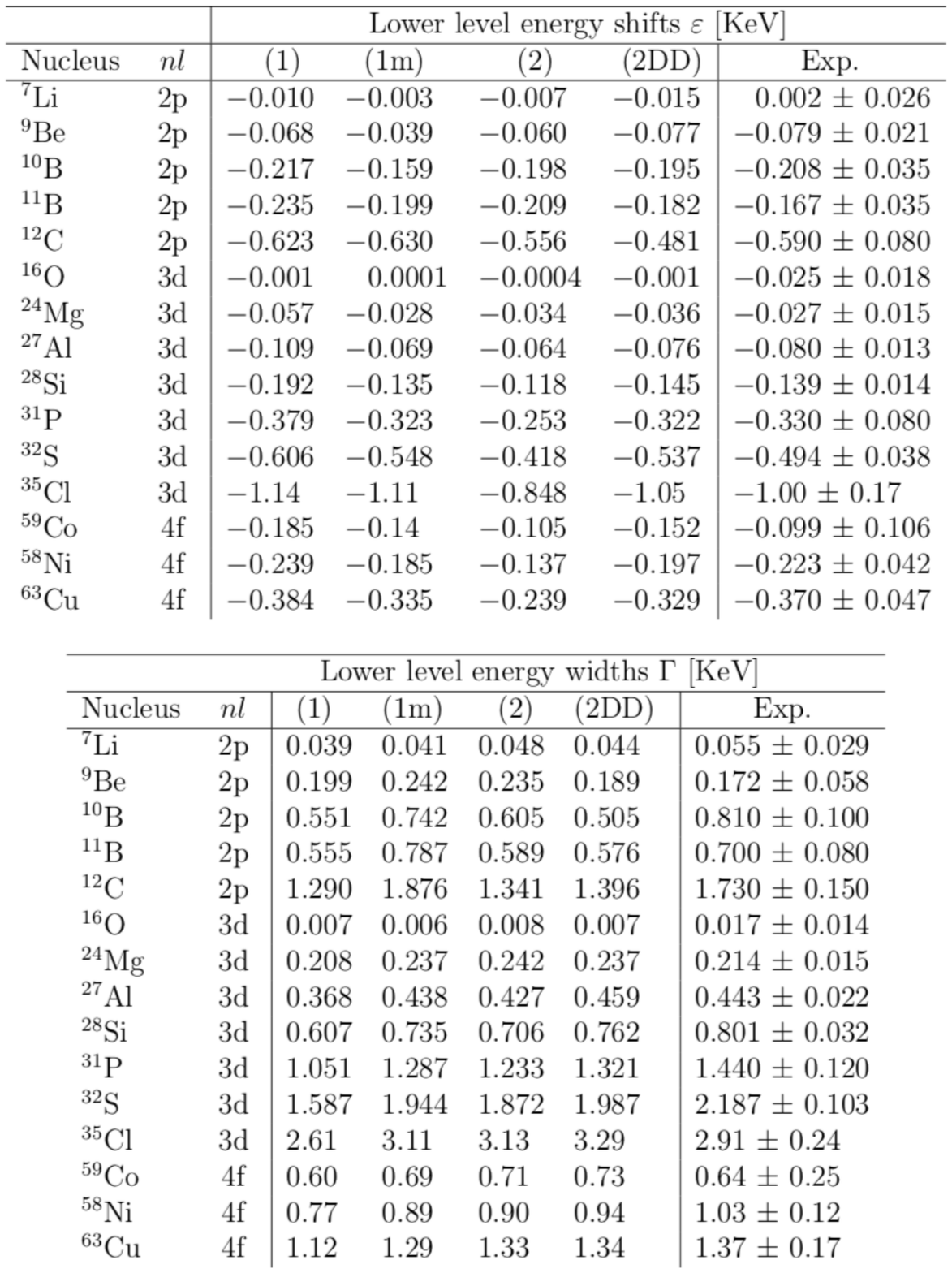}
 \caption{ Left plot: Energy shifts and widths as
function of the nucleus atomic number for $2p$, $3d$, and $4f$ for kaonic atom states. Figure taken from \cite{Hirenzaki:2000da}. Right table:  Energy shifts (upper table) and widths (lower table) of different kaonic atom levels up to $^{63}$Cu $4f$. The (1), (1m), (2) and (2DD) optical potentials are shown together with the experimental data.   Table adapted from \cite{Baca:2000ic}.}
\label{fig:oset-table:hirenzaki}
 \end{center}
\end{figure}

\subsubsection{Kaonic atoms} 

Kaonic atoms are atoms in which an electron is replaced by a negatively charged antikaon. By using X-ray spectroscopy techniques, energy shifts and widths of kaonic atoms levels have been measured for the whole periodic table (see, for example, the compilation of kaonic atoms in Ref.~\cite{Batty:1997zp}).

Kaonic atoms give us complementary information on the interaction of antikaons with nucleons, as the antikaon-nucleus potential can be extracted from best-fit analysis of kaonic-atom data. Some solutions, which use phenomenological potentials (that included an additional non-linear density dependent term)  or hybrid models (that combined a RMF in the nuclear interior and a phenomenological density dependent potential at the surface that was fitted to $K^-$ atomic data) are in agreement with very strongly attractive well depths of the order of -150 to -200 MeV at normal saturation density (see \cite{Friedman:2007zza} for a review).  

However, studies of kaonic atoms using the chiral ${\bar K}N$ amplitudes \cite{Ramos:1999ku} have shown that it is indeed possible to find a reasonable reproduction of the data with a relatively shallow antikaon-nucleus potential \cite{Hirenzaki:2000da}, albeit adding an additional moderate phenomenological piece \cite{Baca:2000ic}. This is seen in the left plot of Fig.~\ref{fig:oset-table:hirenzaki}, where the shifts and widths of kaonic atoms are shown in comparison with the chiral amplitudes. Also, in the right table of Fig.~\ref{fig:oset-table:hirenzaki} of \cite{Baca:2000ic},  the experimental results for shifts and widths of the lower lever states of kaonic atoms up to $^{63}$Cu $4f$ are shown and compared to calculations obtained with shallow and deep phenomenological potentials, obtained from best fits to the data. The former ones are labelled with "1", when only chiral amplitudes are used, and with "1m" when also taking into account a phenomenological contribution. The phenomenological models are indicated with "2" and "2DD".  The reasonable fit to data of the chiral amplitudes has been indeed corroborated by a calculation \cite{Cieply:2001yg}, where a good fit to both scattering $K^- p$ data and kaonic-atom data has only required to modify slightly the parameters of the chiral meson-baryon interaction model of Ref.~\cite{Kaiser:1995eg}.

\begin{figure}[htb]
\begin{center}
\includegraphics[width=0.5\textwidth]{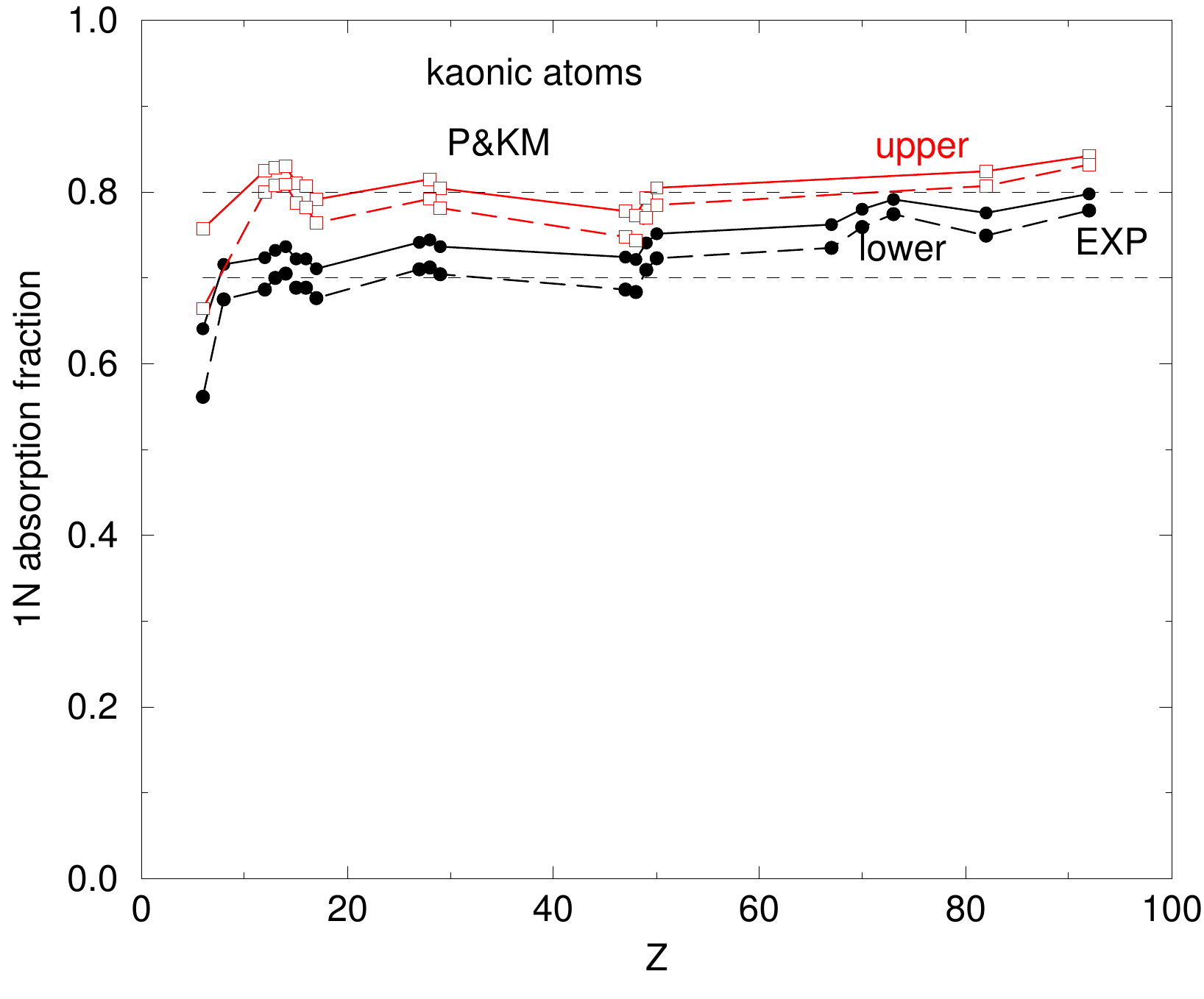}
\caption{\textit{(Color online)} Fractions of single-nucleon absorption for amplitudes 
Prague (P) and Kyoto-Munich (KM): solid circles are for lower states, open squares for upper states. Horizontal dashed lines indicate the experimental values of the single-nucleon 
absorption fraction. Figure taken from \cite{Friedman:2016rfd}.} 
\label{fig:PKMfrac}
\end{center}
\end{figure}

\begin{figure}[htb]
\begin{center}
\includegraphics[width=0.45\textwidth]{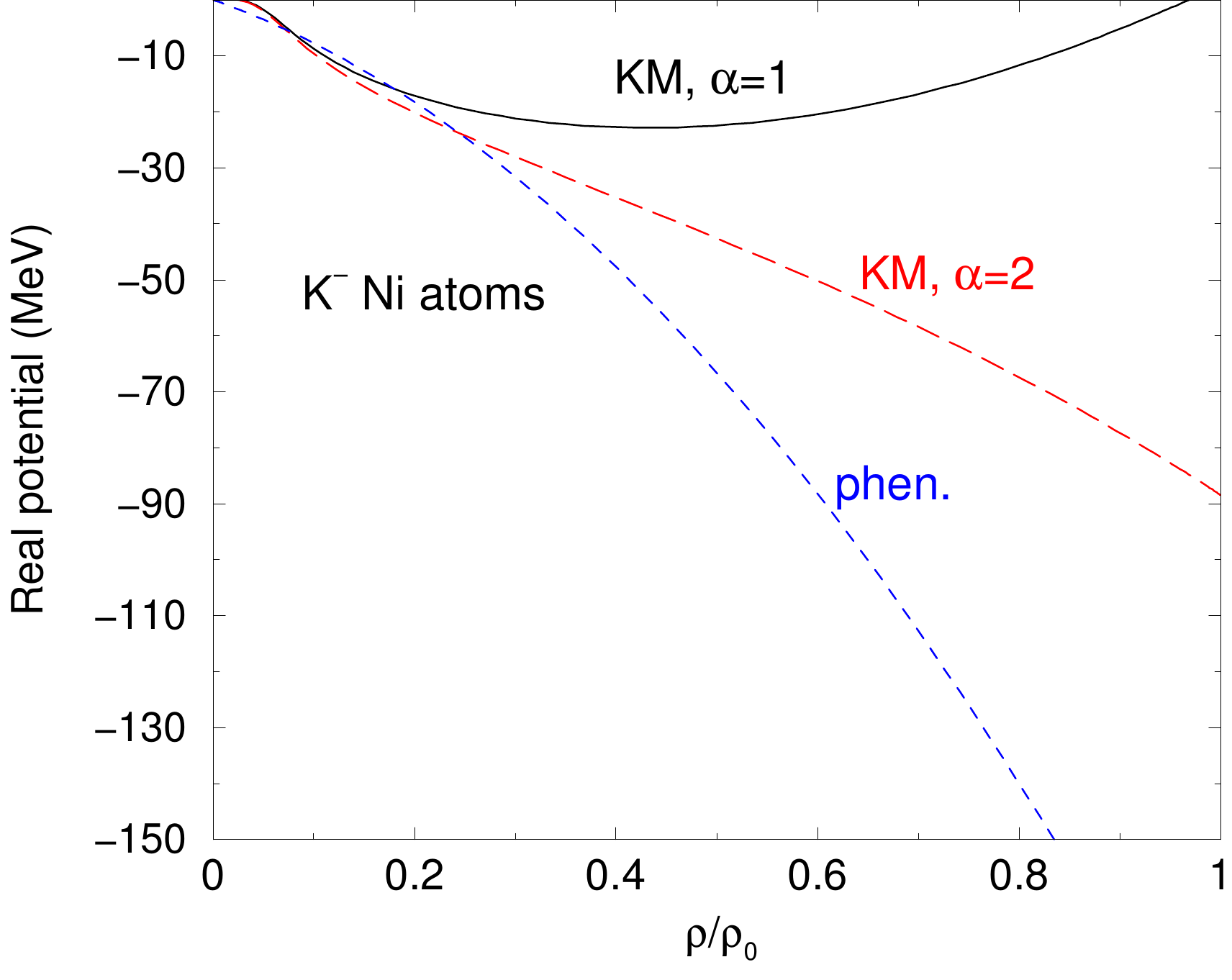}
\includegraphics[width=0.45\textwidth]{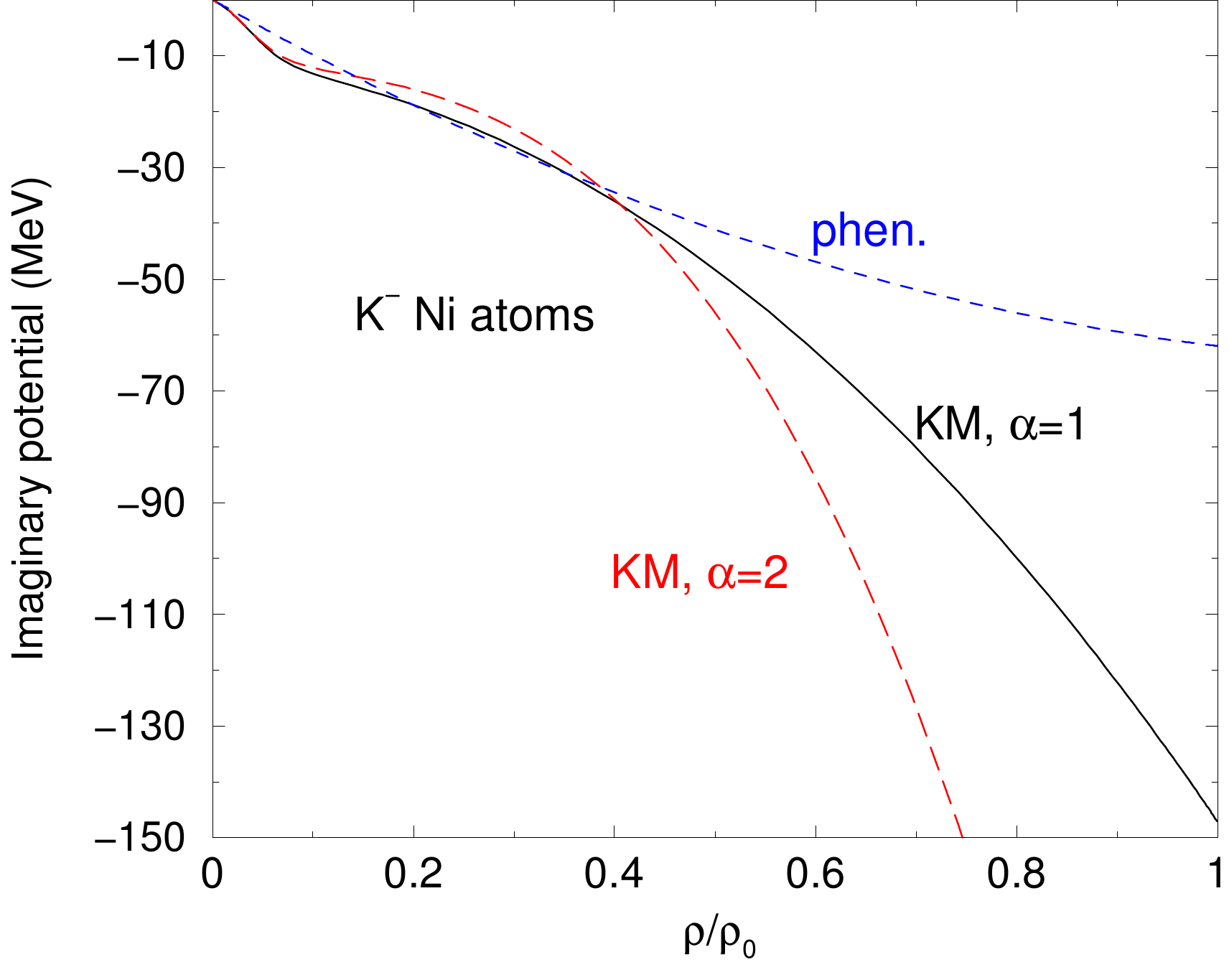}
\caption{\textit{(Color online)} Real part (left panel) and imaginary part
(right panel) of best-fit $K^-$ optical potentials for Ni-kaonic atoms.
Solid curves and long-dashed curves represent the KM single-nucleon
amplitudes plus a phenomenological term $B(\rho /\rho_0)^\alpha$ (with $\alpha=1$ or 2), whereas dashed
curves are for a purely phenomenological density-dependent best-fit
potential. All three potentials lead to equally good fits to the 65 data points in kaonic atoms. Figures taken from \cite{Friedman:2016rfd}.} 
\label{fig:comp}
\end{center}
\end{figure}

More recently, the Jerusalem-Prague Collaboration has reanalyzed kaonic atoms by using the $K^-$ optical potentials derived from state-of-the-art chirally--motivated meson-baryon coupled-channel models, which were supplemented by a phenomenological term representing $K^-$ multi-nucleon interactions \cite{Friedman:2016rfd,Hrtankova:2017zxw}.  Fig.~\ref{fig:PKMfrac} shows the calculated fractions of single-nucleon absorption for lower states (solid circles) and upper states (open squares) of kaonic atoms, when using the chiral-model amplitudes from the Prague (P) and Kyoto-Munich (KM) groups, supplemented by the phenomenological term \cite{Friedman:2016rfd}. These fractions are compared to the best estimates of single-nucleon absorption fractions ($0.75 \pm 0.05$), deduced from Bubble-Chamber studies, as indicated by the two dashed horizontal lines. The agreement of the calculations with the estimated single-nucleon absorption fractions is very good for both models, both producing very similar in-medium $K^- N$ amplitudes down to about 40 MeV below threshold, which is the region of energies relevant for kaonic atoms. In fact, by comparing two KM versions (differing in the phenomenological term) with a purely phenomenological model, as seen in Fig.~\ref{fig:comp}, the Jerusalem-Prague Collaboration found out that the real and imaginary parts of the $K^-$ optical potential are constrained by kaonic data only for densities up to $\sim$ 25 \% (50\%) of saturation density. 

Therefore, the lesson learned from all these works is that kaonic atom data do not really provide a suitable constraint on the antikaon-nucleus potential at densities close to normal nuclear matter density. The strength of the $K^-$ potential might be then found from comparison of available $K^-$-nucleus scattering data for $^{12} {\rm C}$, $^{40} {\rm Ca}$ and $^{208} {\rm Pb}$ to the theoretical predictions \cite{GarciaRecio:2002wh}, from in-flight ($K^-N$) reaction \cite{Magas:2009dk}, from the analysis of the $K^-$  multi-nucleon processes in lighter (few-body) nuclear systems \cite{Hrtankova:2017zxw} or from pion-, proton- or nucleus-nucleus reactions.

\subsubsection{Experimental measurements of antikaons in matter}
\label{AntiKaonMatterExp}

In order to investigate the antikaons properties within nuclear matter, we normally study $K^-$ instead of $\bar{K^0}$, since the detection of the dominant weak decay into three pions, that characterizes the neutral antikaon, is experimentally very challenging.

A treatment of antikaons must consider the  following processes: i) the antikaon-nucleus interaction  is linked to the properties of the $\Lambda(1405)$ resonance and $\phi$ meson; ii) strangeness exchange processes as $\pi \Lambda  \rightarrow \bar{K}N$ link the $\bar K$ production  to the $K$ production; iii) the reverse absorption process $\bar{K}N \rightarrow \pi \Lambda$ also occurs; and iv) a sizeable attractive real part of the $\bar{K}$-nucleus interaction is expected and, if demonstrated, this would have consequences for the equation of state of dense nuclear matter and also neutrons stars.

Experimentally, one can not disentangle all these processes by looking directly at HICs and, so far, there are very scarce quantitative measurements on the properties of antikaons.

\begin{figure}[htb]
\centering 
 \includegraphics[width=
 6cm, height= 9.5 cm]{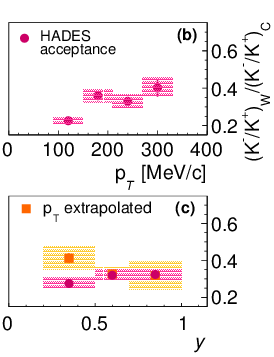}
  \includegraphics[width=11cm]{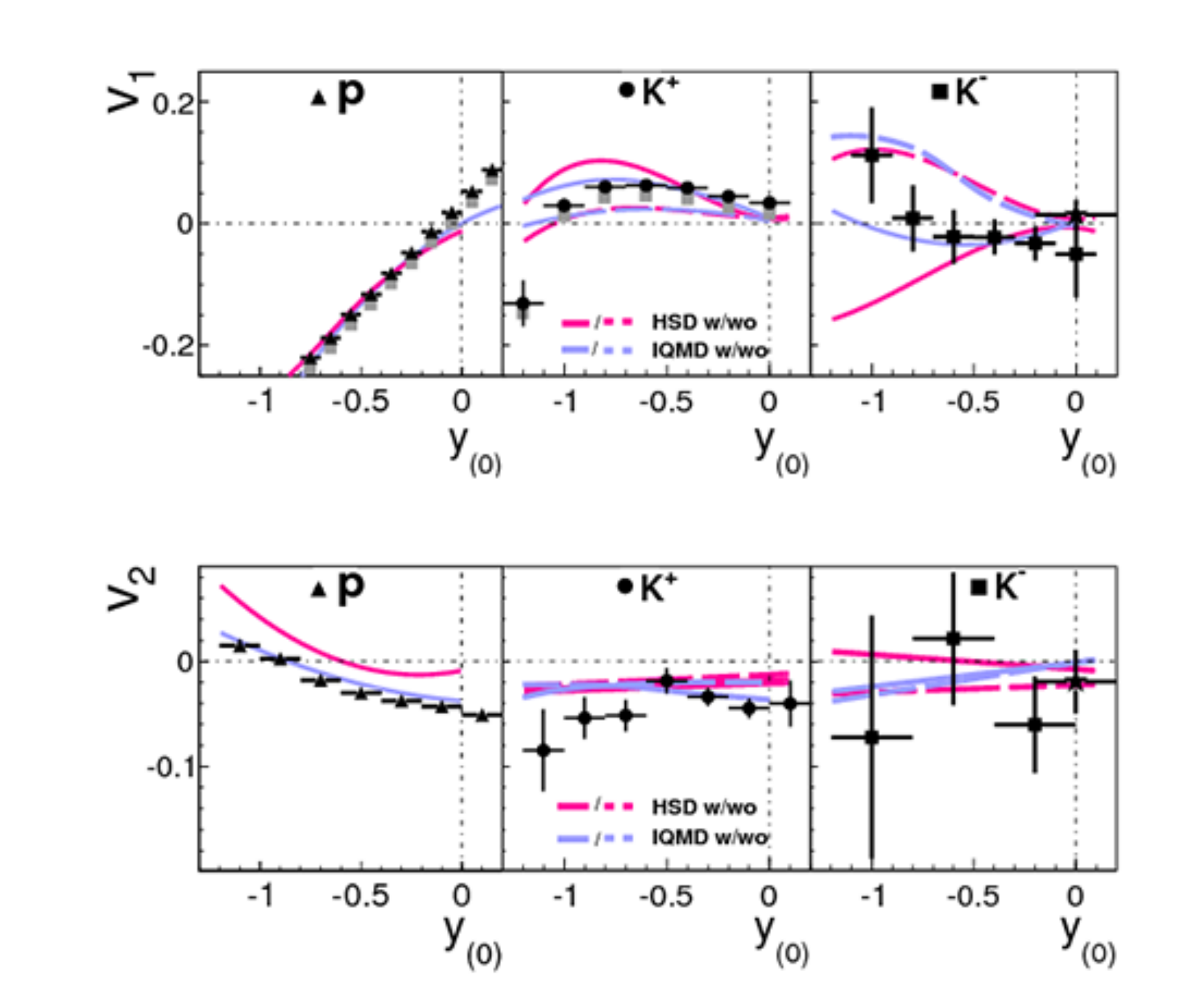}
  \caption{\textit{(Color online)} Left panel: Double ratio $(K^-/K^+)_W/(K^-/K^+)_C$ measured in pion-induced  reactions off  two  different targets  as  a  function of $p_T$ and y \cite{Adamczewski-Musch:2018eik}. Right panel: $v_1$ and $v_2$ flow parameters for protons,  $K^+$ and $K^-$ as a function of y measured in $Ni+Ni$ reactions at 1.9 AGeV by the FOPI collaboration. The experimental  values are compared to theoretical predictions obtained with the HSD and IQMD models \cite{Zinyuk:2014zor}. }
  \label{fig:K-Abs}
\end{figure}

Recent measurements, however, have allowed to quantify the absorption cross sections of $K^-$ off two-, three- and several-nucleons in a quantitative way. The AMADEUS collaboration has published several works on the measurement of low energy $K^-$ absorption in exclusive channels, as $K^-n \rightarrow \Lambda \pi^-$ \cite{Piscicchia:2018rez}, or in semi-inclusive analyses, where either the $\Lambda p$ or the $\Sigma^0 p$ final states are selected \cite{Doce:2015ust,DelGrande:2018sbv}. In the last two works, the multi-nucleons absorption of $K^-$ have been quantified for the first time. Although the results are model dependent, the almost complete phase-space coverage of the measurements allowed to get the quantitative conclusion that multi-nucleons absorption processes are dominant in comparison to two-nucleon absorption. 

Further information on the $K^-$ absorption has been obtained by studying $K^-$ production in $\pi$-induced reactions off light (C) and heavy (W) targets. Thanks to the small mean free path of pions within nuclear matter ($\approx 2$ fm), in $\pi$-induced reactions $K^-$ are produced mainly on the nuclear surface and can, hence, travel the whole nucleus, if emitted in the forward direction. In the analysis published by the HADES collaboration in $\pi^- +C$ and $\pi^- +W$ reactions at a beam  momentum of 1.75 GeV/c \cite{Adamczewski-Musch:2018eik},
the fact that the measurement is conducted at an energy well below threshold allows to study the absorption of $K^-$ in great detail, since the spectra are dominated by primary processes.
In this analysis, the difference in the $K^-$ absorption in $\pi^- +C$ and $\pi^- +W$ reactions has been quantified by taking the $K^+$ as a reference. Indeed, kaons can be re-scattered but can not be absorbed and, by comparing the $K^-/K^+$ ratio measured in the two colliding system, a model independent measurement of the $K^-$ absorption off heavy nuclei with respect to the light nuclei can be extracted. The left panel of Fig.~\ref{fig:K-Abs} shows the double ratio as a function of $p_T$ and $y$. A value around 0.4 is measured while the reference, obtained considering only the different production channels, is $0.96$.  It is clear that if absorption processes are so important, they will tend to shadow the possible modification of the $K^-$ within nuclear matter due to the attractive interaction. 

If we consider HICs, the inclusive momentum spectra of $K^-$ has been measured in several reactions (e.g.: $Ni+Ni$ collisions measured  by FOPI \cite{Piasecki:2018psj}), and since all these measurements are below the $NN$ threshold for $K^-$ production, secondaries processes enter the description of the yields and spectra. The measurement of the flow coefficients $v_1$ and $v_2$
has been carried out as for the $K^+$ (see Sec. \ref{kaonInMatterExp}) to study the sensitivity of this observable to the in-medium potential of $K^-$, but the measurement can not discriminate among different models.

The most recent measurement of kaon and antikaon flow presented in the right panel of Fig.~\ref{fig:K-Abs} shows that transport models can not deliver a solid interpretation of the data due to the limited statistics and also to the not expected trend of the measured flow coefficients. So far, the absorption of $K^-$ is not currently accounted for in transport models, and this is a fundamental ingredient necessary for future studies.

%% file: Content/Phi.tex
\section{$\phi$ Meson in Nuclear Matter}
\label{sec:phi}

\subsection{Experimental detection of $\phi$ in nuclear matter}
\label{phiM}

The investigation of the properties of vector mesons, such as $\rho$, $\omega$ and $\phi$, via direct dilepton decays is one of the most suited way to investigate the spectral shape of these particles, since leptons do not interact strongly with the medium in which they are produced and, hence, carry a direct information about the properties of the vector meson at the decay time. The downside of dilepton decays is, however, the very small branching ratios of vector mesons in these direct channels (BR$ \sim 10^{-5}$).

\begin{figure}[htb]
\centering 
 \includegraphics[width=
 8cm, height= 9.5 cm]{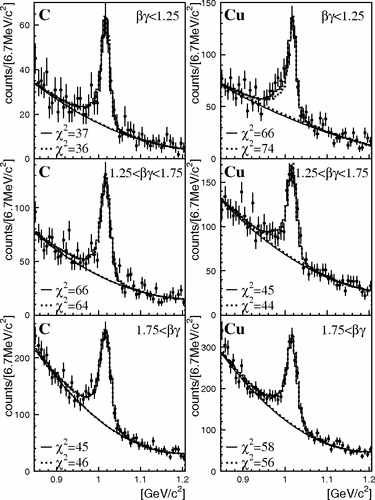}
  \includegraphics[width=9cm]{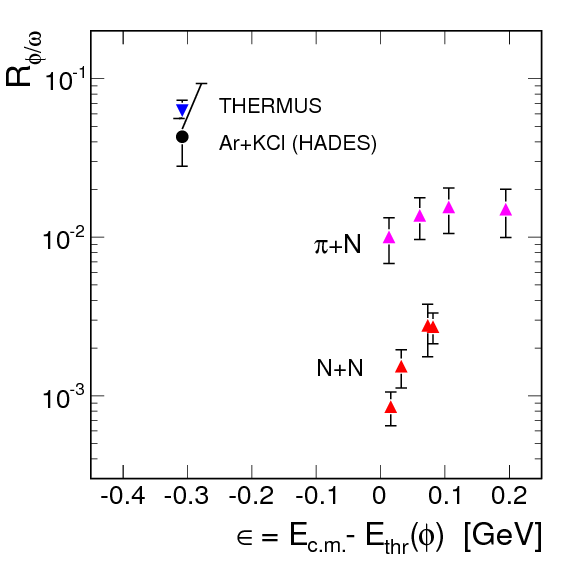}
  \caption{\textit{(Color online)} Left panel: 
  Dilepton invariant mass measured in $p+$C 
  and $p+$Cu reactions at 12 GeV for 
  different dielectron-pair velocities 
  \cite{PhysRevLett.98.042501}. Right panel: $\phi/\omega$ ratio measured in different colliding systems as a function of the $\phi$ excess energy \cite{Agakishiev:2011vf}.}
  \label{fig:phi1}
\end{figure}

Additionally, if we compare the $\phi$ meson life time of $46$ fm with the average fireball lifetime of $10-20$ fm for HICs at different relativistic (GeV) and ultra-relativistic energies ($100$ GeV - TeV), most of the decays will occur outside the fireball, where the vacuum properties of the $\phi$ characterize the spectral shape of the latter. 
The only experimental claim of the observation of a modification of the $\phi$ spectral shape was published in \cite{PhysRevLett.98.042501}. 
There, the $e^+e^-$ invariant mass, measured in $p+$C and $p+$Cu reactions at 12 GeV by the KEK-PS E325 collaboration, 
was used to reconstruct the $\phi$ spectral shape. The
left panel of Fig.~\ref{fig:phi1} shows the
$e^+e^-$ invariant mass measured for reactions off the two targets with different selections on the velocity of the $\phi$ candidate. The shoulder on the left side of the nominal $\phi$ mass, visible in the second upper panel from the left, is interpreted in this analysis as the evidence of the $\phi$ in-medium modification. 
 The slight shift in the invariant mass distribution is interpreted as a mass shift of 3.4\% and a width
increase by a factor of 3.6 at density $\rho_0$ for $\phi$ momenta around 1 GeV/c.
 These results are statistically limited and also model dependent, but so far unique.

Also, the HADES collaboration tried to measured the $\phi$ spectral function using the dilepton 
decay channel in Ar+KCl reactions at 1.76 AGeV, but the collected statistics was not sufficient to allow for such an analysis. On the other hand, the $\phi/\omega$ ratio in the
dilepton channel could be determined and was found to be much larger than the Okubo-Zweig-Iizuka (OZI) rule would allow for. The right panel of Fig.~\ref{fig:phi1} shows the $\phi/\omega$ ratio measured with the HADES spectrometer in comparison with the ratios measured in $\pi N$ and $NN$ reactions as a function of the excess energy $\epsilon=\, E_{c.m.}-E_{thr}$ together with the prediction by the THERMUS code, computed for the colliding system Ar+KCl at 1.76 AGeV. One can see that the measured ratio by HADES agrees with the THERMUS prediction, where the OZI suppression is not accounted for, while it overshoots the systematic measurements in elementary reactions. This observation indicates a violation
of the OZI rule and supports a non negligible in-medium effect for the $\phi$ properties.

A complementary technique with respect to the dilepton decays was pursued by the Spring8-LEPS, CLAS and COSY-ANKE collaborations. There, $\gamma$ and proton-induced reactions off different targets
were used to evaluate to the $\phi$ production and absorption.
The Spring8-LEPS collaboration measured the $\phi\rightarrow K^+K^-$ decay and extracted the production
cross section for photon beams with energies $E_{\gamma}= 1.5-2.4$ GeV, impinging on the following nuclear targets: $Li$, $C$, $Al$ and $Cu$ \cite{Ishikawa:2004id}. The decay
channel into two charged kaons offers the advantage
of a much larger ($\approx$ 50\%) branching ratio with respect to dileptons.
If the $\gamma$+A cross section is measured as a function of A$^\alpha$, the scaling
parameter $\alpha$ is found to be equal to $\alpha= 0.72\pm 0.07$. The analysis relies on a Glauber Monte-Carlo
model to subtract the coherent part of the production. 

This modelling of the cross section is used to extract the $\phi N$ cross section in nuclear
matter. The cross section is found to be equal to $\sigma_{\phi N}=\, 35 ^{+17}_{-11}$mb, which is much larger than the free space estimate of the cross section in $\phi$ photo-production experiments off proton targets, that is equal to $\sigma_{\phi N}^{\rm free}=\, 7.7-8.7$ mb. This enhanced cross section can be also interpreted as an increased in-medium width of about $110$ MeV.

The in-medium $\phi N$ interaction was also analysed by the CLAS collaboration from transparency ratio measurements at JLab \cite{Wood:2010ei}. In this experiment, the $\phi$ mesons were photo-produced (with beam energies up to 4 GeV) on $^2H$, $C$, $Ti$, $Fe$ and $Pb$ targets, and detected via their $e^+e^-$ decay mode.
The transparency ratio $T_A \equiv P_{\rm out}$ is given by 
\begin{equation}
    T_A \equiv P_{\rm out}=\frac{\sigma_{ A}}{A \sigma_{N}}
    \label{eq:transparency},
\end{equation}
that is, the ratio of the nuclear $\phi$ production cross section divided by the mass number A times the cross section into a free nucleon. This can be interpreted as the probability
of a $\phi$ meson getting out of the nucleus, and it is related to the absorptive part of the $\phi$-nucleus potential and, thus, to the $\phi$ width in the nuclear medium.

From an analysis of the transparency ratios normalized to Carbon within
a Glauber model, values of $\sigma_{\phi N}$ in the range of 16 -
70 mb were extracted for an average $\phi$ momentum of $\approx$ 2 GeV/c. Also, in this case, the extracted
$\sigma_{\phi N}$ is much larger than the free one.

The $\phi$ production was also analyzed
in proton-induced reactions off different nuclear targets by the ANKE collaboration. The $p+A$ collisions
at $E_{\rm beam}$= 2.83 GeV, with $A= C, Cu, Ag$, and $Au$,
were measured at the COSY accelerator by the ANKE detector, and the final state $\phi\rightarrow K^+ K^-$
was studied \cite{PhysRevC.85.035206}. The discussion of these results is presented in the next Sec.~\ref{sec:phi-theory}.

\begin{figure}[htb]
\centering 
 \includegraphics[width=
 8cm, height= 10cm]{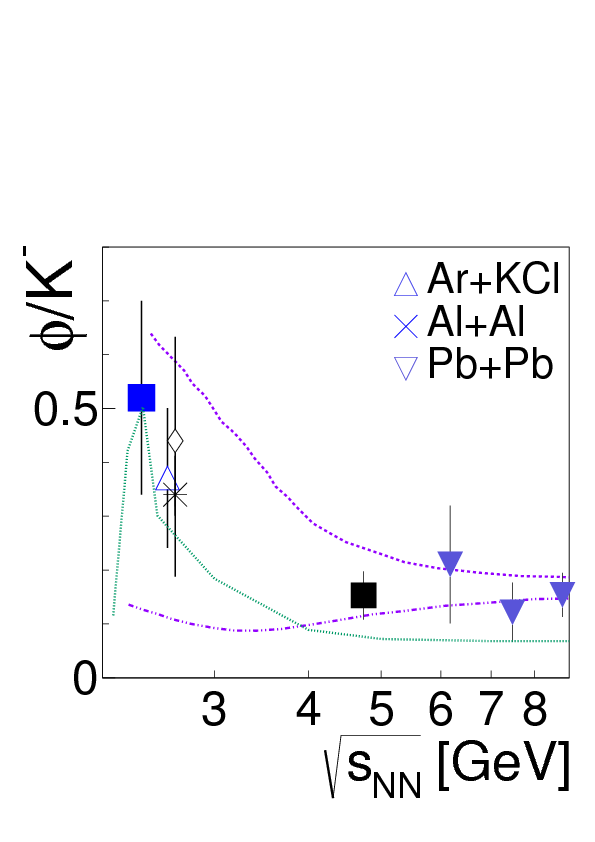}
 \includegraphics[width=8cm, height=8cm]{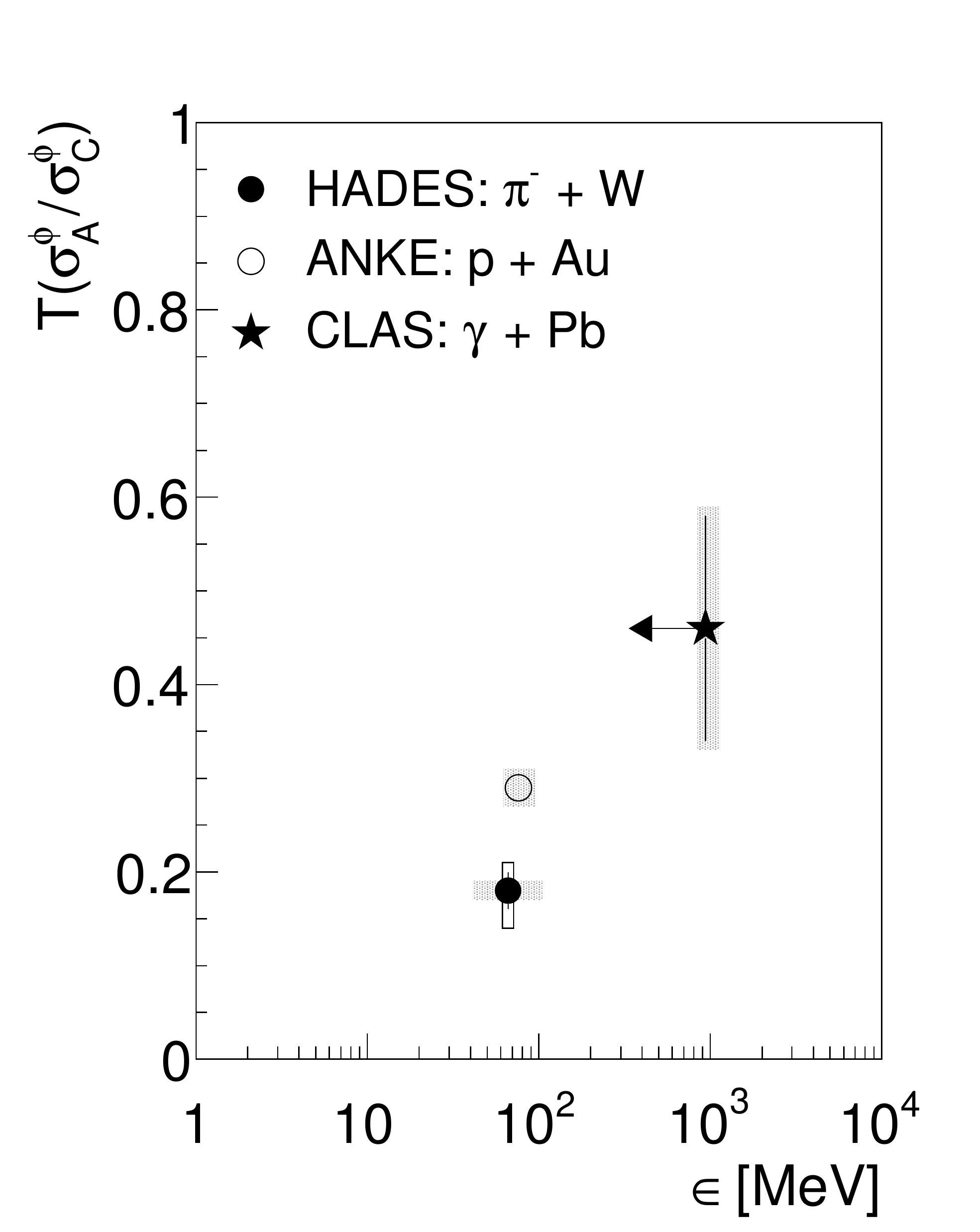}
  \caption{\textit{(Color online)} Left 
  panel: $\phi/K^-$ ratio measured in 
  HICs as a function 
  of the center-of-mass energy 
  \cite{Adamczewski-Musch:2017rtf}. Right panel: $\phi$ transparency 
  ratio normalized to Carbon, measured for different targets and different colliding systems.}
  \label{fig:phi4}
\end{figure}

The absorption of $\bar{K}$ in nuclear matter via one- or multi-nucleons processes was discussed in Sec. ~\ref{AntiKaonMatterExp}, and many experiments demonstrated that these processes do play an important role in the understanding of the in-medium properties of antikaons.
Since the branching ratio of the decay $\phi\rightarrow K^+ K^-$ is rather large ($\sim$ 50\%), the measurement of the $\phi/K^-$ ratio was first used in HICs and then in pion-nucleus collisions to quantify the feed-down from the $\phi$ into the $K^-$ inclusive production.

The left panel of Fig.~\ref{fig:phi4} shows the measured $\phi/K^-$ ratio in different HICs as a function of the center-of-mass energy. One can see that a rather small ratio around 0.2 is measured for $\sqrt{s_{NN}}$ $>$ 4 GeV. This can be explained considering the larger abundance of $u$ and $d$ quarks w.r.t  $s$ quarks in the pre-hadronisation phase formed in HICs by statistical hadronization models \cite{Cleymans:2000ck}.

 For smaller energies, the $\phi/K^-$ ratio increases and the squared symbols in the left panel of Fig.~\ref{fig:phi4} represent the measurements for Au+Au collisions. This enhancement of the ratio was interpreted by the HADES \cite{Lorenz:2010zz} and FOPI collaborations as the evidence of a strong feed-down from the $\phi$ production to the $K^-$ production, down-scaling the importance of processes driven by strangeness-exchange, such as $\Lambda +\pi \rightarrow K^- + N$ for the $K^-$ production \cite{Piasecki:2014zmsF,Agakishiev:2009ar}.
This interpretation was also supported by the fact that, by looking at the $p_T$ distributions obtained for $K^+$, $K^-$ and $\phi$ in $Au+Au$ collisions at 1.23 AGeV measured by HADES and interpreting the slope of the spectra as kinematic temperatures, the $K^-$ slope resembles a mixture of $K^+$ and $\phi$ slopes, with weights consistent with the $\phi/K^-$ ratio \cite{Adamczewski-Musch:2017rtf}.

On the other hand, $\pi$-nucleus reactions at beam energies of 1.65 GeV showed that the same $\phi/K^-$ ratio is obtained when comparing $\pi^- +C$ and $\pi^- +W$ reactions \cite{Adamczewski-Musch:2018eik}. Since a strong difference in the absorption of $K^-$ is observed in the two targets, the constant $\phi/K^-$ ratio speaks for a strong $\phi$ absorption as well. This observation does not match completely the interpretation of the ratio in HICs mentioned before. Indeed, it demonstrates that the in-medium properties of $\phi$ and $K^-$ are connected and the feed-down process together with possible in-medium modifications of the yields are both contributing.

The right panel of Fig.~\ref{fig:phi4} shows the $\phi$ transparency ratio for heavy targets ($W$, $Pb$ and $Au$) normalized to Carbon for $p$-, $\gamma$- and $\pi$-induced reactions as a function of the excess energy. One can see that the ratio measured in $\pi$-induced  is lower than for p-induced reactions, because in the latter secondary production mechanisms within the nucleus shorten the path length of the $\phi$ within it. The discrepancy is, however, very small, and the scaling of the transparency ratio with the excess energy is expected from simple phase space considerations.

In general, although the theoretical interpretation of the transparency ratios measured for $\phi$ in different targets and using different beams is model dependent (as we will discuss in  Sec.~\ref{sec:phi-theory}), we can conclude that the transparency ratio is the most favored experimental scenario, since all the measurements are consistent with a strong absorption of the $\phi$ meson already at saturation density.

\subsection{Theory of $\phi$ in nuclear matter}
\label{sec:phi-theory}

The interest in the modification of the $\phi$ properties in dense matter started with the earlier predictions on the shift of the $\phi$ mass due to the famous Brown-Rho scaling \cite{Brown:1991kk}. Later on, the $\phi$ in matter was studied by means of QCD sum rules  \cite{Asakawa:1994tp,Zschocke:2002mn,Gubler:2018ctz}. In these calculations, a correlator function matched to the $\phi$ spectral function is treated in the framework of the operator product expansion, which incorporates the short (perturbative contribution) and long-distance quark-gluon interactions (non-perturbative terms, such as quark condensates). In this way, the modifications of the spectral function in dense nuclear matter would be correlated to the modification of the quark condensates in medium and, hence, to the restoration of chiral symmetry and to the origin of the hadron masses. However, a clean interpretation of the chiral restoration is shadowed by the possible hadron-hadron scattering processes in matter that are reflected in the spectral function. 

Several theoretical approaches have analyzed the in-medium $\phi$ properties in terms of hadronic processes. They range from effective chiral models in matter \cite{Kuwabara:1995ms} or at finite temperature \cite{Song:1996gw} to NJL models in hot matter \cite{Blaizot:1991af}. In some of these analyses, it has been concluded that the self-energy of the $\phi$ coming from the $K \bar K$ decay is of fundamental importance for understanding its mass shift. Indeed, the $\phi$, having a hidden strangeness content, strongly couples to the $\bar K K$ and, hence, its in-medium dynamics is governed by its decay into the light pseudoscalars, whereas its coupling to the nucleon is forbidden by the OZI rule.

However, soon after, it was realised that the medium effects are more important for the  width of the $\phi$ meson, coming from the dressing of the  $\bar K K$ decay channel. In Ref.~\cite{Ko:1992tp}, within the vector meson dominance model and taking into account the density-dependent kaon effective mass given by linear $\chi PT$,  a slight change of mass was found while a drastic change in the width was observed for large densities. Later on, the full spectral features of the $K$ and $\bar K$ were incorporated confirming the important broadening of the $\phi$ in dense medium \cite{Klingl:1997tm,Oset:2000eg,Cabrera:2002hc}. These works treat the $K$ meson as a quasiparticle with a mass corrected by the $T \rho$ approximation. As for the $\bar K$, the self-energy incorporates the $S$- and $P$-wave contributions, although differences arise in the determination of the $\bar K$ self-energy due to the treatment of the $P$-wave contribution, recoil corrections, and others. The estimate of Ref.~\cite{Oset:2000eg} is in a band between 20-30 MeV for the $\phi$ width at normal nuclear matter density, smaller than the value of 45 MeV of Ref.~\cite{Klingl:1997tm}. In Ref.~\cite{Cabrera:2002hc} some technical novelties were introduced with respect to Ref.~\cite{Oset:2000eg}, such as introducing additional mechanisms required by gauge invariance as well as performing a finer study of the relativistic recoil corrections considered in the $P$-wave $\bar K$ self-energy. A small mass shift of around 8 MeV and a width of about 30 MeV at normal nuclear density were reported, close to the one in Ref.~\cite{Oset:2000eg}.

In view of the modification of the properties of the $\phi$ in matter, several ways of testing them were presented. Reactions such as $\pi^- A \rightarrow \phi X$ \cite{Klingl:1997tm} and $\gamma A \rightarrow \phi X$ \cite{Oset:2000na} were proposed, where small $\phi$ momenta should be accessible with the help of Fermi motion and the rapidity distribution of the $\phi$ could be connected to rescattering processes. However, in Ref.~\cite{Muhlich:2002tu} it was argued that $\gamma-$induced reactions would present the following problems: the small photon-nucleus cross section because of low momenta; the small kaon mean free path that only probes small nuclear densities and distorts the $K^+ K^-$ invariant mass spectrum due to quasielastic $K^+N$ and $K^-N$ scatterings; and the presence of the nuclear Coulomb potential that does not allow to gather information on the in-medium $\phi$ properties.   Similar arguments hold also true for proton-induced reactions.

It was then determined that a more convenient scenario would be to study the imaginary part of the $\phi$ self-energy in nuclei by analysing the transparency ratio. Predictions of $\phi$ photoproduction had been put forward in Ref.~\cite{Cabrera:2003wb}, and it had been found that the $\phi$ survival rate is drastically reduced for increasing values of A. This is due to a modified $\phi$ width at normal nuclear saturation density up to six times larger than the vacuum width. 

This theoretical calculation of the transparency ratio \cite{Cabrera:2003wb} was carried out to match the forthcoming set up of the LEPS at Spring8 experiment \cite{Ishikawa:2004id}. However, a posterior analysis in Ref.~\cite{Muhlich:2005kf} within the semi-classical BUU transport approach showed that the experimental transparency ratio can not be fully explained just by the attenuation of the $\phi$ in nuclei due to the $K \bar K$ channel. A value of $\phi N \approx$ 27 mb  is needed to reproduce the experimental nuclear transparency ratio as well as the $A-$dependence of the total $\phi$ meson yield. This value, however, is considerably higher than the expected $\phi N$ cross section in vacuum according to the OZI rule. 


Apart from $\phi$ photoproduction, the $A$ dependence of the cross section of $\phi$ mesons in nuclei at energies just above threshold in proton-nucleus collisions was also analyzed in Ref.~\cite{Magas:2004eb}, in view of the experimental performance at facilities such as the forthcoming COSY-ANKE \cite{PhysRevC.85.035206}. A strong $A$ dependence was  found, being linked to the distortion of the incident proton and to the absorption of the $\phi$ as it leaves the nucleus.  This latter process reduces the cross section by a factor two in heavy nuclei, thus showing that the $A$ dependence of the cross section can be a good probe to extract information on the $\phi$ width in matter.

\begin{figure}[htb]
\centering 
 \includegraphics[width=0.35\textwidth,angle=-90]{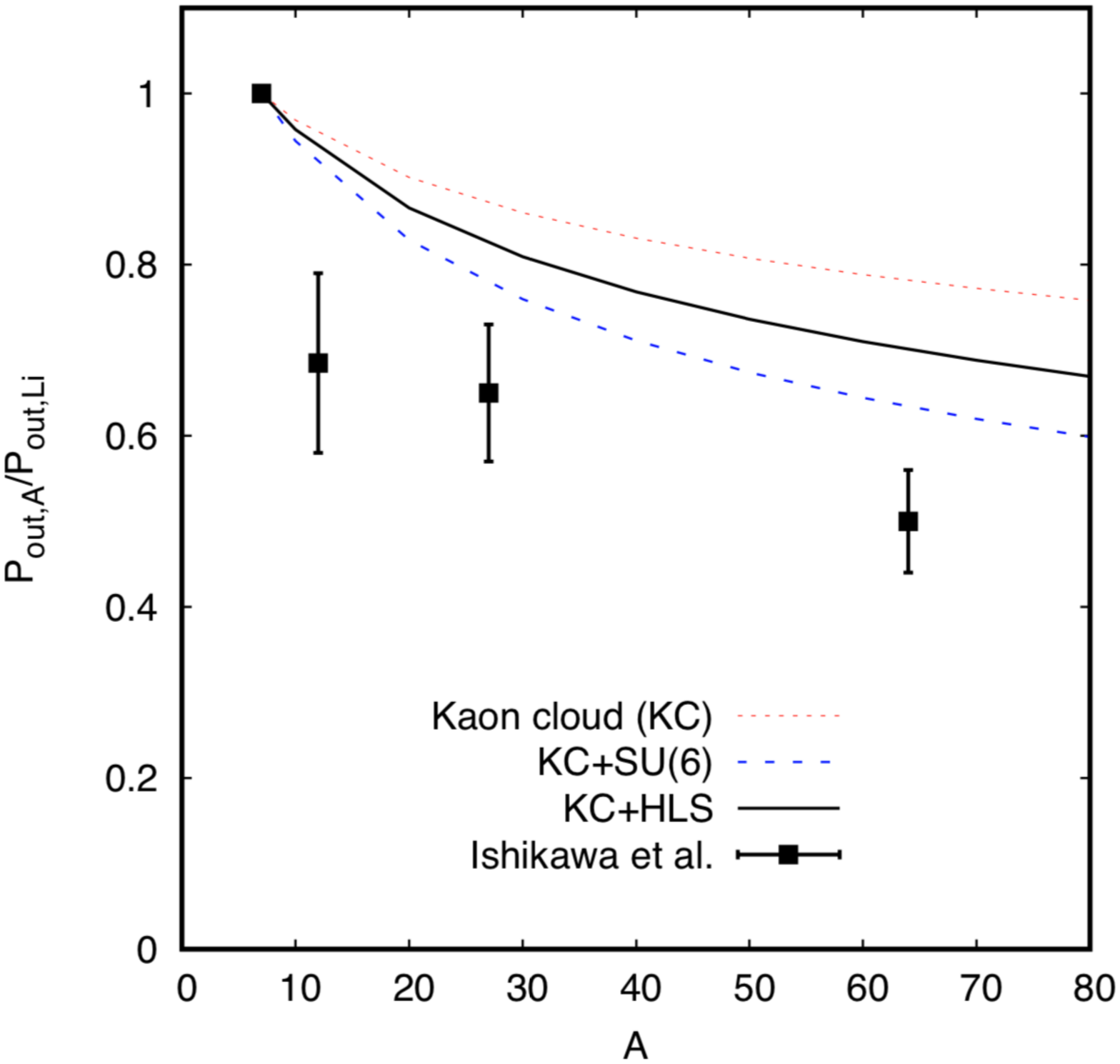}
 \includegraphics[width=0.35\textwidth,angle=-90]{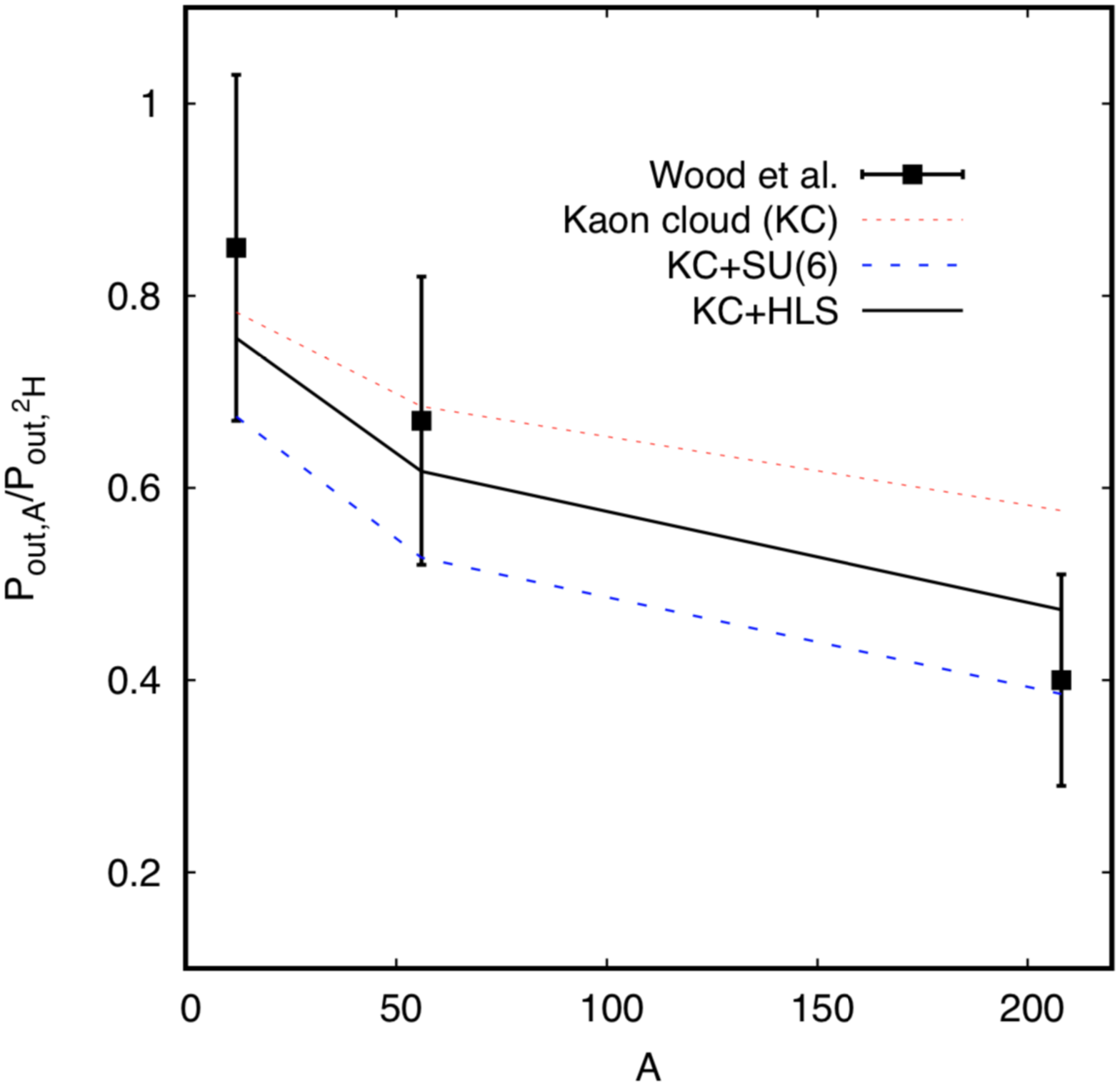}
 \includegraphics[width=0.35\textwidth,angle=-90]{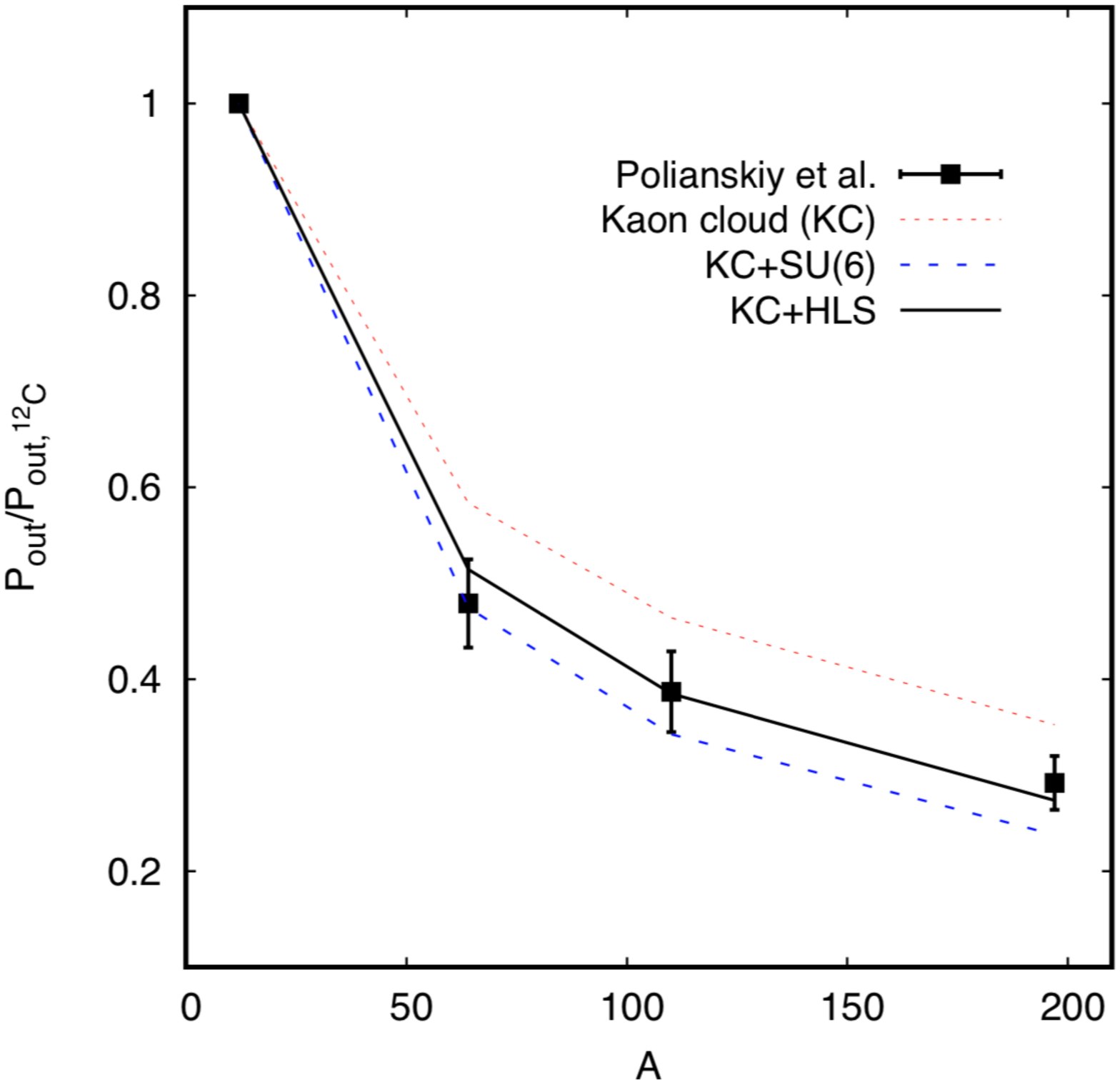}
  \caption{ \textit{(Color online)} $\phi$ transparency ratio compared to Spring8-LEPS photoproduction data normalized to the $^7$Li target \cite{Ishikawa:2004id} (upper left panel), CLAS photoproduction data normalized to the $^2$H target \cite{Wood:2010ei} (upper right panel) and results from COSY-ANKE proton induced reactions normalized to $^{12}$C target \cite{Polyanskiy:2010tj} (lower panel). Three theoretical scenarios are considered: only the $\bar K K$ cloud (KC), the KC and the resonant $\phi N$ scattering from the SU(6) model (KC+SU(6)), and the KC and the resonant $\phi N$ scattering from the hidden-gauge scheme (KC+HLS). These plots are taken from Ref.~\cite{Cabrera:2017agk}.}
  \label{fig:phicomp}
\end{figure}

In view of the more recent experiments on proton-induced experiment by COSY-ANKE \cite{PhysRevC.85.035206} and photoproduction by CLAS \cite{Wood:2010ei} and given that the previous works could not explain the Spring8-LEPS experimental data \cite{Ishikawa:2004id}, there has been a recent revision of the theoretical models. For example, QCD sum rules schemes \cite{Gubler:2016itj} and the QMC model \cite{Cobos-Martinez:2017vtr} have been revisited in order to perform an improved analysis of the $\phi$-$K \bar K$ channel. However, it has been argued that the interaction of $\phi$ with nucleons could account for the missing broadening. Indeed, the fact that the hidden strangeness in the $\phi$ meson can be exchanged with the nucleon by the coupling to $K^*\Lambda$ and $K^*\Sigma$ without violation of the OZI rule has opened a new possibility of accounting for the missing attenuation of the $\phi$ meson in matter \cite{Cabrera:2016rnc}. In Ref.~\cite{Cabrera:2016rnc} two different coupled-channel approaches for the $\phi N$ interactions have been considered: one based on the hidden local symmetry approach (HLS), and a second one where the interaction of vector mesons with baryons has been calculated using a SU(6) spin-flavor symmetry extension of the SU(3) meson-baryon chiral Lagrangian, in the form of a generalized $S$-wave WT interaction.
 
The presence of several broad dynamically generated $N^*$ states in the vicinity of the $\phi N$ threshold, with a non-negligible coupling to it, leads to a sizable contribution to the $\phi$ self-energy, specially for the second approach based on the SU(6) extended WT interaction. Together with the contribution of the $\bar K K$ channel, the in-medium $\phi$ decay width increases up to $\sim 70$ MeV at saturation density, growing with momentum up to 100 MeV for 400-500 MeV/c, and decreasing slowly for higher momenta. An attractive mass shift of up to $\sim 35$ MeV has been also obtained. As a consequence, the $\phi$ spectral function exhibits an important broadening with a shift of the quasiparticle peak to lower energies than the bare mass, whereas developing a second mode above the bare $\phi$ mass coming from the $N^*$ resonant-hole excitations.

The effect of the width of the $\phi$ meson in the $\phi$ transparency ratio due to the direct resonant $\phi N$ scattering together with the kaon decay mechanisms has been explored in the follow-up paper of Ref.~\cite{Cabrera:2017agk}, confronting the results with the CLAS and Spring8-LEPS data on photoproduction, and COSY-ANKE results on proton induced reactions in nuclei. The $\phi$ transparency ratio for Spring8-LEPS (normalized to $^7$Li target), for CLAS (normalized to $^2$H target) and for COSY-ANKE (normalized to $^{12}$C target) are shown in Fig.~\ref{fig:phicomp}, where the Spring8-LEPS, CLAS and COSY-ANKE data are compared to three theoretical scenarios: including only the $\bar K K$ cloud (KC), taking into account the KC and the resonant $\phi N$ scattering from the SU(6) model (KC+SU(6)), and considering the KC and the resonant $\phi N$ scattering from the hidden-gauge scheme (KC+HLS). The authors have used a $\phi$ meson average momentum of 1.8 GeV/c to confront their theoretical results with the Spring8-LEPS experimental data (the photon beams at  Spring8-LEPS have momenta from 1.5 to 2.4 GeV/c), whereas a similar average momentum of 2 GeV/c is used by the theoretical work to analyze CLAS data. A good agreement with the experimental photoproduction data of CLAS is found, as seen in the upper right panel of Fig.~\ref{fig:phicomp}. However, an even stronger $\phi$ absorption will be required in order to reproduce the Spring8-LEPS photoproduction data, as seen in upper left panel of Fig.~\ref{fig:phicomp}. 
The proton induced transparency ratio from COSY-ANKE has been well reproduced (lower panel of Fig.~\ref{fig:phicomp}). However, in this case, only small changes in the transparency ratio are obtained even if the authors substantially change the $\phi$ self-energy, thus,  hinting that the transparency in proton induced reactions is dominated by other effects, such as shadowing.

\begin{figure}[htb]
\centering 
 \includegraphics[width=
 13cm, height= 11.5cm cm]{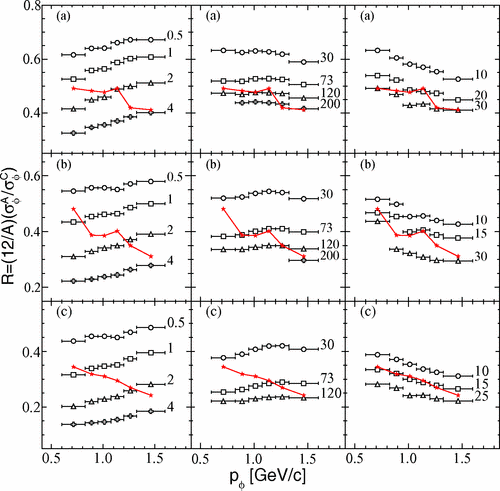}
  \caption{\textit{(Color online)} $\phi$ transparency ratio with respect to the Carbon target for the $Cu$, $Ag$ and $Au$ targets, shown as a function of the $\phi$ momentum in comparison with predictions obtained from three different models \cite{PhysRevC.85.035206}. See text for details.}
  \label{fig:phi3}
\end{figure}

The ANKE collaboration has moreover provided the missing information on the momentum dependence of the $\phi$ absorption below 1.5 GeV/c. For these data, it was possible to study the momentum dependence of the transparency ratio for $\phi$ momenta between $0.5$ and $1.7$ GeV/c, and the results are shown in Fig.~\ref{fig:phi3}. There, the transparency ratio with respect to the Carbon target is shown for the $Cu$, $Ag$ and $Au$ targets as a function of the $\phi$ momentum in comparison with predictions obtained from three different models. In the left panels, curve denotes with 1 results from using the predictions of \cite{Cabrera:2003wb} for the imaginary part of the $\phi$ self-energy in nuclear matter. The other curves correspond to calculations with this self-energy multiplied by factors of 0.5, 2 and 4. In the middle panels, the calculations are performed with different values of a momentum-independent $\phi$ width in its rest frame at density $\rho_0$, and the values are noted next to the curves (in MeV). And,  in the right panels, the computations are produced using the indicated absorption cross section $\sigma_{\phi N}$ (in mb).  One can see that two models fail to reproduce the momentum dependence of the transparency ratio, while one model (right panels) manages to match the data. 

All these results support the idea that the direct $\phi N$ processes are relevant together with the coupling to $\bar K K$ cloud for the description of the $\phi$ meson width in matter. Nevertheless, there are still substantial uncertainties in the description of the $\phi N$ interaction, as seen in the panels of Figs.~\ref{fig:phicomp} and \ref{fig:phi3}. Thus, apart from the need of a better description from the theoretical side, also other experimental scenarios to test $\phi$ in matter are welcome, such as the possible formation of $\phi$-mesic nuclei, as pointed out in Refs.~\cite{YamagataSekihara:2010rb,Cobos-Martinez:2017vtr}.


%% file: Content/Hyperon.tex
\section{Hyperon-Nucleon and Hyperon-Hyperon Interactions}
\label{sec:YN-YY}
\subsection{Experimental searches}
\label{YN_YY_Exp}

The hyperon-nucleon ($YN$) interaction can be tested experimentally via scattering experiments
employing secondary hyperon beams \cite{Eisele:1971mk,Alexander:1969cx,SechiZorn:1969hk},
impinging on hydrogen targets, or by means of the measurement of the correlation in the momentum space 
for hyperon-proton ($Yp$) pairs produced at colliders exploiting the femtoscopy technique \cite{Mihaylov:2018rva,Acharya:2018gyz,FemtopXi,Adamczewski-Musch:2016jlh}.

Scattering experiments were performed at the CERN PS, using the Saclay hydrogen bubble chamber exposed to a secondary $K^-$ beam. The stopped $K^-$
particle and the resulting charged products can be identified looking at the recorded photographs
of the bubble chamber reactions. Typically, hyperons are produced in the following reactions: i) $K^-+p\rightarrow \Lambda +\pi^0$, ii) $K^-+p\rightarrow \Sigma^0 (\rightarrow \Lambda +\gamma ) +\pi^0$, iii) $K^-+p\rightarrow \Sigma^{\pm} +\pi^{\mp}$. The following scattering of hyperons off protons and their weak decays are then searched and tagged.

\begin{figure}[htb]
\centering 
 \includegraphics[width=
 8cm, height= 10cm ]{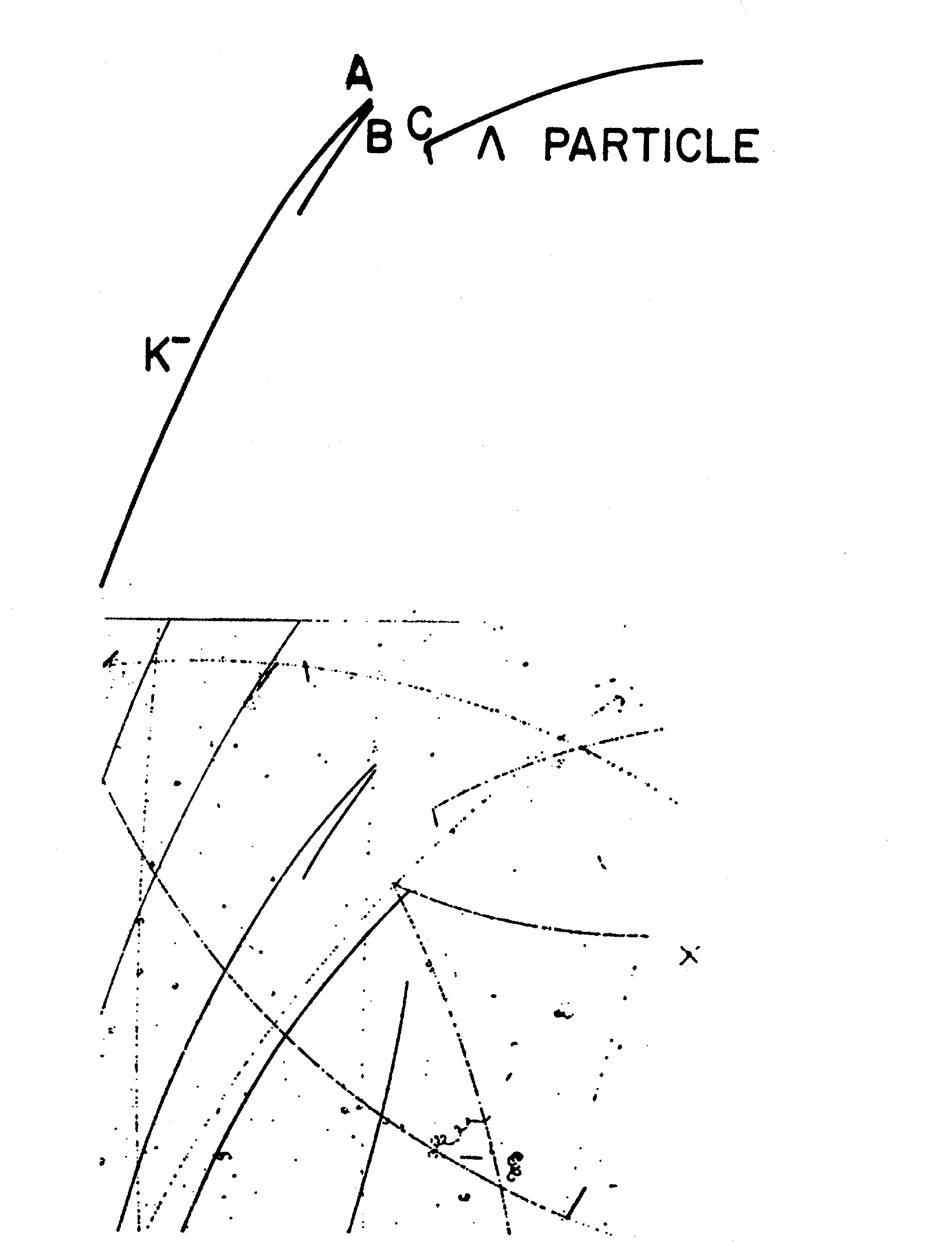}
 \includegraphics[width=8cm, height=7cm]{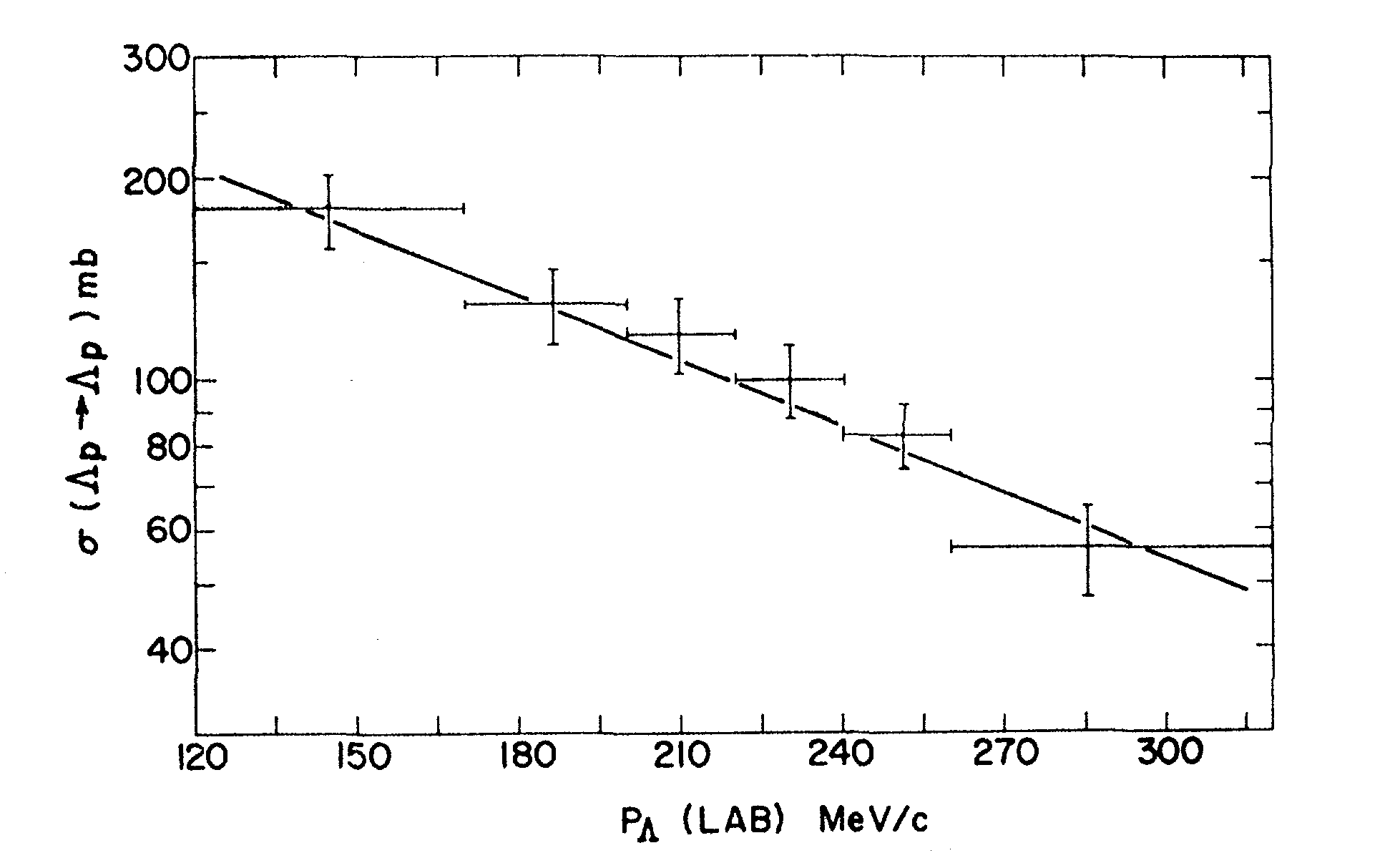}
  \caption{\textit{(Color online)} Left panel: Photograph containing a $\Lambda p$
elastic scattering event and a line drawing of the same event \cite{SechiZorn:1969hk}. Right panel: $\Lambda p$ elastic scattering cross sections as a function of the beam momentum \cite{Alexander:1969cx}.}
  \label{fig:hyp1}
\end{figure}

The left panel of Fig. \ref{fig:hyp1} shows an example of the emulsion photograph for the reaction $K^- + p \rightarrow \Lambda + n$ at the vertex A, followed by a $\Lambda +  p \rightarrow \Lambda + p $
scattering in B and the decay $\Lambda \rightarrow \pi^- +p$ in C \cite{SechiZorn:1969hk}. This is obtained by retaining the imprint of charged tracks in the liquid hydrogen target on an emulsion plate.

This method allows to measure the differential elastic cross section as a function of the hyperon momentum in the laboratory reference system.  
One example of the obtained differential cross sections for the $\Lambda p$ case is shown in the right panel of Fig.~\ref{fig:hyp1} as a function of the laboratory momentum of the scattered hyperon. One can see that the statistics of the data is rather limited. Such cross section measurements could be carried out for the $\Lambda/\Sigma^{\pm}+p$ elastic scattering channels and constitute so far the most stringent constraints for the available theoretical models. Indeed, the J-PARC E40 experiment is going to asses $\Sigma^{\pm} p$ elastic scattering by means of the $p(\pi^{\pm},K^+)\Sigma^{\pm}$ reaction,
followed by an elastic scattering of the charged hyperons with the protons in the extended $LH_2$
target.

Alternatively to scattering experiments, as discussed for the $K^- p$ case in Sec.~\ref{sec:KNexperiment}, the femtoscopy technique can be exploited to investigate different $Yp$ and hyperon-hyperon ($YY$) channels, even those that are not accessible via scattering experiments. 

\begin{figure}[htb]
\centering 
 \includegraphics[width=
 9.cm, height= 8cm ]{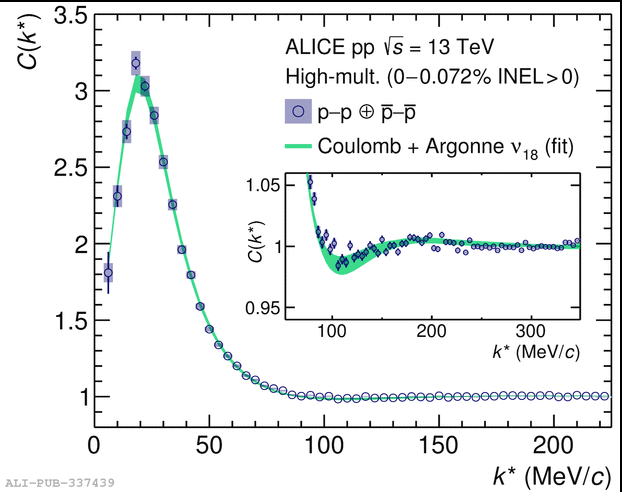}
 \includegraphics[width=9.cm, height=8cm]{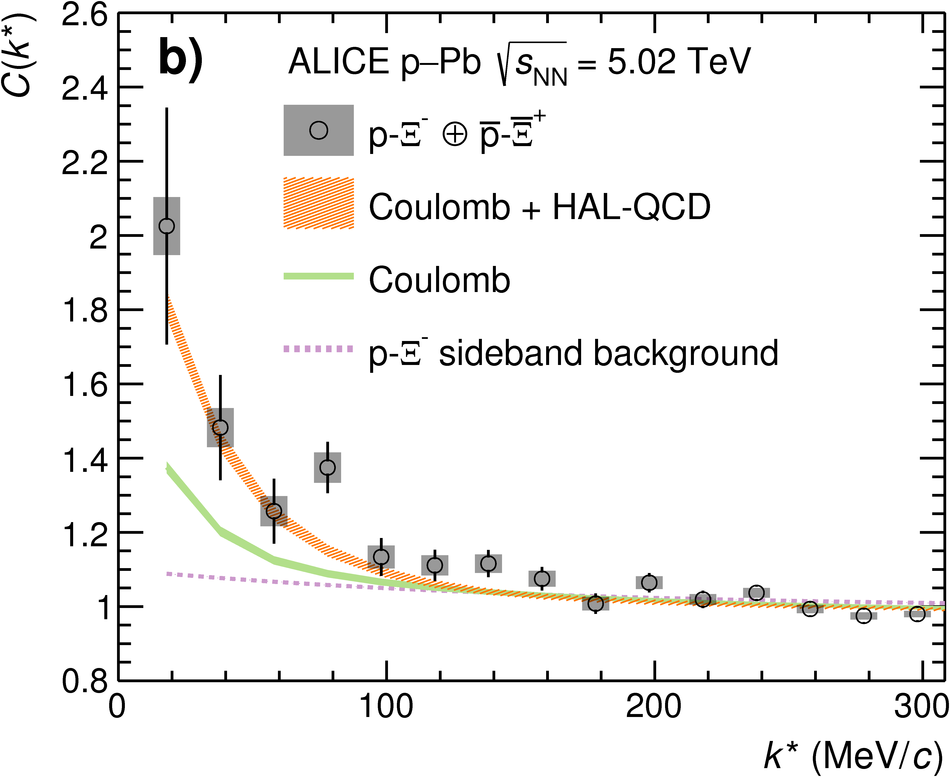}
  \caption{\textit{(Color online)} Left panel: $pp$ correlation measured in $pp$ collisions at $\sqrt{s}= 13$ TeV at the LHC by the ALICE collaboration \cite{femtoSigma0}. Right panel: $p \Xi^-$ correlation function measured in pPb  collisions at $\sqrt{s}= 5.02$ TeV at the LHC by the ALICE collaboration \cite{FemtopXi}.}
  \label{fig:hyp2}
\end{figure}

The sensitivity of the correlation function to the interaction for baryon-baryon pairs is demonstrated by the $pp$ correlation measured in different colliding systems by ALICE. The
left panel of Fig.~\ref{fig:hyp2} shows the measured correlation function in $p+p$ collisions at 13 TeV \cite{femtoSigma0}. 
The experimental correlation function is compared to the 
prediction obtained by using the Coulomb and the AV18 strong
potentials to solve a non-relativistic Schr\"odinger equation and obtain the $pp$ relative wave function, while leaving the radius of the source as a free parameter.
The obtained radius for the inclusive correlation function is equal to $1.197^{+0.035}_{-0.038}$ fm, demonstrating the small size of the source from which hadrons are emitted. 

The right panel of Fig.~\ref{fig:hyp2} shows the correlation 
function of the $p\Xi^-$ final state measured in $pPb$ 
collisions at $\sqrt{s_{NN}}$= 5.02 TeV, together with the 
predictions obtained assuming only the Coulomb interaction 
(green histogram) and considering the Coulomb and strong
interactions predicted by the HAL QCD group with a 
lattice-based approach \cite{HAL1}. One can see
that the experimental correlation function is significantly 
above the Coulomb-only prediction and this demonstrates, in a model-independent way, the presence of an attractive strong $p\Xi^-$ interaction. One can also see that the lattice predictions are  compatible with the measured correlation.

This is the first and also only possible way to measure the 
two body $p\Xi^-$ interaction precisely, since $\Xi^-$ beams are very difficult to handle. The $p\Lambda$ correlation has also been measured and the current available statistics will allow to improve by a factor ten the precision of the measurement with respect to the above-mentioned scattering data.

\begin{figure}[htb]
\centering 
\includegraphics[width=9.cm, height=8cm]{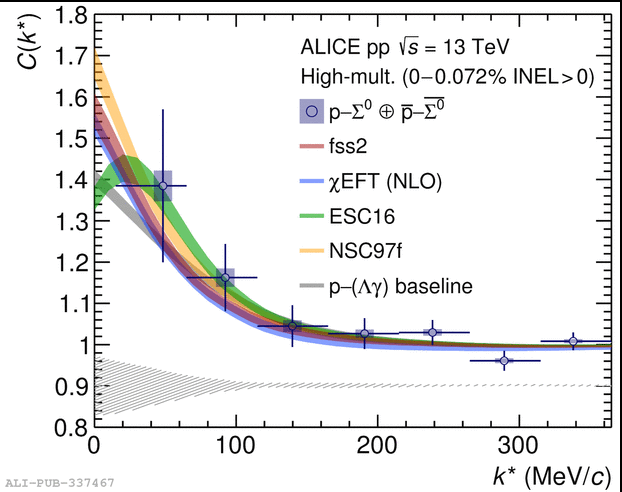}
 \includegraphics[width= 9.cm, height= 8cm ]{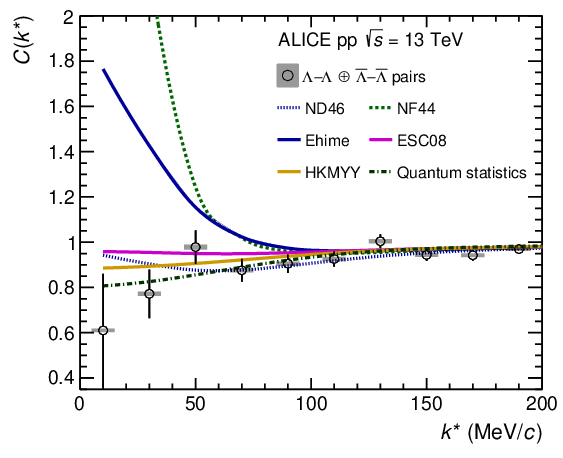}
  \caption{\textit{(Color online)} Left panel: $ p \Sigma^0$ correlation function measured in $pp$ collisions at $\sqrt{s}= 13$ TeV at the LHC by the ALICE collaboration \cite{femtoSigma0}.
  Right panel:  $\Lambda\Lambda$ correlation measured in $pp$ collisions at $\sqrt{s}= 13$ TeV at the LHC by the ALICE collaboration \cite{FemtoLambdaLambda}.}
  \label{fig:hyp3}
\end{figure}

Two further examples of rare $pY$ and $YY$ combinations, which interactions can be studied via correlation 
functions, are the $p\Sigma^0$ and the $\Lambda\Lambda$. The two panels  of Fig.~\ref{fig:hyp3} show the measured 
$p\Sigma^0$ \cite{femtoSigma0} and the $\Lambda\Lambda$ \cite{FemtoLambdaLambda}
correlations together with different theoretical predictions.

The dominant (BR= 100\%) $\Sigma^0\rightarrow \Lambda + \gamma$ 
decay can be reconstructed within the ALICE detector, but the 
signal to background ratio is rather moderate. For this 
reason, a correlation function due to the background must be 
computed as baseline. The gray histogram indicated in the left 
panel of Fig.~\ref{fig:hyp3} shows the resulting background 
contribution to the correlation function. The fact that the
total  $p\Sigma^0$ correlation lays above this histogram
implies a shallow, probably attractive, interaction.
It is clear that the current statistics does not allow 
to discriminate among the different models, but new measurements planned during the Run3 and Run4 data taking at the LHC should improve consistently the situation.

\begin{figure}[htb]
\centering 
 \includegraphics[width=
 11.cm, height= 7cm ]{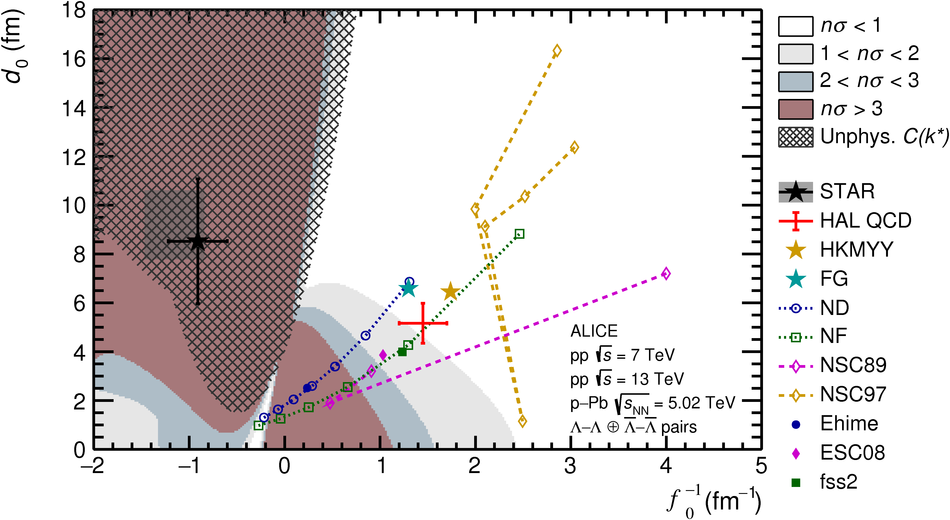}
 \caption{\textit{(Color online)}  Exclusion plot for the $\Lambda\Lambda$ scattering parameters obtained comparing the
 predictions for the correlation function using the Lednicky formalism to the measured $\Lambda\Lambda$ correlations in $pp$ at 7 and 13 TeV and $pPb$ collisions at the LHC \cite{FemtoLambdaLambda}.}
  \label{fig:hyp4}
\end{figure}

The bare $\Lambda\Lambda$ interaction is particularly interesting, 
because of the prediction of the H-dibaryon as possible $\Lambda\Lambda$ bound state \cite{Jaffe:1976yi}.
 Several experimental collaborations have been involved in the search for this state in the decay channels $\mathrm{H\rightarrow \Lambda p \pi}$ and $\mathrm{H\rightarrow \Lambda \Lambda}$, in nuclear and elementary ($\mathrm{e^-e^+}$) collisions, but no evidence has been found \cite{Adam:2015nca,Kim:2013vym,Chrien:1998yt}. 
 
The investigation of the $\Lambda\Lambda$ interaction by means of the femtoscopy method was initiated by the STAR collaboration employing
Au+Au collisions at $\sqrt{s_{NN}}=200$ GeV \cite{Adamczyk:2014vca}, and the scattering parameters extracted in this analysis were 
consistent with a repulsive $\Lambda\Lambda$ interaction. A reanalysis of the data \cite{PhysRevC.91.024916} showed that the background was not correctly accounted for 
in \cite{Adamczyk:2014vca}. After a proper correction, new scattering parameters consistent with a shallow attractive interactions
were found.

A similar analysis was repeated in $pp$ collisions at 7, 13 TeV and $pPb$ collisions at 5 TeV measured by ALICE, and 
the $\Lambda\Lambda$ correlation was analysed. The right panel of Fig. \ref{fig:hyp3} shows the extracted correlation as a function 
of the relative momentum $k*$ in the pair reference frame, together with the predictions obtained by taking the scattering parameters
suggested by different theoretical models, while using the Lednicky-Lyuboshits parametrisation of the correlation function \cite{Lednicky:1981su}.

One can see that the $\Lambda\Lambda$ correlation is rather shallow, close to the baseline (dashed histogram in Fig.~\ref{fig:hyp3}).
Instead of fitting the scattering parameters 
to the measured correlation function, a scan of the whole phase-space for the scattering length and the effective range was carried out.
The prediction is compared to the experimental correlation and a $n \sigma$ value characterises the agreement.
Figure~\ref{fig:hyp4} shows the obtained exclusion plot, where the scattering parameters predicted by different models are indicated.
In particular, we notice the point representing the Nagara  $\Lambda\Lambda$-hypernucleus and the new lattice prediction.
The white region on the exclusion plot represents the allowed parameter space. One can see that mostly a shallow attractive interaction is favoured (large values of $1/f_0$), but a non-excluded region of small negative scattering lengths and small effective ranges is present.

A possible bound state is investigated by computing the binding energy for negative $f_0$ values using the effective-range expansion relation \cite{Gongyo:2017fjb,Naidon:2016dpf}
\begin{equation}\label{eq:BE}
    B_{\Lambda \Lambda}=\frac{1}{m_\Lambda d_0 ^2}\left ( 1-\sqrt{1+2 d_0f^{-1}_0}\right)^2. \nonumber
\end{equation}
Following this prescription, the most probable value of  $B_{\Lambda \Lambda}=3.2-2.4+1.6(stat)-1.0+1.8(syst)$ MeV was obtained for the binding energy of the possible $\Lambda\Lambda$ bound state \cite{FemtoLambdaLambda}.
This measurement should be considered  as an upper limit, since further statistics could completely exclude the region of scattering parameters allowing for a bound state. To this end, this direct measurement of the $\Lambda\Lambda$ interaction is the most precise available.

\subsection{Theoretical approaches for hyperon interactions}
\label{sec:hyper-theory}

There has been a continuous theoretical effort in describing the $YN$ and $YY$ interactions, from meson-exchange models, approaches based on $\chi$EFT, lattice QCD (LQCD) calculations, low-momentum interactions or quark model potentials. 

The interaction between two baryons in the meson-exchange models is mediated by the exchange of scalar, pseudoscalar and vector mesons. The $NN$ meson-exchange model is the starting point to build up the $YN$ and $YY$ interactions, by imposing $SU(3)_{\rm flavor}$ symmetry. Presently, the most used models are those built by the J\"ulich and Nijmegen groups.

The J\"ulich meson-exchange $YN$ potential  \cite{Holzenkamp:1989tq,Haidenbauer:2005zh} is based on the $NN$ Bonn one. It is given in momentum space, containing full-energy dependence and non-localities. The potential is constructed taking into account the exchange of single mesons and also higher-order processes, that includes $\pi$ and $\rho$-exchange processes as well as processes involving $K$ and $K^*$ with $N$, $\Delta$, $\Lambda$, $\Sigma$ and $\Sigma(1385)$ intermediate states. Whereas the coupling constants and cutoff masses at $NN$ and $N \Delta$ vertices are taken from the Bonn model, the couplings involving strange particles are fixed by relating them to the $NN$ and $N\Delta$ values assuming SU(6) symmetry. The cuttoff masses at the strange vertices are adjusted to the existing $YN$ data. The form factors at the vertices are of monopole form, or dipole type when the vertex involves both a spin $3/2$ baryon and a vector.

As for the Nijmegen meson-exchange $YN$ potentials  \cite{Maessen:1989sx,Rijken:1998yy,Stoks:1999bz,Rijken:2006ep,Rijken:2010zzb,Nagels:2015lfa}, the starting point is the Nijmegen $NN$ potential. Those potentials are based in one-meson exchanges and the use of gaussian form factors for the vertices.
In particular,  the Nijmegen Soft-Core 89 (NSC89) $YN$ model \cite{Maessen:1989sx} is a straightforward extension of the Nijmegen $NN$, relating all the coupling constants at the vertices with strange particles to the $NN$ ones using SU(3) symmetry. The Nijmegen Soft-Core 97 (NSC97a-f) \cite{Rijken:1998yy,Stoks:1999bz} and Extended Soft-Core (ESC) \cite{Rijken:2006ep,Rijken:2010zzb,Nagels:2015lfa} are extensions of the Nijmegen potentials for $NN$ and the NSC89 $YN$ model. The coupling constants are determined by using SU(3) relations and fitting to the available $NN$ and $YN$ scattering data. The different models appear due to the existing freedom in fitting the parameters.

\begin{figure}[htb]
\begin{center}
\includegraphics[width=0.325\textwidth]{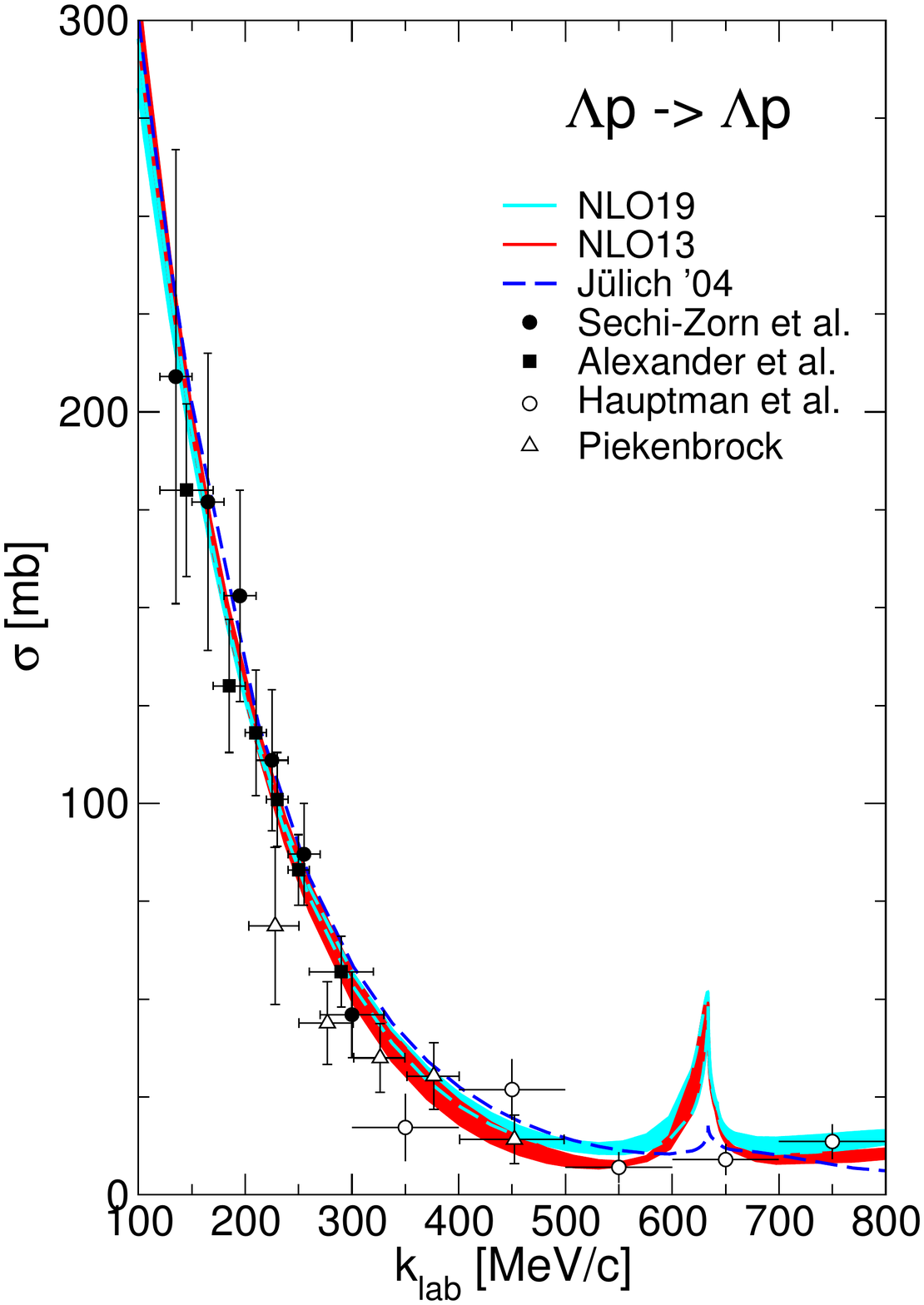}
\includegraphics[width=0.325\textwidth]{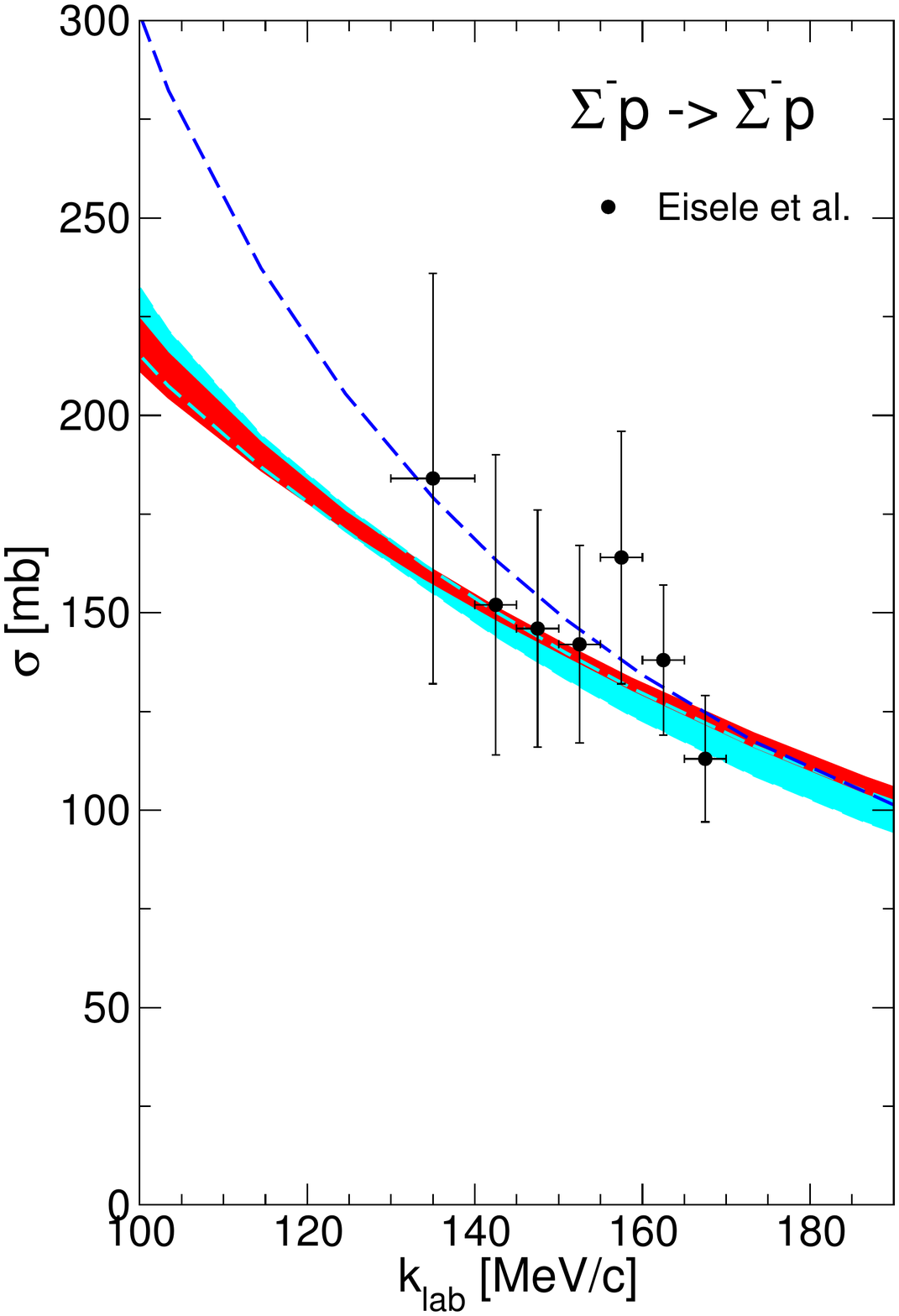}
\includegraphics[width=0.325\textwidth]{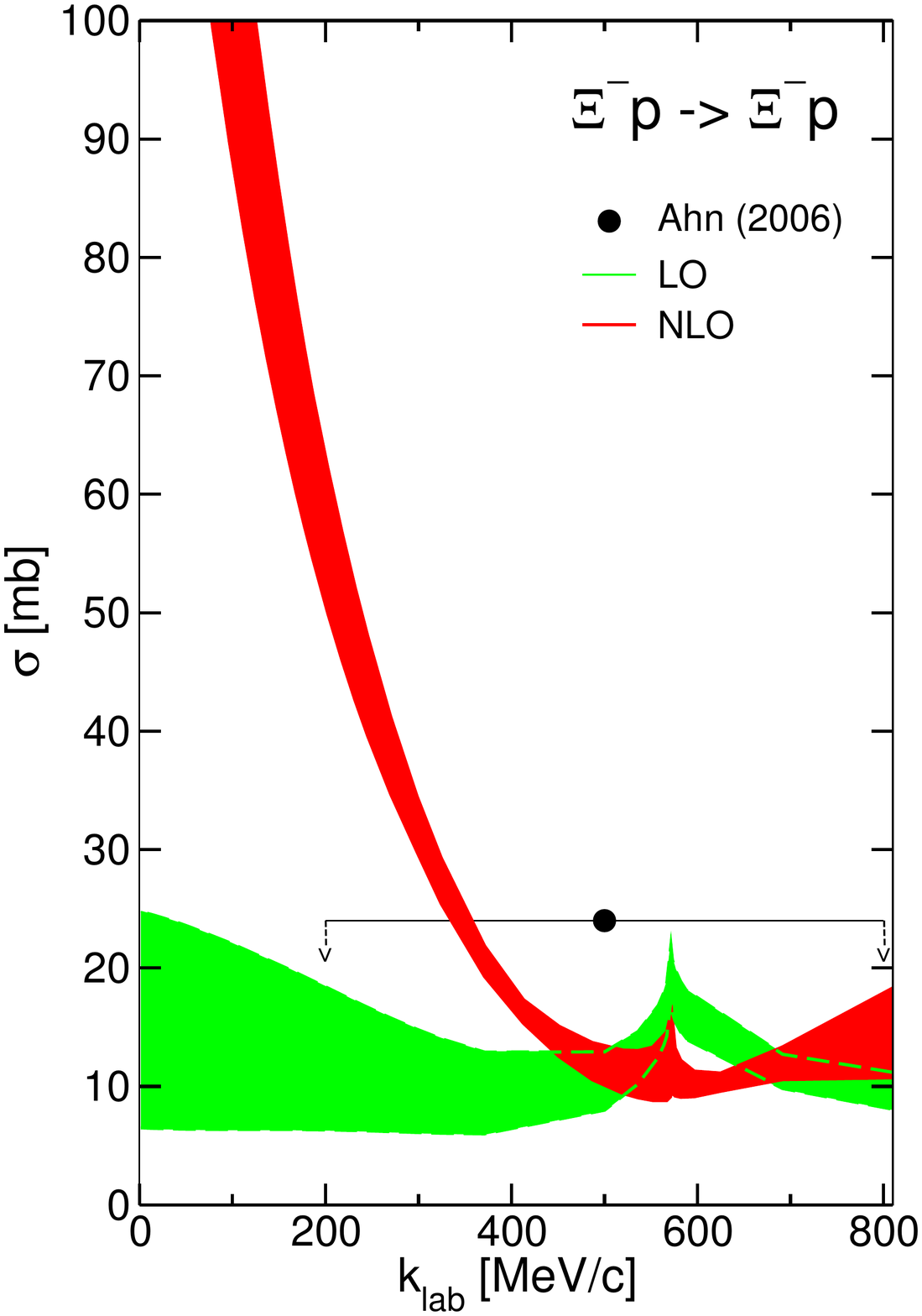}
\caption{
\textit{(Color online)} Left and middle panels: Total cross sections $\sigma$ for the elastic $\Lambda p$ (left panel) and $\Sigma^- p$ (middle panel) channels as a function of $k_{{\rm lab}}$ compared to data, for NLO13 (red band) and NLO19 (cyan band) approaches. The results for the J\"ulich04 meson-exchange model (dashed lines) are also displayed. Adapted from \cite{Haidenbauer:2019boi}. Right panel: The $\Xi^- p$ elastic cross sections for LO (green band) and NLO (red band) are shown. Adapted from \cite{Haidenbauer:2018gvg}.}
\label{fig:cross-hyp}
\end{center}
\end{figure}

The scattering amplitudes and, hence, scattering observables such as cross sections, can be then obtained by solving a regularized Lippmann-Schwinger equation starting from the $YN$ and $YY$ potentials coming from meson-exchange models (Eq.~\ref{bethe-free}). In Fig.~\ref{fig:cross-hyp} the results for the  $\Lambda p \rightarrow \Lambda p$ and $\Sigma^- p \rightarrow \Sigma^- p$ cross sections are shown using the J\"ulich04 (dashed lines) potential as compared to the available scattering data discussed in previous Sec.~\ref{YN_YY_Exp}. In this plot,  the results using $\chi$EFT are also displayed.

Over the past years, approaches based on $\chi$EFT have been also developed so as to construct the $YN$ and $YY$ interactions. The $\chi$EFT is a systematic approach that implements the chiral symmetry of the QCD at low energies, whereas providing a power counting to make consistent calculations order by order and estimating the corrections. During the past years, the $NN$ interaction has been described to high precision, while, more recently, the $YN$ and $YY$ interactions have been built from chiral effective Lagrangians in a similar manner as done for the $NN$ by the J\"ulich-Bonn-Munich collaboration \cite{Polinder:2006zh,Haidenbauer:2013oca,Haidenbauer:2016vfq}. 

 The $YN$ interaction has been obtained within the SU(3) $\chi$EFT, deriving the different orders in the chiral expansion in a consistent way following the Weinberg power counting. At LO in the power counting, the $YN$ potential is built from one pseudoscalar-meson exchanges and non-derivative four-baryon contact terms \cite{Polinder:2006zh}. The NLO takes into account two pseudoscalar-meson exchanges and contact interactions with two derivatives \cite{Haidenbauer:2013oca}.

The scattering observables can be obtained using the LO and NLO contributions to the $YN$ potential as kernel.  Apart from using the available standard set of 36 $YN$ data points for the fitting procedure, the baryon-baryon-meson coupling constants have been fixed by SU(3) symmetry. This symmetry has been also used to derive the relations between the various low-energy constants. However, the SU(3) symmetry has been broken due to the use of the hadron physical masses.  As displayed in the left and middle panels of Fig.~\ref{fig:cross-hyp}, the available $\Lambda N$ and $\Sigma N$ data are described up to NLO, with the so-called NLO13 interaction of Ref.~\cite{Haidenbauer:2013oca}.

More recently, the NLO13 potential for the  $\Lambda N$ and $\Sigma N$ systems has been revisited in \cite{Haidenbauer:2019boi}, by exploring a new option for the low-energy constants that determine the strength of the contact interactions, so giving rise to the NLO19 version. The NLO19 is constructed trying to reduce the number of low-energy constants needed in the fit to the $\Lambda N$ and $\Sigma N$ data, by inferring some of them from the $NN$ sector using SU(3) symmetry. As a consequence, the contribution of the $^3S_1$-$^3D_1$ partial wave is enhanced in the NLO19 with respect to the NLO13. The  NLO13 of Ref.~\cite{Haidenbauer:2013oca} and NLO19 of Ref.~\cite{Haidenbauer:2019boi} yield equivalent results for $\Lambda N$ and $\Sigma N$ scattering observables, as seen in the left and middle panels of Fig.~\ref{fig:cross-hyp}.

Regarding the $\Xi N$ interaction, the calculation within $\chi$EFT up to NLO \cite{Haidenbauer:2018gvg} is in line with empirical constraints on the $\Lambda \Lambda$ $S$-wave scattering length as well as published upper bounds of the $\Xi^- p$ cross sections (see right panel in Fig.~\ref{fig:cross-hyp}). 

Apart from meson-exchange models or $\chi$EFT schemes, the $YN$ and $YY$ interactions can be constructed from solving QCD on the lattice. LQCD is a technique in which space-time is discretized into a four-dimensional grid, and the path integral over the quark and gluon fields at each point in the grid is obtained in Euclidean space-time using Monte-Carlo methods. In the past decade, there has been a lot of progress in deriving baryon-baryon interactions from LQCD calculations.  In the strangeness sector, this work has been carried out by the HAL QCD \cite{Ishii:2006ec,HALQCD:2012aa} and the NPLQCD \cite{nplqcd} collaborations. 

The starting point of the HAL QCD  method is the  Nambu-Bethe-Salpeter wave function within the range of the hadron interaction to define a non-local potential, so that it obeys the Schr\"odinger type equation in a finite box. With an appropriate extrapolation to infinite volume, one can calculate the scattering phase shifts and bound states by solving the Schr\"odinger equation and compare to the experimental results (see \cite{Ishii:2006ec,HALQCD:2012aa} for more details). 

\begin{figure}[htb]
    \centering
    \includegraphics[height=90mm,width=\textwidth]{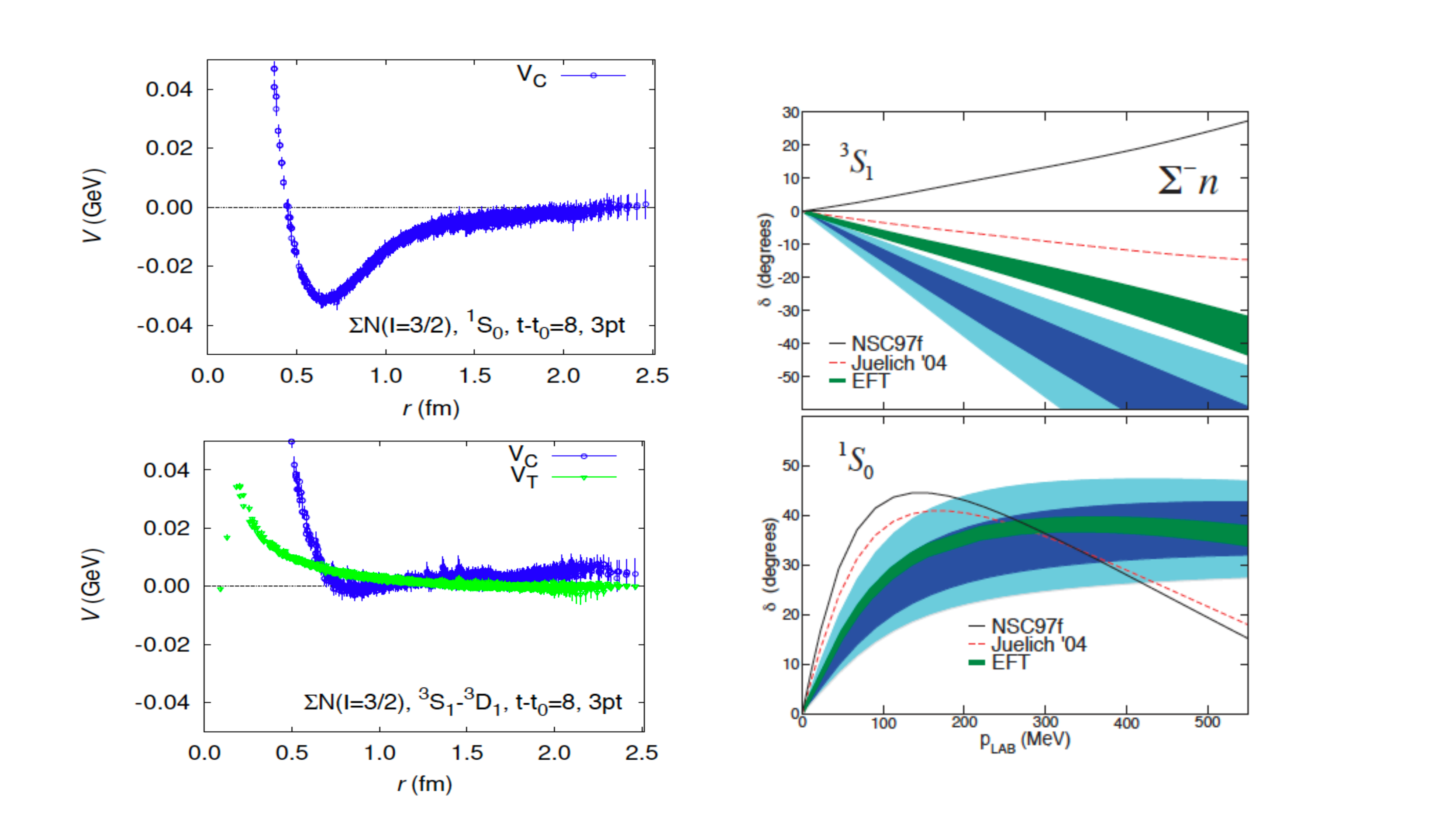}
    \caption{\textit{(Color online)} Left panels: The central and tensor potentials for $\Sigma N(I=3/2)$ in the $^1S_0$ (upper panel) and $^3S_1-^3D_1$  (bottom panel) channels obtained with the HAL QCD method \cite{Aoki:2012tk}. Right panels: The LQCD-predicted $^3S_1$ and $^1S_0$ $\Sigma^- n$ phase shift versus laboratory momentum at the physical pion mass (blue bands) obtained with the L\"uscher method \cite{Beane:2012ey}, compared with the determinations of NSC97f and J\"ulich04  models, as well as the $\chi$EFT calculation.}
    \label{fig:lattice}
\end{figure}

The $S=-1$ $\Lambda N$ and $\Sigma N$ systems have been studied by the HAL QCD collaboration in Ref.~\cite{Aoki:2012tk,Nemura:2017bbw}. The $\Sigma N(I=3/2)$ results for the central and tensor potentials are shown in the left panels of Fig.~\ref{fig:lattice} for $^1S_0$ channel (upper panel) and $^3S_1-^3D_1$ channel (bottom panel). In this work, the $S=-2$ $\Lambda \Lambda$, $\Lambda \Sigma$, $\Sigma \Sigma$ and $\Xi N$ have been also analyzed, with more recent updates in Refs.~\cite{Sasaki:2015ifa,Sasaki:2018mzh}. As we showed in Sec.~\ref{YN_YY_Exp}, the HAL QCD predictions for the $\Lambda \Lambda$ and $\Xi N$ are in good agreement with the recent correlation data.
Moreover, the HAL QCD collaboration has reported results on the  $\Xi \Xi$ \cite{Iritani:2015dhu} and, more recently, on the $\Omega N$ \cite{Iritani:2018sra} and $\Omega \Omega$ \cite{Gongyo:2017fjb} systems. 

The NPLQCD collaboration extracts information on hadron-hadron scattering amplitude following the L\"uscher method.  In the case of the determination of the hadron-hadron scattering amplitude below an inelastic threshold, the deviation of the energy eigenvalues of the two-hadron system in the lattice volume from the sum of the single-hadron energies is linked to the scattering phase shift at the measured two-hadron energies. For energy eigenvalues above kinematic thresholds where multiple channels contribute, a coupled-channels analysis is required as a single phase shift does not parametrize the scattering amplitude. Then, the energy shift for two particles can be determined from the correlation functions for systems containing one and two hadrons, and connected to the scattering amplitude (see \cite{Beane:2010em} for more details).

The NPLQCD collaboration performed the first LQCD calculations of $\Lambda n$ and $\Sigma^- n$ interactions at unphysical pion masses \cite{Beane:2006gf}. Low signal-to-noise ratios are intrinsic in the study of baryons with LQCD. In Ref.~\cite{Beane:2009py} the authors performed a high-statistics LQCD calculation with the aim of reducing the noise, and extract with much more precision the energy of the two-baryon systems in the strangeness 0, -1, -2 and -4. Later on, the $\Sigma^- n$ was revisited in Ref.~\cite{Beane:2012ey} for pion masses of $\sim 390$ MeV and at the physical pion mass by means of a extrapolation using $\chi EFT$. The  $^3S_1$ and $^1S_0$ $\Sigma^- n$ phase shift versus laboratory momentum at the physical pion mass (blue bands) are shown in the  right panel of Fig.~\ref{fig:lattice}, compared to the determinations of NSC97f and J\"ulich04  models, as well as the $\chi$EFT calculation. And, more recently, the low-energy $S$-wave scattering amplitudes of two octet baryons at the SU(3) flavor symmetric point, corresponding to the physical strange quark mass, have been studied, with a focus on the underlying symmetry structures that are expected to emerge in the large-$N_c$ limit of QCD \cite{Wagman:2017tmp}.

Also the NPLQCD has reported the possible formation of bound states, such as the $^1S_0$ $\Xi^- \Xi^-$ state \cite{Beane:2010hg} or the H-dibaryon. In fact, the H-dibaryon, with quantum numbers of the $\Lambda \Lambda$ system, has caught the attention of the $\chi$EFT and LQCD communities. The  NPLQCD Collaboration provided evidence of a H-dibaryon bound state for $m_{\pi} \sim 390$ MeV  \cite{Beane:2010hg,Beane:2011iw}, whereas the HAL QCD group determined also a bound state for $m_{\pi} \sim 469-1171 $ MeV  \cite{Inoue:2010es,Inoue:2011ai}. Extrapolations to the physical light-quark masses, however, pointed to a weakly bound H-dibaryon or a near-threshold state by the NPLQCD Collaboration \cite{Beane:2011iw,Beane:2011zpa,Shanahan:2011su,Haidenbauer:2011ah}. In a later analysis by the NPLQCD Collaboration, the H-dibaryon was found to be deeply bound in the limit of SU(3) flavor symmetric point at the physical strange quark mass \cite{Beane:2012vq}, whereas the HAL QCD has reported very recently that at  the almost  physical  point  ($m_{\pi}=146$MeV  and $m_K=525$MeV)  no bound  or  resonant  di-hyperon  around  the  $\Lambda \Lambda$  threshold  can be produced \cite{Sasaki:2019qnh}. Thus, the existence of the H-dibaryon is still controversial from the point of view of LQCD. 

And last but not least, there are other alternative ways of constructing the $YN$ and $YY$ interactions, such as those considering low-momentum interactions or quark model potentials. The aim of the former one is to obtain a universal effective low-momentum potential for $YN$ and $YY$ using renormalization-group methods \cite{Schaefer:2005fi}, while the latter builds the $YN$ and $YY$ interactions by means of constituent quark models \cite{Fujiwara:2006yh}. We refer the reader to the corresponding references for a more detailed description of these methods.

\subsection{Hypernuclei}
\label{sec:hypernuclei}
\subsubsection{Experimental detection}
\label{sec:hyper-exp}

Hypernuclei \cite{Gal:2016boi,Tamura:2013lwa,Feliciello:2015dua} are formed when a nucleon is replaced by a hyperon inside the nucleus and can occupy any of the bound or unbound nuclear levels, being the hyperon  distinguishable from all the nucleons. From there, the hyperon may escape the nuclear potential, cascade downward in energy, or become trapped in an
isomeric level. A bound $\Lambda$ eventually reaches the ground state and then decays weakly. The hypernucleus
can then be described as the combination of a core nucleus together with a single hyperon, and the 
quantum number of the hypernucleus can be obtained combining the quantum numbers of the core-nucleus and hyperon.
The measurement of the binding energy of hypernuclei quantifies the attractive interaction and, if the hyperon properties inside the nucleus would differ from the vacuum properties, some information about the behaviour of hyperons as a function of the baryonic density could be pinned down.

\begin{figure}[htb]
\begin{center}
\includegraphics[height=70mm,width=0.43\textwidth]{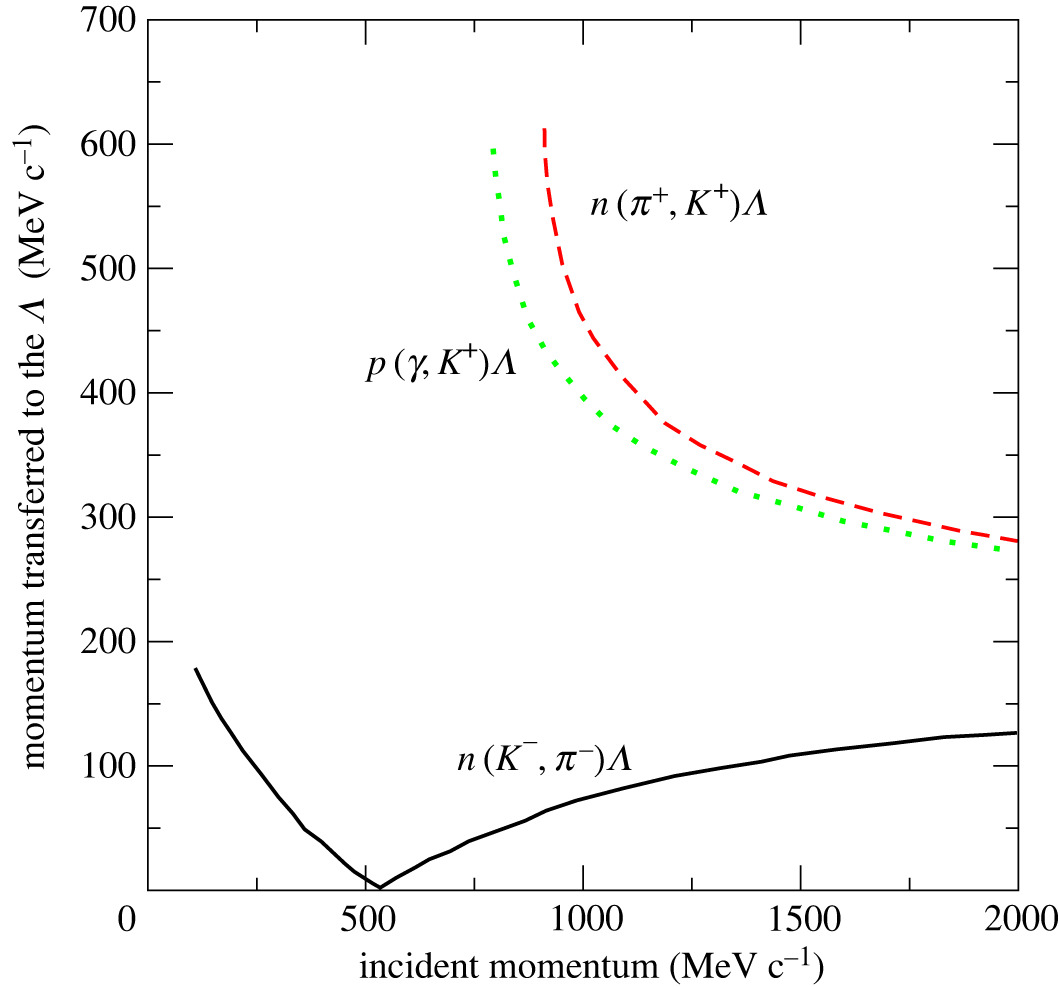}
\includegraphics[height=70mm,width=0.43\textwidth]{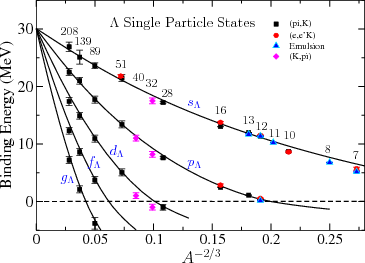}
\caption{\textit{(Color online)} Left panel: Momentum transferred to the $\Lambda$ hyperon for different reactions as a function of the beam momentum. Right panel: Binding energy of hypernuclei measured for different reactions as a function of the mass number $A^{-2/3}$. The full lines represent the predictions obtained from a single-particle potential model \cite{Gal:2016boi}.
}
\label{fig:hypernucleiBE}
\end{center}
\end{figure}

The acquisition of hypernuclear binding energies, well-depths, and positions of the hypernuclear levels began in the 1960s \cite{David63,Juric:1973zq},
using $K^-$ beams, and exploiting the strangeness-exchange reaction: $K^- + ^A \rightarrow ^A_{\Lambda}Z + \pi^- $.
This process can occur for kaon stopped or in-flight. The probability of forming an hypernucleus is the largest for a kaon momentum
of about 500 MeV/c, since in that case the hyperon is produced at rest within the nucleus and, hence, has an high probability 
to form a hypernucleus. The left panel of Fig.~\ref{fig:hypernucleiBE} shows the momentum transferred to the $\Lambda$ hyperons in different formation processes. Experiments with stopped $K^-$ beams have been carried out 
 at DA$\phi$NE \cite{Feliciello:2015dua}, while in-flight $K^-$ have been 
 employed at CERN \cite{Bonazzola:1974ig,Povh:1980hg}, BNL \cite{Chrien:1979wu,May:1981er}, and KEK \cite{Hashimoto:2006aw}.


Another method to produce hypernuclei relies on the associated production mechanism, possible with pion beams in processes as
 $\pi^+ + ^A Z \rightarrow ^A_{\Lambda} Z + K^+$. Such beams have been available at the BNL-AGS \cite{Milner:1985kr,Pile:1991cf} and KEK\cite{Fukuda:1994hi,Nagae:2000sp} accelerators and, although the production cross section
for hypernuclei are smaller than for the $K^-$ beams, pion-induced reactions are more effective in 
producing states with large angular momenta.

In all these measurements, one can determine the binding energy of the formed hypernuclei by measuring the momentum 
of the ejected $\pi^-$ or $K^+$, and employing the missing mass method to evaluate the mass of the hypernucleus. The binding energy is then evaluated by using the formula $BE=[M(\Lambda)+ M(^{A-1}Z)-M(^A_{\Lambda}Z)] c^2 $, where $M(^{A-1}Z)$ is the mass of the core-nucleus and  $M(^A_{\Lambda}Z)$ is the mass of the hypernucleus. Alternatively 
to the missing mass method, the hypernucleus mass can be reconstructed from the invariant mass of the decay products. The missing mass method allowed for a precision of 3-4 MeV for the BNL results and 1-2 MeV for the KEK ones with in-flight pion and kaon beams. The FINUDA experiment reached precisions of 1.4 MeV for the determination of the $BE$ of light hypernuclei (A$<$16) \cite{Botta_2017}.

Also electron beams can be used to produce hypernuclei exploiting reactions such as 
$e^- + {}^AZ \rightarrow {e^-}^{'} + ^A_{\Lambda}(Z-1) + K^+$. This mechanism leads to production cross sections of an
order of magnitude smaller than for in-flight $K^-$ beams, but allows for a better energy resolution ($0.5-0.8$ MeV)
and also to study neutron-rich hypernuclei. Moreover, the electro-production of free $\Lambda$ and $\Sigma$ hyperons
off a proton target allows for an absolute energy calibration, while for the methods mentioned previously only a relative 
calibration among different nuclei is possible. The experiments with electron beams were extensively carried out at JLab. The reached precision in  the determination of the binding energy is due to the fact that both the scattered 
electron and the produced kaon are fully reconstructed. Also, the beam momentum determination is much more accurate than 
for the secondary $K^-$ and $\pi^+$ beams. Electrons beams have been also employed at MAMI to produce hypernuclei \cite{Esser:2015trs}. There, both the outgoing production of $K^+$ and the $\pi^-$ emitted in the weak decay of the hypernucleus can be reconstructed.

The binding energies measured for different hypernuclei are summarized in the right panel of Fig.~\ref{fig:hypernucleiBE} \cite{Gal:2016boi} as a function of the mass number $A^{-2/3}$.
The different colors refer to different methods employed to produce the hypernuclei, and the lines correspond to the predicted binding
energy according to a single particle potential model with a Wood-Saxon potential of depth $V_0$ = -30 MeV.
One can see that the binding energy saturates to values of about -28 MeV for large nuclei, and also that the single-particle model reproduces the measurements in a satisfactory way.

The study of hypernuclei can also provide information on the $YN$ weak interactions by their non-mesonic weak decay (NMWD: $\Lambda n\rightarrow nn$ or $\Lambda p  \rightarrow np$) \cite{Botta_2017,Feliciello:2015dua}. In light hypernuclei, the mesonic decay modes are favored and the decay process is basically the same as that of a free $\Lambda$ particle ($\Lambda \rightarrow p\pi^-$ or $\Lambda \rightarrow n\pi^0$). On the other hand, in medium-heavy systems, due to the Pauli blocking and the increased overlap of wave functions between $\Lambda$ and nucleons, the non-mesonic channels, in which a $\Lambda$ decays via weak interactions with neighboring nucleon(s), dominate the decay process \cite{Alberico:2001jb}.  Until now, NMWD is the only means to study the strangeness change baryon-baryon weak interaction. An important observable of the nonmesonic decay of hypernuclei is the ratio of the neutron-induced ($\Lambda n \rightarrow  nn$) over the proton-induced ($\Lambda p \rightarrow np$) decay rate, labelled as $\Gamma_n/\Gamma_p$. The experimental ratios had been reported to be close or greater than unity, indicating the dominance of the neutron-induced channel, while the theoretical ratios based on one-meson-exchange model were as small as 0.1 \cite{Kim:2009zzp}. This puzzle has been solved recently through the progress made in theoretical and experimental works by well taking account of the final state interaction and the three-body process ($\Lambda NN\rightarrow nNN$) \cite{Kim:2009zzp}.

A more detailed study of the $\Lambda$ interaction is based on the measurement of the $\gamma$-spectroscopy of $\Lambda$
hypernuclei \cite{Tamura:2013lwa}. Indeed, the $\gamma$ spectroscopy allows to study the fine-structure of the interaction and elucidate the spin and angular momentum-dependent effective interaction. One can express the $\Lambda N$ interaction for nuclei, assuming that the $\Lambda$ hyperon is in the s-orbit, as \cite{Gal:2016boi} 
\begin{equation}
 V_{\Lambda N}= \, V_0(r) + \Delta \vec{s}_N\cdot \vec{s}_{\Lambda} + S_{\Lambda}\vec{l}_N \cdot \vec{s}_{\Lambda} +S_N \vec{l}_N \cdot \vec{s}_N +TS_{12}, \nonumber   
\end{equation}
where $\vec{s}_{\Lambda, N}$ are the spins of the $\Lambda$  and the valence nucleon(s), $\vec{l}_N$ is the relative angular momentum between the core nucleus and the $\Lambda$, $S_{12}$ is the tensor component of the interaction, and $\Delta, \,S_{\Lambda},\,S_N,\,T$ parameters are to be determined by fitting experimental spectra. The energy levels of the low-lying states are primarily determined by the strength of the spin-dependent term. The measured transitions of p-shell hypernuclei allow to determine the parameters \cite{Tamura:2013lwa,Gal:2016boi}, that are found to be equal to $\Delta=\, 0.43$ MeV $(A=7)$ or $0.33$ MeV$ (A>10)$, $S_{\Lambda}= -0.001$ MeV, $S_N=\, -0.4$ MeV and $T=\, 0.03$ MeV.

Such corrections have been quantified 
analyzing so far only the $\gamma$-spectra for light hypernuclei (A $<$ 16). An improved shell-model has been employed in the fitting of the p-shell $\gamma$-ray measurements in \cite{Millener:2012zz} and the effect of the $\Lambda-N$ to  $\Sigma-N$ coupling has been found to be non negligible. 

Future measurements planned at JPARC should extend these studies to different nuclei species to test the models also against the contribution of three-body interactions such as $\Lambda NN$, that are currently not considered.
Presently, the hypernuclei data do not provide any quantitative  information on $\Lambda NN$, because if $\Sigma N$-$\Lambda N$ coupling can be projected and, hence, absorbed in a two-body interaction, the three body interaction can not. 

\begin{figure}[htb]
\begin{center}
\includegraphics[height=95mm,width=0.73\textwidth]{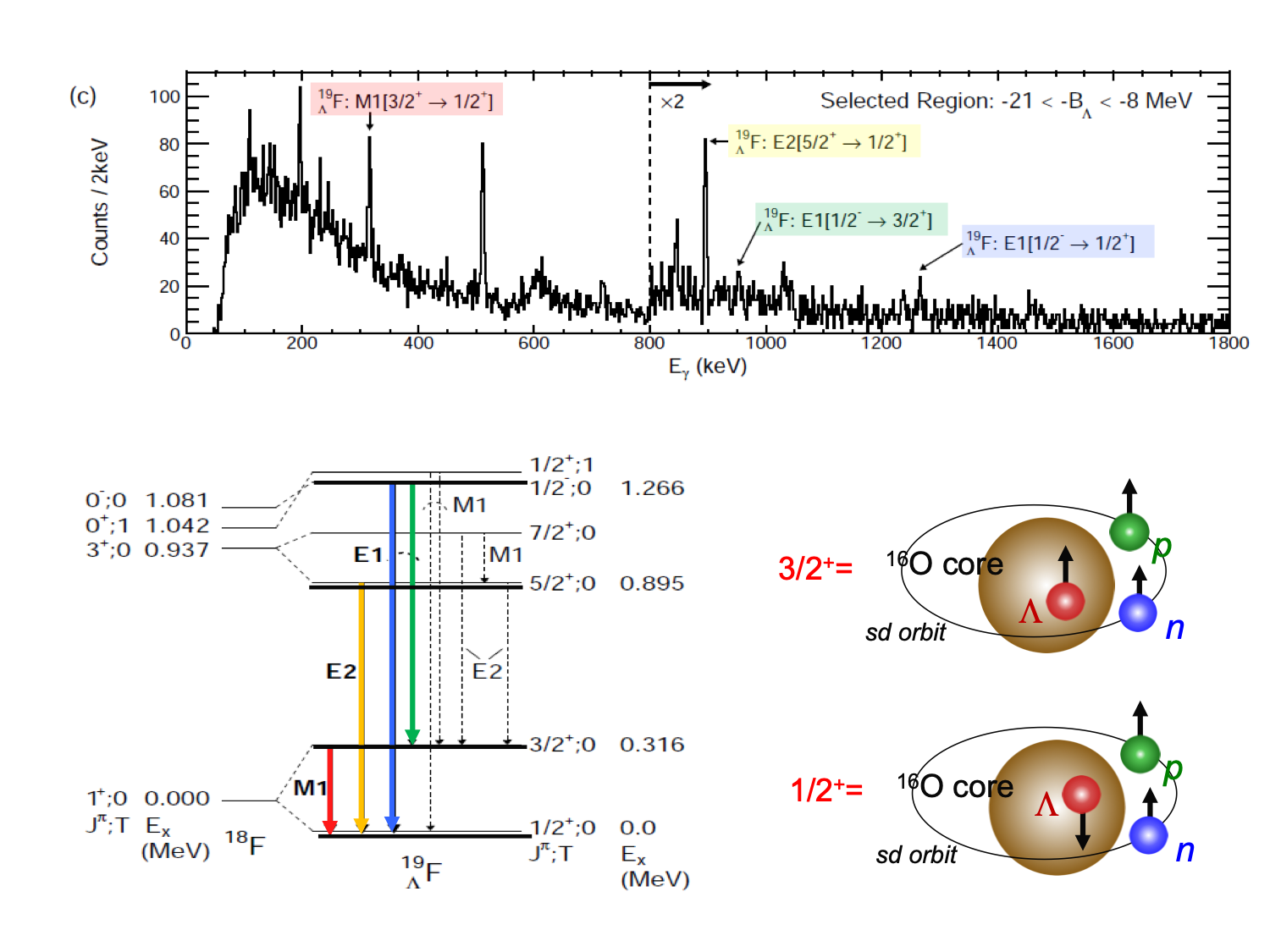}
\caption{\textit{(Color online)} $\gamma$-ray spectra obtained from the $^{19}F(K^-,\pi^-)$ at at a beam momentum of 1.8 GeV/c, and level diagram for the $^{19}_{\Lambda}F$ with the assigned $\gamma$ ray transitions \cite{PhysRevLett.120.132505}.
}
\label{fig:Fe19gamma}
\end{center}
\end{figure}

More recently, measurements of the $\gamma$-ray spectra obtained in $^{19}F(K^-,\pi^-)$ at a beam momentum of 1.8 GeV/c allowed to pin down, for the first time, sd hypernuclei\footnote{Hypernucleus where the $\Lambda$ is in the s-orbit and the two nucleons are in the d-orbit.} \cite{PhysRevLett.120.132505}, that can extend the study of the $\Lambda N$ spin-spin interactions to a wide A-range of hypernuclei. Figure \ref{fig:Fe19gamma} shows the $\gamma$ ray spectrum obtained gating the $\Lambda$ binding energy with a selection of the missing mass for the $^{19}F(K^-,\pi^-)$ reaction. The case showed here refers to the low-lying energy states ($-21 <- B_{\Lambda} < 5 $ MeV), and the peak highlighted in red shows the $3/2^+ \rightarrow 1/2^+$ transition that provides the first experimental measurement of sd hypernuclei. The schematic representation of the spin configuration shown in Fig.~\ref{fig:Fe19gamma} evidences that the measured energy spacing of $315\pm0.4\,(stat)^{+0.6}_{-0.5}\,(syst)$ keV is primarly determined by the $\Lambda N$ spin-spin interaction. 
This energy line is reproduced to better than 20\% by shell-model analyses that use its fitted strength in the p-shell \cite{PhysRevLett.120.132505}.
This new measurement is of significant interest to test the theoretical framework for heavier hypernuclei.

The precise measurement of the $\gamma$ spectra of heavier hypernuclei and the understanding of
their structure is considered important also to get information on the $\Lambda NN$ three-body force (or nuclear-density dependence of $\Lambda$N interaction). Indeed, the existence of the (short-range)
repulsive $NNN$ force has been found by combining various data of
ordinary nuclei and precise pp/pn scattering data,
through ab-initio calculations.
The same approach will be followed for hypernuclei. 

An additional topic investigated experimentally with hypernuclei is the charge symmetry breaking (CSB).
This effect deals with the invariance of the strong interaction to a rotation in the isospin space (substituting protons with neutrons), and it is almost completely maintained in standard nuclei (70 keV difference between $^3$H and $^3$He).

The situation is different for the 
strangeness sector, where the separation 
of the binding energies of the $\Lambda$ 
in the mirror nuclei $^4_{\Lambda}$H and 
$^4_{\Lambda}$He has been measured  \cite{Yamamoto:2015avw,Esser:2015trs,Schulz:2016kdc} and the most recent value was found to be equal to $233 \pm 92$ keV \cite{Schulz:2016kdc}. This effect is, hence, much more sizable for hypernuclei than for normal nuclei.
This enhanced CSB effect for hypernuclei might play
an important role in the computation of the 
hyperon interactions in a neutron-rich environment.


The measurement of neutron-rich $\Lambda$ hypernuclei can constrain the $\Lambda$-$\Sigma$ mixing that occurs within nuclei. This mixing is not 
present in free space, but  the $\Sigma N$-$\Lambda N$  coupling becomes relevant for nuclei  with a large isospin value ($N \gg Z$). 
The neutron-rich environment is also closely related to the study of the equation of state of high-density neutron matter, as present within neutron stars.
Although the density tested in the hypernuclei is at most $\rho_0$, a detailed knowledge of the $\Sigma N$-$\Lambda N$ coupling is essential to build a realistic equation of state.

The most efficient way to produce neutron-rich hypernuclei are double charge exchange reactions, such as the ($\pi^-$, $K^+$) and ($K^-$, $\pi^+$). In these reactions, two protons are converted to a $\Lambda$ and a neutron. 
The FINUDA collaboration claimed the observation of a $^6_{\Lambda}{\rm H}$ hypernucleus \cite{PhysRevLett6HLambda} produced in the reaction $^6{\rm Li}(K^{-}_{\rm stopped},\pi^+)^6_{\Lambda}{\rm H}$, followed by the decay $^6_{\Lambda}{\rm H}\rightarrow ^6{\rm He} +\pi^-$, where the secondary $\pi^-$ was identified.  The reconstructed invariant mass reports a binding energy of 2.31 MeV w.r.t. the ${\rm t}+2n+\Lambda$ threshold, which would indicate a rather small contribution from $\Sigma N$-$\Lambda N$ coupling. On the other hand, the J-PARC E10 experiment searched for the same $^6_{\Lambda}$H state in the reaction $^6{\rm Li}(\pi^-,K^+)X$ and could only determine an upper limit \cite{PhysRevC.96.014005}. This means that new experiments are called for to confirm the existence of such neutron-rich hypernuclei.
In this context, also the investigation of neutron-deficient hypernuclei plays an important role. These hypernuclei are not measured at the moment, but future experiments planned at the FAIR facility by the R3B collaboration employing radioactive beams will study such hypernuclei \cite{Sun:2017unn}.

The existence of $\Sigma$-hypernuclei is not completely ruled out. Indeed, evidence of a relatively narrow $\Sigma$-hypernucleus was suggested in the 1980s using in-flight (CERN \cite{Bertini:1979qg} and  BNL \cite{Piekarz:1982ud}) and stopped (KEK \cite{Yamazaki:1984hi}) $K^-$ beams.
These data supported a $\Sigma$-nuclear attractive  interaction  of 25-30 MeV.
On the other end, more recent measurements of the $\Sigma^-$ at KEK \cite{PhysRevLett.89.072301} have established that the $\Sigma$-nuclear interaction is strongly repulsive.
Considering the rather shallow correlation measured for the $p\Sigma^0$ correlation discussed in Sec. \ref{YN_YY_Exp}, a sizeable interaction seems excluded, but more precise data are needed to clarify the situation.

Hypernuclei allows also to investigate the interaction in the S=-2 sector since $\Lambda\Lambda$- and $\Xi$- and have been studied as well. Before the advent of the direct $\Lambda\Lambda$ correlation measurements described in Sec.~\ref{YN_YY_Exp}, double-$\Lambda$ hypernuclei have been employed to investigate the $\Lambda\Lambda$ interaction. 
The $\Lambda\Lambda$ interaction strength in $^A_{\Lambda\Lambda}Z$ can be determined from the measurement of the masses of the double- and single-$\Lambda$ hypernuclei. The following equation is used,
\begin{equation}
\Delta B_{\Lambda\Lambda}(^ A_{\Lambda\Lambda}Z)= B_{\Lambda\Lambda}(^ A_{\Lambda\Lambda}Z) -2B_{\Lambda}(^ {A-1}_{\Lambda}Z),
\end{equation}
where $B_{\Lambda\Lambda}$ and  $B_{\Lambda}$ are the binding energies of the two $\Lambda$ hyperons in the double-$\Lambda$ hypernucleus and a single $\Lambda$ hyperon in the single-$\Lambda$ hypernucleus, respectively.

Among the several double-$\Lambda$ hypernuclei so far reported in emulsion experiments \cite{Danysz:1963zza,Prowse:1966nz,Aoki:1991ip,Takahashi:2001nm}, the clearest evidence was shown by the NAGARA event discovered by the KEK-PS E373 experiment \cite{Takahashi:2001nm}. In this work, the $^6_{\Lambda\Lambda}$He state was identified unambiguously in the reaction $\Xi^- +^{12}{\rm C} \rightarrow ^6_{\Lambda\Lambda}{\rm He} + ^4{\rm He} +{\rm t}; \,\,^6_{\Lambda\Lambda} {\rm He} \rightarrow ^5_{\Lambda}{\rm He} + p + \pi^- $,  and the value of $\Delta B_{\Lambda\Lambda}(^6_{\Lambda\Lambda}He)=\, 0.67 \pm 0.17$ MeV was determined \cite{Ahn:2013poa}. 

The assignment of this event to a double-$\Lambda$ hypernucleus is unique, because all the involved hypernuclei have no particle-stable excited states.
This very small number questions the existence of the H-dibaryon state, but, on the other hand, the number is greatly affected by the structure of the core nucleus. Hence, it is necessary to measure different double-$\Lambda$ hypernuclei with different core sizes and p-shell hypernuclei to extract more precise constraints on the genuine $\Lambda\Lambda$ interaction.

Future perspectives for the production of $\Lambda\Lambda$-hypernuclei are offered by the $\bar{P}$ANDA collaboration at FAIR, exploiting the large cross section of hyperon-antihyperon  pairs, produced via antiprotons to form single- and double-$\Lambda$ hypernuclei \cite{Singh:2016hoh}. A high resolution spectroscopy system will complement the detection system. 

For $\Xi$-hypernuclei, the KISO event was measured by the KEK E373 experiment using emulsion exposures \cite{Nakazawa15}.
This event is uniquely tagged as $\Xi^- + ^{14}{\rm N} \rightarrow ^{10}_{\Lambda}{\rm Be} + ^5_{\Lambda}{\rm He}$, but it is not clear whether
the final state of $^{10}_{\Lambda}$Be is in a ground or excited state.  The resulting B$_\Xi$ binding is rather shallow 
and found to be equal to $1.11\pm 0.25$ MeV. 
This result is consistent with the hypothesis of an initial $\Xi^-$ absorption 
on a 3D level or deeper, as it is also assumed 
in the analysis of the $\Lambda\Lambda$-hypernuclei.
This demonstrates how
these analyses are complementary to each others and both necessary to pin down the different interactions precisely. 

If we compare this rather small number to the rather attractive two-body potential predicted by HAL QCD calculations \cite{HAL1} and validated by the p$\Xi^-$ correlation measured by ALICE \cite{FemtopXi} (see Sec.~\ref{YN_YY_Exp}), one can see rather clearly that the sensitivity of hypernuclei to two-body interactions is limited. Indeed, the clear difference between the p$\Lambda$ and p$\Xi^-$ correlation functions reflect directly the difference in the interaction, but within hypernuclei one can test only the average value of the interactions because of the presence of many nucleons.

\begin{figure}[htb]
\begin{center}
\includegraphics[height=100mm,width=0.9\textwidth]{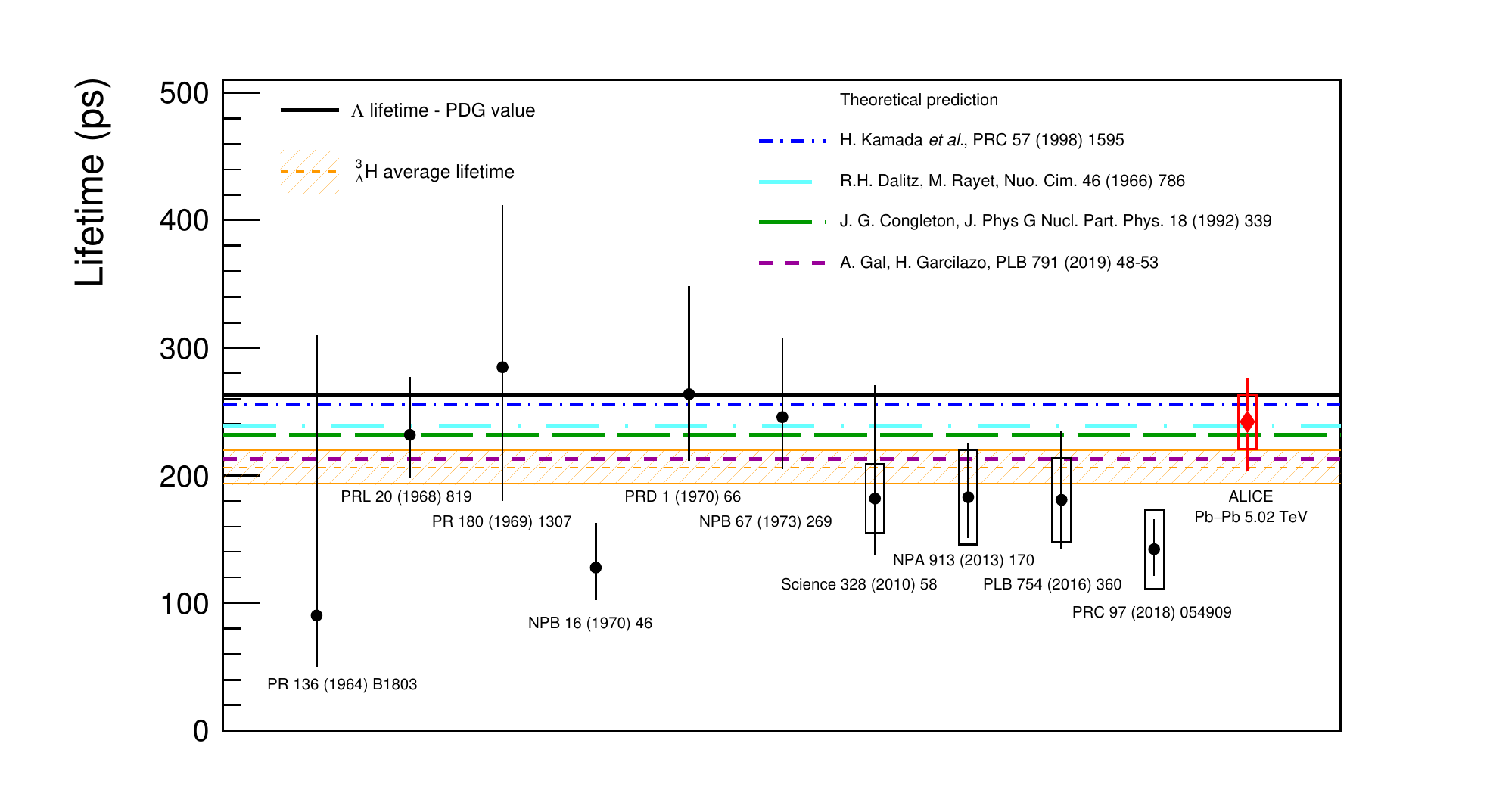}
\caption{Collection of the hypertriton lifetime measurements obtained with different experimental techniques. The orange band represents the average of all the lifetime measurements and the black horizontal line is the free $\Lambda$ lifetime \cite{Acharya:2019qcp}.
}
\label{fig:HypTrit}
\end{center}
\end{figure}

A complementary technique to produce hypernuclei is to employ HICs, either using projectile fragmentation reactions in fixed target experiments at GSI (HYPHI collaboration \cite{Rappold:2013fic}) or with HICs at colliders (STAR \cite{Adamczyk:2017buv} and ALICE \cite{Acharya:2019qcp} collaborations). In particular, light hypernuclei have been successfully measured in the last years with focus on the determination of the hypertriton ($^3_{\Lambda}$H) lifetime.
The separation energy of the $\Lambda$ in this hypernucleus is only about $130$ keV, and this results in an RMS radius of $10.6$ fm for the hypertriton. A very low binding energy suggests a small change of the $\Lambda$ wave function in a nucleus and, hence, one can expect the lifetime of the hypertriton to be very close to that of the free $\Lambda$.
Several measurements systematically reported a shorter lifetime for $^3_{\Lambda}$H, as shown in Fig. \ref{fig:HypTrit} and this generated the so called 'hypertriton lifetime puzzle'. 
Indeed, a possible modification of the lifetime would have consequences on the $\Lambda N$ and $\Lambda NN$ interactions.
The most recent measurement of the hypertriton lifetime performed by ALICE is compatible with the $\Lambda$ free lifetime (see red symbol in Fig. \ref{fig:HypTrit} in comparison with the black horizontal line).
This measurement goes in the direction of solving the hypertriton puzzle.

\subsubsection{Theory of hypernuclei: from chiral effective field theory to lattice QCD}
\label{sec:hypernuclei-theory}

Several approaches have been used over the years to determine the properties of hyperons in finite nuclei and to relate them with the underlying $YN$ and $YY$ interactions \cite{Gal:2016boi}. Those range from mean-field potentials of Woods-Saxon type \cite{Motoba:1988ww}, non-relativistic Hartree-Fock calculations using Skyrme-type $YN$ interactions \cite{Millener:1988hp} to relativistic approaches, based on Dirac phenomenology \cite{Brockmann:1980ar} or mean-field theory \cite{Boguta:1980wi} (see  Ref.~\cite{Vidana:2018bdi} for a review and an extensive list of references for all previous approaches). 
Also, microscopic hypernuclear structure analyses derived from diagrammatic expansions using the bare $YN$ have been performed (see \cite{Vidana:1998ed} and references therein), as well as quantum Monte-Carlo calculations of single- and double- $\Lambda$ hypernuclei using two- and three-body forces between the $\Lambda$ and nucleons \cite{Lonardoni:2013gta,Lonardoni:2013rm}. The quality of the description of hypernuclei in most of these previous works relies on the validity of the mean-field picture.  However, the correlations induced by the $YN$ interaction can substantially change this picture, as seen in Ref.~\cite{Vidana:2016ayd}.

Whereas most of the microscopic calculations on hypernuclear structure have been done using $YN$ meson-exchange models, in the recent years there has been some progress in the study of light hypernuclei with $YN$ derived from $\chi$EFT (see \cite{Gazda:2015qyt,Haidenbauer:2019boi,Wirth:2019cpp} for recent calculations and references therein). In particular, in Ref.~\cite{Gazda:2015qyt} the large CSB observed in the binding energies of $^4_{\Lambda}$He and $^4_{\Lambda}$H hypernuclei was reproduced within ab-initio no-core shell model calculations using the LO of the $\chi$EFT $YN$ potential plus a charge symmetry breaking $\Lambda$-$\Sigma^0$ mixing vertex.   

More recently, the $\Lambda$ single-particle states for various hypernuclei from $^5_{\Lambda} {\rm He}$ to $^{209}_{\Lambda}{\rm Pb}$ have been calculated using the NLO13 and NLO19 versions of the NLO potentials, finding a qualitatively good agreement with the data for the NLO19 interaction \cite{Haidenbauer:2019thx}. As for the hypertriton and its binding energy, a new measurement by the STAR collaboration suggests a value of $E_{\Lambda}= 0.41 \pm 0.12$ MeV \cite{Adam:2019phl}, which is about three times larger than the original benchmark of $E_{\Lambda}= 0.13 \pm 0.05$ MeV \cite{Juric:1973zq}. This has stimulated the reanalysis of the hypertriton binding energy in the context of $\chi$EFT. The use of the recent $YN$ NLO13 and NLO19 potentials within Faddeev and Yakubovsky equations has shown that it is possible to reconcile a larger binding energy of the hypertriton, whereas keeping the overall description of the $\Lambda p$ and $\Sigma N$ data. This can be done by increasing the attraction in the $\Lambda p$ $^1S_0$ partial wave, while correspondingly reducing the attraction in the $^3S_1$ channel, with the caveat of breaking the strict SU(3) symmetry for the contact interactions in the $\Lambda N$ and $\Sigma N$ channels \cite{Le:2019gjp}. 
On the other hand,  as discussed in Sec.~\ref{sec:hyper-exp}, recent measurements of the hypertriton lifetime by ALICE \cite{Acharya:2019qcp} indicates consistency with the free $\Lambda$ lifetime, thus hinting to a small value of the hypertriton binding energy, that could lead to a possible reanalysis of the theoretical predictions.  

\begin{figure}[htb]
    \centering
    \includegraphics[width=0.8\textwidth]{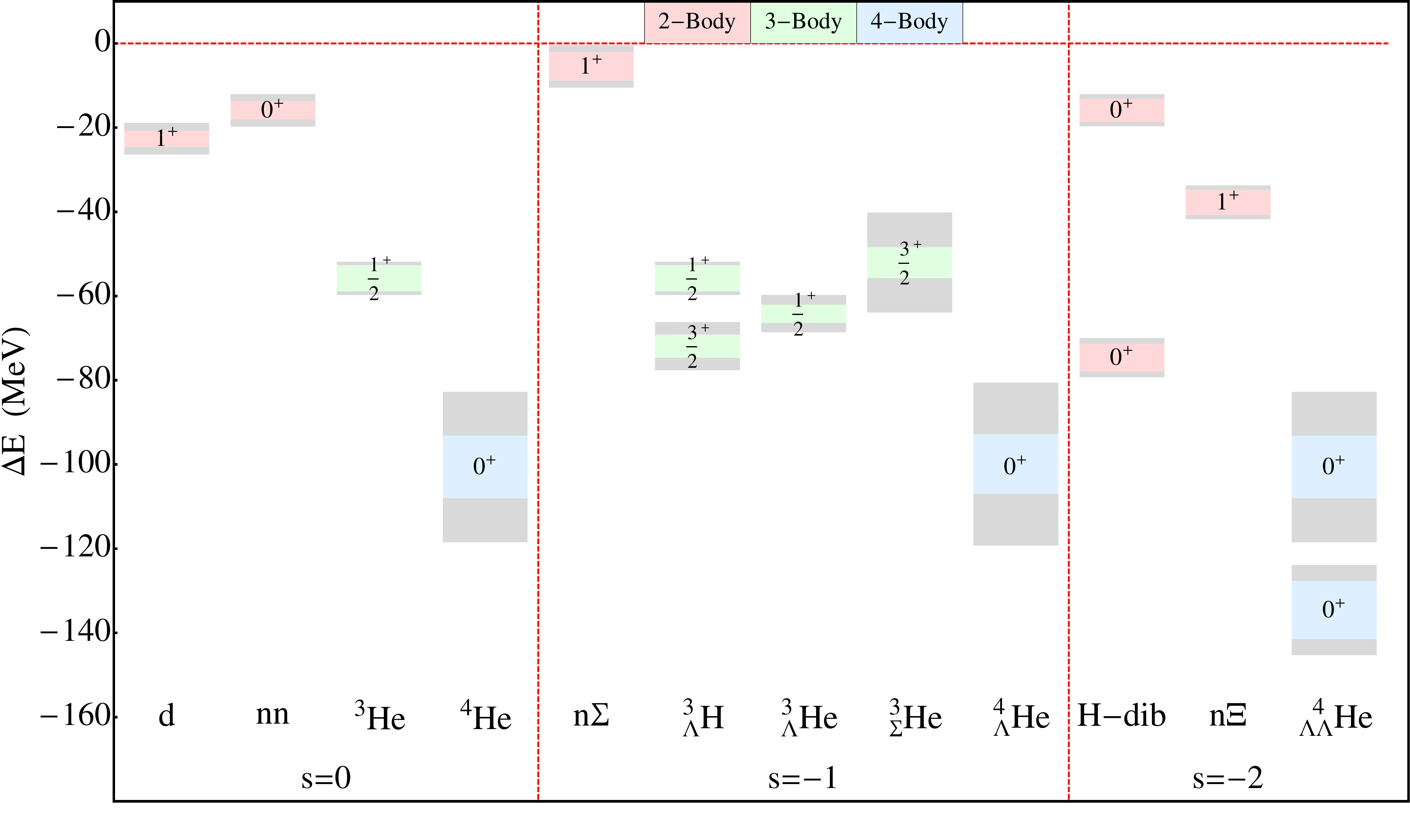}
    \caption{\textit{(Color online)} Results of the NPLQCD collaboration on the low-lying energy levels for different nuclei and hypernuclei, with the spin and parity $J^{\pi}$. Revised figure from \cite{Beane:2012vq}.}
    \label{fig:hyper-lattice}
\end{figure}

LQCD collaborations have also addressed the study of hypernuclei. In Ref.~\cite{Beane:2012vq} the lightest nuclei and hypernuclei were analyzed in the absence of electromagnetic interactions, showing the first results on LQCD for a number of s-shell nuclei and hypernuclei with $A\leq 5$, including the hypertriton $^3_{\Lambda} {\rm H}$ as well as $^3{\rm He}$, $^3_{\Lambda}{\rm He}$, $^4_{\Lambda}{\rm He}$, $^4_{\Lambda \Lambda}{\rm He}$ and the H-dibaryon in the limit of exact SU(3)-flavor symmetry at the physical strange-quark mass, with $m_{\pi}=m_{K}=m_{\eta} \sim 800$ MeV.  The binding energies of the extracted nuclear and hypernuclear states with their corresponding spin and parity $J^{\pi}$ are shown in Fig.~\ref{fig:hyper-lattice}. At this unphysical value of the quark mass, all the baryon systems are bound. The same kind of calculations at lighter quark-mass values are then needed  to determine how the binding energies evolve as the physical point is approached. Work alone this line is in progress.

In Ref.~\cite{Hiyama:2019kpw}, on the other hand, the authors have searched for  $\Xi N N $ and $\Xi NNN$ bound systems using phenomenological potentials and the $\Xi N$ potential extracted from LQCD \cite{Sasaki:2018mzh}. The $\Xi NNN$ system with isospin 0 and  $J^{\pi}$ = $1^+$ appears to be bound, being deeply bound for the Nijmegen potential and shallow for the HAL QCD one, below the $^3{\rm H}$/$^3{\rm He}$ + $\Xi$ threshold.

\subsection{Hyperons in dense matter}
\label{sec:hyperon-dense}

\subsubsection{Hyperon production in hadron-hadron collisions}

Hyperon properties within nuclear matter have been also
investigated by measuring the hyperons momentum and rapidity 
distributions in different hadron-hadron collisions.
The main motivation for this kind of studies is that, in HICs at intermediate energies (kinetic energies of few GeV maximally),
baryonic densities up top $2-3$ $\rho_0$ should be reached and, hence, the kinematic of the produced hyperons might be modified by the hyperon
rescattering within the dense medium. Such measurements have been
carried out for $p+p/A$ \cite{Adamczewski-Musch:2016vrc,Agakishiev:2014kdy}, $\pi^-+A$ \cite{Benabderrahmane:2008qs,Adamczewski-Musch:2018eik} and $A+A$ \cite{Adamczewski-Musch:2018xwg} collisions at intermediate energies to study elementary reactions as a reference, light colliding systems to test the properties of hyperons at saturation densities, and finally HICs to infer on the hyperon propagation within dense environments.
So far, the only way to investigate possible in-medium modifications of hyperons is the comparison to transport models, where the interaction can be varied. As mentioned before, first elementary reactions are investigated to calibrate the production cross sections within the transport models.

\begin{figure}[htb]
\begin{center}
\includegraphics[height=95mm,width=0.73\textwidth]{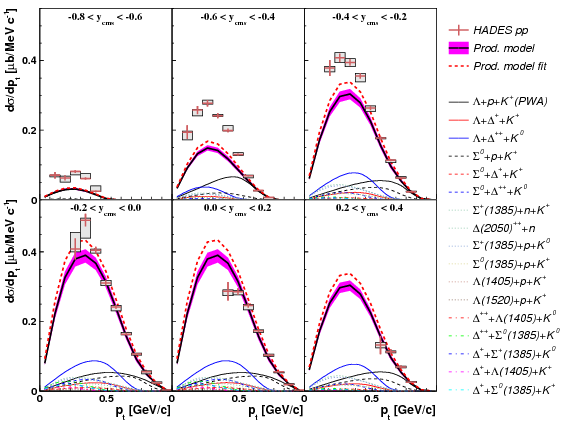}
\caption{\textit{(Color online)} $\Lambda$ differential cross sections as a function of $p_T$ for different rapidity intervals measured in $pp$ collisions at 3.5 GeV by HADES \cite{Adamczewski-Musch:2016vrc}. The magenta histogram represents the sum of all simulated production channels. The other curves listed in the legend correspond to the decomposition of the model in the single channels.
}
\label{fig:pp3.5Lambda}
\end{center}
\end{figure}

The $pp$ collisions at 3.5 GeV measured by HADES provided precision $p_T$ spectra for a rather broad rapidity interval \cite{Adamczewski-Musch:2016vrc}. Figure~\ref{fig:pp3.5Lambda} shows the measured differential production cross section for $\Lambda$ as a function of $p_T$ compared to a model that includes all known, and partially exclusively measured channels contributing to the inclusive final state. The model also accounts for the contributions of broad $N^*$ resonances and their interferences \cite{Agakishiev:2014dha}. One can see that the model underestimates the measured cross section for backward rapidities. Since this represents the reference for transport calculations, it is then seen that the reference for elementary reactions is not under control.

This result anticipates the limits of a direct comparison of data and models for the hyperon kinematic distributions in heavier colliding systems. It, hence, does not come as a surprise that the $\Lambda$ spectra measured in $Au+Au$ collisions at 1.23 AGeV by HADES \cite{Adamczewski-Musch:2018xwg} can not interpreted quantitatively relying on transport models.

\begin{figure}[htb]
\begin{center}
\includegraphics[height=85mm,width=0.45\textwidth]{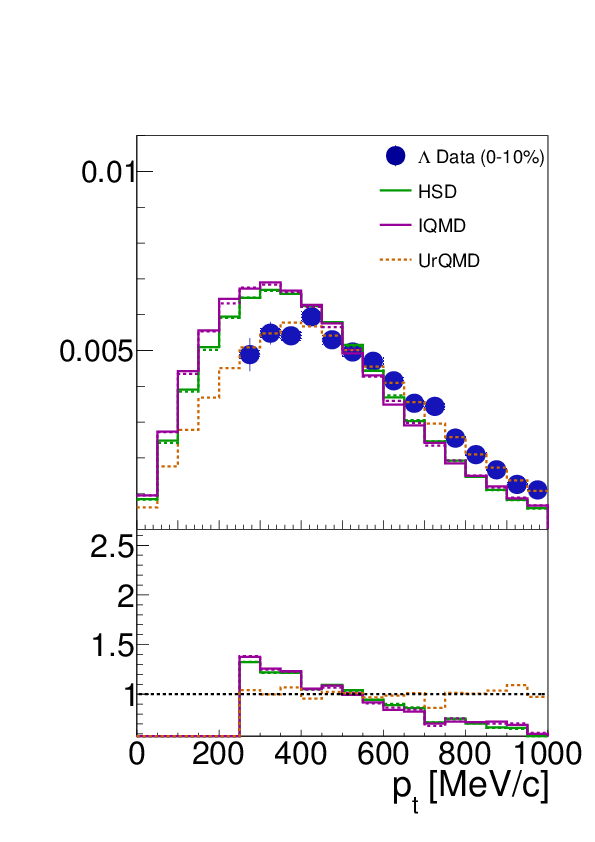}
\includegraphics[height=75mm,width=0.4\textwidth]{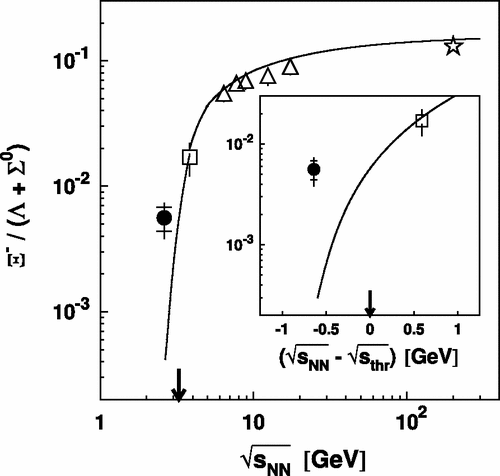}
\caption{\textit{(Color online)} Left panel: Inclusive $\Lambda$ $p_T$ distribution measured in $Au+Au$ collisions at 1.23 AGeV by HADES \cite{Adamczewski-Musch:2018xwg}, compared to normalized distributions obtained from different transport models. Right panel: $\Xi^-/(\Lambda +\Sigma^0)$ ratio  as function of $\sqrt{s_{NN}}$ or $\sqrt{s_{NN}}- \sqrt{s_{thr}}$ (inset) \cite{Agakishiev:2009rr}. The full circle represents the HADES measurement in $Ar+KCl$ at 1.76 AGeV, the arrow gives the threshold in free $NN$ collisions. The open star, triangles and square represent data for central $Au+Au$ and $Pb+Pb$ collisions measured at RHIC. The solid line refers to the prediction from a statistical hadronization model.
}
\label{fig:LambdaAuHADES}
\end{center}
\end{figure}

Figure \ref{fig:LambdaAuHADES} shows the inclusive $\Lambda$ $p_T$ distribution compared to the normalized distributions obtained from three different transport models. One can see that only the UrQMD model reproduces the shape of the measured distribution, but an absolute comparison shows that the UrQMD overpredicts the measured $\Lambda$ yield by 30\%  and does not describe the $K^0_s$ at the same time. Hence no conclusion on the hyperon in-medium properties can be drawn. 

Additional measurements of $\Lambda$ hyperons have been carried out by the FOPI collaboration \cite{PhysRevC.76.024906}, but also for this data set not conclusive information about the behaviour of hyperons in matter could be extracted.

The $\Sigma^0$ hyperons have also been measured in $p +Nb$ reactions a 3.5 GeV by HADES \cite{Adamczewski-Musch:2017gjr} by exploiting the decay $\Sigma^0 \rightarrow \Lambda + \gamma $. However, only an inclusive  production cross section could be extracted from this measurement, because the reconstruction of the $80$ MeV photon via its decay into $e^+e^-$ pairs has a very low efficiency. 
 
 Moreover, $\Xi^-$ hyperons have been measured in $Ar+KCl$ collisions at 1.76 AGeV also with HADES  \cite{Agakishiev:2009rr}. The weak decay $\Xi^- \rightarrow \Lambda + \pi^-$ was exploited and also, in this case, the statistics is too limited to allow for a differential analysis. The measured $\Xi^-/(\Lambda +\Sigma^0)$ yield can be, however, compared to other measurements at higher energy and to the predictions from a statistical hadronization model. Although the process is deep sub-threshold, the measured yield is found about a factor 24 higher than expected according to the statistical model. This result indicates that the conditions for the applicability of statistical models are not fulfilled for cascade production in small systems far below threshold, and that the unexpectedly high $\Xi^-$ yield might be due to modifications of the hyperon properties within nuclear matter. On the other hand, so far no quantitative explanation of this effect could be found.

\subsubsection{Hyperons in dense matter from  baryon-baryon interactions}
\label{sec:hypmatter-theory}

 Starting from the realistic $YN$ and $YY$ interactions (see Sec.~\ref{sec:hyper-theory}), the properties of hyperons in dense matter can be obtained by incorporating medium corrections coming from the interaction with the surrounding many-body medium. This so-called microscopic formulation has been attempted following various many-body approaches, such as variational calculations \cite{Akmal:1998cf},  renormalization group methods \cite{Djapo:2008au}, LQCD calculations \cite{Inoue:2018axd,Beane:2012ey} or diagrammatic expansions within the Brueckner-Hartree-Fock approach (BHF) \cite{Schulze:1995jx,Schulze:1998jf,Baldo:1998hd,Vidana:1999jm,Vidana:2000ew,Schulze:2011zza} or Dirac-Brueckner-Hartree-Fock (DBHF) \cite{Sammarruca:2009wn, Katayama:2015dga}.

One of the most widely used method is the BHF approach to compute single-particle potentials of hyperons within nuclear matter. The  fundamental  input  of  the BHF  calculations are  the  potentials  in  the  $NN$, $YN$,  and  $YY$  sectors, supplemented  by  three-body  forces,  which  in  the  $NN$  sector  are  required  in  order  to  ensure  a  correct saturation point of nuclear matter.  One of the first BHF calculations was performed using  $YN$ interactions adopted from the Nijmegen  potentials \cite{Schulze:1998jf}. The calculation did not take into account, however, the  $YY$ interactions. Later on, the single-particle potentials of nucleons and hyperons were obtained using the  $NN$, $YN$ and $YY$ interactions from the Nijmegen potentials NSC89 and Nijmegen97(e) \cite{Vidana:1999jm}. A further step forward was made within BHF by employing the version ESC08b of the Nijmegen $YN$ potential ~\cite{Schulze:2011zza}. One interesting feature of the ESC08b $YN$ is the repulsive $\Sigma^- N$ interaction, which was  attractive in the previous Nijmegen models.

In more recent works, the single-particle potentials of the $\Lambda$ and $\Sigma$ in nuclear matter have been calculated within the same BHF approach, but using the LO and NLO (NLO13) $YN$ interactions of \cite{Haidenbauer:2014uua}. The $\Lambda$ single-particle potential turns out to be in good qualitative agreement with the empirical values extracted from hypernuclei \cite{Gal:2016boi}, whereas the $\Sigma$-nuclear potential, on the other hand, has been found to be repulsive, although some independent studies of $\Sigma^-$ atoms conclude that the $\Sigma^-$-nucleus potential should be attractive \cite{Batty:1978sba,Oset:1989ey}. 
The analysis of Ref.~\cite{Haidenbauer:2014uua} has been subsequently revisited in Ref.~\cite{Petschauer:2015nea} by a) employing also the $NN$ from $\chi$EFT, b) implementing the continuous choice for intermediate-state spectra in the BHF approach, and c) investigating isospin-asymmetric matter. The $\Sigma$-nuclear potential turns out to be moderately repulsive for both LO and NLO13, and the $\Lambda$-nuclear potential becomes repulsive from two-to-three times saturation density.

\begin{figure}[htb]
\begin{center}
\includegraphics[width=0.6\textwidth,angle=-90]{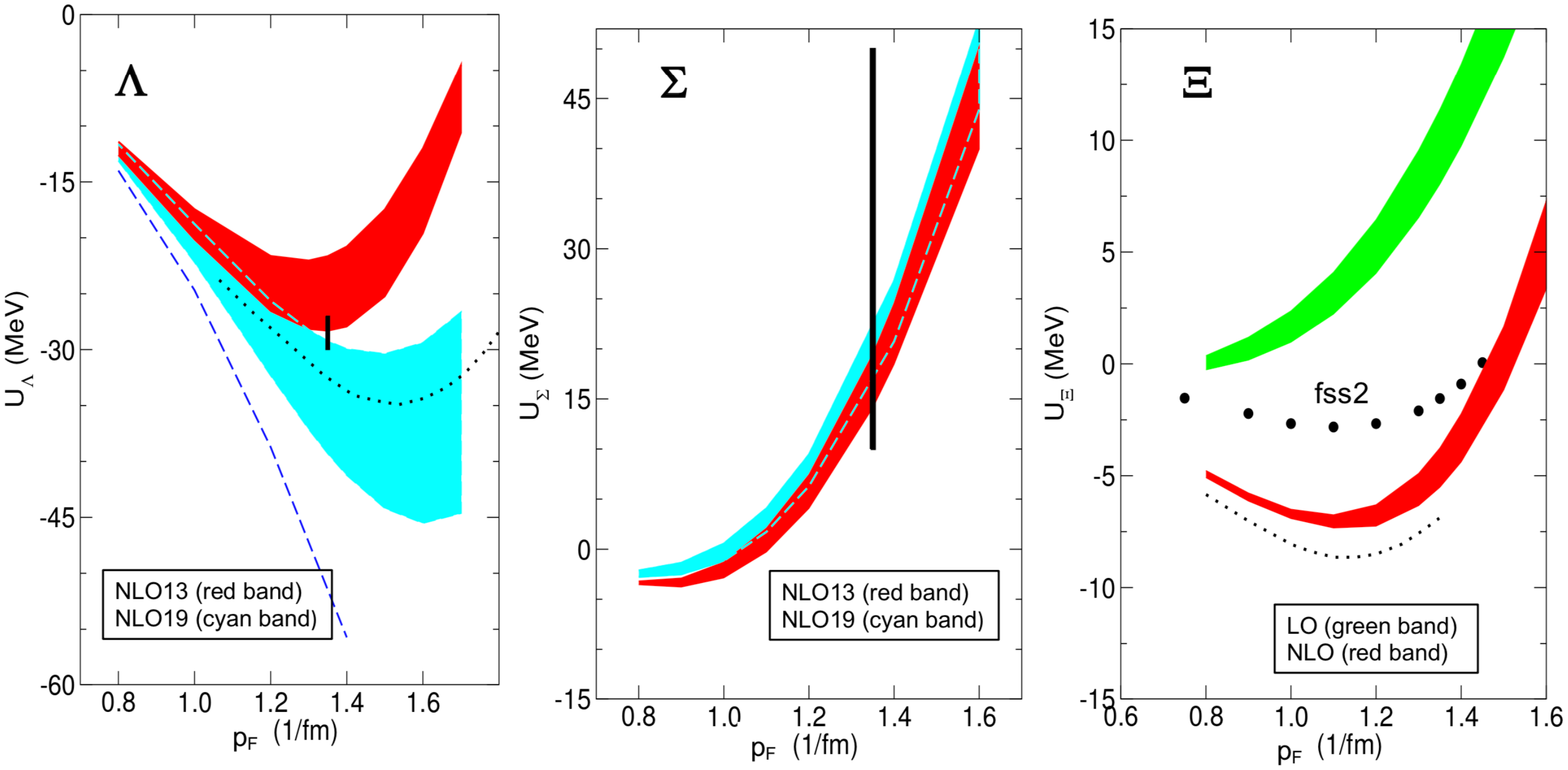}
\caption{
\textit{(Color online)} Left and middle panels: $\Lambda$ and $\Sigma$ single-particle potentials as a function of the Fermi momentum $p_F$ in symmetric nuclear
matter for NLO13 (red band) and NLO19 (cyan band). The results for the Nijmegen NSC97f potential (dotted lines) and the J\"ulich04 meson-exchange potential (dashed lines) are also shown.  The vertical bar indicates the empirical value. Figures adapted from \cite{Haidenbauer:2019boi}. Right panel:  The  $\Xi$ single-potential as a function of the Fermi momentum $p_F$ in symmetric nuclear
matter for LO (green band) and NLO (red band). The circles correspond to the fss2 scheme and the dotted line to the ESC08c Nijmegen potential. Figure adapted from \cite{Haidenbauer:2018gvg}.}
\label{fig:la-sig-xi-pot}
\end{center}
\end{figure}

Figure~\ref{fig:la-sig-xi-pot} shows the single particle potentials for $\Lambda,\,\Sigma$ and $\Xi$ as a function of Fermi momenta. In first panel, one can see that the older calculations based on meson-exchange interactions (dashed lines) lead to a mere attractive single-particle potential for $\Lambda$ at saturation density. The BHF based on NLO13 interactions (red band) shows also attraction at saturation, with a repulsive component dominating at higher densities. The $\Sigma$ single-particle potential is repulsive in all the considered calculations, as seen in the middle panel of Fig.~\ref{fig:la-sig-xi-pot}.

Very recently, the NLO19 $\chi EFT$ $YN$ has been used for the computation of the hyperon single-particle energies, differing from the previous NLO13 as described in Sec.~\ref{sec:hyper-theory}. As discussed in Sec.~\ref{sec:hyper-theory}, it has been found that the NLO13 and NLO19 variants yield equivalent results for $\Lambda N$ and $\Sigma N$ scattering observables \cite{Haidenbauer:2019boi}. However, the in-medium $\Lambda$ and $\Sigma$ properties in matter are different when using NLO13 or NLO19, as the strength of $\Lambda N$-$\Sigma N$ transition potential changes.

From the left panel of Fig.~\ref{fig:la-sig-xi-pot}, it is observed that the $\Lambda$-nuclear potential using the new NLO19 is much more attractive in the medium than for NLO13. Comparing the results from the J\"ulich04 model and NSC97f meson-exchange potentials to NLO13 and NLO19, the $\Lambda$ single-potential is more attractive for the J\"ulich04 potential (dashed lines), whereas the NSC97f (dotted lines) shows a similar behaviour as the NLO19 (cyan band). As for the $\Sigma$-nuclear potential, the NLO19 interaction provides slightly more repulsion that for the other interactions, although there is very little difference with respect to NLO13 (middle panel of Fig.~\ref{fig:la-sig-xi-pot}). The dominant contribution of the $^3S_1$ partial wave of the $\Sigma^+ p$ channel is the responsible for the repulsion. The authors indicate that their result for $\Sigma$ single-particle potential is in accordance with the observation from studies of the level shifts and widths of $\Sigma^-$ atoms, and
measurements of ($\pi^-, K^+$) inclusive spectra for the $\Sigma^-$-formation in heavy nuclei, that show that the $\Sigma$-nuclear potential is repulsive.

The effect of three-body forces has also been addressed in the case of the $\Lambda$-nuclear interaction in Ref.~\cite{Haidenbauer:2016vfq}. Three-body forces are necessary to reproduce few-nucleon binding energies, scattering observables and the nuclear saturation properties in non-relativistic many-body approaches, such as BHF. Adding a density-dependent effective $\Lambda N$ interaction constructed from the chiral $\Lambda N N$ force, the $\Lambda$ single-particle potential becomes more repulsive when three-body forces are included.

As for the $\Xi$-nuclear potential (right panel of Fig.~\ref{fig:la-sig-xi-pot}), the interaction strength varies between -3 to -5 MeV, smaller than the usually reported values of -14 MeV \cite{Gal:2016boi}, but in line with another BHF outcome using the J\"ulich $\Xi N$.

\begin{figure}[htb]
    \centering
    \includegraphics[angle=-90, width=0.9\textwidth]{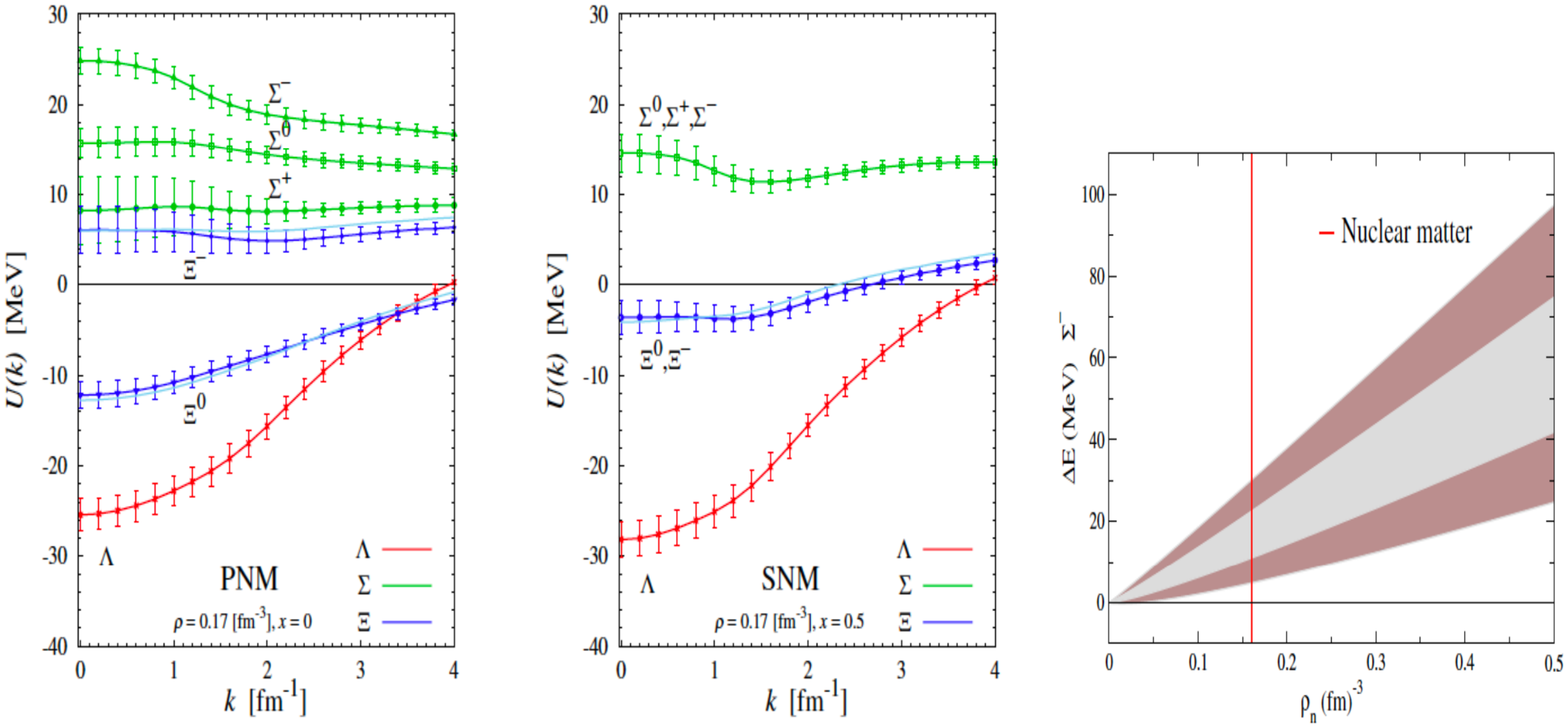}
    \caption{\textit{(Color online)} Left and middle panels: Hyperon single-particle potentials in pure neutron matter (left) and normal nuclear matter (middle), based on the interaction potentials extracted from the lattice by the HAL QCD collaboration. Figures taken from \cite{Inoue:2018axd}. Right panel: The energy shift of $\Sigma^-$ versus the neutron density, with the inner (outer) bands encompassing the statistical (systematic) uncertainties, obtained by the NPLQCD collaboration. Figure taken from \cite{Beane:2012ey}.}
    \label{fig:pot-lattice}
\end{figure}

Interestingly, the single-particle potentials for the $\Lambda$, $\Sigma$ and $\Xi$ have been also calculated within the BHF approach using the potentials extracted from the lattice by the HAL QCD collaboration \cite{Inoue:2018axd}. The authors have obtained the single-particle potentials for hyperons in pure neutron matter and, also, in symmetric nuclear matter (see left and middle panels of Fig.~\ref{fig:pot-lattice}). The values for saturation density in symmetric nuclear matter are $U_{\Lambda}$=-28 MeV, $U_{\Sigma} =$ 15 MeV, and $U_{\Xi} =$ -4 MeV, with a statistical error about $\pm$ 2 MeV associated with the Monte-Carlo simulation. As discussed in Sec.~\ref{YN_YY_Exp}, the measured $p\Xi^-$ correlation at the LHC has validated the HAL QCD predictions. The resulting slightly repulsive single-particle potential for $\Xi^-$ in pure neutron matter disfavors the $\Xi$ appearance within neutron stars. 

The NPLQCD collaboration, on the other hand, has been able to estimate the $\Sigma^-$ energy shift in pure neutron matter. 
Fumi's theorem relates the energy shift due to a static impurity in a non-interacting Fermi system to an energy integral over the phase shifts. Using this result together with the LQCD determinations of the phase shifts (and allowing for a 30$\%$ uncertainty), the authors find that the repulsion in the $\Sigma^-n$ system does not exclude the presence of $\Sigma$ hyperons in neutron star matter. The energy shift of $\Sigma^-$ versus the neutron density, with the inner (outer) bands encompassing the statistical (systematic) uncertainties, is shown in the right panel of Fig.~\ref{fig:pot-lattice}. 

Last but not least, quark models \cite{Kohno:2009sc} or renormalization group schemes in matter \cite{Djapo:2008au} have been also used to determine the properties of hyperons in medium. As an example, we should mention the $YN$ potential low-momentum interaction deduced from the available $YN$ bare potentials. It has been found that, due to the limited experimental data on $YN$, the $YN$ bare potentials are not well constrained, and thus there is no-universal $YN$ potential low-momentum interaction \cite{Djapo:2008au}, so no clear determination of the hyperon properties in dense matter is possible.

%% file: Content/EOSandNeutronStars.tex
\section{Connection to Neutron Stars}
\label{sec:NS}

\subsection{Brief description of Neutron Stars}
\label{sec:briefNS}

Neutron stars (NSs) are a unique laboratory to test the fundamental properties of matter under strong gravitational and magnetic fields as well as at extreme conditions of density, isospin asymmetry and temperature \cite{Lattimer:2006xb,Watts:2016uzu,Watts:2018iom}.

\begin{figure}[htb]
    \centering
    \includegraphics[width=0.6\textwidth]{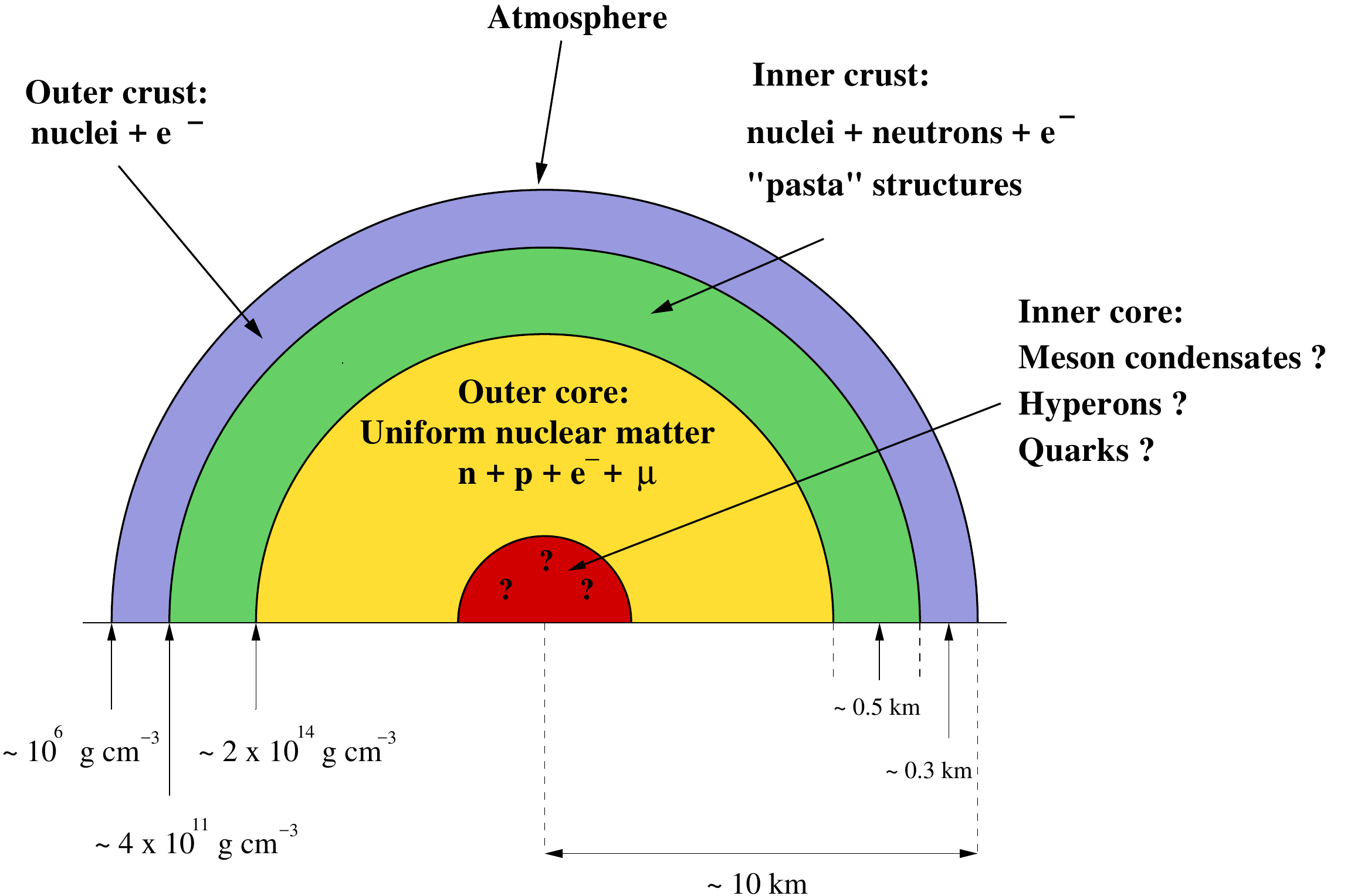}
    \caption{\textit{(Color online)} Schematic picture of the interior of a neutron star. Figure adapted from \cite{Vidana:2018lqp}. }
    \label{fig:NSscheme}
\end{figure}

NSs are supported against gravitational collapse mainly by the neutron degeneracy pressure. NSs have typically masses of the order of 1-2M$_{\odot}$ and radii within the range 10-12 km. Such masses and radii yield an averaged density for NSs of the order of $\sim$ $10^{14} {\rm g/cm}^3$. Nevertheless, the expected densities span over a wide range as the internal structure can be described as an onion-like, as depicted in Fig.~\ref{fig:NSscheme}. In this figure, a schematic picture of the internal structure of NSs is shown, where one can distinguish several layers: the atmosphere, the thin outermost layer; the outer and inner crust with about $\approx$ 1 km and a mass of only a few percent of the total mass of the star; and the core, divided in the outer and the inner core, with a radius of about $\approx$ 10 km and almost the total mass of the star. The composition of this inner region is though not known, and several hypothesis have been postulated, such as pion or kaon condensates, baryonic matter made of nucleons and hyperons, and/or deconfined quark matter. In this review we assume that baryonic matter occupies the inner core of NSs, as we are interested in the properties of nucleons and hyperons in neutron star matter.

NSs are in equilibrium against weak interaction processes, that is, 
\begin{eqnarray}
b_1 \rightarrow b_2+l+\bar{\nu_l}, \hspace{1cm} b_2 + l  \rightarrow b_1 + \nu_l,
\end{eqnarray}
 where $b_i$ refers to a certain type of baryon, $l$ represents a lepton, and $\nu_l$ and $\bar \nu_l$ the corresponding neutrino and antineutrino, respectively. The composition of $\beta-$equilibrated charged neutral matter can be determined by analyzing all the possible reactions among the different components of matter, and expressed them in terms of the different chemical potentials $\mu_i$, so that one arrives to 
\begin{eqnarray}
\mu_{b_i}&=&B_{b_i} \mu_n - q_{b_i} \mu_e , \nonumber \\ 
\mu_{l_j}&=&-q_{l_j} \mu_e ,
\end{eqnarray}
where $B_{b_i}$ is the baryonic number of the baryon $b_i$, $q_{b_i(l_j)}$ the charge of $b_i$ baryon ($l_j$ lepton), and $\mu_n$ and $\mu_e$ the chemical potential of the neutron and electron, respectively. Note that since the mean free path of the (anti-)neutrinos is much larger than the typical size of a NS, they freely escape without contributing to the energy balance.
The charged neutrality is ensured by 
\begin{eqnarray}
\sum_{b_i} q_{b_i} \rho_{b_i} + \sum_{l_j} q_{l_j} \rho_{l_j}=0 ,
\end{eqnarray}
where $\rho_{b_i(l_j)}$ is the density of $b_i (l_j)$, whereas the total baryonic density is given by
\begin{eqnarray}
\rho=\sum_{b_i} B_{b_i} \rho_{b_i}.
\end{eqnarray}
 
The structure of NSs can be determined using Einstein's general relativity theory.  Einstein's field equations for a spherical static star take the form of the  Tolman-Oppenheimer-Volkoff (TOV) structure equations, that using $G=c=1$ units read
\begin{eqnarray}
&&\frac{dP(r)}{dr}=-\frac{G}{r^2} \left[\varepsilon (r) + P(r)\right] [M(r)+4 \pi r^3 P(r)] \left[1- \frac{2GM(r)}{r} \right]^{-1}, \label{TOVeq1} \\
&& \frac{dM(r)}{dr}=4 \pi  r^2 \varepsilon(r). \label{TOVeq2}
\end{eqnarray}
The TOV equations are the equations of the hydrostatical equilibrium in general relativity. They can be interpreted as follows. Taking a shell of matter of radius $r$ and thickness $dr$, we can see that Eq.~(\ref{TOVeq2}) gives the mass energy in this shell. The left-hand side of Eq.~(\ref{TOVeq1}) is the net force acting outwards on the surface of the shell by the pressure difference between the interior and the exterior,  $dP(r)$, whereas the right-hand side is the force of gravity acting on the shell by the mass in the interior. 

As earlier mentioned, the TOV equations are valid for a spherical static compact objects. However, NSs are rotating stars, detected as they pulsate (pulsars). Thus,  the spherical symmetry is broken and only the axial symmetry remains as the NSs flatten with rotation. In this case, the usual approach to address rotation in NSs is based on a perturbative method developed by Hartle and Thorne \cite{Hartle:1968si}. 

The equation of state (EoS), i.e., the relation between the pressure $P$ and the energy density $\epsilon$, is the manner in which matter and its composition enters the equations of stellar structure. For a given EoS, the TOV equations can be integrated by specifying the initial conditions, that is, the enclosed mass and the pressure at the center of the star, $M(r=0)=0$ and $P(r=0)=P_c$, with $P_c$ taking an arbitrary value. The integration of the TOV equations over the radial coordinate $r$ ends when $P(r=R)= 0$, where $R$ is the radius of the star and $M(R)$ the total mass $M$.

\subsection{The nuclear Equation of State}
\label{sec:nucEoS}

As previously indicated, to solve the structure equations of NSs, one needs to first determine the  EoS. This task is, however, complicated given the fact that the EoS for NSs span for a wide range of densities, temperatures and isospin asymmetries.

Models of the EoS in the NS crust are based on  atomic nuclei experimental data, nucleon scattering results, and theoretical models for strongly coupled Coulomb systems. As we go deeper in the star and for densities $\approx  10^{14} {\rm g/cm^3}$, matter becomes a uniform fluid of neutrons, protons and electrons. The EoS in the outer core of the NS can be obtained by using models for nuclear structure calculations. However, as we enter the unknown inner core, we rely on theoretical calculations of the nuclear EoS that could  be only tested by astrophysical observations.

The nuclear EoS describes an idealised infinite uniform system of nucleons, where the Coulomb interaction is switched off. Symmetric nuclear matter, with an equal number of neutrons and protons, is the simplest approximation to bulk matter in heavy atomic nuclei, whereas pure neutron matter is the simplest approach to the matter in the NS core.

It is convenient to express the energy per nucleon of the nuclear system of density $\rho$ as
\begin{equation}
    \frac{E}{A}(\rho,\delta) = \frac{E}{A}(\rho,0)+S(\rho) \delta^2 + ...,
\end{equation}
with $\delta=(N-Z)/A$, $N(Z)$ being the neutron (proton) number and $A=N+Z$. The $(E/A)(\rho,0)$ term describes the energy per nucleon of symmetric nuclear matter ($\delta=0$), whereas $S(\rho)$ is the symmetry energy that measures the energy cost involved in changing the protons into neutrons in nuclear matter. Expanding both terms around saturation density for symmetric matter, $\rho_0$, one obtains
\begin{eqnarray}
\frac{E}{A}(\rho,0)= \frac{E}{A}(\rho_0)+\frac{1}{18}K_0 \epsilon^2 + ... , \nonumber \\
S(\rho)=S_0+\frac{1}{3}L \epsilon + \frac{1}{18} K_{\rm sym} \epsilon^2 +...,
\end{eqnarray}
where $\epsilon=(\rho-\rho_0)/\rho_0$. The isoscalar parameters are the binding energy per nucleon at saturation $(E/A)(\rho_0)$ and the incompressibility at the saturation point $K_0$. The isovector parameters are the symmetry energy coefficient at saturation $S_0$, and $L$ and $K_{\rm sym}$ that characterize the density dependence of the symmetry energy around saturation. The $K_0$, $S_0$, $L$ and $K_{\rm sym}$ parameters are defined as
\begin{eqnarray}
K_0 \equiv 9 \rho_0^2 \left( \frac{\partial^2 (E/A)(\rho,\delta)}{\partial \rho^2}\right)_{\rho_0,\delta=0}, \hspace{1cm}
S_0 \equiv \frac{1}{2} \left( \frac{\partial^2 (E/A)(\rho,\delta)}{\partial \delta^2}\right)_{\rho_0,\delta=0}, \nonumber \\
L \equiv 3 \rho_0\left( \frac{\partial S(\rho)}{\partial \rho}\right)_{\rho_0}, \hspace{1cm}
K_{\rm sym} \equiv 9 \rho_0^2 \left( \frac{\partial^2 S(\rho)}{\partial \rho^2}\right)_{\rho_0}.
\label{eq:eosparam}
\end{eqnarray}

Note that the energy density of the system is given by
\begin{equation}
\varepsilon(\rho,\delta)=\rho \frac{E}{A}(\rho,\delta),
\end{equation}
and the pressure is then easily obtained using thermodynamic relations
\begin{eqnarray}
P=\rho^2\frac{\partial (E/A)(\rho,\delta)}{\partial \rho}=\rho \frac{\partial \varepsilon(\rho,\delta)}{\partial \rho}- \varepsilon(\rho,\delta). 
\end{eqnarray}

\subsubsection{Theoretical models on the nuclear Equation of State}

The theoretical description of nuclear matter in the core of NS is based on the use of different theoretical many-body approaches. These are usually divided in two main categories, microscopic ab-initio approaches and phenomenological schemes. 

 Microscopic ab-initio approaches obtain the nuclear EoS by solving the many-body problem starting from two-body and three-body meson-exchange or chiral interactions, that are fitted to experimental scattering data and to the properties of finite nuclei (see discussion in Sec.~\ref{sec:hyper-theory}). These approaches deal with the difficulty of the treatment of the short-range repulsive core of the nuclear force. The ab-initio approaches include schemes based on the variational method \cite{Akmal:1998cf}, quantum-montecarlo techniques \cite{Wiringa:2000gb,Carlson:2003wm,Gandolfi:2009fj}, the correlated basis function formalism \cite{Fabrocini:1993eaz}, diagrammatic methods  (Brueckner-Bethe-Goldstone expansion \cite{Day:1967zza}, the DBHF approach \cite{TerHaar:1986xpv,Brockmann:1990cn} and the self-consistent Green's function \cite{2005mbte.book.....D}), renormalization group methods \cite{Bogner:2003wn}, and LQCD calculations \cite{Beane:2010em,Ishii:2006ec}. The advantage of most of these approaches is the systematic addition of higher-order contributions allowing for a controlled description of the EoS. However, the main disadvantage is the applicability to high densities, as the inclusion of higher-order contributions make the calculations more difficult and tedious.

Phenomenological schemes rely on density-dependent interactions that are adjusted to nuclear observables and NS observations. Among them, one can find energy-density functional schemes of non-relativistic type, such as the Skyrme \cite{Skyrme:1959zz} or Gogny \cite{Decharge:1979fa} schemes, or relativistic models, usually derived from a Lagrangian with baryon and meson fields, at the mean-field or Hartree-Fock level \cite{Boguta:1977xi,Serot:1984ey}. The advantage of these schemes is the possible application to high densities beyond the saturation density, whereas the disadvantage lies on the fact that those approaches are built in a non-systematic manner. 
 
For a more detailed explanation of the different models and an exhaustive list of references, we refer the reader to recent reviews, such as Refs.~\cite{Oertel:2016bki,Burgio:2018mcr}.

\subsection{Constraints on the nuclear Equation of State}
\label{sec:constraints}

Constraints on the nuclear EoS can be inferred from nuclear physics experiments and astrophysical observations. However, in most cases, these constraints are extracted after theoretical modelling and/or after extrapolations to regions not accessible by experiments and observations. Hence, these constraints have to be taken with caution.

\subsubsection{Experimental constraints}
\label{sec:expconst}

In order to study the EoS of dense nuclear matter, several experimental methods can be exploited. With respect to the  connection to the physics of NSs, it is particularly relevant to test experimentally the role played by neutrons in the EoS. Hence, a measurement of the symmmetry energy $S(\rho)$ for different densities is aimed for.

\begin{figure}[htb]
    \centering
    \includegraphics[width=0.45\textwidth]{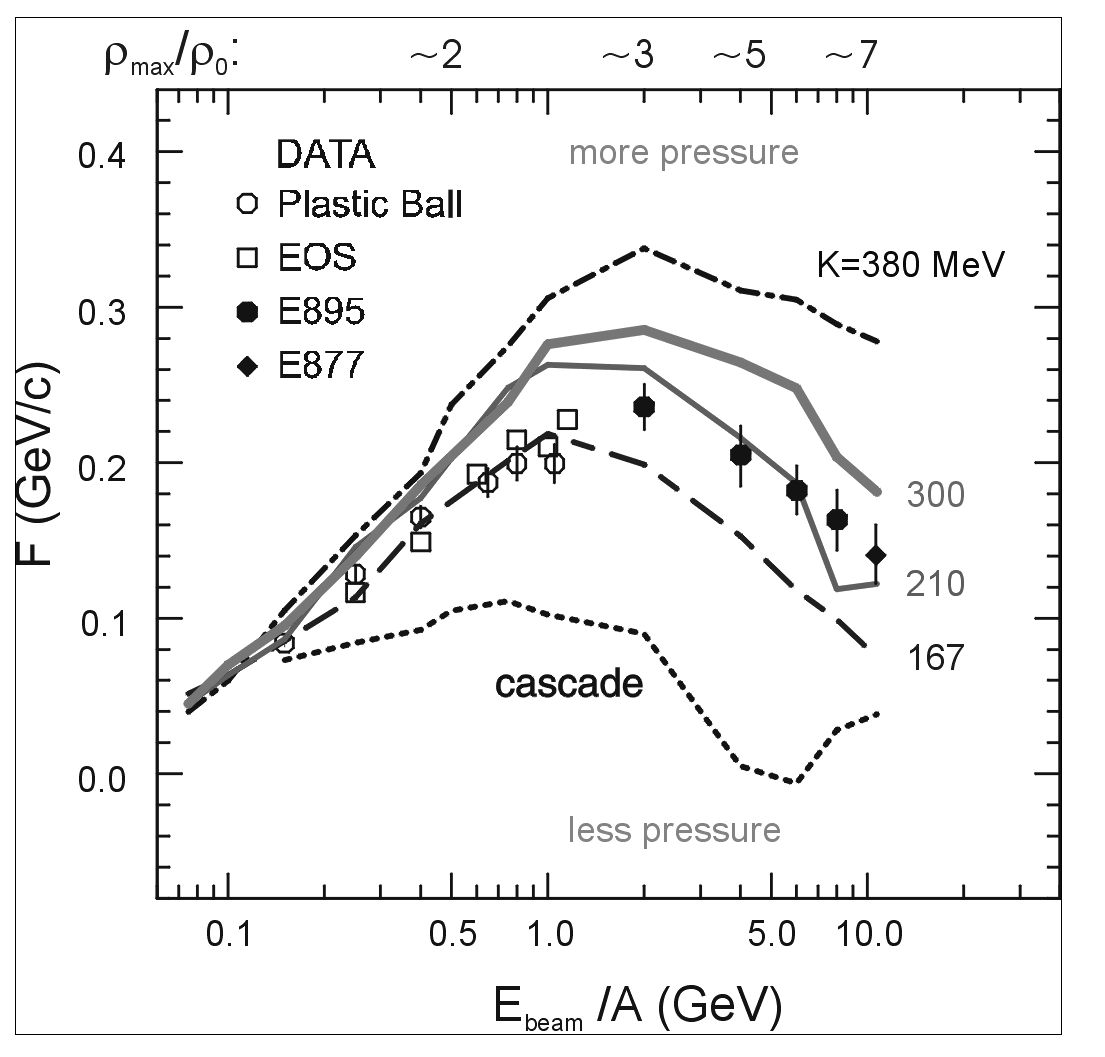}
      \includegraphics[width=0.5\textwidth]{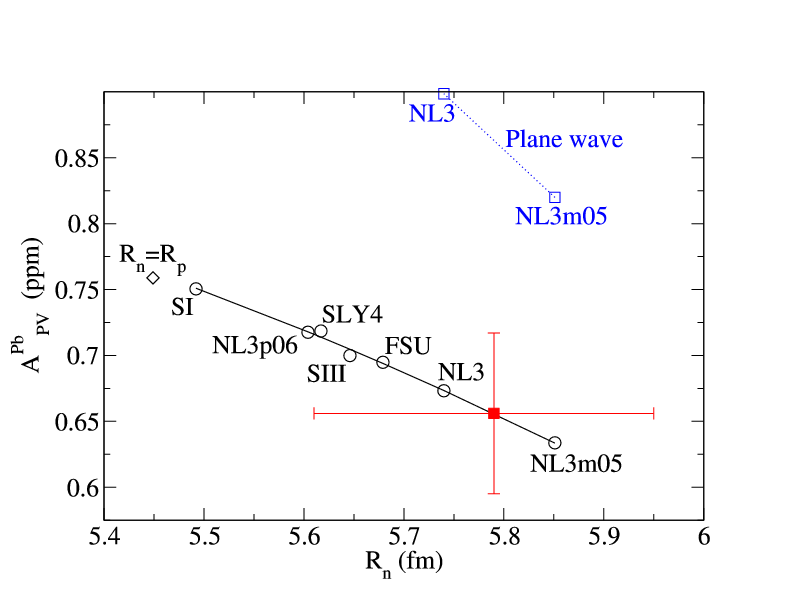}
    \caption{\textit{(Color online)} Left panel: The solid and open points show experimental values for the transverse flow as a function of the incident energy per nucleon for HICs in fixed target experiments.  The various lines represent the predictions from transport code calculations assuming different incompressibility (parameter $K_0$) of the nuclear EoS. The maximal densities reached in the collisions are also estimated via transport calculations \cite{Danielewicz:2002pu}. Right panel: Measured asymmetry of electron scattering versus the neutron point radius $R_n$  in $^{208}{\rm Pb}$ \cite{Abrahamyan:2012gp}.
}
   \label{fig:EOSHIC}
\end{figure}

HICs at different energies have been exploited to quantify the EoS of symmetric nuclear matter and the transverse flow measurement\footnote{Collective flow perpendicular to the beam axis} of charged particles emitted in collisions \cite{Danielewicz:2002pu} has been employed.
The left panel of Fig.~\ref{fig:EOSHIC} shows the transverse flow of charged particles measured for collisions with $^{197}$Au projectiles, impinging on target nuclei at incident kinetic energies $E_{\rm beam}$/A ranging from about 0.15 to 10 GeV per nucleon (29.6 to 1970 GeV total beam kinetic energies). 
The experimental data are compared to the predictions by transport model calculations assuming different values of the incompressibility of nuclear matter $K_0$ (see Eqs.~(\ref{eq:eosparam})). This parameter quantifies the curvature of the energy per nucleon as a function of the system density at saturation density. Hence, it is crucial for the extrapolation at high densities. In the case of low values of $K_0$ (160-210) we speak of a soft EoS, while larger values of $K_0$ ($>$ 300) indicate a stiff EoS. Using these analyses, a strongly repulsive nuclear EoS and weakly repulsive EoS with phase transitions at densities less than three times that of stable nuclei are ruled out, and soft EoSs at higher densities are favoured.

Moving to neutron rich matter, the symmetry energy $S(\rho)$ is related to the pressure exerted by the neutron excess. One experimental method to evaluate the symmetry energy within nuclei is to study
the response of a nucleus to an external electric field. Indeed, if protons are displaced by
the field, regions of asymmetry are created that are not energetically favorable due to the 
symmetry energy. One can study the collective excitation resulting after the external field has
been switched off, that normally results in a Giant Dipole Resonance. The frequency of the
oscillation is connected to the symmetry energy. 
Additionally, in case of a neutron excess inside the nucleus, a neutron skin could develop and also 
a low-energy dipole strength due to the oscillation of the excess neutrons with respect to the symmetric core appears. Such an oscillation is called Pigmy Dipole Resonance, and it is correlation to the thickness of the neutron skin \cite{Thiel:2019tkm}.

This neutron pressure can also be quantified at saturation density by measuring the thickness of the neutron skins of neutron-rich nuclei. Indeed, the larger the pressure is, the thicker is the resulting skin. A recent review on the measurement of neutron skins can be found in \cite{Thiel:2019tkm}. Here, we want to discuss an example of such measurements. The PREX collaboration was pioneering in exploiting parity-violating measurements of neutron densities by means of electron scattering \cite{Abrahamyan:2012gp}. This measurement is sensitive to the neutron density, because the neutron coupling to the weak-neutral $Z^0$ boson is much larger than for the proton. Longitudinally polarised electrons are scattered off $^{208}$Pb nuclei, and the difference in the scattering of left- and right-handed electrons is related to the weak and charged form nuclear factor.
Since the charged form factor is accurately determined, the weak form factor can be extracted for a given value of the transferred momentum and a matter radius can be extracted. The right panel of Fig.~\ref{fig:EOSHIC} shows the measured asymmetry as a function of the neutron radius $R_n$.
The PREX experiment reported a parity-violating asymmetry equal to A$^{{\rm Pb}}_{PV}= 656 \pm 60 (stat) \pm 14$ (syst) ppb and a derived neutron skin thickness of $R^{{\rm Pb}^{208}}_{{\rm skin}}= 0.33 ^{+0.16}_{- 0.18}$ fm.

This pioneering result does not yet provide a sufficiently stringent constraint on the EoS of neutron-rich matter, because of the still sizable error of the extracted $R_n$. Future measurements at Jlab (PREX2) aim at achieving a precision of $\pm 0.06$ fm, and the upcoming MESA facility will enable a new parity-asymmetry experiment (P2) \cite{Becker:2018ggl}, with higher beam intensities and a full azimuthal coverage. This will allow for a precision of $\pm 0.03$ fm in the matter radius determination.

Alternative measurements of neutron-removal cross sections in high-energy nuclear collisions of 0.4 to 1 GeV/nucleon have been proposed in this context \cite{Aumann:2017jvz}. Indeed, the measurement of neutron-removal cross sections on the neutron skin of medium-heavy neutron-rich nuclei could be correlated to the symmetry energy, and even a better precision  envisaged for the future asymmetry in parity-violating experiment is claimed. Future experiments in this direction offer a complementary tool to investigate the properties of neutron-rich matter.

The measurement of isospin diffusion in $^{112}{\rm Sn} + ^{124} {\rm Sn}$ and $^{124} {\rm Sn} + ^{112}{\rm Sn}$ reactions at a kinetic energy of 50 AMeV was used to estimate the symmetry energy at sub-saturation densities ($0.4 \leq \rho/\rho_0\leq 1.2$) \cite{Tsang:2004zz,Tsang:2008fd}. These measurements have been performed at MSU, and they consist in studying the isotopic distribution near the projectile rapidity to see if neutrons can diffuse from one nucleus to another, because of the effect of the nuclear mean-field.

\begin{figure}[htb]
    \centering
    \includegraphics[width=0.5\textwidth]{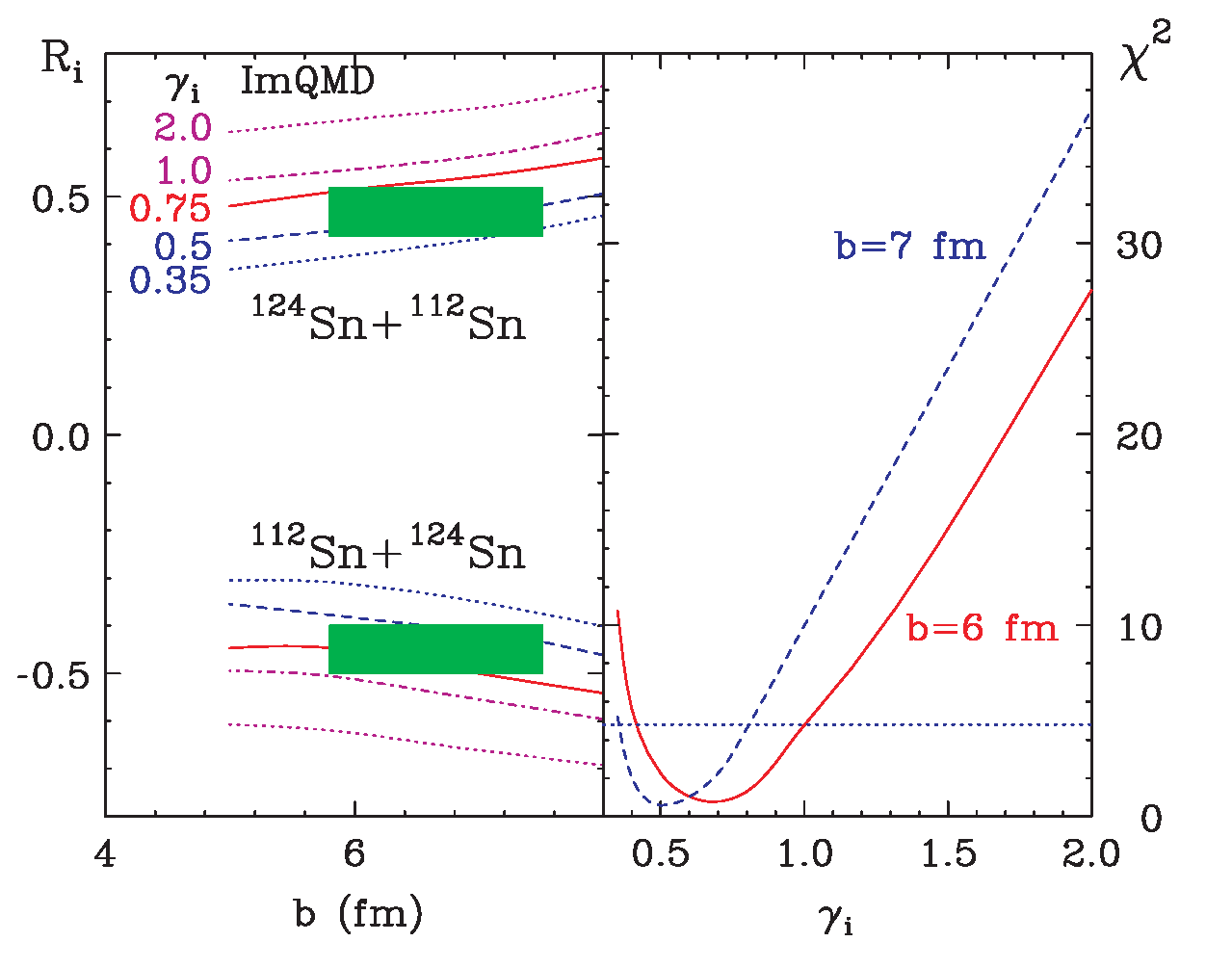}
      \includegraphics[width=0.45\textwidth]{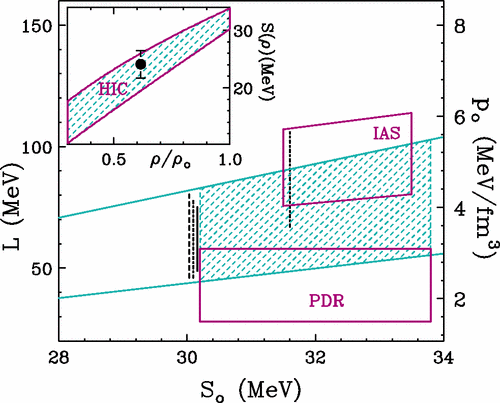}
    \caption{\textit{(Color online)} Left panel: Comparison of experimental isospin transport ratios (shaded regions) to ImQMD results (lines) for different impact parameters and different values of the $\gamma$ parameter. $\xi^2$ analysis for b=6 fm (solid curve) and b=7 fm (dashed curve) as a function of rapidity \cite{Tsang:2008fd}. Right panel: Representation of the extracted constraints for $S_0$ and $L$. The right axis corresponds to the neutron matter symmetry pressure at saturation density. The region bounded by the diagonal lines represents the constraints obtained by the isospin diffusion experiments \cite{Tsang:2008fd}.
}
   \label{fig:IsospinDiff}
\end{figure}

The left panel of Fig.~\ref{fig:IsospinDiff} shows the measured ratio $R_i$ for the two asymmetric colliding systems as a function of the measured impact parameter. In case of no diffusion, $R_i$ should be equal to $-1$ for the $^{112}{\rm Sn} + ^{124} {\rm Sn}$ and  equal to $1$ for the  $^{124} {\rm Sn} + ^{112}{\rm Sn}$ colliding system. One can see that in both cases the absolute value is smaller than one, indicating the diffusion of neutrons from the neutron-rich to the neutron-deficient nucleus. The experimental values are compared with transport calculations based on the ImQMD model assuming different values of the $\gamma$ parameter. The latter defines the EoS from very soft ($\gamma=\, 0.35$) to very stiff ($\gamma=\, 2$). 

The right panel of Fig. \ref{fig:IsospinDiff} shows the extracted constraints on the $L$ and $S_0$ parameters (see Eqs. ~(\ref{eq:eosparam})) with the diagonal light blue lines. 
These results overlap with recent constraints obtained from Giant Dipole Resonances, Pygmy Dipole Resonances, and mass data.

HICs at intermediate energies, such as the $^{197}{\rm Au}+^{197}{\rm Au}$ reaction at 400 MeV/nucleon, have been also employed to study the isospin dependence of the EoS \cite{Russotto:2011hq,Russotto:2016ucm}. In the collisions of heavy nuclei at moderate energies, a density up to 2 or 3 times the saturation density can be reached within a short time scale ($\approx$ 20 fm/c).
The resulting pressure induces a collective motion of produced particles towards the outer direction, with a strength that depends on the symmetry energy within the asymmetric systems.

\begin{figure}[htb]
    \centering
    \includegraphics[width=0.45\textwidth]{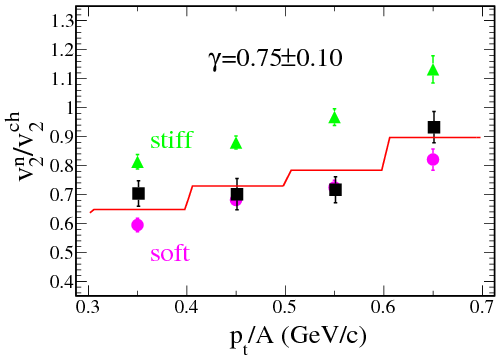}
      \includegraphics[width=0.4\textwidth]{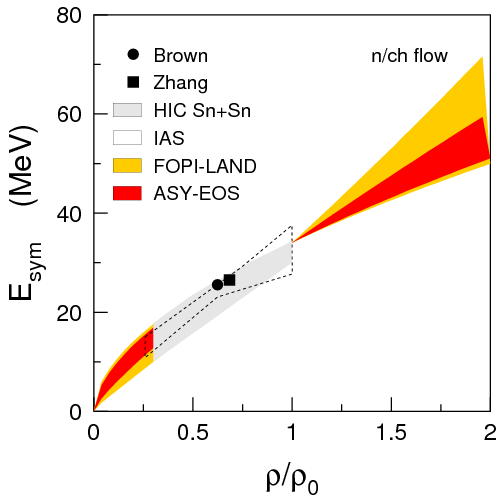}
    \caption{\textit{(Color online)} Left panel: Elliptic flow ratio of neutrons over all charged particles measured in central $^{197}{\rm Au}+^{197}{\rm Au}$ at 400 MeV/nucleon as a function of the transverse momentum per nucleon \cite{Russotto:2016ucm}. The data are shown by the black symbols, while the colored symbols represent the results of two transport calculations assuming a stiff ($\gamma=\, 1.5$) and soft ($\gamma=\, 0.5$) EoS. Right panel: Constraints from the ASY-EOS and FOPI-LAND experiments for the density dependence of the symmetry energy  as a function of  $\rho/\rho_0$ \cite{Russotto:2016ucm}.
}
   \label{fig:ASYEOS}
\end{figure}

The observables in this case are the directed and elliptic flow of neutrons compared to light charged particles. In the case of a sizable symmetry energy for increasing densities, a difference in the flow of neutrons and charged particles should be observed. 
First, the FOPI-LAND \cite{Russotto:2011hq} and then the ASY-EOS \cite{Russotto:2016ucm} collaborations carried out such measurements, by means of a rather complex experimental set-up that allow to reconstruct neutral and charge particles, and the reaction plane as well as to determine rather precisely the impact parameter of the reaction.
The UrQMD transport model was used to interpret the measured data. In such a model, it is possible to emulate an asymmetric collisions and the flow of all emitted particles for soft and stiff nuclear EoSs. 

The left panel of Fig.~\ref{fig:ASYEOS} shows the measured elliptic flow  ratio of neutrons over all charge particles as a function of the transverse momentum per nucleon from $^{197}{\rm Au}+^{197}{\rm Au}$ collisions at kinetic energies of 400 MeV/nucleon. The black symbols represent the experimental data, and the green and magenta symbols the simulated flow ratio according to a soft and stiff EoS.
The soft EoS provides a better agreement with the data in accordance with the analysis discussed in
\cite{Danielewicz:2002pu}. 

The right panel of Fig.~\ref{fig:ASYEOS} shows the symmetry energy as a function of the density extracted from the flow measurements with the yellow and red histograms.
The constraints obtained for the FOPI-LAND and ASY-EOS experiments are compatible, and the more precise ASY-EOS measurement shows an energy dependence as a function of the density at most linear.

In this context, one should mention that also the flow of charged kaons and the 
$\pi^-/\pi^+$ ratio  could be sensitive to the EoS.
The pion ratio was investigated and the variation
expected for this observable for a soft versus stiff EoS was found very small \cite{Reisdorf:2006ie}.
The ratio $K^+/K^0$ is even less sensitive to the EoS \cite{PhysRevC.75.011901}.

\begin{figure*}[hbt]
  \centering
  \includegraphics[width=0.7\textwidth,height=140mm]{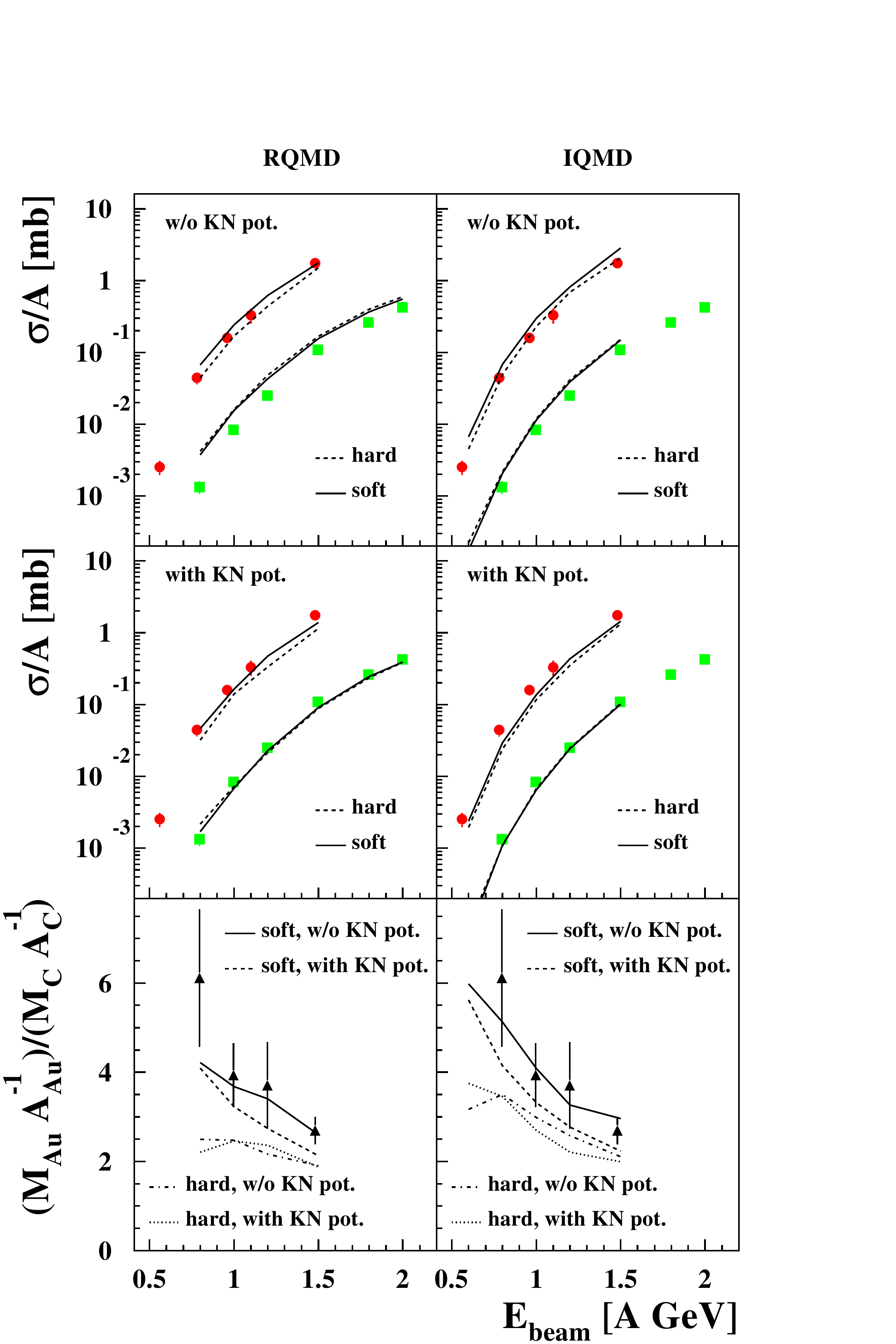}
\caption{\textit{(Color online)} Upper and middle panels: Comparison
of the $K^+$ production cross section per mass
number ($\sigma(K^+)$/A)  as a function  of the beam energy for $Au+Au$ (red
circles) and for $C+C$ collisions (green squares), compared to  RQMD
(left-hand side) and to IQMD calculations (right-hand
side), assuming soft or hard EoSs and considering a repulsive $KN$ potential \cite{Forster:2007qk}. 
Lower panels: Double ratio of the $K^+$ multiplicities per mass
number $M/A$ in $Au+Au$ divided by the one in $C+C$, and the
comparison to the various transport-model calculations. }
  \label{fig:KaonMat6}
\end{figure*}

Since kaons are produced in $NN$ or $N\Delta$
collisions and, hence, in the first stage of the HICs 
they test the largest density of the system, that is reached during this initial stage.
Moreover, the large kaon mean free path 
of about 5 fm renders absorption processes rather improbable, so that the measured kaon yields
in different colliding system are thought to be sensitive to the EoS. The upper and middle panels of Fig.~\ref{fig:KaonMat6} shows the excitation  functions of $K^+$ for a light ($C+C$) and heavy ($Au+Au$) colliding system, indicated by the green  and red symbols, respectively, measured by the KaoS experiment. The measurements are compared with two versions of transport models (RQMD \cite{PhysRevLett.86.1974} and  IQMD  \cite{PhysRevLett.96.012302}), where different incompressibility  of nuclear matter can  be implemented. Figure~\ref{fig:KaonMat6} shows that only the ratio of the kaon multiplicity per participant in the heavy and light colliding system is sensitive  to the variation of incompressibility of the system and kaon-nucleus  potential.
Also this finding speaks for a soft EoS for symmetric nuclear matter \cite{PhysRevLett.86.1974,PhysRevLett.96.012302}.

\subsubsection{Observational constraints}
\label{sec:obsconstraints}

Apart from constraints on the EoS from nuclear experiments, one can extract information on the EoS from  astrophysical observations on masses and radii.  

The mass of a NS can be inferred directly from observations in binary systems. There are five Keplerian or orbital parameters that can be measured precisely in binary systems. They are the binary orbital period ($P_b$), the eccentricity of the orbit ($e$), the projection of the semi-major axis on the line of sight ($x \equiv a_1 {\rm sin} \, i/c$, where $i$ is inclination angle of the orbit), and the time ($T_0$) and longitude ($\omega_0$) of the periastron. From Kepler's laws, one can obtain the so-called mass function that relates the masses of both stars in the binary system with the observed orbit parameters, that is, $f=m_c^3 \ {\rm sin^3} i/(m_P + m_C)= 4\pi^2/P_b^2 x^3$, with $m_P$ being the mass of the pulsar and $m_C$ the mass of the companion. 

As we can see, if only the mass function is measured, we cannot go any further in the determination of the masses. However, accounting for the deviations from the Keplerian orbit due to general relativity effects helps in this endeavour. The relativistic effects can be parametrized in terms of the so-called post-Keplerian parameters, that is, the advance of the periastron of the orbit ($\dot{\omega}$), the combined effect of changes in the transverse Doppler shift and gravitational redshift around an elliptical orbit ($\gamma$), the orbital decay ($
\dot{P_b}$), and the range ($r$) and shape ($s$) of the Shapiro time delay of the pulsar signal as it propagates through the gravitational field of its companion. The post-Keplerian parameters depend on the Keplearian ones and the two masses of the binary system. Thus, the determination of at least two of them together with the mass function allows for the determination of the masses of the two stars in the binary system. The measurement of a third post-Keplerian parameter results in a test of general relativity.

With more than 2000 pulsars known up to date, one of the best determined pulsar masses is that of the Hulse-Taylor of 1.4$M_{\odot}$ \cite{Hulse:1974eb}. The detection and measurement of the masses of this binary system gave Hulse and Taylor the Nobel Prize in 1993, as it served as a test for Einstein's general relativity. Accurate values of almost $2M_{\odot}$ have been recently reported, such as for the PSR J1614-2230 \cite{Demorest:2010bx,Fonseca:2016tux}, by measuring the post-Keplerian Shapiro delay as well as for the PSR J0348+0432 \cite{Antoniadis:2013pzd}, using optical techniques in combination with pulsar timing. More recently, the mass of PSR J0740+6620 has been measured to be $2.14_{-0.09}^{+0.10} M_{\odot}$  \cite{2019NatAs.tmp..439C}, by determining the shape and range of the Shapiro delay. These measurements are, however, in tension with the soft behaviour of the nuclear EoS found in HiCs, as described in Sec.~\ref{sec:expconst}, because a rather stiff EoS is needed in order to obtain these large masses.

As for radii, they have been extracted in the past from the analysis of X-ray spectra emitted by the NS atmosphere. The X-ray spectra strongly depends on the distance to the source, its magnetic field and the composition of its atmosphere, hence making the measurement of the radius a rather difficult task.  With space missions such as 
NICER (Neutron star Interior Composition ExploreR) \citep{2014SPIE.9144E..20A} and the future eXTP (enhanced X-ray Timing and Polarimetry)  \cite{Watts:2018iom}, 
high-precision X-ray astronomy, based on pulse-profile modeling X-ray spectral-timing event data, will offer precise measurements of masses and radii simultaneously. Indeed, the first precise measurement of the size and mass of the millisecond pulsar PSR J0030+0451 has been recently reported by the NICER collaboration, with a mass of  $M =1.34^{+0.16}_{-0.15}M_{\odot}$ and equatorial radius of $R=12.71_{-1.19}^{+1.14}$ km \cite{Riley:2019yda}. 
 
\begin{figure} [htb]
    \centering
    \includegraphics[width=0.85\textwidth]{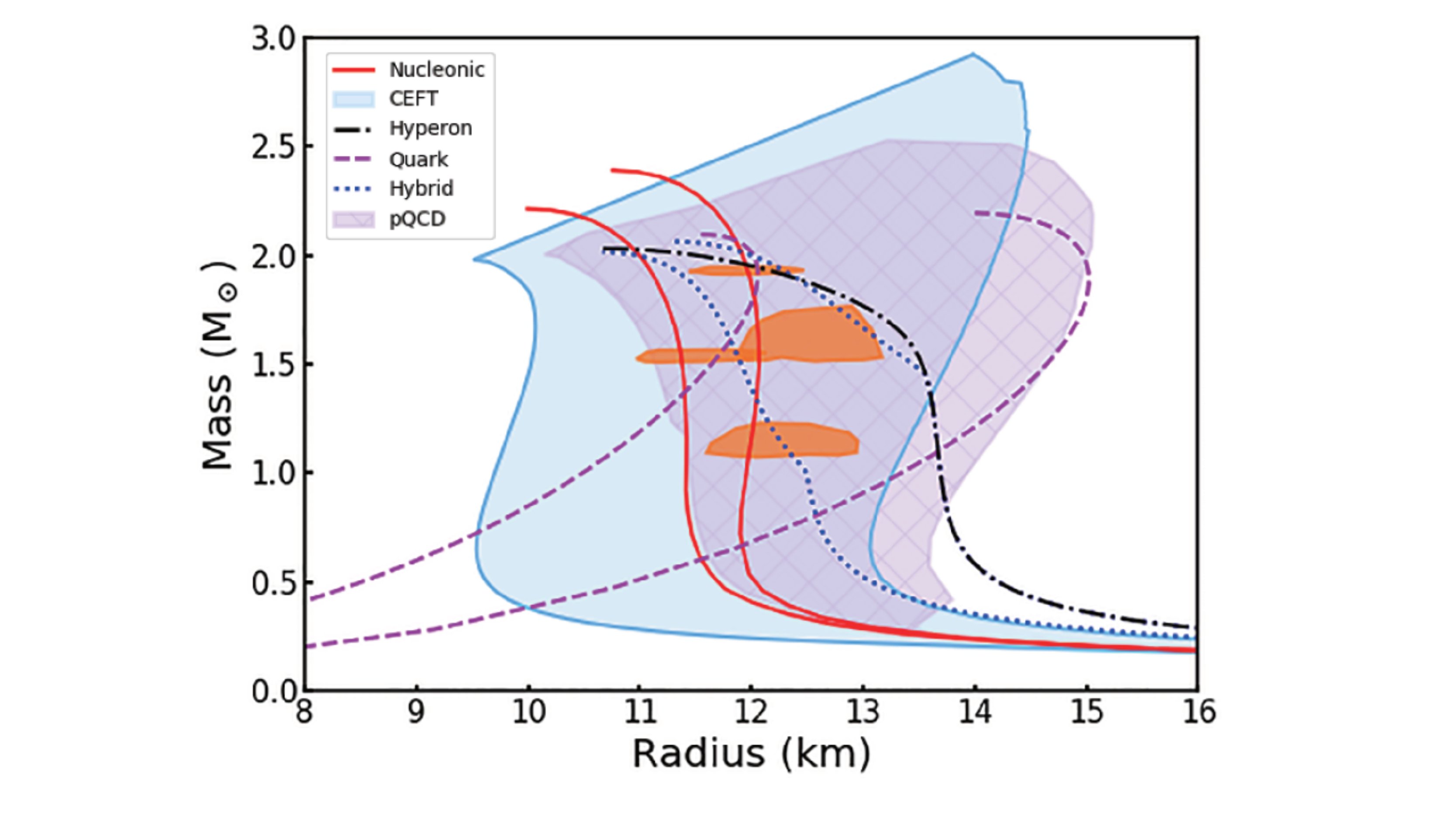}
    \caption{\textit{(Color online)} Constraints from pulse profile modelling of rotation-powered pulsars with eXTP. Figure adapted from \cite{Watts:2018iom}. }
    \label{fig:mass-radius}
\end{figure}

The expected constraints for the mass and radius of NSs coming from the new advances in X-ray astronomy are shown in  Fig.~\ref{fig:mass-radius}. There, the constraints from pulse profile modelling of rotation-powered pulsars with eXTP are indicated with the orange error contours for four millisecond pulsars with masses known precisely (PSR J1614-2230 \cite{Demorest:2010bx,Fonseca:2016tux}, PSR J2222-0137 \cite{Kaplan:2014mka}, PSRJ0751+1807 \cite{Desvignes:2016yex} and PSR J1909-3744 \cite{Desvignes:2016yex}). The underlying model assumed in the simulations is the variational AP3 nucleonic EoS (red) \cite{Akmal:1998cf}. The EoS models indicated in the figure are nucleonic (models AP3 and AP4) \cite{Akmal:1998cf}, but also schemes where the constituent species are quarks (u,d,s quarks) \cite{Li:2016khf,Bhattacharyya:2016kte}, nucleons and hyperons (inner core of hyperons, outer core of nucleonic matter) \cite{Bednarek:2011gd}, or quarks and nucleons forming hybrid stars (inner core of uds quarks, outer core of nucleonic matter) \cite{Zdunik:2012dj}. The label CEFT indicates the range of a nucleonic EoS based on $\chi$EFT \cite{Hebeler:2013nza}, whereas the pQCD is the range of nucleonic EoS that results from interpolating from CEFT at low densities and matching to perturbative QCD (pQCD) calculations at higher densities \cite{2014ApJ...789..127K}. 

Interestingly, the recent discovery of gravitational waves emitted from two merging NSs from the LIGO and
VIRGO collaborations \cite{TheLIGOScientific:2017qsa,Abbott:2018wiz} has opened a new era in astrophysics. Gravitational waves from the late inspirals of NSs are sensitive to EoS, through the so-called tidal deformability of the star. Since the tidal effects are strongly dependent on the stellar compactness, a measurement of the tidal deformability offers insights into the EoS and the possibility to discriminate among EoSs that predict similar masses but different radii.

Moreover, aside from mass and radius determinations, one of our best windows to the interior of NSs is through observations of their luminosities or temperatures as function of age, the so-called NS cooling curves \cite{Yakovlev:2004iq}. The crucial question is whether the cooling through neutrino emission is rapid or slow, and this is intimately related to the composition and phases inside NSs, given by the EoS.

\subsection{Equation of State with strangeness}
\label{sec:eos-strange}

The conditions of matter inside NSs are very different from those on Earth, thus permitting the existence of new phases of matter. In particular, the presence of antikaons and hyperons in the interior of NSs is a possibility that has been explored extensively over the years.

\subsubsection{Antikaons in neutron stars: Kaon condensation}
\label{sec:kaoncond}

\begin{figure}[htb]
    \centering
    \includegraphics[width=0.37\textwidth, angle=-90]{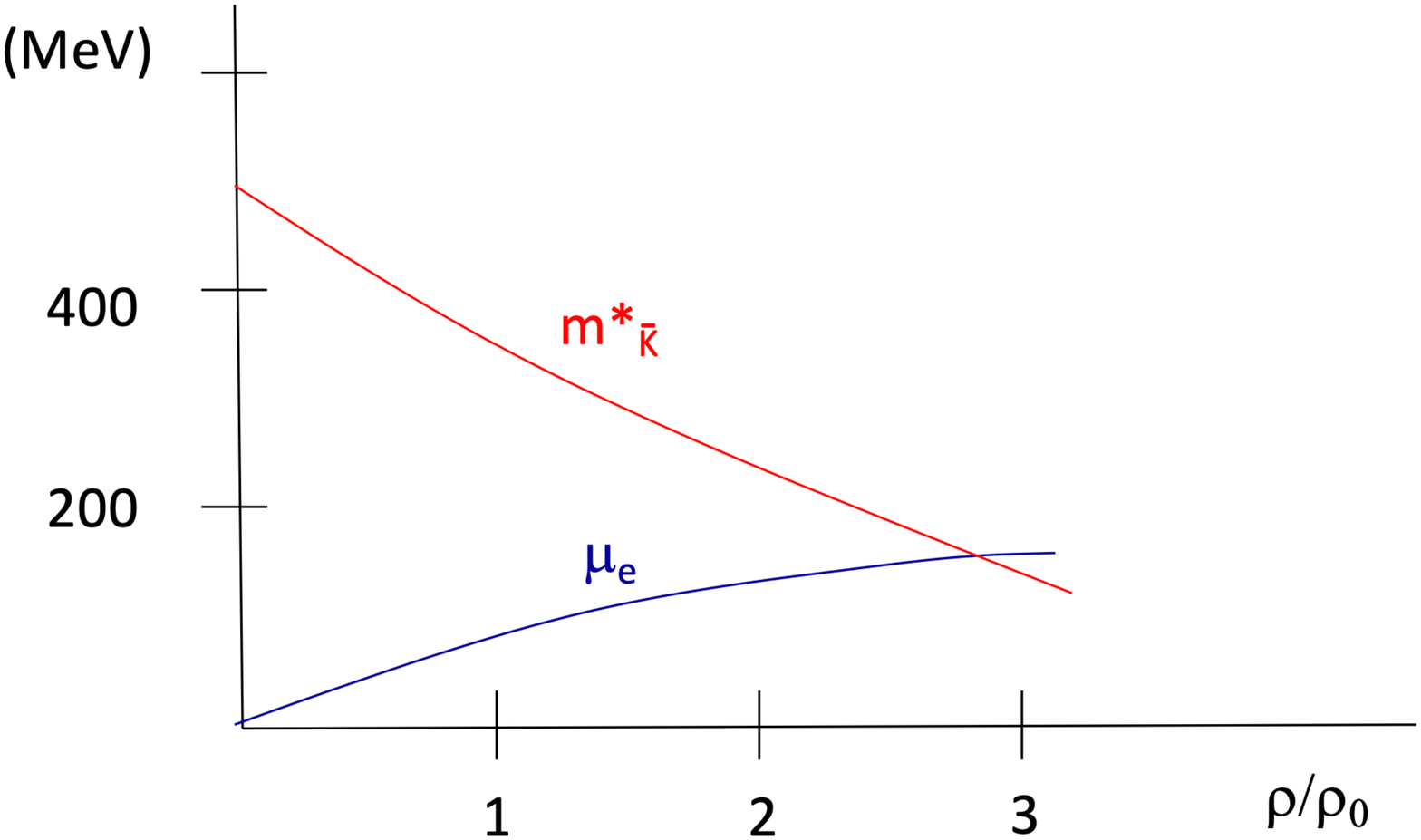}
    \caption{\textit{(Color online)} Schematic drawing depicting the evolution of the electron chemical potential $\mu_e$ and the antikaon effective mass $m^*_{\bar K}$ with baryon density in the interior of NSs.}
    \label{fig:kaoncond}
\end{figure}

The appearance of strange particles, such as antikaons, is one of the possible scenarios inside the core of NSs. As mentioned in Sec.~\ref{sec:briefNS}, the composition of matter in NSs is found by demanding equilibrium against weak interaction processes. Assuming NS matter made of neutrons, protons and electrons, the weak interaction processes that take place are  $n \rightarrow p e^- \bar \nu_e$ and $e^- p \rightarrow n \nu_e$, implying $\mu_p=\mu_n+\mu_e$ and $\rho_p=\rho_e$, with $\rho=\rho_p+\rho_n$. However, if the chemical potential of the electron increases dramatically with density in the interior of NSs, it might become energetically more favourable to produce antikaons instead of electrons, via reactions of the type $n \leftrightarrow p + \bar{K}$. In order for this to happen, the chemical potential of the electron for a given density should overcome the effective mass of antikaons, that is, $\mu_e \ge m^*_{\bar K}$, as depicted in Fig.~\ref{fig:kaoncond}. If this would be the case and given that antikaons are bosons, the phenomenon of kaon condensation would arise. 

Since the pioneer work of Ref.~\cite{Kaplan:1986yq}, this possibility has been largely debated.  The discussion is based on whether the mass of antikaons could be largely modified by the interaction with the surrounding nucleons.  Whereas $\chi$EFT approaches have shown moderate changes in the mass of antikaons, some phenomenological models tend to agree with this scenario. We refer the reader to the previous discussion in Sec.~\ref{sec:kaons-matter} on antikaons in dense matter.

\subsubsection{Hyperons in neutron stars: The hyperon puzzle}

\begin{figure*}[h]
\begin{center}
\resizebox{0.6\textwidth}{!}
{
\includegraphics[width=1.1\textwidth, height=0.7\textwidth]{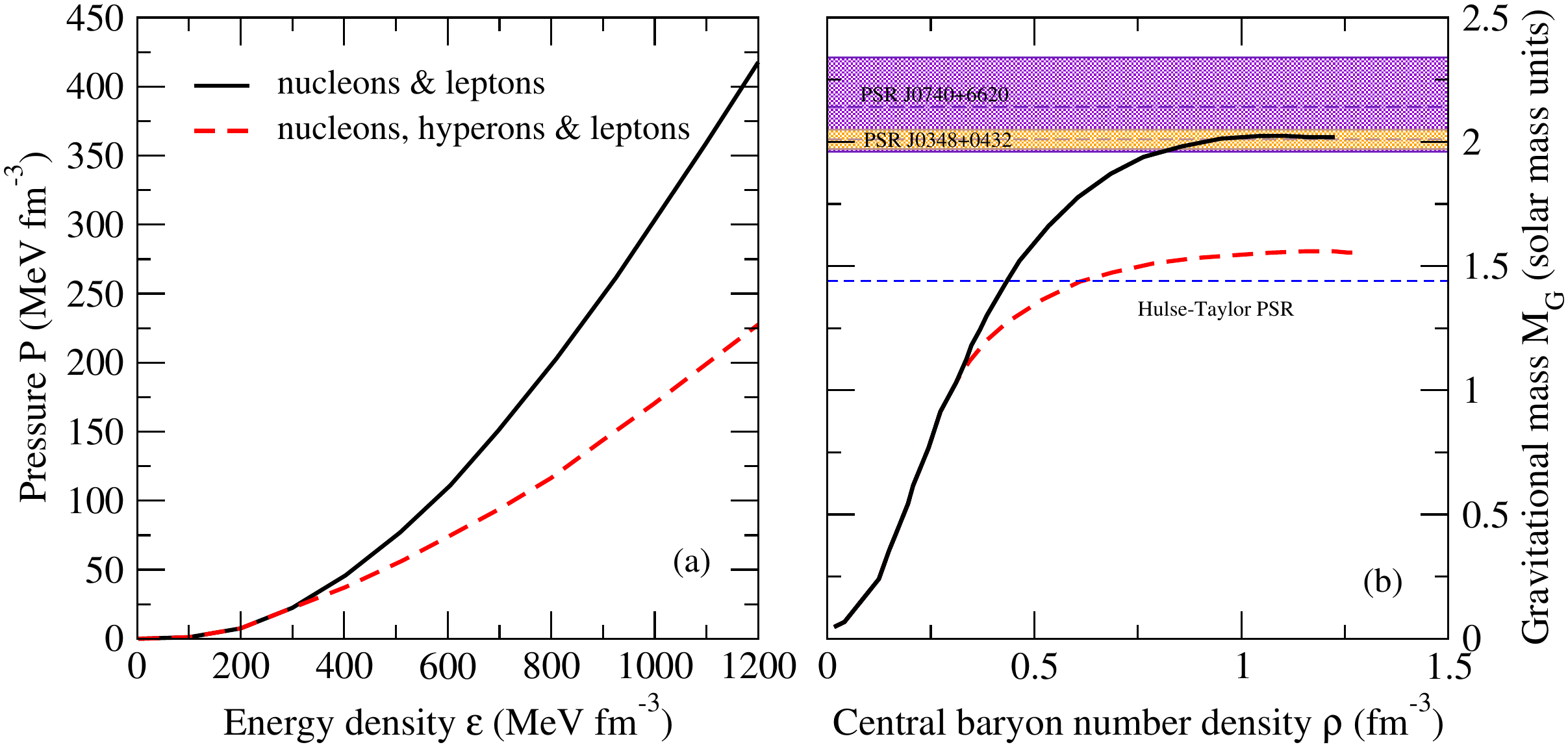}
}\caption{\textit{(Color online)} The EoS (panel (a)) and the corresponding NS mass (panel (b)). An example with (black solid line) and without (red dashed line) hyperons has been considered. The mass of the Hulse-Taylor pulsar and the observed PSR J1614-2230 \cite{Demorest:2010bx,Fonseca:2016tux} and PSR J0348+0432 \cite{Antoniadis:2013pzd} are shown with horitzontal lines. Also, the new mass measurement of PSR J0740+6620 is shown \cite{2019NatAs.tmp..439C}. Figures taken from \cite{vidana}.}
\label{fig:EOS}       
\end{center}
\end{figure*}

Another hypothesis is the appearance of hyperons in the core of NSs \cite{Vidana:2018bdi}. Due to the high value of density at the center of a NS and the rapid increase of the nucleon chemical potential with density, the appearance of hyperons might be energetically favourable. 

The presence of hyperons in NSs was first considered in the pioneering work of \cite{1960SvA.....4..187A}. Since then, the study of the properties of hyperons inside NSs have been studied by many authors using phenomenological schemes (see for example Refs.~\cite{Balberg:1997yw,Glendenning:1984jr,Weber:1989uq,Knorren:1995ds,Schaffner:1995th,Tolos:2016hhl,Tolos:2017lgv}), or microscopical models (see  \cite{Schulze:1995jx,Schulze:1998jf,Baldo:1998hd,Vidana:1999jm,Vidana:2000ew,Schulze:2006vw,Djapo:2008au,Sammarruca:2009wn,Schulze:2011zza,Lonardoni:2013gta,Katayama:2015dga} as examples). These approaches agree that hyperons may appear in the inner core of neutron stars at densities of $\approx 2$-$3 \rho_0$. At those densities, the nucleon chemical potential is large enough to make the conversion of nucleons into hyperons energetically favourable. This conversion relieves the Fermi pressure of the system, making the EoS softer when hyperons are present, as seen in the left panel of Fig.~\ref{fig:EOS}. The softer the EoS, the less pressure there is inside a NS and, hence, the less mass that the NS can sustain, as shown in the right panel of Fig.~\ref{fig:EOS}.

Although the presence of hyperons in NSs seems to be energetically inevitable, the strong softening of the EoS leads (mainly in microscopic models) to maximum masses not compatible with the $2M_{\odot}$ observations, as seen in the right panel of Fig.~\ref{fig:EOS}. In the literature this fact is often referred as 'the hyperon puzzle'. Several solutions have been advocated in order to have hyperons in the interior of $2M_{\odot}$ NSs, that we comment here briefly. 

\begin{figure}[htb]
    \centering
    \includegraphics[width=0.5\textwidth, height=0.4\textwidth]{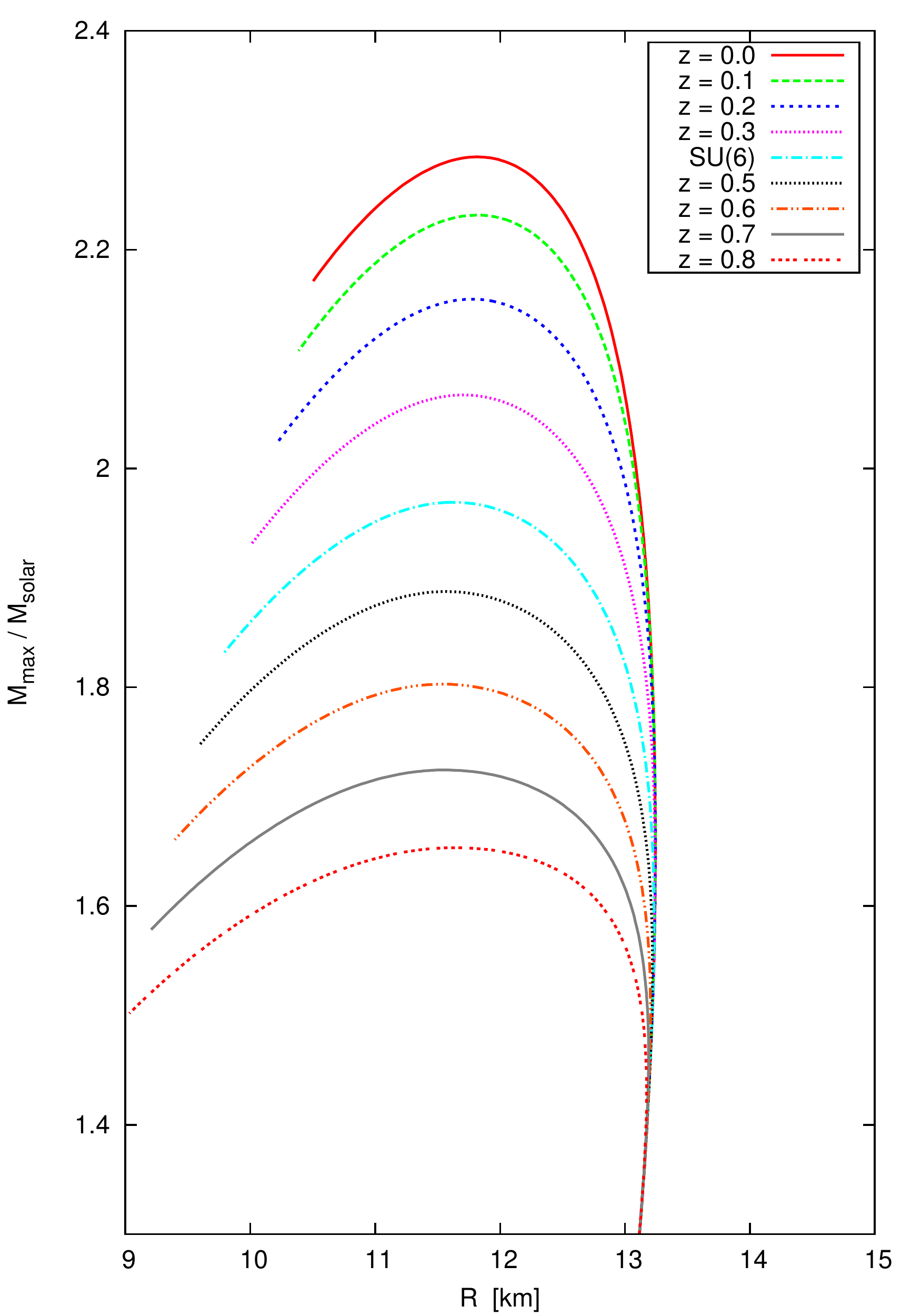}
    \caption{\textit{(Color online)} Mass-radius relation for various EoS within the RMF model of Ref.~\cite{Weissenborn:2011ut}. The different lines correspond to different $\phi$ coupling strengths.}
    \label{fig:schaffner}
\end{figure}

One possible way of solving the hyperon puzzle lies in the use of stiff $YN$ and $YY$ interactions (see \cite{Bednarek:2011gd,Weissenborn:2011ut,Oertel:2014qza,Maslov:2015msa} for some examples), so as to overcome the softening induced by the presence of hyperons, thus reaching $2M_{\odot}$.  The stiffening of $YN$ and $YY$ has been extensively explored in the RMF models, as it is well known that the exchange of vectors mesons among baryons leads to a repulsive behaviour of the baryon-baryon interaction. In particular, the role of the $\phi$ coupling to hyperons has become crucial to shift the hyperon onset to higher energies so as to reach the 2$M_{\odot}$ limit. To illustrate this, in Fig.~\ref{fig:schaffner} the mass-radius relation is shown for various EoSs as a $z$ parameter is changed. This $z$ parameter accounts for the singlet and octet meson coupling constants to baryons, that is, it is correlated with the $\phi$ coupling to baryons. The $z$ value is varied around the SU(6) value and, as seen in this figure, the decreasing of $z$ below its SU(6) value leads to larger maximum masses, as the EoS becomes stiffer.
 
 \begin{figure}[htb]
    \centering
    \includegraphics[width=0.55\textwidth, angle=-90]{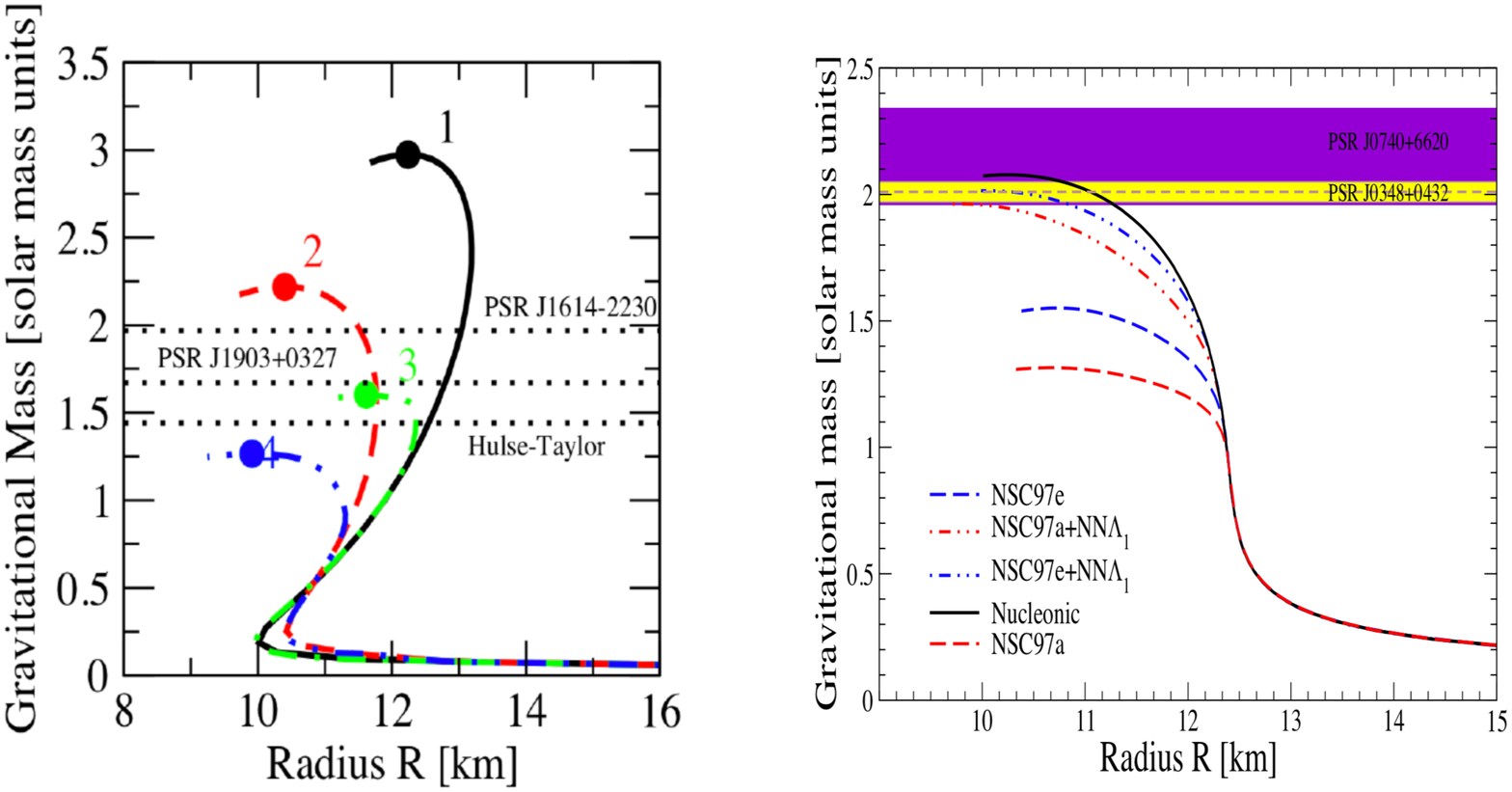}
    \caption{\textit{(Color online)} Left panel: Mass-radius relation for various models of the EoS including nucleons (lines 1 and 2), and nucleons and hyperons (lines 3 and 4). The hyperonic models, in this case, do not reach the 2$M_{\odot}$ limit. Figure adapted from \cite{Vidana:2010ip}. Right panel: Mass-radius relation for various models of the EoS including nucleons (solid line), and nucleons and hyperons (dashed and dotted lines). The hyperonic models with 3BF $NN \Lambda$, in this case, reach the 2$M_{\odot}$ limit.  Figure adapted from \cite{Logoteta:2019utx}. }
    \label{fig:TBF}
\end{figure}
 
The stiffening induced by hyperonic three-body forces has also been studied so as to solve the hyperon puzzle. The hyperonic three-body forces, as in the case of three-nucleon forces, might induce an additional repulsion at high densities so as to make the EoS stiff enough to reach 2$M_{\odot}$  \cite{Takatsuka2002,Takatsuka:2004ch,Vidana:2010ip,Yamamoto:2013ada,Yamamoto:2014jga,Lonardoni:2014bwa,Haidenbauer:2016vfq,Logoteta:2019utx,Gerstung:2020}. There is, however, not a general consensus regarding whether the hyperonic three-body forces will solve the hyperonic puzzle. While some models do not reach the 2$M_{\odot}$ limit, as seen in the left panel of Fig.~\ref{fig:TBF} for curves 3 and 4 that include hyperons \cite{Vidana:2010ip}, other schemes do \cite{Takatsuka:2004ch,Logoteta:2019utx,Lonardoni:2014bwa}, as reported in the right panel of Fig.~\ref{fig:TBF} for the EoS models including the 3BF $NN \Lambda$. We should also indicate  that the results of the first Quantum-Montecarlo calculation of Ref.~\cite{Lonardoni:2014bwa} with neutron and $\Lambda$ matter are not conclusive enough, as they strongly depend on the $\Lambda nn$ force used. Moreover, the effect of three-hyperonic forces from $\chi$EFT on the properties of hyperonic matter has been recently addressed in Ref.~\cite{Haidenbauer:2016vfq,Kohno:2009sc}, hinting again at a possible solution of the hyperon puzzle.

\begin{figure}[htb]
    \centering
    \includegraphics[width=0.5\textwidth,height=0.5\textwidth]{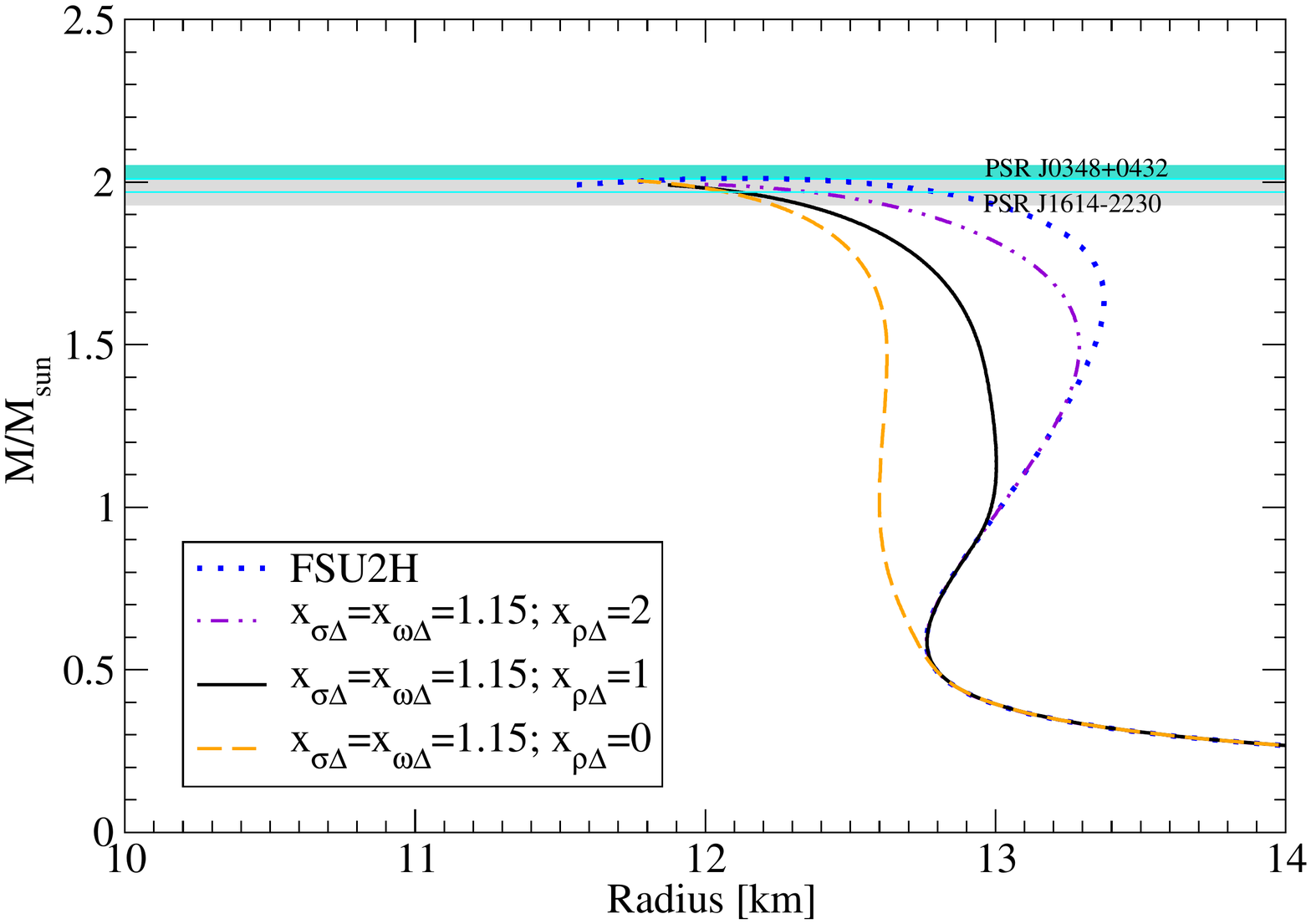}
    \caption{\textit{(Color online)} Mass-radius relation for various EoS adapted from Ref.~\cite{Ribes:2019kno}. The  FSU2H model without $\Delta$ baryons and three variations of the model including $\Delta$s, that depend on the coupling of $\Delta$ to the $\rho$ meson, are shown.}
    \label{fig:ribes}
\end{figure}

Other solutions are based on the appearance of new hadronic degrees of freedom that push the hyperon onset to higher densities, such as $\Delta$ baryons or a kaon condensate. Whereas we refer the reader to Sec.~\ref{sec:kaoncond} for the formation of kaon condensates in neutron stars, here we discuss the possibility of $\Delta$ baryons in NSs. It has been shown recently that $\Delta$ isobars might appear before hyperons \cite{Schurhoff:2010ph,Drago:2014oja,Ribes:2019kno}. As in the case of hyperons, the presence of a new hadron in $\beta$-stable matter leads to the softening of the EoS, and that might imply a reduction of the maximum mass below the observational limit of 2$M_{\odot}$ \cite{Drago:2014oja}. However, the expected softening of the EoS for densities around the onset of appearance of $\Delta$ might be overcome at higher densities, thus still reproducing 2$M_{\odot}$ observations \cite{Ribes:2019kno}. In Fig.~\ref{fig:ribes} the mass-radius relation is displayed for the hyperonic FSU2H model without $\Delta$ baryons and three variations of the model including $\Delta$s, that depend on the coupling of $\Delta$ to the $\rho$ meson. All these three models reach the 2$M_{\odot}$ limit, whereas the radii become smaller. The reduction in the value of the radius is due to the softening of the EoS at intermediates densities of 1-2$\rho_0$, as the $\Delta$ baryons appear.  Indeed, small  values  for  radii  favour  smaller  tidal deformabilities, more consistent with the value derived from  the  recent GW170817 event  \cite{Ribes:2019kno}.

Also solutions based on the appearance of non-hadronic degrees of freedom have been invoked. The presence of an early phase transition to quark matter below the hyperon onset could lead to the formation of hybrid stars. The EoS of the inner quark matter core should be stiff enough so as to sustain 2$M_{\odot}$. In order for this to happen, deconfined quark matter should be energetically favoured in the inner core while providing the sufficient repulsion required to reach the 2$M_{\odot}$ limit \cite{Zdunik:2012dj}. This possibility has been indeed exhaustively studied in the literature (see Refs.~\cite{Weissenborn:2011qu,Bonanno:2011ch,Klahn:2013kga,Lastowiecki:2011hh,Zdunik:2012dj} for recent papers). 

And, finally, alternative solutions to the hyperon puzzle have been discussed, such as the use of modified gravity models in order to present a consistent description of the maximal mass of neutron star while accommodating hyperons in the interior of NSs \cite{Astashenok:2014pua}.

%% file: Content/Conclusions.tex
\section{Conclusions}
\label{sec:conclusions}

Considering hadrons as the relevant degrees of freedom  within nuclear matter at low- and high- densities, two- and three-body interactions among hadrons should be pinned down to extract information about the equation of state of dense systems, ranging from nucleus to neutron stars. 
In the strangeness sector, many of these interactions are not known. The purpose of this review was to summarize the state-of-the-art of the experimental and theoretical efforts on strangeness in nuclei and neutron stars, thus taking into account kaons, antikaons, $\phi$ and hyperons and their interactions with nucleons and nuclear matter,  while trying to elucidate possible future directions.
%


Concerning strange mesons, the $K^-p$ interaction is dominated by the presence of the $\Lambda(1405)$. Theoretically, this state is understood as the superposition of two poles, one strongly coupled to the $\bar KN$ state, and a second one with a dominant coupling to the coupled-channel $\pi \Sigma$. Several experiments on photon-, pion-, kaon-induced reactions as well as proton-proton have been carried out. New experiments planned at JPARC 
will provide even more precise information on the structure of the $\Lambda(1405)$.

Above the $\bar{K}N$ threshold, correlation measurements have shown to be a complementary tool to elastic scattering  in order to study the $\bar{K}N$ interaction. The contribution of coupled channels influences the  correlations, since only the final state and not the initial state is fixed, and this contribution should be quantified more precisely in the  future. Only in this way, the correlation data can be useful to better constrain the theoretical calculations. Future measurements of kaonic-deuterium planned both at Da$\phi$NE and at JPARC will also help to disentangle the $I=0$ and $I=1$ components of the $\bar{K}N$ interaction at threshold.

Moving to three-body interactions containing kaons, kaonic bound states were also discussed in this review. A very heterogeneous landscape of predictions and measurements are available for the smallest of the kaonic bound states, that is,  $K^-pp$.  Theoretical determinations for the $K^-pp$ bound state show a broad variety of results for the binding energies, going from a shallow to a deeply bound state with binding energies of $\sim$ 100 MeV, while the decay widths could be comparable to the binding. 

Experimentally, it has been demonstrated that low energy $K^-$-nucleus interactions are dominated by $1N$, $2N$ and $3N$ absorption processes, so that it is difficult to impossible to isolate a broad $K^-pp$ state. For example, the analysis of the AMADEUS data on $K^-$ absorption did not confirm the claim of the observation of a $K^-pp$ state by FINUDA based on similar reactions. 
A similar situation occurred for $p+p$ reactions, where the DISTO collaboration claimed the observation of a broad $K^-pp$ state. However,  a partial wave analysis of the same data set interpreted it as a coherent sum of different known $N^*$ resonances. 
 The only promising experimental method is at the moment the employment of energetic (order of GeV) $K^-$ beams on light nuclear targets,  where a large momentum transfer can facilitate the $K^-pp$ creation. The E15 experiment seems indeed to have a solid signal for the kaonic bound state.

In this review we also addressed the interaction of kaons with nucleons and nuclear matter. 
It is of importance to understand $KN$ as the in-medium properties of kaons are linked to the ones of antikaons and hyperons, and these particles are strongly connected to the physics of nuclei and neutron stars. De facto, $K^0_S$ and $K^+$ measurements are much easier to interpret w.r.t. other strange particles, because of the absence of coupled channels. The repulsive interaction predicted theoretically and measured in $KN$ scattering experiments has been also quantified within nuclear matter. Different observables (flow, momentum and cross-section analyses) validate a repulsive  interaction with a value  of $V=\, +20 - +40$ MeV at saturation density and zero kaon momentum. Although the strength of the interaction is at most 10\% of the kaon mass, the kinematics of kaons is influenced by this mass change and, hence, relevant observables could be pinned down.

As for antikaons, the presence of the $\Lambda(1405)$ resonance and its behaviour in dense nuclear matter determine the strength of the $\bar KN$ interaction also in dense matter. 
While kaons behave almost as a good quasiparticle, antikaons have a broad spectral function due to the strong interaction with the medium. As a consequence, antikaons get primarily absorbed when they meet nucleons, as seen in the context of kaonic bound states. This means that besides the attractive real part of the optical potential extracted from heavy kaonic atoms, additional effects within nuclear matter are completely shielded by absorption. Nevertheless, in the last years the $K^-$ absorption could be quantified in several experiments. This could have an impact for the physics of neutron stars if antikaons are allowed.

The $\phi$ in matter was also a matter of discussion in this review, as $\phi$ mesons are strongly absorbed in the nuclear medium as $K^-$. The absorption is due to the coupling $\phi \rightarrow \bar{K}K$ and the probable increase of the $\phi N$ cross section in the nuclear medium. This would imply a violation of the OZI rule that could be circumvented with an enhanced coupling of the $\phi N$ to the $K^*\Lambda$ and $K^*\Sigma$ channels.
Nevertheless, there are still substantial uncertainties in the description of the $\phi N$ interaction that do not allow to a simultaneous description of the experimental data on photoproduction and proton-proton reactions.


Last but not least, we reviewed the hyperon-nucleon and hyperon-hyperon interactions in vacuum and in matter, the latter connected to the phases of dense matter in the interior of neutron stars.
The appearance of hyperons within neutron stars is under debate, and it depends on the interaction of $\Lambda$, $\Sigma$ and $\Xi$ with nucleons at different densities.  
Two-body interactions can profit now and in the future of correlation data, that should substitute scattering data soon. In particular, channels as $p\Xi^-$, $p\Sigma^0$ and $\Lambda\Lambda$ could be observed directly for the first time. Future measurements at colliders will provide brand new data for such two-body interactions, whereas lattice calculations have shown to be also a new venue worth exploring to understand hyperon-nucleon and hyperon-hyperon interactions.

For the study of hyperon properties in matter,
a lot of theoretical effort has been developed using phenomenological and microscopic ab-initio models. Starting from two- and three-body interactions, the microscopic approaches range from variational calculations, renormalization group methods, lattice QCD calculations to diagrammatic expansions. The theoretically predictions have tried, among others, to explain the physics of hypernuclei.

Indeed, hypernuclei represent the only tool able to deliver quantitative information on the $\Lambda$ properties within nuclear matter. Despite of the rich set of measurements of $\Lambda$ in different heavy-ion collisions, the hyperon kinematics does not provide observables clearly connected to the in-medium properties. In this context, one also has to mention that transport models, that could be instrumental in the interpretation of the experimental data, do fail also in describing elementary collisions. The legacy of future hypernuclei experiments is to achieve a data base for spectroscopy measurements more close to the one available within nuclear physics. Only in this way, the contribution of three- and four- body interactions, which are needed to determine the equation of state with strange hadrons, will be quantified.

In order to quantify the $\Sigma N $ interaction, new scattering experiments at JPARC are planned. It is, however, unclear if the measurement of $\Sigma$-hypernuclei will ever achieve a good precision. Similar arguments hold for $\Xi$-hypernuclei. However, in this case, the attractive interaction, predicted by lattice QCD and  confirmed by correlation measurements, supports the hypernuclei formation. On the other hand, the data base for $\Lambda \Lambda$ hypernuclei will also improve, but the impact of these measurements for the search of an H-dibaryon will always be shielded by the presence of several nucleons. In this context, the genuine $\Lambda\Lambda$ correlations offer a more direct measurement of the two-body interaction.

An additional aspect that involves all strange hadrons is the observation of a universal scaling of their production yield as a function of the number of participating nucleons in the reaction. This feature is observed in heavy-ion collisions at intermediate energies (GeV), and suggests that hadron formation could proceed differently than through the superposition of nucleon-nucleon collisions.  A new theoretical approach is then necessary to evaluate the possible impact on the equation of state of dense systems.

In fact, the equation of state of nuclear matter without strangeness was investigated in this review.  Neutron-rich systems are relevant for neutron stars, and several experimental efforts focus on the determination of the symmetry energy as a function of the system density. Nowadays, the most promising tools in this field are either the neutron-skin measurements or the analyses of neutron flow and ratio of isospin partners produced in heavy-ion collisions at low (50 MeV) and intermediate (400 MeV) energies.

Heavy-ion measurements have also delivered constraints on the symmetry energy up to two times the saturation density, and the results speak for a linear dependence of the symmetry energy with density. This seems to exclude a stiff equation of state for neutron rich matter. Neutron skin measurements are still hampered by large errors, so that they do not yet impose stringent constraints on the symmetry energy, but future campaigns at Jlab and MAMI will significantly improve on the precision.

Apart from experimental constraints on the nuclear equation of state, astrophysical observations are also of relevance, such as mass observations, radius determinations, cooling measurements or the recent gravitational wave emission signals from two merging neutron stars. Whereas masses are well constrained from observations in binary systems, the determination of the radii of neutron stars is rather difficult. It is expected that in the near future high-precision X-ray astronomy, with missions such as NICER and eXTP, will offer simultaneous precise measurements of masses and radii.  On the other hand, the expected measurement of new gravitational wave events coming from the merging of two neutron stars will help to to discriminate among equations of state that predict similar masses but different radii.

The inclusion of the strange degree of freedom on the equation of state of neutron stars is also the focus of a continuous effort. The presence of antikaons and hyperons in the inner core is under an intense debate. Due to the high value of density at the center of a neutron star and the rapid increase of the nucleon chemical potential with density, the appearance of strange hadrons might be energetically favourable. Whereas the possible presence of antikaons in its condense form is disfavoured by most of the recent microscopic models, there is not a consensus about the presence of hyperons in the inner core of neutron stars. The onset of hyperons in neutron stars seems to be energetically unavoidable, but the induced softening of the equation of state due to their existence might lead to maximum masses not compatible with the recent $2M_{\odot}$ observations. This is the so-called hyperon puzzle, whose solution has not been found yet.